\documentclass[12pt,a4paper]{book}
\usepackage{slashed,cite}
\usepackage{latexsym}
\usepackage{amssymb}
\usepackage{lscape}
\usepackage{colordvi}
\usepackage{epsf}
\usepackage{epsfig}
\usepackage{times}
\usepackage{fancyhdr}

\def\nn{\nonumber}

\def\rmi{{\,\rm i\,}}
\newsavebox{\uuunit}
\sbox{\uuunit}
    {\setlength{\unitlength}{0.825em}
     \begin{picture}(0.6,0.7)
        \thinlines
        \put(0,0){\line(1,0){0.5}}
        \put(0.15,0){\line(0,1){0.7}}
        \put(0.35,0){\line(0,1){0.8}}
       \multiput(0.3,0.8)(-0.04,-0.02){12}{\rule{0.5pt}{0.5pt}}
     \end {picture}}
\newcommand {\unity}{\mathord{\!\usebox{\uuunit}}}

\def\a{\alpha}

\def\g{\gamma}

\def\d{\delta}
\def\D{\Delta}
\def\e{\epsilon}
\def\ve{\varepsilon}
\def\f{\phi}

\def\p{\psi}

\def\k{\kappa}
\def\l{\lambda}

\def\m{\mu}
\def\n{\nu}
\def\r{\rho}
\def\s{\sigma}

\def\o{\omega}

\newcommand{\SU}{\mathop{\rm SU}}
\newcommand{\SO}{\mathop{\rm SO}}
\newcommand{\Sp}{\mathop{\rm Sp}}
\newcommand{\Spin}{\mathop{\rm Spin}}
\newcommand{\U}{\mathop{\rm {}U}}
\newcommand{\USp}{\mathop{\rm {}USp}}

\newcommand{\Symp}{\mathop{\rm {}Sp}}
\newcommand{\Sl}{\mathop{\rm {}S}\ell }
\newcommand{\Gl}{\mathop{\rm {}G}\ell }
\newcommand{\su}{\mathop{\mathfrak{su}}}
\newcommand{\so}{\mathop{\mathfrak{so}}}
\newcommand{\symp}{\mathop{\mathfrak{sp}}}
\newcommand{\spin}{\mathop{\mathfrak{spin}}}
\newcommand{\un}{\mathop{\mathfrak{u}}}

\newcommand{\spl}{\mathop{\mathfrak{sl}}}
\newcommand{\gl}{\mathop{\mathfrak{gl}}}

\def\Ric{R}
\newcommand{\covder}{\mathfrak{D}}
\def\rmi{{\,\rm i\,}}
\newtheorem{defi}{Definition}

\def\rmd{{\rm d}}
\fancypagestyle{plain}{%
\fancyhead{} 
}
\begin{document}
\setcounter{page}{-5}
\thispagestyle{empty}
\parbox{4.3cm}{
\epsfig{file=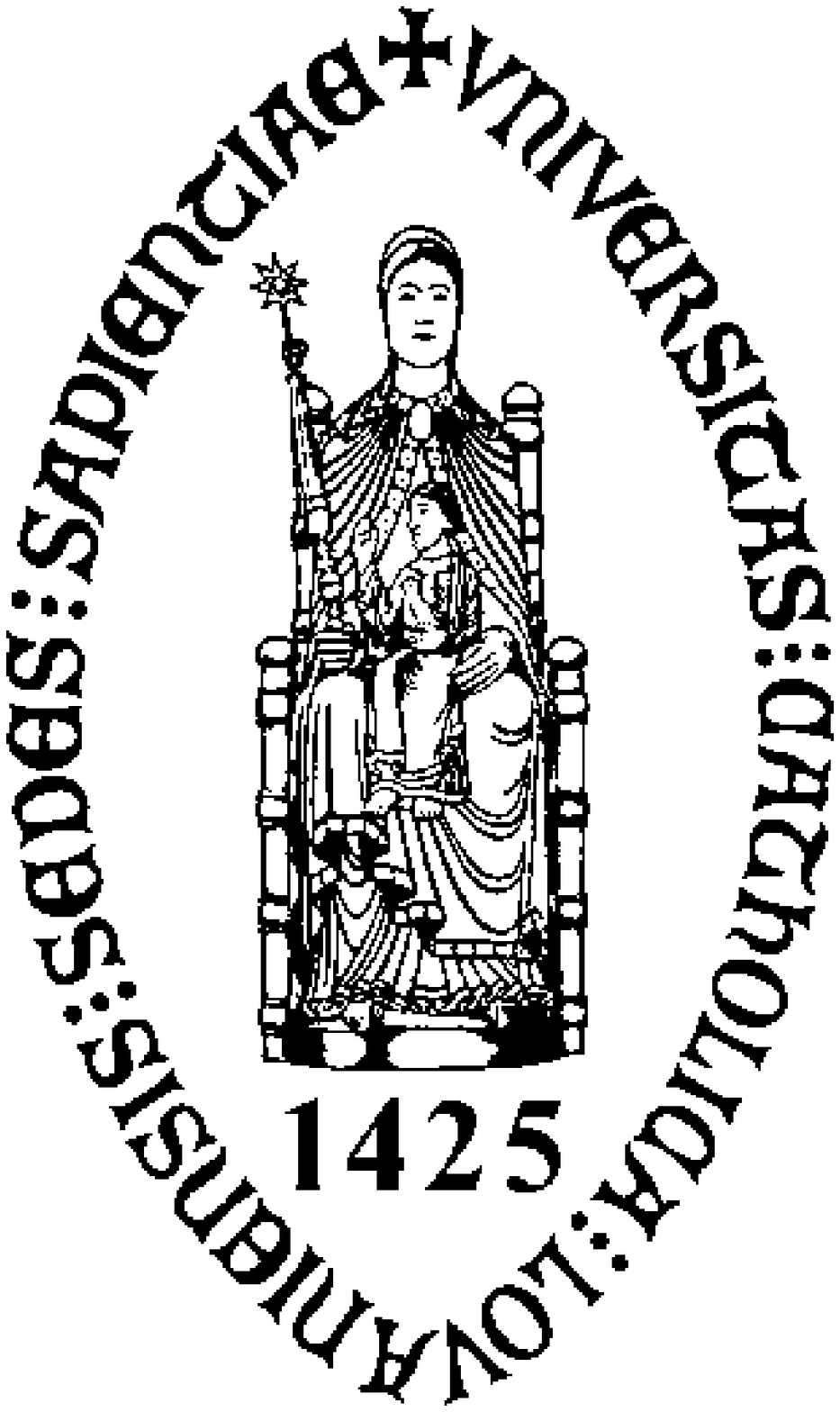,height=2cm}\hfill}\hfill
\parbox{6.9cm}{
\begin{flushright}
\hfill Katholieke Universiteit Leuven\newline
\hfill Faculteit der Wetenschappen\\
\hfill Instituut voor Theoretische Fysica\\[.5cm]
\hfill KUL-TF 04/34\\
\hfill hep-th/0411126\\
\end{flushright}
}\\\vspace{.5cm}
\vspace{4cm}
\begin{center}
{\huge \textsc{Aspects of on-shell supersymmetry}}\\[2.5cm]
Jos Gheerardyn\footnote{Present address: Dipartimento di Fisica Teorica,
Universit\` a di Torino, Via P. Giuria 1, I-10125 Torino, Italy. Email:
gheerard@to.infn.it}\\[5.5cm]
\end{center}
\parbox[b]{7cm}{
Promotor:\hfill\\
Prof. Dr. Antoine Van Proeyen\hfill\\}\hfill
\parbox[b]{6.8cm}{
\begin{flushright}
\hfill Proefschrift ingediend tot\\
\hfill het behalen van de graad van\\
\hfill Doctor in de Wetenschappen\\
\end{flushright}}

\begin{center}
(2004)
\end{center}
\newpage ~
\thispagestyle{empty}
\pagebreak
\section*{Dankwoord}
Als ik nu vertelde dat ik citroenen zou gaan kweken op de zuidelijke flank van de berg van Koekelare, dan zouden velen zich verbazen over zoveel optimisme. Hetzelfde gebeurde enkele jaren geleden toen ik als ingenieur besliste het boek van Polchinski te lezen, getiteld `String Theory'. Daarom dank ik in de eerste plaats Henri Verschelde, Jan-Willem Rombouts, Jeroen De Pessemier en Karel Van Acoleyen omdat zij mij tijdens het schrijven van mijn ingenieurs-eindwerk in contact brachten met de wondere wereld van de hoge energie fysica.

Voor het schrijven van een thesis is origineel werk nodig, en origineel zijn als beginnend wetenschapper is een zware taak. Daarom dank ik natuurlijk mijn promotor, Antoine Van Proeyen, om mij hierbij te helpen. Niet alleen de vele discussies zijn mij bijgebleven, vooral ook zijn dynamisme en no-nonsense werkmethode hebben diepe indruk op mij gemaakt. Bovendien ben ik hem dankbaar omdat hij mij de kans gaf deel te nemen aan vele conferenties.

I would also like to thank Bert Janssen, Eric Bergshoeff, Joke Adam, Patrick Meessen, Sorin Cucu, Stefan Vandoren, Tim de Wit and Yolanda Lozano for fruitful collaborations. Verder was de sfeer op het instituut altijd aangenaam, en dit niet enkel omwille van de persistente geur van koffie. Daarom ook mijn dank aan het adres van mijn collega's en de leden van de staf. Ook dank ik het Fonds voor Wetenschappelijk Onderzoek - Vlaanderen, omdat ze mij het mandaat van Aspirant hebben toegekend.

De overgang van de wereld van de theorie naar de harde realiteit werd altijd verzacht door de aanwezigheid van mijn vrienden. Vandaar ook enkele zoete woordjes voor Bart, Eun-Jung \& Jan-Willem, Isaac, Kaat, Klaas, Koen, Leander, Lennaert, Michel, Rozelien \& Andy, Sofie \& Frederik en vele anderen.

Ook mijn broer dank ik van harte voor zijn kritische kijk. Mijn ouders wil ik bedanken omdat ze mij de kans en de vrijheid gaven te studeren. Daarom draag ik dit werk aan hen op.

Mijn laatste woord van dank is uiteraard voor Marjoleine, bedankt om zoveel te begrijpen. 
\thispagestyle{empty}
\newpage ~
\thispagestyle{empty}
\pagebreak
\vspace*{16cm}
\begin{flushright}
{\it Sapere aude \hspace*{1cm}}
\end{flushright}
\thispagestyle{empty}
\newpage
\thispagestyle{empty}
\pagebreak
\tableofcontents

\chapter{Introduction}\label{intr}
\section{Purpose of this work}
In this thesis, we wish to study classical field theories that do not admit an action principle, implying that their formulation is entirely in terms of equations of motion. All kinds of dynamical field equations could obviously be written down, but we would like to study only a subset that satisfies a certain number of restrictions. First of all, we want the equations of motion to be covariant under the Poincar\'e group, locally when the theory is coupled to general relativity or rigidly when it is not. Secondly, we ask the theory to be supersymmetric, which is the property that it is invariant under a certain interchange of fields corresponding to bosonic and fermionic particles. Finally, we will only consider equations of motion with at most two derivatives.

In the theories that we want to consider, supersymmetry is realized on the mass-shell, meaning that the closure of the algebra is only guaranteed on all the fields when they satisfy the equations of motion. Hence, we can use supersymmetry as an organizing principle to build field equations governing the dynamics, prior to the construction of an action. We will find that the supersymmetry algebras previously considered in the literature, allow for a larger set of theories than the ones that can be formulated in terms of a Poincar\'e-invariant and supersymmetric least action principle. These theories can first of all admit a broader class of potentials. A second important generalization lies in the fact that these theories can describe more general geometries. More precisely, the scalar fields of a so-called nonlinear sigma model can be interpreted as coordinate functions on a certain manifold, called the target space. Supersymmetry restricts the allowed spaces, but we will find that for a theory that does not admit a Lagrangian description, these target spaces can be more general. Hence, we can use techniques from theoretical physics (like e.g. superconformal tensor calculus) to deduce new geometrical results. 

The existence of such theories raises some conceptual problems, the most important being the embedding in string theory, which is believed to be the fundamental theory governing all physics. A first (partial) answer to this question will also be presented in this thesis, as we will show in an easy example how dimensional reduction of a theory with action can yield a theory without. More specifically, we will discuss the construction of a massive nonchiral (i.e. type IIA) supergravity in ten dimensions via a Scherk-Schwarz reduction on the equations of motion of eleven-dimensional supergravity (i.e. classical M-theory). Moreover, we will present a simple method to construct supersymmetric solutions to this IIA theory.
\section{The origin of field theories and supersymmetry}
In classical mechanics, the motion of a particle subjected to conservative forces, generating a potential $V$, can be described using the least action principle. This means that the orbit of the particle will be such that the kinetic minus the potential energy computed along the trajectory is minimized. More explicitly, parametrizing a trajectory as $\vec x:\mathbb{R}\to \mathbb{R}^3:t\mapsto \vec  x(t)$, the action $\mathcal S$ of a particle with mass $m$ and potential $V$ reads
\begin{equation}
\mathcal S[\vec x]=\int dt \; \Big[\frac12 m \left(\dot{\vec x}(t)\right)^2-V\big(\vec x(t)\big)\Big]\,,
\end{equation}
with $\dot{\vec x}(t)\equiv \rmd \vec x(t)/\rmd t$.
The orbit should therefore satisfy the following differential equation in order to extremize the action,
\begin{equation}
-\frac{\delta \mathcal S[\vec x]}{\delta\vec x(t)}=m\frac{\rmd^2 \vec x(t)}{\rmd t^2}+\vec\nabla V(\vec x(t))=0\,.
\end{equation}
This formulation of the classical mechanics of point particles is important in itself.  Moreover, the least action principle plays a central role in modern theoretical physics, due to the existence of a conceptually very appealing quantization method, called path integral quantization. The central idea, originally from Dirac~\cite{Dirac} and elaborated by Feynman~\cite{Feynman:1948ur} is that the transition amplitude ${<\vec x_f,t_f\vert\vec x_i, t_i>}$ of the above mentioned particle can symbolically be written as 
\begin{equation}
<\vec x_f,t_f\vert\vec x_i, t_i>=\int \mathcal{D}\vec x(t)\exp \left(\frac{i}{\hbar}\int_{t_i}^{t_f}dt \; \Big[\frac12 m \left(\dot{\vec x}(t)\right)^2-V\big(\vec x(t)\big)\Big]\right)\,,
\end{equation}
where the integration is over all possible trajectories starting at $\vec x_i$ on $t=t_i$ and ending at $\vec x_f$ on $t=t_f$.

Lagrangians that are functionals of fields can similarly be used to describe the classical mechanics of extended systems with continuously distributed degrees of freedom, like e.g. strings. Moreover, in case we want to combine the special theory of relativity with quantum mechanics, we are naturally led to quantum field theories, see e.g.~\cite{Peskin:1995ev}. Since the path integral quantization method carries over to the quantization of fields, the least action principle for field theories gained even more importance due to this quantization process. An important class of field theories that is widely used for quantum mechanical purposes is that of the so-called (non-Abelian) gauge theories (Yang-Mills theories) that are invariant under the local action of a (non-Abelian) group.

By now, the `standard model' in quantum field theory, which is a gauge theory with gauge group $\SU(3)\times \SU(2)\times \U(1)$, is used to describe the electro-weak and the strong interactions. Although the theory has been tested to a remarkable precision (hence its name), there are still some conceptual problems to be unravelled. In this introduction, we will mainly focus on the problem of mass scales in the standard model, as it can be used to motivate the introduction of a new invariance, called supersymmetry. 

The $\SU(2)\times \U(1)$ symmetry in the standard model has to be broken to a diagonal $\U(1)$, and the conventional way to do this is by giving a charged scalar field a nonzero vacuum expectation value. In this Brout-Englert-Higgs mechanism, the fluctuations around the expectation value of the scalar (the so-called Higgs field) yield a mass for the gauge field. However, the problem with this is that in order to give the W- and Z-bosons their observed masses, the bare mass of the Higgs has to be remarkably small, i.e. many orders of magnitude below its renormalized (natural) value. This is referred to as the gauge hierarchy problem. 

A very elegant way to make quantum field theories more tractable and to introduce more mathematical structure is the inclusion of a symmetry between bosons and fermions, called supersymmetry. A generator of such a symmetry has to commute with the Hamiltonian and has to transform bosonic into fermionic states (and vice versa). Due to the spin-statistics theorem~\cite{Pauli1}, these generators have to combine in a spinorial representation of the Lorentz group. Moreover, the representation has to have spin $1/2$ as higher spin generators would lead to trivial theories~\cite{Coleman:1967ad}.\footnote{The reason is that the Coleman-Mandula theorem~\cite{Coleman:1967ad} rules out any higher spin conservation laws, which would appear in the right-hand side of~(\ref{ordefsusy}).} The defining relation for supersymmetry generators is the following anticommutator
\begin{equation}\label{ordefsusy}
\{Q_\alpha^i,Q^\beta_{j}{}^\dagger\}=2\delta^i_j\sigma^\mu_{\alpha}{}^\beta P_\mu\,.
\end{equation}
In this equation, $Q$ denotes the supersymmetry annihilation operator, while $Q^\dagger$ is the hermitian conjugated creation operator. The indices $\alpha,\beta$ are chiral spinor indices, while $i,j=1,\dots,\mathcal N$ refer to the number of spinors of supersymmetry generators. The matrices $\{\sigma^\mu\}$ denote $\{\unity,\sigma^{\mathfrak i}\}$ where $\{\sigma^{\mathfrak i}\}$ are the Pauli matrices. The (conserved) vector quantity $P$ should be the energy-momentum 4-vector as otherwise, the theory would again become trivial~\cite{Coleman:1967ad}. 

Consider now as an example a field theory of a complex scalar field $\varphi$. Due to supersymmetry, the self interaction would be related to interactions of its supersymmetric fermionic partner $\psi$. The same holds for gauge theories. For every gauge field $A_\mu$, supersymmetry instructs us to introduce a (chiral) fermion $\lambda$, called the gaugino. Due to these stringent relations between bosons and fermions, the renormalization of supersymmetric theories simplifies drastically. To see this, note e.g. that in a supersymmetric theory, every scalar particle $|\varphi>$ with a mass $m\neq0$ has a fermionic partner $Q|\varphi>$ of the same mass. As the mass of a fermion can only diverge logarithmically, while that of a scalar can run to infinity in a quadratic way, the latter divergences have to cancel in a theory with supersymmetry realized at the quantum level. These cancellations occur at every order in perturbation between diagrams involving bosons and diagrams involving virtual fermions. 

Another simplification has to do with the vacuum energy. As the vacuum has zero three-momentum $P^{\mathfrak i}=0$,~(\ref{ordefsusy}) implies that a supersymmetric vacuum state $|0>$ has zero energy as~(\ref{ordefsusy}) yields
\begin{equation}
<0|H|0>=0\,.
\end{equation}
Hence, in a supersymmetric theory, the positive contributions to the vacuum energy of the bosonic fields are cancelled exactly by the negative contributions of the fermions. 

To conclude, let us briefly comment on the rather obvious fact that theories with more supersymmetry are more restricted. For instance, $\mathcal N=4$ Yang-Mills theory has vanishing $\beta$ functions, implying that the theory appears to be exactly conformally invariant. This observation has profound consequences in string theory~\cite{Maldacena:1998re}. Moreover, the Montonen-Olive conjecture that stated originally~\cite{Montonen:1977sn} that the Lagrangian field theory `dual' to the Georgi-Glashow model would posses electric monopole solutions analogous to the BPS monopole, was proven in the context of this $\mathcal N=4$ super Yang-Mills theory~\cite{Vafa:1994tf}.

The introduction of supersymmetry moreover yields a solution to the hierarchy problem. In a supersymmetric standard model, the Higgs would be one of many scalar fields present in the theory. As already noted, the masses of these scalars would only be renormalized multiplicatively (as they diverge logarithmically). If supersymmetry would be broken in such a way as to give mass differences of a few hundred GeV between the quarks and leptons and their (still unobserved) bosonic partners, this would yield a Higgs particle with the correct mass~\cite{Witten:1981nf,Nilles:1984ge,Haber:1985rc}.

In the case of massive particles in four dimensions, the spin is the value of the Casimir of the little group, while the helicity of a massless particle is the charge under the only generator of that group. It is a basic fact about supersymmetry~\cite{Lykken:1996xt} that acting on a state of definite spin (helicity) changes the spin (helicity) by $1/2$. Hence, it turns out that in the absence of gravity (i.e. where the maximum spin (helicity) is $1$), the maximum number of supersymmetric generators is $16$, while in the presence of gravity (thus with maximum spin (helicity) equal to $2$), this number is $32$. In the former, supersymmetry is a global symmetry, while in the latter, it appears to be gauged. Hence, the maximal dimension for a rigid supersymmetric theory is $d=10$, while in the case of local supersymmetry (called supergravity), the maximal dimension turns out to be $d=11$. The reason why a supersymmetric theory with gravity admits local supersymmetry is that the graviton, which is the gauge field for coordinate transformations, always forms a multiplet\footnote{A multiplet is a set of fields transforming into each other under supersymmetry.} with at least one fermionic field of helicity $3/2$, called the gravitino. And, it turns out that this field is the gauge field for supersymmetry.

Although supersymmetry improves the renormalization properties of a theory, supergravities still are not renormalizable. Hence, in order to unify all elementary particle interactions, we might need a new principle. One of the best candidate theories to yield a quantum description of all forces is called string theory. There, elementary particles are considered to be extended in one dimension. More precisely, the theory is defined through a conformal nonlinear sigma model defined on a two-dimensional Lorentzian surface, called the world-sheet of the string. The target space in that theory is interpreted as our space-time. In quantizing the theory, supersymmetry again plays an important role as it is used to eliminate a nonphysical (tachyonic) mode from the spectrum.

String theory thus differs fundamentally from supergravity, but it still yields effective field theories describing the low-energy behaviour. Let us briefly explain this remark~\cite{Polchinski}. For simplicity, we will
consider the bosonic string but the construction can be extended to the supersymmetric case. As we already mentioned, the scalars in a nonlinear sigma model are coordinate functions on the target space and mostly, the fundamental object defined on that space is the metric $g_{XY}$, where $X$ and $Y$ label the scalar fields.
However, in string theory it turns out that it is necessary that we also introduce on the target space an antisymmetric 
two-tensor $B_{XY}$ called the Neveu-Schwarz two-form and a real scalar function $\Phi$ called the dilaton. The
role of $\hbar$ is played by the only free parameter in string theory, named $\alpha '$, which is inversely proportional to the tension of the string. The string action then reads
\begin{eqnarray}\label{bosonicstring}
\mathcal S&=&\frac1{4\pi \alpha '} \int_M d^2 \sigma \, \sqrt{-h}\big[h^{\mu \nu}g_{XY}(\varphi) \partial_\mu \varphi^X\partial_\nu
\varphi^Y+\rmi \varepsilon^{\mu \nu} B_{XY}(\varphi)\partial_\mu \varphi^X\partial_\nu \varphi^Y +\alpha ' R(\varphi) \Phi(\varphi)\big]\,.\nonumber\\
\end{eqnarray}
This nonlinear sigma model is defined on a two-dimensional surface $M$ with Lorentzian metric $h$, while $\varepsilon$ is the two-dimensional Levi-Civita symbol (vo\-lu\-me-form) with curved indices, see Appendix~\ref{appconventions}. $R$ denotes the target space Ricci scalar (computed from the target space Levi-Civita connection corresponding to the metric $g$). The two-form admits an Abelian gauge symmetry, with a target space one-form $\xi$ as gauge parameter,
\begin{equation}
\delta B_{XY}(\varphi)=2\partial_{[X}\xi_{Y]}(\varphi)\,,
\end{equation}
which adds a total derivative to the Lagrangian. The corresponding gauge-invariant field strength reads
$H_{XYZ}=3\partial_{[X}B_{YZ]}$. It is well-known that the theory~(\ref{bosonicstring}) admits a classical conformal invariance, generated by the trace of the energy-momentum tensor, which is however anomalous. In the simplest case of a flat target space
(i.e. $g_{XY}$ equals the Minkowski metric $\eta_{XY}$, and $B_{XY}=\Phi=0$), the conformal anomaly disappears at leading order in $\alpha'$ when the dimension of the target space is 26. For the quantum theory of~(\ref{bosonicstring}) in a general background, the trace of the energy-momentum tensor can be expanded as follows
\begin{equation}
T_\mu{}^\mu=-\frac1{2\alpha'}\beta^g_{XY}h^{\mu \nu}\partial_\mu \varphi^X\partial_\nu \varphi^Y-\frac1{2\alpha'}\rmi
\beta^B_{XY}\varepsilon^{\mu \nu}\partial_\mu \varphi^X\partial_\nu \varphi^Y-\frac12 \beta^\Phi R\,.
\end{equation}
Hence, up to quadratic order in $\alpha'$, we find~\cite{Polchinski}
\begin{eqnarray}
\beta^g_{XY}&=&\alpha ' R_{XY}+2\alpha ' \mathfrak{D}_X\mathfrak{D}_Y \Phi -\frac{\alpha '}4
H_{XVW}H_Y{}^{VW}+\mathcal{O}(\alpha'{}^2)\,,\nonumber\\
\beta^B_{XY}&=&-\frac{\alpha '}2\mathfrak{D}^ZH_{ZXY}+\alpha'\mathfrak{D}^Z \Phi H_{ZXY}+ \mathcal{O}(\alpha'{}^2)\,,\\
\beta^\Phi&=&\frac{d-26}6-\frac{\alpha '}2 \Box \Phi +\alpha ' \mathfrak{D}_X\Phi \mathfrak{D}^X\Phi
-\frac{\alpha '}{24}H_{XYZ}H^{XYZ}+\mathcal{O}(\alpha'{}^2)\,,\nonumber
\end{eqnarray}
where $d$ is the dimension of the target space and the derivatives are covariant with respect to the target space Levi-Civita connection. If we want to have conformal invariance, we thus need to set the
above expressions to zero,
\begin{equation}
\beta^g_{XY}=\beta^B_{XY}=\beta^\Phi=0\,.
\end{equation}
These functionals (when truncated at order $\alpha'^2$) are the equations of motion of gravity in 26
dimensions coupled to a two-form gauge potential and a scalar field. Hence, quantizing string theory leads in a natural way to equations of motion for the background fields and these are the equations of a regular field theory, when we can trust the $\alpha'$ truncation. Similar considerations for local $\mathcal N=1$ two-dimensional nonlinear sigma models, i.e. superstring theory, lead to the equations of motion of ten-dimensional supergravities. Therefore, it is often stated that supergravity is the low-energy effective theory that describes string theory. Note that the equations of motion of IIB supergravity cannot be generated from a standard action principle, due to the appearance of a self-dual four-form gauge field.
\section{On-shell realizations of supersymmetry}
The standard representation of a set of symmetries on (classical) fields is a so-called off-shell realization. This means that the algebra of generators closes without having to impose dynamical constraints (equations of motion) on the fields. In supersymmetric theories, an on-shell realization is also possible. This implies that in the right-hand side of the anticommutator~(\ref{ordefsusy}), a nonclosure functional of the fields appears that is second order in the derivatives.\footnote{More specifically, it is second order on the independent components of the fields.} Hence, by computing the supersymmetry algebra, we can construct classical equations of motion. 

It turns out that these dynamical constraints are not always integrable, meaning that it is not always possible to find a supersymmetric and Poincar\'e-invariant action functional for which these field equations would describe the stationary points. At first sight, one could think of discarding such theories as they cannot be quantized in a standard fashion. We think however that because these generalizations of known classical field theories appear naturally in the framework of supersymmetry, we should first study their consequences before deciding on their fate. Moreover, we will show that at least some of these theories can consistently be embedded into string theory, 
which makes the need for a quantization scheme less urgent or even unnecessary. 

Let us be more precise about this point. Superstring theory is only well-defined in ten space-time dimensions and its low-energy behaviour is described by ten-dimensional supergravity. However, as our universe appears (at least locally) to be four-dimensional, we have to think of ways to confine physics to these four dimensions. One possibility is the Kaluza-Klein dimensional reduction process, which boils down to the idea that the six other dimensions are compact implying that (four-dimensional) low-energetic fluctuations are constant along the compact directions. Examples exist of a generalization of such a procedure that yields lower-dimensional equations of motion without an action~\cite{Lavrinenko:1998qa,Bergshoeff:2003ri}. Hence, such theories without actions describe the low-energy behaviour of a supergravity, that itself describes a particular sector of string theory.

Something similar is encountered in the Randall-Sundrum scenarios, where we are thought of as living on a four-dimensional defect of a higher-dimensional space \cite{Randall:1999vf,Randall:1999ee}. The four-di\-men\-sio\-nal theory found by integrating out the bulk fields in the path integral is highly nonlocal and is used as an effective theory, while the higher-dimensional theory is local and serves as the more fundamental theory, see e.g.~\cite{Porrati:2004yi}. These Randall-Sundrum scenarios are moreover historically one of the main reasons why the theories discussed in Chapters~\ref{onshellsusy} and~\ref{suptencalc} are five-dimensional. 

The study of these on-shell generalizations of supersymmetry can also be valuable in a mathematical context.
To see this, let us first return to the standard case in which an action exists. If a multiplet contains one or more scalar fields, this sector of the theory might be described by a nonlinear sigma model. In such a theory, the kinetic energy in the Lagrangian is field-dependent in the following way
\begin{equation}\label{form}
\mathcal{S}= -\frac12\int d^dx \;g_{XY} (\varphi)\partial_\mu \varphi^X \partial^\mu \varphi^Y\,.
\end{equation}
Similar to our discussion of the string theory action, $g$ denotes the target space metric. The equations of motion originating from~(\ref{form}) then read
\begin{equation}\label{form2}
\Box \varphi^X \equiv \partial^\mu \partial_\mu \varphi^X+\Gamma_{YZ}{}^X\partial_\mu \varphi^Y \partial^\mu \varphi^Z=0\,,
\end{equation}
in which $\Gamma$ denotes the Levi-Civita connection. If we directly consider the above equations of motion, the torsionless connection appearing in~(\ref{form2}) can also be affine, meaning that it does not have to preserve a metric under parallel transport. Hence, a nonlinear sigma model with action admits a target space with a metric, while a similar theory without a Lagrangian description can have a more general target space. 

Supersymmetry moreover constrains the possible target spaces of the theory. It turns out that in theories with more than eight supercharges, only a discrete number of allowed target manifolds exists. Subtleties aside, it can be stated that such a theory is completely fixed once the number of multiplets is known. In theories with eight supersymmetries, there is more freedom in the sense that the metric can continuously be deformed without breaking the supersymmetry. In theories with even less supersymmetry, there is still more freedom for the choice of the target space. Hence, theories with eight supercharges are special in the sense that the target space geometry can be continuously deformed, although the huge amount of symmetries is very restrictive. In the present work, we will show that such theories can admit more general target spaces than previously considered in the literature~\cite{Alvarez-Gaume:1981hm,Bagger:1983tt} if the equations of motion are not derivable from an action principle. This will be similar in spirit to the generalization by~(\ref{form2}) of the theory in~(\ref{form}). Moreover, these remarks will allow us to use physical methods to derive new mathematical relations between the target space manifolds. Note that these spaces are used as well in conformal quantum mechanics, describing e.g. the moduli space of certain black hole configurations~\cite{Michelson:1999zf}.
\section{Content}
In the present work, we will thus consider supersymmetric classical field theories that are formulated in terms of equations of motion rather than in terms of actions. We will discuss how these theories can be constructed in a systematic way and will point out precisely what are the generalizations with respect to the corresponding theories that can be described using a Lagrangian.

The content of this thesis is as follows. In Chapter~\ref{prelim} the necessary mathematical and physical concepts will be introduced. In Chapter~\ref{onshellsusy} we will discuss two examples of on-shell realizations of rigid $\mathcal{N}=2$ super-Poincar\'e theories. We will discover that these theories allow for more general target spaces and can have more general potentials~\cite{Bergshoeff:2002qk,Gheerardyn:2003rf}. Chapter~\ref{suptencalc} will be a similar discussion in the context of local $\mathcal{N}=2$ supersymmetry. However, since these theories are much more elaborate than the rigid ones, we will use the so-called superconformal tensor calculus to construct them. This will lead us to a few new results. We will again show that the nonlinear sigma models can be made more general, we will point out new relations between different types of so-called quaternionic-like spaces and we will present the most general $\mathcal N=2$ supergravity with vector- and hypermultiplets in five dimensions~\cite{Bergshoeff:2002qk,Bergshoeff:inprep1,Bergshoeff:inprep2}. In Chapter~\ref{dimred}, we will show how similar theories without action can be constructed from generalized dimensional reduction. More precisely, we will point out that we can use Scherk-Schwarz dimensional reduction to retrieve a certain ten-dimensional supergravity that does not possess a Lagrangian formulation. We will discuss how the supersymmetry transformation rules can be derived from this dimensional reduction process and we will construct supersymmetric solutions~\cite{Gheerardyn:2002wp}. We will conclude in Chapter~\ref{concel} by listing some possible directions for future research. We have added three Appendices. In Appendix~\ref{spinors}, we will list the basic properties of spinors in diverse dimensions.  The Appendix~\ref{appconventions} contains our conventions while Appendix~\ref{samenvatting} is a r\'esum\'e in Dutch. 

From the above, it must be clear that we will mostly be considered with the construction of these on-shell theories. We will point out what are the generalizations, but will most often not go into any detail concerning the possible applications. We will however make two exceptions. In Chapter~\ref{suptencalc}, we will discover new relations between hypercomplex and quaternionic manifolds and between hyperk\"ahler and quaternionic-K\"ahler manifolds. In Chapter~\ref{dimred}, we will show that because of the fact that theories without action allow for more general potentials, we can find a cosmologically interesting de Sitter solution to this ten-dimensional supergravity, a fact that has proven to be very hard in more `regular' supergravities~\cite{Maldacena:2000mw}, although other exceptions exist~\cite{Fre:2002pd,Fre':2003gd}. 

Let us conclude by stating that it is our believe that theories that do not obey an action principle can only  be discarded as unphysical if string theory prohibits their existence. We can only hope of discovering such a conclusive answer if we are prepared to study the consequences of these generalized on-shell supersymmetric theories.

\chapter{Preliminaries}\label{prelim}
In this chapter, we will discuss the necessary background to study supersymmetric theories. In the starting Section~\ref{s:difgeom}, we will repeat some standard results from basic differential geometry. We will set the stage by giving the definition of a manifold and of a bundle. We will subsequently introduce the concept of a connection to define parallel transport, and use it to discuss the integrability of G-structures, which will prove to be very useful in characterizing different types of manifolds. Next, we will introduce complex manifolds, quaternionic-like manifolds, spin manifolds and manifolds of special holonomy, which will all play a prominent role in supersymmetric theories.
In Section~\ref{s:susy}, we will discuss the $\mathcal{N}=2$ super-Poincar\'e algebra, since this is the superalgebra we will most of the time be concerned with. To give a flavour of the subject, we will subsequently give an easy example of how supersymmetry constrains the target space geometry of two-dimensional nonlinear sigma models. Finally, we will introduce the Batalin-Vilkovisky formalism in Section~\ref{s:BV}, since it is a very elegant way to keep track of the requirements posed by the introduction of an action.
\section{Differential geometry}\label{s:difgeom}
\subsection{Manifolds}\label{ss:manifolds}
For completeness, we will start this section by giving some basic definitions.
\subsubsection{Definitions}
\begin{defi}Consider a set X, an index set $I$, let $i\in I$, a set $A$ of subsets $U_i$ of $X$ called open subsets. $X$ is then called a topological space if it admits the following properties:
\begin{enumerate}
\item $X \in A$ and $\emptyset \in A$.
\item For $J \subset I$, $\bigcup_{j \in J} U_j \in A$.
\item For $K \subset I$ a finite subset, $\bigcap_{k \in K}U_k \in A$.
\end{enumerate}
\end{defi}
To be able to introduce the definition of a manifold in a rigorous way, we need the more restrictive notion of a topological manifold. This requires some basic concepts from topology, which can be found in e.g.~\cite{McCarty}.
\begin{defi}
A topological manifold is a topological space that is Hausdorff, satisfies the second axiom of countability and is locally homeomorphic to $\mathbb{R}^n$ for some value of $n$.
\end{defi}
The value of $n$ is called the (real) dimension of the topological manifold. 
\begin{defi}Consider an index set $I$, a topological manifold $X$ of dimension $n$, $U_i$ are (some) open sets of $X$, and a set of bijective maps $x_i$ between $U_i$ and an open set in $\mathbb{R}^n$. X is called a $C^r$ n-dimensional manifold if it exhibits the following properties:
\begin{enumerate}
\item $U_i$ cover M, i.e. $\bigcup_{i \in I} U_i=M$.
\item If $U_i \cap U_j \neq \emptyset$, then the map $x_ix_j^{-1}:x_j(U_i \cap U_j)\to x_i(U_i \cap U_j)$ is a $C^r$ map\footnote{This means that the map is r times continuously differentiable.} between two open sets in $\mathbb{R}^n$.
\end{enumerate}
\end{defi}
The functions $x_i$ are called coordinates or coordinate functions while $x_ix_j^{-1}$ are named transition functions.
The set of sets $\{U_i\}$ is called an open covering of $M$, also known as an atlas, while $U_i$ is called a chart. 
We will only consider $C^{\infty}$ manifolds, call them (differentiable) manifolds and we will denote them by $M$.\footnote{Unless otherwise stated, all maps, functions, etc. will also be taken to be $C^\infty$. Such maps, functions etc. may also be called smooth.} The set of functions on $M$ is called $\mathcal{F}(M)$. Later on, we will introduce more structure on the manifold, like e.g. a metric, a complex structure, a connection, etc. Hence, if we will use the word `manifold', we will refer to the above definition together with all necessary structure on the space. 

A curve $c$ on $M$ is a map from the real numbers to the manifold, i.e. $c:I\subset\mathbb{R} \to M:t\mapsto c(t)$.\footnote{We will not consider self-intersecting curves.} An open curve corresponds to $I$ being an open interval, and by convention we will in this case always choose $0\in I$. Let $p$ be a point of M. We can now define an equivalence class of open curves going through $p$. If $c_1^p$, $c_2^p$ are two open curves in $M$, $f$ any function on $M$ and $c_1^p(0)=c_2^p(0)=p$, these two curves are called equivalent in $p$ if
\begin{equation}
\frac{df\big(c_1^p(t)\big)}{dt}\Big|_{t=0}=\frac{df\big(c_2^p(t)\big)}{dt}\Big|_{t=0}\,. 
\end{equation}
The equivalence is denoted by $c_1^p\equiv c_2^p$, while the equivalence class is represented as $[c_1^p]$. These equivalence classes of curves span an $n$-dimensional vector space $T_pM$, called the tangent space at $p$ and we will denote a set of basis vectors by $\{\partial/\partial x^\mu(p)\}\equiv\{\partial_\mu(p)\}$ with $\mu=1,\dots,n$. An element $X(p)\in T_pM$ corresponding to $[c^p]$ is called a tangent vector at $p$ and we may write 
\begin{equation}\label{defvector}
X[f](p)=\frac{df\big(c^p(t)\big)}{dt}\Big|_{t=0}\,.
\end{equation} 
The vector space dual to $T_pM$ is called the cotangent space at $p$ and is denoted by $T_p^*M$. The basis dual to $\{\partial/\partial x^\mu(p)\}$ is called $\{dx^\mu(p)\}$ and defined in the usual way 
\begin{equation}
dx^\mu(p):\; T_pM\to \mathbb{R} :\; dx^\mu(p)\big(\partial/\partial x^\nu(p)\big)=\delta_\nu^\mu\,.
\end{equation}
An element of this vector space is called a one-form at $p$.
A (p,q) tensor is a multi-linear object which maps q elements of $T_p^*M$ and p vectors of $T_pM$ to a real number. 
Writing the local expression for the vector defined in~(\ref{defvector}) in two different sets of coordinates $\{x^\mu\}$ and $\{y^\mu\}$, we can deduce that a vector transforms under coordinate transformations in the following way
\begin{equation}\label{transfvect}
X(p)=X'{}^\mu(p)\frac{\partial\,}{\partial y^\mu}(p)=X^\nu(p) \frac{\partial y^\mu}{\partial x^\nu}(p)\frac{\partial \,}{\partial y^\mu}(p)\Rightarrow X'{}^\mu(p)=X^\nu (p)\frac{\partial y^\mu}{\partial x^\nu}(p)\,,
\end{equation}
and similarly for other tensors.

Unless otherwise stated, we will suppose that a manifold is orientable. This is explained in the following definition.
\begin{defi}
A connected\footnote{See~\cite{McCarty} for a proper definition.} manifold $M$ is called orientable if for any two overlapping charts $U_i$ and $U_j$ there exist local coordinates $\{x^\mu\}$ on $U_i$ and $\{y^\mu\}$ on $U_j$ such that $\det (\partial x^\mu/\partial y^\nu)>0$.
\end{defi}

Having associated a vector space $T_pM$ to every point of the manifold, we now consider the union of all these spaces. The tangent bundle $TM$ to the manifold $M$ is the set of vector spaces $\bigcup_{p\in M}T_pM$. A vector field on $M$ assigns to every point $p$ a vector in $T_pM$ in a smooth way, i.e. if $X$ is a vector field, then for every function $f\in \mathcal{F}[M]$ we have that $X[f]\in \mathcal{F}[M]$. The space of all vector fields on $M$ is called $\chi(M)$. Similarly, we can introduce the cotangent bundle and hence, more general tensor bundles. A (p,q) tensor field will then similarly assign to every point a (p,q) tensor. The most prominent example of such tensor field is the metric, which is a symmetric $(2,0)$ tensor field that is nowhere singular.\footnote{The metric defines a regular symmetric matrix at every point.} A manifold with a metric is called Riemannian, while a manifold with a positive definite metric is called Euclidean.

In~(\ref{defvector}) we have defined a vector $X$ in $T_pM$ as an equivalence class of curves through $p$. This local correspondence can now be made global in the following way. Note that although we will only explain the construction in a certain coordinate frame, it can easily be shown that the correspondence is actually coordinate-independent.
\begin{defi}\label{defflow}
Let $X$ be a vector field on $M$. The flow of $X$ through $p\in M$ is a curve $\sigma^p:I\subset \mathbb R \to M:t\mapsto \sigma^p(t)$  for which both $\sigma^p(0)=p$ and locally
\begin{equation}
\frac{d\sigma^{p\mu}(t)}{dt}=X^\mu\big(\sigma^p(t)\big)
\end{equation}
holds.
\end{defi}

Let $M$ and $N$ be two $n$-dimensional manifolds, let $p\in M$, $X \in T_pM$, let $f$ be a smooth map from $M$ to $N$ and let $g \in \mathcal{F}[N]$. Using the map $f$ between these two manifolds, we can construct a map between the (co)tangent bundles. 
\begin{defi}
The induced map $f_*:T_pM\to T_{f(p)}N$, called the push-forward, is given by
\begin{equation}
(f_*X)[g]\big(f(p)\big)=X[gf](p)\; . 
\end{equation} 
The map $f$ also induces a map $f^*:T_{f(p)}^*N\to T_p^*M$ called the pull-back. Let therefore be $\omega$ any element in $T_{f(p)}^*N$, then
\begin{equation}
<f^*\omega,X>=<\omega,f_*X> \,.
\end{equation}
\end{defi}
Using the above concepts, we can now introduce the notion of the Lie-de\-ri\-va\-ti\-ve, which measures the change of a tensor field along the flow of a vector field in a coordinate-invariant fashion.
\begin{defi}
Let $X\in \chi(M)$, and consider its flow $\sigma^p(t)$ through $p \in M$. Let $Y$ be a vector field, $\omega$ a one-form and $f$ a function, all defined at least in a neighbourhood of $p$. The Lie-derivative along $X$ of $Y$, $\omega$ and $f$ in $p$ is respectively
\begin{eqnarray}
\mathcal{L}_X Y(p)\equiv[X,Y](p)&=&\lim_{\epsilon\to 0} \frac{\big(\sigma^p(-\epsilon)\big)_*Y\vert_{\sigma^p(\epsilon)}-Y\vert_p}{\epsilon}\,,\nonumber\\
\mathcal{L}_X \omega(p)&=&\lim_{\epsilon\to 0} \frac{\big(\sigma^p(\epsilon)\big)^* \omega \vert_{\sigma^p(\epsilon)}-\omega\vert_p}{\epsilon}\,,\nonumber\\
\mathcal{L}_X f(p)&=&\lim_{\epsilon\to 0} \frac{f\big(\sigma^p(\epsilon)\big)-f(p)}{\epsilon}=X[f](p)\,,\label{deflieder}
\end{eqnarray}
\end{defi}
where $[X,Y]$ is called the Lie-bracket. Using these definitions, the Lie-de\-ri\-va\-ti\-ve of any tensor field can be computed. For convenience, we now list the different expressions of~(\ref{deflieder}) in a given set of coordinates,
\begin{eqnarray}
\mathcal{L}_X Y(p)&=&\big(X^\nu(p) \partial_\nu Y^\mu(p)- Y^\nu(p) \partial_\nu X^\mu(p)\big)\frac{\partial\,}{\partial x^\mu}(p)\,,\nonumber\\
\mathcal{L}_X \omega(p)&=&\big(X^\nu(p) \partial_\nu \omega_\mu(p)+\omega_\nu(p) \partial_\mu X^\nu(p)\big)dx^\mu(p)\,,\nonumber\\
\mathcal{L}_X f(p)&=&X^\mu (p)\partial_\mu f(p)\,.\label{compLieder}
\end{eqnarray} 
Consider now in general the Lie-derivative of some tensor field $T$ along a vector field $X$. If $X(p)\neq 0$, then there exists a neighbourhood of $p$ in which it is possible to change coordinates such that $X(p)=\partial_z(p)$~\cite{Nakahara}. In that case, the Lie-derivative reduces to a plain derivative along $z$. For instance, if $T$ is a vector field, we have in components
\begin{equation}\label{proplieder}
\mathcal{L}_XT^\mu(p)=\partial_zT^\mu(p)\,.
\end{equation}
Suppose that $M$ is a Riemannian manifold with metric $g$. We will often encounter vector fields for which the Lie-derivative of the metric along that vector field satisfies special properties. A Killing vector $X$ is a vector field on $M$ such that
\begin{equation}
\mathcal{L}_X g=0\,.
\end{equation}
A conformal Killing vector satisfies a more general identity,
\begin{equation}\label{confkilvect}
\mathcal{L}_X g=2fg,
\end{equation}
where $f$ is a function on the manifold. If the function is a constant $m$, i.e.
\begin{equation}\label{homkilvect}
\mathcal{L}_X g=2mg,
\end{equation}
the vector field is called homothetic Killing.
\subsubsection{Bundles and connections}
We start this section by introducing the different types of bundles (with connections) on a manifold. Subsequently, we will discuss the frame bundle and we will close by discussing G-structures.
\begin{defi}
Let G be a Lie-group and M a manifold. The left action of G on M is a differentiable map $\sigma:G\times M\to M$ which satisfies
\begin{enumerate}
\item $\sigma(e,p)=p$ for any $p\in M$ and with $e$ the unit element of $G$,
\item $\sigma(g_1,\sigma(g_2,p))=\sigma(g_1g_2,p)$ for any $p\in M$ and $g_1,g_2\in G$
\end{enumerate}
\end{defi}
The right action is defined in a similar way, but the second condition becomes $\sigma(g_1,\sigma(g_2,p))=\sigma(g_2g_1,p)$. 
\begin{defi}
A fibre bundle $(E,\pi,M,F,G)$ consists of the following data~\cite{Nakahara}:
\begin{enumerate}
\item Differentiable manifolds $E,M$ and $F$ respectively called the total space, the base space and the typical fibre.
\item A Lie-group $G$, called the structure group acting on $F$ from the left.
\item A surjection $\pi:E\to M$ called the projection, where the image of the inverse map $\pi^{-1}(p)$ is isomorphic with $F$ and is called the fibre at $p\in M$.
\item An open covering $\{U_i\}$ of $M$ with a set of diffeomorphisms $\{\phi_i\}$, called local trivializations, for which $\phi_i:U_i\times F\to \pi^{-1}(U_i)$ and $\pi \phi_i(p,f)=p$ with $f\in F$.
\item For every $U_i \cap U_j\neq \emptyset$, a smooth map $t_{ij}$ can be introduced by  $\phi_j(p,f)=\phi_i(p,t_{ij}(p)f)$ which should take values in $G$. The maps $\{t_{ij}\}$ are called transition functions.
\end{enumerate}
\end{defi}
In order to be able to glue the local pieces of the fibre bundle together, the transition functions need to satisfy three consistency conditions (no sum on repeated indices):
\begin{equation}\label{consist}
t_{ii}=1\,,\; \; t_{ij}=t_{ji}{}^{-1}\,,\; \;t_{ij}t_{jk}=t_{ik}\,.
\end{equation} 
Note that the set of local trivializations is not unique. To see this, let $\{g_i\}$ be a set of homeomorphisms mapping $U_i$ to $G$ and let $p\in U_i\cap U_j$. We can now define new local trivializations as $\tilde \phi_i(p,f)=\phi_i(p,g_i(p)f)$, while the corresponding transition functions are $\tilde t_{ij}(p)=g_i(p)^{-1}t_{ij}(p)g_j(p)$.

A special case is that of a trivial fibre bundle which is just the direct product of the base with the fibre. Otherwise stated, a trivial bundle is a fibre bundle for which there exists a set of transition functions being identity maps. 
A basic object on a bundle is a section $s$ which is a fibrewise and smooth map from the base $M$ to $E$, i.e. $\pi s=1$. An important theorem states that if a bundle with a typical fibre isomorphic to the structure group, admits a global section, the bundle is trivial~\cite{Nakahara}. A local section $s_i$ is a section defined on a chart $U_i$. 
\paragraph{Vector bundles}
We now specify to a particular type of fibre bundles.
\begin{defi}
A vector bundle is a fibre bundle where the typical fibre is a vector space.
\end{defi}
A prominent example of the latter is the tangent bundle $TM$ of a manifold $M$ with dimension $n$, where the fibre is $T_pM\equiv\mathbb{R}^n$ and the structure group equals $\Gl(n,\mathbb{R})$, a fact which can be deduced from the transformation property~(\ref{transfvect}). Vector fields are sections of this bundle. A basic object, defined on the tangent bundle is the affine connection.
\begin{defi}
An affine connection $\mathfrak{D}$ is a map $\mathfrak{D}:\chi(M)\times \chi(M)\to \chi(M):(X,Y)\mapsto \mathfrak{D}_X Y$ that satisfies the following conditions.
\begin{eqnarray}
\mathfrak{D}_X(Y+Z)&=&\mathfrak{D}_X Y+\mathfrak{D}_X Z\,,\nonumber\\
\mathfrak{D}_{X+Y}Z&=&\mathfrak{D}_X Z+\mathfrak{D}_Y Z\,,\nonumber\\
\mathfrak{D}_{fX}Y&=&f\mathfrak{D}_XY\,,\nonumber\\
\mathfrak{D}_X(fY)&=&X[f]Y+f\mathfrak{D}_XY\,,\nonumber
\end{eqnarray}
with $f\in \mathcal{F}(M)$ and $X,Y,Z \in \chi(M)$.
\end{defi}
We can use this definition to deduce how the connection acts on more general tensor fields.
Locally, we can define a set of functions called connection coefficients (or simply call the set a connection) $\Gamma_{\mu \nu}{}^\lambda$ which can be defined as
\begin{equation}
\mathfrak{D}_{\partial_\mu }\partial_\nu=\Gamma_{\mu \nu}{}^\rho \partial_\rho\,.
\end{equation}
In physics, the connection $\mathfrak{D}$ acting on a tensor field, is called a covariant derivative. 
For example, if $T$ is a $(1,1)$ tensor field, we may write locally 
\begin{equation}
\mathfrak{D}_\mu T_{\nu}{}^{\rho}=\partial_\mu T_{\nu}{}^{\rho}-\Gamma_{\mu \nu}{}^\sigma T_{\sigma}{}^{\rho}+\Gamma_{\mu \sigma}{}^\rho T_{\nu}{}^{\sigma}\,.
\end{equation}
Note that the covariant derivative of a $(p,q)$ tensor field transforms as a $({p+1},q)$ tensor field.
Another important concept is the notion of parallel transport. Consider a curve $c$ and a vector field $X$ defined at least in a neighbourhood of $c$. Let $Y$ be a vector field generating $c$ as its flow. We say that $X$ is parallel transported along $c$ if the condition 
\begin{equation}
\mathfrak{D}_YX\vert_c=0\,,
\end{equation}
is met. If a vector field $X$ is parallel transported along any $Y\in \chi(M)$, $X$ is said to be preserved by $\mathfrak D$ or to be covariantly constant or parallel.\footnote{The same nomenclature will be used for any other tensor field or even for a spinor field, see Section~\ref{spinnen}.}

It is well-known that the affine connection does not transform as a tensor, a fact that follows directly from its definition. Therefore, it cannot have any intrinsic meaning implying that it is necessary to introduce the torsion $T:\chi(M)\otimes \chi(M) \to \chi(M)$ and the Riemannian curvature $R:\chi(M)\otimes \chi(M)\otimes \chi(M)\to \chi(M)$, which do transform as tensors.
\begin{eqnarray}
T(X,Y)&\equiv &\mathfrak{D}_XY-\mathfrak{D}_YX-[X,Y]\,,\nonumber\\
R(X,Y,Z)&\equiv&\mathfrak{D}_X\mathfrak{D}_YZ-\mathfrak{D}_Y\mathfrak{D}_XZ-\mathfrak{D}_{[X,Y]}Z\,,
\end{eqnarray}
with $X,Y,Z \in \chi(M)$. For our purposes, it will most of the time be sufficient to use local expressions.
\begin{eqnarray}
T_{\mu \nu}{}^\rho&=&2\Gamma_{[\mu \nu]}{}^\rho\,,\nonumber\\
R_{\mu \nu \rho}{}^\sigma&=&2\partial_{[\mu}\Gamma_{\nu]\rho}{}^\sigma+2\Gamma_{\tau[\mu}{}^\sigma\Gamma_{\nu]\rho}{}^\tau\,.
\end{eqnarray}  
The curvature tensor always satisfies the second Bianchi identity
\begin{equation}
\mathfrak{D}_{[\mu}R_{\nu \rho ]\sigma}{}^\tau=0\,.
\end{equation}
In case of a torsionless connection (i.e. $T\equiv 0$), the first Bianchi identity is also satisfied.
\begin{equation}
R_{[\mu \nu \rho]}{}^\sigma=0\,.
\end{equation}
\paragraph{Principal and induced bundles}
Another type of fibre bundles is often encountered in gauge theories.
\begin{defi}
A principal bundle is a bundle where the structure group coincides with the typical fibre.
\end{defi}
Such a bundle $(P,\pi,M,G,G)$ will also be denoted by $P(M,G)$. As in the case of the tangent bundle $TM$, we will also introduce a connection on principal bundles, but this requires some extra work.

We start by introducing the left action of $G$ on a principal bundle, via its local trivialization. Suppose that $\phi_i(p,h)=u$, we can define $L_g u:u\mapsto gu\equiv\phi_i(p,gh)$ with $g,h\in G$ and similarly for the right action $R_g$. 
Let $A\in \mathfrak g$, the Lie-algebra of $G$. The fundamental vector field $A^\sharp$ generated by $A$ is defined as
\begin{equation}
A^\sharp[f](u)=\frac{d\;}{dt}f\big(u \exp (tA)\big)\vert_{t=0}\,,
\end{equation}
with $u\in P(M,G)$ and $f\in \mathcal F(P)$. All $A^\sharp (u)$ span a subspace of $T_uP$, tangent to the fibre $G$. Therefore, this space is called the vertical subspace $V_uP$, which is isomorphic to $\mathfrak g$. To specify its complement in $T_uP$, called the horizontal subspace $H_uP$, in a unique way, we have to introduce a connection on $P(M,G)$.
\begin{defi}\label{defcononeform}
A connection one-form $\omega\in \mathfrak g\otimes T^*P$ is a projection of $T_uP$ onto $V_uP$, $\forall u\in P(M,G)$.
\begin{enumerate}
\item $\omega(A^\sharp)(u)=A$,
\item $\omega(R_{g^*}X)(ug)=g^{-1}\omega(X)(u)g$, with $X\in T_uP$ and $g\in G$.
\end{enumerate}
\end{defi}
The horizontal subspace is then given by the kernel of $\omega(u)$.

Given a local section $\sigma_i$ of $P(M,G)$ on $U_i\subset M$, we can define the local connection one-form as 
\begin{equation}\label{defloccon}
\mathcal A_i\equiv \sigma_i^*\omega \in \mathfrak g \otimes T^*U_i\,.
\end{equation}
Conversely, given such a local one-form, we can construct a connection $\omega_i$ on $P(M,G)$ such that~(\ref{defloccon}) is satisfied on $U_i$~\cite{Nakahara}. However, if a local section and a local connection one-form are given in every chart, a unique connection $\omega$ on $P(M,G)$ compatible with all the local one-forms is only possible if $\omega_i=\omega_j$ on $U_i\cap U_j$, and this connection then satisfies $\omega|_{U_i}=\omega_i$. The compatibility condition needed to be able to construct this $\omega$ is
\begin{equation}\label{contrans}
\mathcal A_j=t_{ij}{}^{-1}\mathcal A_it_{ij}+t_{ij}{}^{-1}dt_{ij}\,.
\end{equation}
Note that since the principal bundle is not necessary trivial, the pull-back $\mathcal A_i=\sigma_i^*\omega$ may only exist locally. Therefore, the local connection one-form cannot carry any global information. It is the complete set of local one-forms $\{\mathcal A_i\}$ or equivalently $\omega$ that encodes the nontrivial information of the bundle.
Finally, on a trivial bundle there exists a set of vanishing local one-forms $\mathcal A_i\equiv 0$ defined on a set $\{U_i\}$ that covers $M$.

The counterpart of the Riemannian curvature is the field strength of the connection one-form of a principal bundle $P(M,G)$,\footnote{Sometimes, this field strength will be called the curvature tensor.} for which we will only need its expression in terms of the local one-form 
\begin{equation}
F_{i \mu \nu}=2\partial_{[\mu}\mathcal A_{i\nu]}+2\mathcal{A}_{i[\mu}\mathcal{A}_{|i|\nu]}\,,
\end{equation}

In practical computations, we will often work with so called associated vector bundles. 
Suppose that $G$ acts on a vector space $F$ from the left in a certain representation $\rho$, we may define the action of $g\in G$ on $P(M,G)\times F$ by
\begin{equation}
g\big[(u,f)\big]=(ug,\rho[g^{-1}]f)\,,
\end{equation}
where $u\in P(M,G)$ and $f\in F$. The associated fibre bundle $P\times_\rho F/G$ is an equivalence class for which the points $(u,f)$ and $(ug,\rho[g^{-1}]f)$, for any $g\in G$, are identified. Conversely, a vector bundle $(E,\pi,M,F,G)$ can naturally induce a principal bundle $P(M,G)$ by employing the same transition functions~\cite{Nakahara}.

Consider now a principal bundle $P(M,G)$ with connection $\omega$ and an associated vector bundle $E\equiv P\times_\rho F/G$, and restrict attention to the overlap of two charts $U_i\cap U_j$. Suppose we have two corresponding local sections $\sigma_i:U_i\to E$ and $\sigma_j:U_j\to E$. Under a change in the local canonical trivialization~\cite{Nakahara}, the local section transforms as $\sigma_j=\sigma_it_{ij}$. Similar to the covariant derivative on the tangent bundle, we want to introduce a derivative operator $\mathfrak D$ such that $\mathfrak D \sigma_i$ transforms in the same way under local trivializations as $\sigma_i$ itself. Therefore, we define the covariant derivative on a local section of $E$ as
\begin{equation}
\mathfrak D_\mu \sigma_i=\partial_\mu\sigma_i+\sigma_i\mathcal{A}_{i\mu}^\rho \,,
\end{equation}  
where $\mathcal A_{i\mu}^\rho=\rho[\mathcal A_{i\mu}]$ is called the induced local connection one-form.\footnote{We will however often call this the local connection one-form or even briefly the connection.} Using~(\ref{contrans}), we can easily check that $\mathfrak D_\mu \sigma_j=t_{ij}\mathfrak D_\mu \sigma_i$. Hence, the covariant derivative is said to transform covariantly, meaning that it has the same form in all charts. This nomenclature can also be used to denote the transformation properties of other objects than the covariant derivative. As in the case of a vector bundle, a section is said to be parallel transported along a curve $c:I\subset \mathbb R \to M$, if 
\begin{equation}
X^\mu\mathfrak{D}_\mu \sigma_i\vert_c=0\,,
\end{equation}
where $X\in \chi(M)$ is a vector field tangent to $c$, i.e. 
\begin{equation}
X[f]\big(c(t)\big)=\frac{df\big(c(t)\big)}{dt}\,,
\end{equation}
with $f\in \mathcal F(M)$.

Consider now as an example an $n$-dimensional manifold $M$ and its tangent bundle $TM$. With this vector bundle, we can associate a principal bundle, called the frame bundle $LM$ by assigning to every point of $M$ the group $\Gl(n,\mathbb{R})$ in the following way. In a chart $U_i$ with coordinates $\{x^\mu\}$, the bundle $TU_i$ is obviously trivial, and hence, we can choose $n$ linearly independent local sections. A natural basis of sections is $\{\partial/\partial x^\mu\}$. A frame $u$ at $p\in U_i$ is the value of these sections at $p$, hence it is a set of $n$ linearly independent vectors in $T_pM$, and can be written as
\begin{equation}
u=\{X_1(p),\dots,X_n(p)\} \qquad \mbox{with}\qquad X_a(p)=X_a{}^\mu(p) \partial_\mu(p)\, ,
\end{equation}
with $a=1,\dots,n$.
Note that $(X_a{}^\mu(p))$ has to be a $\Gl(n,\mathbb{R})$-matrix since otherwise, the vectors would not be independent. As such, the set of frames in $T_pM$ is isomorphic to $\Gl(n,\mathbb{R})$ defining a principal bundle over $M$. Local trivializations $\phi_i$ are defined by $\phi_i{}^{-1}:LM\to U_i\times \Gl(n,\mathbb{R}):u\mapsto (p,(X_a{}^\mu))$. The projection is obviously $\pi (u)=p$. On an overlap $U_i\cap U_j\neq \emptyset$ with coordinates $\{x^\mu\}$ and $\{y^\mu\}$, the transition functions can be determined easily from the transformation properties of a vector~(\ref{transfvect}).
\begin{equation}
X_a(p)=X_a{}^\mu (p)\frac{\partial}{\partial x^\mu}(p)= X'{}_a{}^\mu(p) \frac{\partial}{\partial y^\mu}(p)\, \Rightarrow \; \;
t_{ij}(p)=\frac{\partial x^\mu}{\partial y^\nu}(p)\in \Gl(n,\mathbb{R})\,.
\end{equation}
We can now associate a new vector bundle to $LM$ with base $M$ and fibre $\mathbb{R}^n$ where the structure group acts in the fundamental representation $\rho$ on the fibre. This bundle will be denoted by  $aLM\equiv LM\times_\rho \mathbb{R}^n/\Gl(n,\mathbb{R})$, and can be shown to be isomorphic to $TM$~\cite{Nakahara}. 

This is a well-known fact in physics. On the manifold $M$, we can introduce (locally) a set of Vielbeine $\{e_\mu{}^a\}$, which are the components of $n$ linearly independent one-forms,\footnote{The Vielbeine define a so-called coframe while their inverses $\{e_a{}^\mu\}$ define a frame.} hence the matrix $(e_\mu{}^a)$ is nonsingular everywhere. These Vielbeine (locally) give the isomorphism between $TM$ and $aLM$, in that the components of a vector field $X\in \chi(M)$ can locally be mapped to the components of a section of $aLM$ by
\begin{equation}
X^\mu \mapsto X^a=e_\mu{}^a X^\mu\,.
\end{equation}
In physics, $\mu$ is called a curved index and $a$ a flat one. 

An affine connection $\Gamma$ induces a connection on $aLM$. In physics, this is shown by stating that the Vielbeine are covariantly constant,
\begin{equation}\label{defspinconnection}
\mathfrak{D}_\mu e_\nu{}^a=\partial_\mu e_\nu{}^a  -\omega_{\mu b}{}^a e_\nu{}^b-\Gamma_{\mu \nu}{}^\rho e_\rho{}^a=0\,\Rightarrow\, \omega_{\mu b}{}^a\equiv e_b{}^\nu \left( \partial_\mu e_\nu{}^a-\Gamma_{\mu \nu}{}^ \rho e_\rho{}^a\right)\,.
\end{equation}
The local expression for the corresponding curvature reads
\begin{equation}\label{Rspinconnection}
R_{\mu \nu a}{}^b=2\partial_{[\mu}\omega_{\nu]a}{}^b+2\omega_{[\mu| a}{}^c\omega_{|\nu]c}{}^b\,.
\end{equation}
\paragraph{Holonomy}
If $c$ is a closed loop going through $p\in M$, we can parallel transport the vector $X(p)\in T_pM$ along that loop using the connection $\mathfrak{D}$ on $TM$. The resulting vector $X'(p)\in T_pM$ might not coincide with $X(p)$, hence this procedure generates an action on $T_pM$. Parallel transporting $X(p)$ along every possible loop, the action on $T_pM$ defines a group, called the holonomy group $\Phi_{\mathfrak{D}}(p)$. If we restrict our attention to loops homothopic\footnote{This means contractible in a continuous way, see~\cite{McCarty}.} to a point, the corresponding group is called the restricted holonomy group $\Phi^0_{\mathfrak{D}}(p)$. 
It turns out that this group is generated by $R_{\mu \nu \rho}{}^\sigma(p)$, seen as a two-form. This is a consequence of the Ambrose-Singer theorem~\cite{Ambrose}.
\paragraph{G-structures}
Note that we still have not introduced a metric on the manifold $M$ in our discussion of connections on the tangent bundle. If we do so, there is a unique torsionless connection on $TM$ that leaves the metric invariant. It is called the Levi-Civita connection and its component expression reads
\begin{equation}
\Gamma_{\mu \nu}{}^\rho=\frac12g^{\rho \sigma}(2\partial_{(\mu}g_{\nu)\sigma}-\partial_\sigma g_{\mu \nu})\,,
\end{equation}
where $g_{\mu \nu}$ is the local expression for the metric and $g^{\mu\nu}$ for its inverse. The components $\Gamma_{\rho,\mu\nu}\equiv \Gamma_{\mu\nu}{}^\sigma g_{\rho \sigma}$ are called the first Christoffel connection coefficients. 

The structure group of the frame bundle can now be reduced to $\SO(t,n-t)$ (supposing that $M$ is orientable), where $t$ denotes the number of negative eigenvalues of the metric evaluated at any point, by restricting attention to so-called `admissible' frames $\{e_a{}^\mu\}$, for which
\begin{equation}
e_a{}^\mu e_b{}^\nu g_{\mu \nu}=\eta_{ab}\,, \qquad \det (e_a{}^\mu)=+1\,,
\end{equation}
where 
\begin{equation}
\eta=\mbox{diag}(\underbrace{-1,\dots,-1}_{\mbox{t}},\underbrace{1,\dots,1}_{\mbox{n-t}})\,.
\end{equation}
The condition~(\ref{defspinconnection}) with $\Gamma_{\mu \nu}{}^\rho$ being the Levi-Civita connection can now moreover be solved for the spin connection
\begin{equation}\label{expromega}
\omega_\mu{}^{ab}=2e^{\nu[a}\partial_{[\mu}e_{\nu]}{}^{b]}-e^{\nu[a}e^{b]\sigma}e_{\mu c}\partial_\nu e_\sigma^c\,.
\end{equation}
This connection now takes values in $\so(t,n-t)$. Moreover, due to the Ambrose-Singer theorem, the restricted holonomy group of the manifold is (contained in) $\SO(t,n-t)$. This reduction of the structure group of the frame bundle thus stems from the fact that we have introduced a new object (the metric) on the manifold. This process can be generalized using the theory of G-structures.\footnote{A review of this theory in the context of Riemannian geometry can e.g. be found in~\cite{Gurrieri:2002wz}.}

\begin{defi}
An $n$-dimensional manifold admits a G-structure if the structure group of the frame bundle can be reduced to $G\subset \Gl(n,\mathbb{R})$.
\end{defi}
An alternative definition of a G-structure is in terms of one or more G-invariant nowhere-vanishing, globally defined tensor fields $\xi$. A tensor field is called G-invariant if it is invariant under $G$ rotations of the frame. Since $\xi$ is globally defined (and nonvanishing), this amounts to a global reduction of the structure group of $LM$. Stated more technically, working in a certain chart $U_i$, we can introduce a frame $\{e_a{}^\mu\}$ and dual coframe $\{e_\mu{}^a\}$. We can now restrict attention to the frames in which the components of $\xi$ have some specific (fixed) form. The subgroup $G\subset \Gl(n,\mathbb R)$ that rotates these frames into each other defines the reduction of the structure group of the frame bundle.
Considering e.g. the above mentioned metric $g$, we have to look for frames for which $g_{\mu \nu}=e_\mu{}^ae_\nu{}^b\eta_{ab}$ and it is obvious that $\SO(t,n-t)$ rotations transform such frames into each other.

Conversely, suppose that the manifold $M$ admits a G-structure. In generic cases, we can then construct a G-invariant nowhere-vanishing tensor field $\xi$ that determines the G-structure in the following way. Looking at the branching rules for $\Gl(n,\mathbb R)\to G$, there will be a certain irreducible $p$-dimensional representation $\rho$ of $\Gl(n,\mathbb R)$ which contains a trivial representation of $G$ in its reduction
\begin{equation}
\Gl(n,\mathbb R)\to G\qquad:\qquad\mathbf{p}\to \mathbf{1}+\dots\,.
\end{equation}
When reducing the structure group, the associated bundle $E$ which equals $LM\times_\rho \mathbb R^p/\Gl(n,\mathbb R)$ will decompose into the direct sum of different bundles,\footnote{See~\cite{Nakahara} for a proper definition of direct sums of bundles and of subbundles.} following the above branching rule. Hence, $E$ will have a subbundle on which all transition functions\footnote{Note that these functions take values in $G$.} can be trivially represented. Hence this subbundle is trivial and it admits a nonzero section, which we call $\xi$. For instance, suppose that an $n$-dimensional manifold admits an $\SO(n)$ structure. Looking at the following branching rule,
\begin{equation}
\Gl(n,\mathbb R)\to \SO(n)\qquad:\qquad \mathbf{n^2}\to \mathbf{1}+\mathbf{\frac{n(n-1)}2}+(\mathbf{\frac{n(n+1)}2-1})\,,
\end{equation}
we see that there is one component of the adjoint representation of $\Gl(n,\mathbb R)$ that transforms trivially under $\SO(n)$. On the level of Lie-algebras, the generator of $\gl(n,\mathbb R)$ that commutes with $\so(n)$ is of course $\delta_{ab}$, which defines the invariant tensor field whose components read $g_{\mu\nu}=\delta_{ab}e_\mu{}^a e_\nu{}^b$ (if we restrict to admissible frames).

Suppose that a manifold admits a G-structure determined by the G-in\-va\-ri\-ant tensor field $\xi$. A G-connection on the frame bundle is a connection one-form that takes values in $\mathfrak g$. Under certain technical assumptions (which will always be valid in the cases we will encounter) such a connection does always exist~\cite{Kobayashi}. It follows that a connection preserves $\xi$, i.e.
\begin{equation}\label{parGstr}
\mathfrak D \xi=0\,,
\end{equation}
if and only if it is a G-connection. Hence, the existence of a G-structure with a connection satisfying~(\ref{parGstr}) implies that the holonomy of the connection is included in $G$. A G-structure is called 1-flat if it possible to find a torsionless G-connection.
\subsection{Complex manifolds}\label{sss:complexmanifolds}
Suppose that a $2n$-dimensional manifold M admits a globally defined $(1,1)$ tensor $J$ with local expression $J_\mu{}^\nu dx^\mu \otimes \partial_\nu$ which enjoys the following properties:
\begin{eqnarray}
J_\mu{}^\mu&=&0\,, \nonumber\\
J_\mu{}^\kappa J_{\kappa}{}^\nu&=&-\delta_\mu^\nu\,,\label{defJ}
\end{eqnarray}
then the tensor is called an almost complex structure and M is called an almost complex manifold.
Since this tensor is nowhere-vanishing, it defines a G-structure. To see this, consider a $2n$ real dimensional vector space $V=\mathbb R^{2n}$ equipped with an endomorphism $J_0$, also called an almost complex structure,  satisfying~(\ref{defJ}). The subgroup of automorphisms of $V$ leaving $J_0$ invariant is $\Gl(n,\mathbb{C})$. This means that we can identify $V$ with $\mathbb{C}^n$, as we can perform a $\Gl(2n,\mathbb R)$ transformation on $J_0$ bringing it in the form 
\begin{equation}
J_0{}_i{}^j=\rmi \delta_i^j\, , \; \;J_0{}_{\bar{i}}{}^{\bar{j}}=-\rmi \delta_{\bar{i}}^{\bar{j}}\, , \; \;J_0{}_{\bar{i}}{}^{j}=J_0{}_{i}{}^{\bar{j}}=0\,,
\end{equation}
where $i,\bar i=1,\dots n$. This makes clear why $J_0$ has to be traceless, since it would otherwise have an unequal number of eigenvalues $\pm \rmi$ and this would of course make the identification of $V$ with $\mathbb C^n$ impossible. A field of such almost complex structures $J_0$ is thus called an almost complex structure on the manifold $M$ and this implies that the structure group of the frame bundle can be reduced to $\Gl(n,\mathbb{C})$ for a $2n$ (real-)dimensional manifold, by restricting attention to frames $\{e_\mu{}^i\,,\; e_\mu{}^{\bar{i}}\}$ for which
\begin{equation}
J_\mu{}^\nu=J_{0j}{}^i e_i{}^\mu e_\nu{}^j+J_{0\bar j}{}^{\bar i} e_{\bar i}{}^\mu e_\nu{}^{\bar j} \, .
\end{equation}
Using the almost complex structure, we can define a mixed three-tensor, called the Nijenhuis tensor $N$, with components 
\begin{equation}
N_{\mu \nu}{}^\rho=\frac16 J_\mu{}^\sigma \partial_{[\sigma}J_{\nu ]}{}^\rho-(\mu \leftrightarrow \nu)\,.
\end{equation}
It can be proven that the Nijenhuis tensor vanishes identically if and only if the almost complex structure is a complex structure (see e.g.~\cite{Candelas:1987is}). The existence of the latter structure means that it is possible to find a holomorphic atlas on $M$, i.e. in every chart, coordinates $\{z^m,\bar z^{\bar m}\}$ with $m,\bar m=1,\dots,n$ exist for which
\begin{equation}\label{defcoords}
J_m{}^n=\rmi \delta_m^n\, , \; \;J_{\bar{m}}{}^{\bar{n}}=-\rmi \delta_{\bar{m}}^{\bar{n}}\, , \; \;J_{\bar{m}}{}^{n}=J_{m}{}^{\bar{n}}=0\,.   
\end{equation}
and moreover, the transition functions are holomorphic. Hence, the coordinates take values in $\mathbb C^n$.
\begin{defi}
An almost complex structure for which the Nijenhuis tensor vanishes, is called a complex structure and the corresponding manifold a complex manifold. 
\end{defi}
Note that the condition for a complex manifold can be reformulated in terms of G-structures as well~\cite{Newlander}, since an almost complex structure is complex if and only if it is 1-flat.

If an almost complex manifold is Riemannian and the metric satisfies 
\begin{equation}\label{hermmetric}
J_\mu{}^\rho J_\nu{}^\sigma g_{\rho \sigma}=g_{\mu \nu}\,,
\end{equation}
the metric is called almost hermitian. This condition is equivalent to $J_{\mu \nu}=J_\mu{}^\rho g_{\nu \rho}$ being antisymmetric, and $J_{\mu \nu}$ is then called the fundamental two-form. The couple $(J,g)$ again defines a G-structure (called an almost-hermitian structure), and the structure group of $LM$ can now be reduced further to $\U(n)$.
An almost hermitian manifold is called hermitian if the Nijenhuis tensor vanishes and there exists a (possibly) torsionful connection that preserves both the complex structure and the metric.
An important class of hermitian manifolds are K\"ahler manifolds. 
\begin{defi}
A hermitian manifold is called K\"ahler if the fundamental two-form is closed, i.e. $\partial_{[\mu} J_{\nu \rho]}=0$.
\end{defi}
This definition is equivalent with the statement that the almost hermitian structure is 1-flat~\cite{Candelas:1987is}. Note that for a K\"ahler manifold, the corresponding connection on the tangent bundle is the Levi-Civita connection.
\subsection{Quaternionic-like manifolds}\label{sss:quatlikestruct}
We will now introduce quaternionic-like manifolds, since they play an important role as allowed target spaces in $\mathcal{N}=2$ supersymmetric theories.
A very thorough reference on the subject is~\cite{AM1996}.
\subsubsection{Quaternionic-like structures}
Suppose that $V=\mathbb{R}^{4n}$. A triple $H=(J^1,J^2,J^3)$ of complex structures with
\begin{equation}\label{defhcstr}
J^\alpha J^\beta=-\unity_{4n}\delta^{\alpha \beta}+\varepsilon^{\alpha \beta \gamma}J^\gamma\,,
\end{equation}
is called a hypercomplex structure on $V$. Denote the space of endomorphisms of $V$ by End $V$. The three-dimensional subspace $Q$ of End $V$, defined by
\begin{equation}
Q=\mathbb{R} J^1+\mathbb{R} J^2+\mathbb{R} J^2\,,
\end{equation}
is called a quaternionic structure, i.e. $Q$ is the set of real linear combinations of the complex structures. A triple $H$ is called an admissible base of $Q$. Any two admissible bases are connected by an $\Sp(1)$ transformation. If we introduce a metric $g$ on $V$, satisfying~(\ref{hermmetric}) for every $J^\alpha$, then the pair $(H,g)$ is called a hypercomplex hermitian structure while $(Q,g)$ is called quaternionic hermitian. From now on, $\mathcal{S}$ will denote any of the four structures introduced ($H$, $Q$, $(H,g)$ or $(Q,g)$) and will be called a quaternionic-like structure. The subset of automorphisms of $V$ that leave $\mathcal{S}$ invariant are denoted in Table~\ref{autquatlikestruct} and obviously $\mbox{Aut}(\mathcal{S}) \subset \Gl(4n,\mathbb{R})$.
\begin{table}[tb]
\begin{center}
\begin{tabular}{|c||c|c|c|c|}
\hline
$\mathcal{S}$&$H$&$Q$&$(H,g)$&$(Q,g)$\\
\hline
$\mbox{Aut}(\mathcal{S})$&$\Gl(n,\mathbb{H})$&$\Sp(1)\cdot \Gl(n,\mathbb{H})$&$\Sp(n)$&$\Sp(1)\cdot \Sp(n)$\\
\hline
\end{tabular}
\end{center}
\caption{\it Automorphism groups Aut($\mathcal S$) preserving the different quaternionic-like structures $\mathcal S$.}\label{autquatlikestruct}
\end{table}  
Note that the $A\cdot B$ in the Table means $(A \times B)/\mathbb{Z}_2$.\footnote{This $\mathbb Z_2$ acts as follows. We let $A$ act from the left and $B$ from the right on $Q$. Hence, multiplying both $A$ and $B$ with $-1$ does not change anything.}

In the context of a complex structure on a vector space, we have pointed out that the tracelessness condition stemmed from the fact that we wanted to be able to introduce complex coordinates. In the case at hand, the trace of $J^\alpha$ vanishes as the hypercomplex structure forms a matrix representation of $\symp(1)$. 
\subsubsection{Connections}
Consider now a $4n$-dimensional manifold $M$ with a field of an $\mathcal{S}$-structure, which we will also denote by $\mathcal{S}$. A manifold with such a G-structure is called almost quaternionic-like and the structure group of the tangent bundle can be reduced to the corresponding groups given in Table~\ref{autquatlikestruct}. As explained in~\cite{AM1996}, an almost quaternionic-like structure is said to define a quaternionic-like structure if it is 1-flat. The holonomy of the corresponding connection is again included in the groups given in Table~\ref{autquatlikestruct}. The index $\alpha$ labelling the $\symp(1)$ index on the hypercomplex structure will not be written any more, and we will adopt vector notation, see Appendix~\ref{appconventions}. 
\paragraph{Hypercomplex manifolds}
Given a manifold with an almost hypercomplex structure, there always exists a unique (possibly torsionful) connection preserving it.
\begin{equation}
\mathfrak{D}_\mu \vec J_\nu{}^\rho=\partial_\mu\vec J_\nu{}^\rho-\Gamma_{\mu\nu}{}^\tau \vec J_\tau{}^\rho+\Gamma_{\mu\tau}{}^\rho \vec J_\nu{}^\tau=0\,.
\end{equation}
If the torsion vanishes (i.e. if the hypercomplex structure is 1-flat), the manifold is called hypercomplex and this is equivalent with the vanishing of the diagonal Nijenhuis tensor defined as 
\begin{equation}\label{Nijenhuisdiag}
N^{\rm d}{}_{\mu \nu}{}^\rho=\frac16 \vec J_\mu{}^\sigma \cdot \partial_{[\sigma}\vec J_{\nu ]}{}^\rho-(\mu \leftrightarrow \nu)\,.
\end{equation} 
In that case, the torsionless affine connection on $TM$ is called the Obata connection~\cite{Obata} and its components are given by the following expression.
\begin{equation}\label{defobata}
\Gamma^{\rm Ob}{}_{\mu\nu}{}^\rho=-\frac16(2\partial_{(\mu}\vec J_{\nu)}{}^ \upsilon+\vec J_{(\mu|}{}^\sigma\times\partial_{\sigma}\vec J_{|\nu)}{}^\upsilon)\cdot \vec J_\upsilon{}^\rho\,.
\end{equation}
More generally, given an almost hypercomplex structure, the unique connection preserving it is given by
\begin{equation}
\Gamma_{\mu \nu}{}^\rho=\Gamma^{\rm Ob}{}_{\mu \nu}{}^\rho+N^{\rm d}{}_{\mu\nu}{}^\rho\,.
\end{equation}
\paragraph{Quaternionic manifolds}
On an almost quaternionic manifold, there again always exist (possibly torsionful) connections preserving the almost quaternionic structure, 
\begin{equation}\label{intJquat}
\mathfrak{D}_\mu \vec J_\nu{}^\rho=\partial_\mu\vec J_\nu{}^\rho-\Gamma_{\mu\nu}{}^\tau \vec J_\tau{}^\rho+\Gamma_{\mu\tau}{}^\rho \vec J_\tau{}^\rho+2\vec \omega_\mu \times \vec J_\nu{}^\rho=0\,.
\end{equation}
The final term in~(\ref{intJquat}) is an $\mathfrak{sp}(1)$ connection on the nontrivial $\Sp(1)$ bundle of admissible bases, which we will elaborate on shortly. Given a field of almost quaternionic structures, the 1-flatness, i.e. the vanishing of the torsion, now boils down to the following condition on the diagonal Nijenhuis tensor,
\begin{equation}\label{condNquat}
(1-2n)N^{\rm d}{}_{\mu\nu}{}^\rho=-\vec J_\upsilon{}^\tau\cdot\vec J_{[\mu}{}^\rho N^{\rm d}{}_{\nu]\tau}{}^\upsilon\,.
\end{equation}
In that case, the manifold is called quaternionic. The Nijenhuis tensor satisfying this condition, can be used to define the so-called Oproiu connection, which preserves the quaternionic structure, as
\begin{equation}\label{defOproiu}
\vec \omega^{\rm Op}{}_\mu=\frac1{1-2n}N^{\rm d}{}_{\mu\nu}{}^\rho\vec J_\rho{}^\nu\,,\qquad\Gamma^{\rm Op}{}_{\mu\nu}{}^\rho =\Gamma^{\rm Ob}{}_{\mu\nu}{}^\rho-\vec J_{(\mu}{}^\rho\cdot \vec \omega_{\nu)}\,.
\end{equation}
One can show that this $\symp(1)$ connection satisfies $\vec J_\mu{}^\nu\cdot \vec \omega_\nu=0$
using the following property of the diagonal Nijenhuis tensor,
\begin{equation}
N^{\rm d}{}_{\mu \nu}{}^\rho=\vec J_\mu{}^\sigma \cdot \vec J_\tau{}^\rho N^{\rm d}{}_{\sigma \nu}{}^\tau\,.
\end{equation}
Note that the torsionless connections on a quaternionic manifold are not unique, since it is easy to check that~(\ref{intJquat}) is left invariant by the following transformation of the connections.
\begin{eqnarray}
\Gamma_{\mu\nu}{}^\rho&\to& \Gamma_{\mu\nu}{}^\rho+S_{\mu\nu}^{\tau\rho}\xi_\tau  \qquad {\rm with}\qquad S_{\mu\nu}^{\tau \rho}=2(\delta_{(\mu}^\tau\delta_{\nu)}^\rho-\vec J_{(\mu}{}^\tau \cdot \vec J_{\nu)}{}^\rho)\,,\nonumber\\
\vec \omega_\mu &\to& \vec \omega_\mu+\vec J_\mu{}^\nu\xi_\nu\label{xitf}\,,
\end{eqnarray}
and with $\xi_\mu$ the components of an arbitrary one-form.
\paragraph{Hyperk\"ahler manifolds}
For hyperk\"ahler manifolds (manifolds with a 1-flat field of hypercomplex hermitian structures), the Levi-Civita connection coincides with the Obata connection. 
\paragraph{Quaternionic-K\"ahler manifolds}
For a quaternionic-K\"ahler manifold (a manifold with a 1-flat field of quaternionic hermitian structures), the Levi-Civita connection selects one of the possible torsionless connections from the family of connections of~(\ref{xitf}).
\paragraph{$\symp(1)$ connection}
To make the role of the $\mathfrak{sp}(1)$ connection introduced on quaternionic(-K\"ahler) manifolds\footnote{From now on, we will only consider 1-flat quaternionic-like structures and their corresponding connections.} more clear, we will define Vielbeine on the manifold, which give the isomorphism between $TM$ and the associated vector bundle $aLM$ over $M$ with typical fibre $\mathbb{R}^{4n}$ and structure group given by the corresponding automorphism group in Table~\ref{autquatlikestruct}. The two possible factors in these groups are reflected in the index structure of the Vielbeine $f_\mu^{\mathfrak i \mathfrak a}$ where $\mu=1, \dots 4n$, $\mathfrak i=1,2$ and $\mathfrak a=1,\dots,2n$. To define the reality of these Vielbeine, we have to introduce two more matrices $\rho_{\mathfrak a}{}^{\mathfrak b}$ and $E_{\mathfrak i}{}^{\mathfrak j}$ that satisfy
\begin{equation}\label{defrho}
\rho \rho^*=-\unity_{2n}\,,\;\; EE^*=-\unity_2\,,
\end{equation}
where $*$ denotes complex conjugation.
Now, complex conjugation is realized on the Vielbeine in the following way:
\begin{equation}\label{complconjViel}
(f_\mu^{\mathfrak i \mathfrak a})^*=E_{\mathfrak j}{}^{\mathfrak i}\rho_{\mathfrak b}{}^{\mathfrak a}f_\mu^{\mathfrak j\mathfrak b}\,.
\end{equation}
Since the Vielbeine give an isomorphism between two bundles, they should be invertible. Hence, the following identities hold
\begin{equation}
f_\mu^{\mathfrak i\mathfrak a}f_{\mathfrak i\mathfrak a}^\nu =\delta_\mu^\nu\,,\;\;f_\mu^{\mathfrak i\mathfrak a}f^\mu_{\mathfrak j\mathfrak b}=\delta^{\mathfrak i}_{\mathfrak j}\delta^{\mathfrak a}_{\mathfrak b}\,.
\end{equation}
On $aLM$ we can introduce connections induced by the affine connection $\Gamma_{\mu\nu}{}^\rho$ by requiring the Vielbeine to be covariantly constant, as in~(\ref{defspinconnection}),
\begin{equation}\label{covcteViel}
\mathfrak{D}_\mu f_\nu^{\mathfrak i\mathfrak a}=\partial_\mu f_\nu^{\mathfrak i\mathfrak a}-\Gamma_{\mu\nu}{}^\rho f_\rho^{\mathfrak i\mathfrak a}+\omega_{\mu \mathfrak j}{}^{\mathfrak i}f_\nu^{\mathfrak j\mathfrak a}+\omega_{\mu \mathfrak b}{}^{\mathfrak a}f_\nu^{\mathfrak i\mathfrak b}=0\,.
\end{equation}
From~(\ref{covcteViel}) we can see that the $\mathfrak{sp}(1)$ connection is part of the connection on the associated bundle. Therefore, if the reduced structure group of $LM$ of the quaternionic-like manifold does not admit an $\Sp(1)$ factor (i.e. for hypercomplex or hermitian hypercomplex structures), the $\mathfrak{sp}(1)$ connection is trivial,\footnote{as it admits global nonvanishing sections (the complex structures)} and we will work in a gauge (i.e. use local trivializations) for which it vanishes, unless stated otherwise. The connection $\omega_{\mu \mathfrak a}{}^{\mathfrak b}$ introduced in~(\ref{covcteViel}), takes values in the Lie-algebra of the other factor of the structure group. Finally, note that we can express the hypercomplex structure in terms of the Vielbeine as
\begin{equation}\label{defJf}
\vec J_\mu{}^\nu=-\rmi f_\mu^{\mathfrak i \mathfrak a}\vec \sigma_{\mathfrak i}{}^{\mathfrak j} f_{\mathfrak j \mathfrak a}^\nu\,.
\end{equation}
\subsubsection{Curvatures}
For every connection introduced in the previous Paragraph, we can of course compute its curvature
\begin{eqnarray}
 R_{\mu\nu\rho}{}^\upsilon &\equiv & 2
\partial_{[\mu}\Gamma_{\nu]\rho}{}^\upsilon + 2 \Gamma_{\tau[\mu}{}^\upsilon \Gamma_{\nu]\rho}{}^\tau
\,,\nonumber\\
R_{\mu\nu \mathfrak b}{}^{\mathfrak a} &\equiv & 2\partial_{[\mu} \omega_{\nu]\mathfrak b}{}^{\mathfrak a}
+ 2\omega_{[\mu|\mathfrak c|}{}^{\mathfrak a} \omega_{\nu] \mathfrak b}{}^{\mathfrak c}\,, \nonumber\\
 \vec  R_{\mu\nu}  &\equiv & 2\partial_{[\mu}\vec \omega _{\nu]}
+2\vec \omega _\mu\times \vec \omega _\nu\,.\label{defcurvs}
\end{eqnarray}
There exist two different ways of splitting the Riemannian curvature on qua\-ter\-ni\-o\-nic-like manifolds. The first one stems from the integrability condition of~(\ref{covcteViel}). For all quaternionic-like manifolds, we can write
\begin{eqnarray}\label{RdecompJ}
R_{\mu\nu\upsilon}{}^\rho&=&{R^{\mathfrak{sp}(1)}{}_{\mu\nu\upsilon}{}^\rho}+
R^{\mathfrak{gl}(n,\mathbb{H})}{}_{\mu\nu\upsilon}{}^\rho  \\
&=&-\vec J_\upsilon{}^\rho\cdot\vec R_{\mu\nu} +
L_\upsilon{}^\rho{}_{\mathfrak a}{}^{\mathfrak b}R_{\mu\nu \mathfrak b}{}^{\mathfrak a} \,,\qquad \mbox{with}\qquad\; \;
L_\upsilon{}^\rho{}_{\mathfrak a}{}^{\mathfrak b}\equiv f^\rho_{\mathfrak i\mathfrak a}f_\upsilon^{\mathfrak i\mathfrak b} \,.\nonumber
\end{eqnarray}
Hence, the curvature is split in a part generating the $\symp(1)$ term of the restricted holonomy algebra (which is of course absent in the hypercomplex and hyperk\"ahler case), while the other part is generating $\mathfrak{gl}(n,\mathbb H)$ or $\symp(n)$. From~(\ref{RdecompJ}) we can isolate the different curvatures in the following way.
\begin{equation}
\vec R_{\mu\nu}=\frac1{4n} R_{\mu\nu\rho}{}^\upsilon\vec J_\upsilon{}^\rho\,,\qquad
R_{\mu\nu \mathfrak a}{}^{\mathfrak b}=\frac12 R_{\mu\nu\rho}{}^\upsilon L_\upsilon{}^\rho{}_{\mathfrak a}{}^{\mathfrak b}\,.
\end{equation}
Furthermore, the Ricci tensor $R_{\mu\nu}\equiv R_{\rho\mu\nu}{}^\rho$ might have both a symmetric and
antisymmetric part. This antisymmetric part can be traced back
to the $\mathfrak{u}(1)$ part in
$\gl(n,\mathbb{H})=\spl(n,\mathbb{H})+ \un(1)$. Indeed, using the first
Bianchi identity, we find
\begin{equation}\label{defu1}
  \Ric_{[\mu\nu]}=R_{\rho[\mu\nu]}{}^\rho=-\frac12R_{\mu\nu\rho}{}^\rho=-R_{\mu\nu \mathfrak a}{}^{\mathfrak a}\equiv -R^{\un(1)}{}_{\mu\nu}\,.
\end{equation}
Therefore, the antisymmetric part of the Ricci tensor follows completely
from this $\un(1)$ part. Moreover, the antisymmetric part of the Ricci-tensor is a closed two-form.
\begin{equation}\label{Ricciclosed}
\partial_{[\mu}R_{\nu \rho]}=\mathfrak{D}_{[\mu}R_{|\sigma|\nu \rho]}{}^\sigma=-\frac12\mathfrak{D}_{[\mu}R_{\nu \rho]\sigma}{}^\sigma=0\,,
\end{equation}
where we have used the second Bianchi identity in the last step.

Note that the separate curvature terms in the first line of~(\ref{RdecompJ}) do not
satisfy the first Bianchi identity (unless the $\symp(1)$ curvature vanishes). We can however consider another splitting of the full curvature
where both terms separately satisfy this cyclicity property,
\begin{equation}\label{RRBRW}
  R_{\mu\nu\upsilon}{}^\rho=R^{\rm Ric}{}_{\mu\nu\upsilon}{}^\rho+R^{(\rm W)}{}_{\mu\nu\upsilon}{}^\rho\,.
\end{equation}
The first part only depends on the Ricci tensor of the full curvature,
and is called the Ricci part. It is defined by~\cite{Musso:1992}
\begin{eqnarray}
  R^{\rm Ric}{}_{\mu\nu\rho}{}^\upsilon&\equiv &\delta _{[\mu}{}^\upsilon B_{\nu]\rho}
  -\delta _\rho{}^\upsilon B_{[\mu\nu]}
  -2\vec J_\rho{}^{(\upsilon}\cdot \vec J_{[\mu}{}^{\tau)}B_{\nu]\tau}\,,\label{defRB}\\
 B_{\mu\nu}&\equiv& \frac{2n+3}{8n(n+2)}R_{(\mu\nu)}-\frac{1}{8n(n+2)}\vec
 J_{(\mu}{}^\rho\cdot \vec J_{\nu)}{}^\upsilon R_{\rho\upsilon}+\frac{1}{4(n+1)}R_{[\mu\nu]}\,.\nonumber
\end{eqnarray}
We can further split the Ricci part as
\begin{equation}
  R^{\rm Ric}{}_{\mu\nu\rho}{}^\upsilon=
  \left( R^{\rm Ric}_{\rm symm}+R^{\rm Ric}_{\rm antis}\right)
  _{\mu\nu\rho}{}^\upsilon\,,
\end{equation}
where the first term is the construction~(\ref{defRB}) using only the
symmetric part of $B$ (the symmetric part of the Ricci tensor), and the
second term uses only the antisymmetric part of $B$ (of the Ricci
tensor).

The Ricci part now does satisfy the cyclicity property, and its Ricci
tensor is just $R_{\mu\nu}$. The second term in~(\ref{RRBRW}) is defined as
the remainder, and its Ricci tensor therefore equals zero. For this reason, it is
called the Weyl part~\cite{AM1996}. Following the
discussion in~\cite{Bergshoeff:2002qk}, we can rewrite the Weyl part in
terms of a symmetric and traceless tensor $\mathcal{W}$, such
that\footnote{The case of four-dimensional quaternionic manifolds must
be treated separately, but~(\ref{RdecompRBWA}) is
still satisfied. See~\cite{Bergshoeff:2002qk} for more details.}
\begin{equation}
  R_{\mu\nu\rho}{}^\upsilon=R^{\rm Ric}{}_{\mu\nu\rho}{}^\upsilon - \frac 12 f^{\mathfrak i\mathfrak a}_\mu
\varepsilon_{\mathfrak i\mathfrak j}f^{\mathfrak j\mathfrak b}_\nu f_\rho^{\mathfrak l\mathfrak c}f^\upsilon_{\mathfrak l\mathfrak d}
 \mathcal{W}_{\mathfrak a\mathfrak b\mathfrak c}{}^{\mathfrak d}\,,
 \label{RdecompRBWA}
\end{equation}
or conversely,
\begin{equation}
\mathcal{W}_{\mathfrak c\mathfrak d\mathfrak b}{}^{\mathfrak a} \equiv \frac12 \varepsilon ^{\mathfrak i\mathfrak j} f^\mu_{\mathfrak j\mathfrak c} f^\nu_{\mathfrak i\mathfrak d}
f_{\mathfrak l\mathfrak b}^\rho f_\upsilon^{\mathfrak l\mathfrak a} {R^{(\rm W)}}_{\mu\nu\rho}{}^\upsilon\,.\label{WfromRW}
\end{equation}
Similarly, the $\gl(r,\mathbb{H})$ curvature can also be decomposed in its Ricci and
Weyl part while the $\symp(1)$ curvature is determined only by the Ricci
tensor:
\begin{eqnarray}
 R_{\mu\nu \mathfrak a}{}^{\mathfrak b} & = & R^{\rm Ric}{}_{\mu\nu \mathfrak a}{}^{\mathfrak b}+ R^{({\rm W})}{}_{\mu\nu \mathfrak a}{}^{\mathfrak b}\,,\nonumber\\
  R^{\rm Ric}{}_{\mu\nu \mathfrak a}{}^{\mathfrak b}&\equiv&\frac12 L_\upsilon{}^\rho{}_{\mathfrak a}{}^{\mathfrak b} R^{\rm
  Ric}{}_{\mu\nu\rho}{}^\upsilon= 2\delta_{\mathfrak a}{}^{\mathfrak b} B_{[\nu\mu]}+4L_{[\mu}{}^\tau{}_{\mathfrak a}{}^{\mathfrak b} B_{\nu]\tau}\,,
  \nonumber\\
  R^{({\rm W})}{}_{\mu\nu \mathfrak a}{}^{\mathfrak b}&\equiv&\frac12
  L_\upsilon{}^\rho{}_{\mathfrak a}{}^{\mathfrak b}R^{({\rm W})}{}_{\mu\nu\rho}{}^\upsilon=-f_\mu^{\mathfrak i\mathfrak c}\varepsilon
  _{\mathfrak i\mathfrak j}f_\nu^{\mathfrak j\mathfrak d}\mathcal{W}_{\mathfrak c\mathfrak d\mathfrak a}{}^{\mathfrak b}\,,\nonumber\\
  \vec {R}_{\mu\nu} &=& 2\vec J_{[\mu}{}^\rho B_{\nu]\rho}\,.\label{R=JB}
\end{eqnarray}

We can now summarize the different curvature decompositions in the
following scheme:
\begin{equation}
  \begin{array}{rccccc}
  R_{\mu\nu\rho}{}^\upsilon= & \big( R^{\rm Ric}_{\rm symm}  & + & R^{\rm Ric}_{\rm antis}  & + &
  R^{({\rm W})}\big)_{\mu\nu\rho}{}^\upsilon
  \\
  &\multicolumn{3}{c}{\phantom{.}\hspace{5mm}\leavevmode \epsfxsize=65mm
 \epsfbox{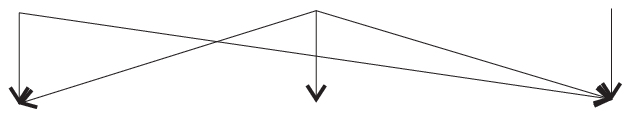}\hspace{-18mm}} \\
  = & \big( R^{\symp(1)} & + & R^{\un(1)} & + &
R^{\spl(n,\mathbb{H})}\big)_{\mu\nu\rho}{}^\upsilon\,.
\end{array}
\end{equation}
The terms in the second line depend only on specific terms of the first
line as indicated by the arrows. This is the general scheme and thus
applicable for quaternionic manifolds. For
specific other quaternionic-like manifolds some parts are absent as can
be seen from Table~\ref{tbl:CurvQuatlikeMan}.
\begin{table}[tb]
\begin{center}
  \begin{tabular}{|c|c|}\hline
   \textit{\textbf{hypercomplex}} & \textit{\textbf{hyperk{\"a}hler}}  \\
    $R^{\rm Ric}_{\rm antis}+ R^{({\rm W})}$
      & $R^{({\rm W})}$ \\
    $R^{\un(1)} + R^{\spl(n,\mathbb{H})}$
      & $ R^{\spl(n,\mathbb{H})}$ \\
 \hline
    \textit{\textbf{quaternionic}} & \textit{\textbf{quaternionic-K{\"a}hler}} \\
    $R^{\rm Ric}_{\rm symm}+R^{\rm Ric}_{\rm antis}+ R^{({\rm W})}$
      & $R^{\rm Ric}_{\rm symm}+ R^{({\rm W})}$ \\
    $R^{\symp(1)} + R^{\un(1)} + R^{\spl(n,\mathbb{H})}$
      & $R^{\symp(1)} + R^{\spl(n,\mathbb{H})}$ \\
\hline
  \end{tabular}
  \caption{\it The curvatures in quaternionic-like manifolds.}\label{tbl:CurvQuatlikeMan}
\end{center}
\end{table}

We end the discussion on the curvatures by noting that the $\xi$ transformation~(\ref{xitf}) on quaternionic manifolds does not leave the curvatures invariant, hence the holonomy of the manifold might change. As a consequence, we do not consider this transformation as a symmetry of the manifold, but merely as a transformation relating different manifolds. It can for instance be shown that in every class of quaternionic manifolds that can be related via a $\xi$ transform, there is at least one space with vanishing $\un(1)$ part of the curvature~(\ref{transfxiu1}). For completeness, we give the $\xi$ transformation of the $\su(2)$ curvature, the Riemannian curvature and the Ricci tensor.
\begin{eqnarray}
\Gamma _{\mu\nu}{}^\rho&=&\tilde \Gamma_{\mu \nu}{}^\rho+S_{\mu\nu}{}^{\rho\sigma}\xi _\sigma\,,\qquad
   \vec \omega _\mu=\tilde {\vec\omega} _\mu+\vec J_\mu{}^\nu\xi _\nu\,,\nonumber\\
\vec R(\vec \omega )_{\mu\nu} & = &  R(\tilde {\vec\omega})_{\mu\nu} +2\vec J_{[\nu}{}^\rho \mathfrak D_{\mu]}\xi
 _\rho +\vec J_{[\nu}{}^\rho S_{\mu]\rho}{}^{\sigma \tau} \xi _\sigma \xi _\tau \,,\nonumber\\ 
R(\Gamma )_{\mu\nu\rho}{}^\sigma & = &R(\tilde \Gamma )_{\mu\nu\rho}{}^\sigma+2S_{\rho[\mu}{}^{\sigma \tau}
  \tilde {\mathfrak D}_{\mu]}\xi _\tau+2S_{\upsilon[\mu}{}^{\sigma\tau}S_{\nu]\sigma}{}^{\upsilon \phi} \xi _\tau\xi _\phi\,, \nonumber\\
R(\Gamma )_{\mu\nu} & = &R(\tilde \Gamma )_{\mu\nu}-4n \mathfrak D_{(\mu}\xi _{\nu)}
 -8\Pi _{(\mu\nu)}{}^{\rho\sigma} \mathfrak D_\rho\xi _\sigma - 4(n+1)\partial _{[\mu}\xi _{\nu]}\nonumber\\
 && -8n\xi_\mu\xi _\nu+16(n+1)\Pi _{(\mu\nu)}{}^{\rho\sigma}\xi _\rho\xi _\sigma\label{transcurvxi}\,,
\end{eqnarray} 
with $S$ as defined in~(\ref{xitf}),
\begin{equation}\label{defpi}
\Pi_{\mu\nu}{}^{\rho\sigma}=\frac14 (\delta_\mu^\rho \delta_\nu^\sigma+\vec J_\mu{}^\rho\cdot \vec J_\nu{}^\sigma)\,,
\end{equation}
and $\mathfrak D$ is a covariant derivative with respect to $\Gamma$ while $\tilde{\mathfrak D}$ is the one with respect to $\tilde \Gamma$.
\subsection{Spin manifolds}\label{spinnen}
Consider an $n$-dimensional Riemannian manifold $M$. The structure group of the tangent bundle $TM$ is then contained in $\SO(n)$. Let $LM$ be the frame bundle. The transition functions of the latter obey the consistency conditions~(\ref{consist}). A spin structure on $M$ is defined by transition functions $\tilde t_{ij}$ which obey~(\ref{consist}) and moreover
\begin{equation}
\mathcal{H}(\tilde t_{ij})=t_{ij}\,,
\end{equation}
where $\mathcal{H}$ is the double covering $\Spin(n)\to\SO(n)$ (see Appendix~\ref{spinors}). A manifold admitting a spin structure is called a spin manifold.

Thus, if a manifold is spin, it is possible to associate a spinor bundle defined via the transition functions $\tilde t_{ij}$ to the frame bundle. The local connection one-form on this spinor bundle is
\begin{equation}
\frac14\slashed{\omega}_\mu=\frac14 \omega_\mu{}^{ab}\gamma_{ab}\,,
\end{equation}
where $\gamma_a$ are Clifford matrices and where $\omega$ is the spin connection introduced in~(\ref{defspinconnection}). A spinor field is a section of that spinor bundle.

On the spinor bundle, we can again try to measure the change of a section along the flow of a vector field via the Lie-derivative. If $X$ is a Killing vector field on a Riemannian spin manifold $M$, we can introduce the Lichnerowicz-Lie-derivative along $X$ at $p\in M$ of a spinor $\epsilon$ defined in at least the neighbourhood of $p$. The component expression reads~\cite{Lichenrowicz,figueroa2001}
\begin{equation}\label{defliespinor}
\mathcal{L}_X \epsilon(p)=X^\mu (p)\mathfrak{D}_\mu \epsilon(p)+\frac14 \partial_\mu X_\nu (p)\gamma^{\mu \nu}\epsilon(p)\,.
\end{equation} 
Note that from now on, we will not explicitly write the dependence on $p$.
To show the necessity of $X$ being Killing, we will prove the property~(\ref{proplieder}). First of all, we can rewrite~(\ref{defliespinor}) as
\begin{equation}\label{rewrdefliespinor}
\mathcal{L}_X \epsilon=X^\mu \partial_\mu\epsilon+\frac14X^\mu\slashed{\omega}_\mu\epsilon +\frac14 e_{\mu b}\mathfrak{D}_a X^\mu \gamma^{ab}\epsilon\,,
\end{equation}
where $\{e_\mu{}^a\}$ are Vielbeine.
Suppose now that $X(p)\neq 0$ such that we can change to coordinates adapted to $X$, i.e. $X=\partial_z$. Since $X$ is Killing, this implies that the metric does not depend on $z$. As a consequence, the $z$ dependence of the Vielbeine reads
\begin{equation}
\partial_z e_\mu{}^a=\Lambda_b{}^a e_\mu{}^b\,,
\end{equation}
where $\Lambda \in \SO(n)$ for an $n$-dimensional manifold. However, we can always choose a gauge for which the Vielbeine are $z$ independent.\footnote{This can easily be seen by defining new Vielbeine $e'{}^a=(\e^{-\Lambda z}e)^a$.} Writing~(\ref{defspinconnection}) in this gauge, 
\begin{equation}
\mathfrak{D}_z e_\mu{}^a\equiv \partial_z e_\mu{}^a-\Gamma_{z\mu}{}^\nu e_\nu{}^a-\omega_z{}^{ab}e_{\mu b}=-\Gamma_{z\mu}{}^\nu e_\nu{}^a-\omega_z{}^{ab}e_{\mu b}=0\,,
\end{equation}
implies that the last two terms in~(\ref{rewrdefliespinor}) cancel. Hence, in these coordinates, we find~(\ref{proplieder}) which we wanted to prove.

The definition~(\ref{defliespinor}) is actually also valid in the case X is a homothetic Killing vector~(\ref{homkilvect}), as was explained in~\cite{Rindler}. In that case, the $z$ dependence of the metric (again in coordinates adapted to $X$) reads
\begin{equation}
g_{\mu \nu}=e^{2mz}h_{\mu \nu}\,,
\end{equation}
where $h$ is a metric that does not depend on $z$ any more. Similarly to the previous case, we can choose a gauge for the Vielbeine such that $\partial_z e_\mu{}^a=m e_\mu{}^a$. Then, $\mathfrak{D}_z e_\mu{}^a=0$ leads in the same way to the fact that the last two terms in~(\ref{rewrdefliespinor}) cancel, again proving~(\ref{proplieder}). 
\subsection{Berger's list}\label{berger}
It is possible to classify all possible holonomy groups of the Levi-Civita connection $\mathfrak D$ on a Riemannian manifold, which amounts to the famous list of Berger~\cite{Berger}.\footnote{A nice review can be found in~\cite{Cleyton}.} Let therefore $M$ be a simply connected,\footnote{This means that every loop is homothopic to a point} orientable manifold of dimension $n$ and let $g$ be a Euclidean metric that is irreducible and nonsymmetric.\footnote{The first demand is the requirement that we do not consider direct product spaces, while the second means that the manifold should not be a coset space with an invariant metric.} Then exactly one of the following cases holds:
\begin{enumerate}
\item $\Phi_{\mathfrak D}=\SO(n)$,
\item $\Phi_{\mathfrak D}=\U(m)$ for $n=2m$ with $m\geq 2$ (=K\"ahler manifold),
\item $\Phi_{\mathfrak D}=\SU(m)$ for $n=2m$ with $m\geq 2$ (=Calabi-Yau manifold),
\item $\Phi_{\mathfrak D}=\Sp(m)$ for $n=4m$ with $m\geq 1$ (=Hyperk\"ahler manifold),
\item $\Phi_{\mathfrak D}=\Sp(1)\cdot\Sp(m)$ for $n=4m$ with $m\geq 1$ (=quaternionic-K\"ahler manifold),
\item $\Phi_{\mathfrak D}=G_2$ for $n=7$ (=$G_2$-manifold),
\item $\Phi_{\mathfrak D}=\Spin(7)$ for $n=8$ (=$\Spin(7)$-manifold).
\end{enumerate}
Due to the requirements imposed on the manifold, the holonomy group is independent of the point considered up to conjugation in $\SO(n)$, hence we have written $\Phi_{\mathfrak D}$ instead of $\Phi_{\mathfrak D}(p)$.  
Note that all these types of manifolds can be described in terms of 1-flat G-structures. We have already discussed the cases 1, 2, 4 and 5 previously. The G-structure on a $2n$ (real) dimensional Calabi-Yau manifold is defined by three nowhere-vanishing parallel tensor fields $(g,J,\omega)$ where $g$ is a hermitian metric with respect to the complex structure $J$ and $\omega$ is a holomorphic $n$-form, i.e.
\begin{equation}
\prod_{i=1}^n \big[\frac12(\unity-\rmi J)_{\mu_i}{}^{\nu_i}\big] \omega_{\nu_1\dots \nu_n}=\omega_{\mu_1\dots \mu_n}\,.
\end{equation}
On a $G_2$-manifold, the G-structure is given by the metric $g$ and a certain three-form $\phi$, while on a $\Spin(7)$-manifold, we need a metric $g$ and a certain self-dual four-form $\varphi$. 

On a manifold with a G-structure, the structure group of the frame bundle can be reduced. If the manifold is spin, we can look at the action of the reduced structure group on the spin bundle. In some cases, there will exist a trivial subbundle, which will amount to the existence of a nowhere vanishing spinor field covariantly constant with respect to the spin connection, i.e. a parallel spinor. Hence, the G-structure can then also be defined in terms of these parallel spinors. Moreover, the G-invariant tensor fields can be constructed using the parallel spinors~\cite{Wang}. Below we list the subset of manifolds in Berger's list where this is possible. In the case of odd-dimensional manifolds, $N$ will denote the number of parallel spinors, while in the even-dimensional case $N_+$ and $N_-$ respectively denote the number of parallel left-and right-handed Weyl spinors. For a definition of chiral spinors, see Appendix~\ref{spinors}.
\begin{enumerate}
\item $\Phi_{\mathfrak D}=\SU(2m)$ implies $N_+=2$ and $N_-=0$,
\item $\Phi_{\mathfrak D}=\SU(2m+1)$ implies $N_+=N_-=1$,
\item $\Phi_{\mathfrak D}=Sp(m)$ implies $N_+=m+1$ and $N_-=0$,
\item $\Phi_{\mathfrak D}=G_2$ implies $N=1$,
\item $\Phi_{\mathfrak D}=\Spin(7)$ implies $N_+=1$.
\end{enumerate}
\section{Supersymmetry}\label{s:susy}
\subsection{The super-Poincar\'e algebra}
We will start by introducing the concept of a Lie-superalgebra as listed in~\cite{Frappat:1996pb}. We will subsequently specify to the super-Poincar\'e algebra, which is of main physical interest.
\begin{defi}
A Lie-superalgebra $\mathfrak g$ over a field $\mathbb K$ is a $\mathbb Z_2$ graded algebra that is a $\mathbb K$ vector space direct sum of two vector spaces $\mathfrak g_0$ and $\mathfrak g_1$ on which a product $[\cdot,\cdot]$ is defined as follows:
\begin{itemize}
\item $\mathbb Z_2$ gradation 
\begin{equation}
[\mathfrak g_a,\mathfrak g_b]\subset \mathfrak g_{(a+b)\; {\rm mod}\; 2}\,,
\end{equation}
\item graded antisymmetry 
\begin{equation}
[X_i,X_j]=(-1)^{\epsilon_i.\epsilon_j+1}[X_j,X_i]\,,
\end{equation}
\item generalized Jacobi identity 
\begin{equation}
(-1)^{\epsilon_i.\epsilon_k}[X_i,[X_j,X_k]] +{\rm permutations}=0\,,
\end{equation}
\end{itemize}
where $X_i$ denotes a generator, $a=0$ or $1$, $i$ labels the generators and $\epsilon_i$ is the degree, i.e. $\epsilon_i=0$ if $X_i\in \mathfrak g_0$ and $\epsilon_i=1$ for $X_i\in \mathfrak g_1$.
\end{defi}
For the superalgebras under consideration, $\mathfrak g_0$ is called the even or bosonic part, while $\mathfrak g_1$ is called the odd or fermionic part.
A sub-superalgebra $\mathfrak k=\mathfrak k_0\oplus \mathfrak k_1$ of $\mathfrak g$ is a subset of elements of $\mathfrak g$ that forms a vector subspace of $\mathfrak g$ that is closed with respect to the Lie-product of $\mathfrak g$ such that $\mathfrak k_0\subset \mathfrak g_0$ and $\mathfrak k_1\subset \mathfrak g_1$.

A super-Poincar\'e algebra in $p$ timelike and $q$ spacelike dimensions is a special case of this. The bosonic part of the algebra has the form $\mathfrak g_0=[\mathbb R^{p+q} \ltimes \so(p,q)]\oplus\mathfrak r$, where $\mathfrak r$ is called the R-symmetry algebra, which will be specified below. The fermionic part $\mathfrak g_1$ is a spinor representation of $\so(p,q)$ tensored with a representation of $\mathfrak r$.  

To explain this in more detail, let us recall the defining anticommutation relation~(\ref{ordefsusy}) for an $\mathcal N=1$ super-Poincar\'e algebra. 
\begin{equation}\label{Q^2=P}
\{Q_\alpha,Q_\beta\}=(\mathcal{C}\slashed{P})_{\alpha \beta}\,.
\end{equation}
Here, $Q$ denotes the generators in $\mathfrak g_1$ transforming in the spinor representation of $\so(p,q)$. We will always take spinors to be Grassmann-valued. The $\alpha, \beta$ are the spinor indices, while $P$ in~(\ref{Q^2=P}) denotes the generators of translation and $\mathcal C$ is the charge conjugation matrix, introduced in Appendix~\ref{spinors}. 

Extended supersymmetry contains $\mathcal{N}>1$ spinorial fermionic generators, labelled by an index $i$. As explained in the Appendix~\ref{spinors}, in five dimensions (which will be the relevant case in our discussion) we can take spinors to satisfy a symplectic Majorana condition. For extended supersymmetry, the defining relation~(\ref{Q^2=P}) becomes 
\begin{equation}
\{Q_\alpha^i,Q_{\beta j}\}=\delta_j^i(\mathcal{C}\slashed{P})_{\alpha \beta}\,,
\end{equation}
while the R-symmetry then acts on the supercharges as
\begin{equation}
[R_A,Q^i_\alpha]=(U_A)_j{}^i Q^j_\alpha\,,
\end{equation}
Using Jacobi identities and the fact that $\mathfrak r$ commutes with the other part of the bosonic algebra, we arrive at the result that the R-symmetry algebra in $d=5$ is $\symp(\mathcal N)$~\cite{VanProeyen:1999ni}. In the cases where the supersymmetry generators transform in another irreducible spinor representation, a similar reasoning leads to other R-symmetry algebras and they are all listed in Table~\ref{Rsymmalg}.
\begin{table}[tb]
\begin{center}
\begin{tabular}{|c|c|}
\hline 
spinor type       & R-symmetry algebra \\
\hline\hline
M and d odd&$\so(\mathcal N)$\\
M and d even&$\un(\mathcal N)$\\
MW&$\so(\mathcal N_-)\times\so(\mathcal N_+)$\\
S&$\symp(\mathcal N)$\\
SMW&$\symp(\mathcal N)\times\symp(\mathcal N)$\\
\hline 
\end{tabular}
\end{center}
\caption{\it The possible R-symmetry algebras classified using the spinor representation of the supercharges. We use $\mathcal N$ for the total number of spinors generating $\mathfrak g_1$, while $\mathcal N_-$ and $\mathcal N_+$ is the number of left-and right-handed ones, $\mathfrak g_0$ contains $\so(p,q)$ and $d=p+q$. `M' denotes Majorana, `S' symplectic and `W' Weyl spinors.}\label{Rsymmalg}
\end{table}

As we do not intend to discuss superalgebras in full generality, we will now give the example of $\mathcal{N}=2$ super-Poincar\'e symmetry in five dimensions. For the conventions on the spinors, the reader is again referred to the Appendix~\ref{spinors}. The extended super-Poincar\'e algebra contains the generators of Lorentz-rotations $M$, translations $P$, the $\symp(1)\equiv \su(2)$ automorphisms $U$ and the supersymmetries $Q^i$ with $i=1,2$. Note that these fermionic generators are symplectic Majorana spinors with respect to $\so(1,4)$. We then have the following commutation relations.
\begin{equation}
\begin{array}{ll}
\lbrack M_{ab},M_{cd}\rbrack=\eta_{a\lbrack c}M_{d\rbrack b}-\eta_{b\lbrack c}M_{d\rbrack a}\,, & \lbrack P_a,M_{bc}\rbrack=\eta_{a\lbrack b}P_{c\rbrack}\, ,\\
\lbrack P_a,P_b\rbrack=0\,,&\lbrack U_i{}^j,U_k{}^l\rbrack=\delta_i^lU_k{}^j-\delta_k^jU_i{}^l\,,\\
\lbrack M_{ab},Q^i_\alpha\rbrack=-\frac14 (\gamma_{ab}Q^i)_\alpha\,,&\lbrack U_i{}^j,Q_\alpha^k\rbrack=\delta_i^kQ_\alpha^j-\frac12 \delta_i^jQ_\alpha^k\,,\\
\{Q_{i\alpha},Q^{j \beta}\}=-\frac12 \delta_i^j(\gamma^a)_\alpha{}^\beta P_a\,.&\\
\end{array}
\label{N2d5poinc}
\end{equation}

A set of fields transforming in an irreducible representation of a supersymmetry algebra is called a multiplet. Transforming the bosonic fields of the multiplet twice under supersymmetry should be equivalent to acting with $\slashed{P}$ on these fields, due to~(\ref{Q^2=P}). In theories where a translation is an invertible operation, this implies that the numbers of fermionic and bosonic degrees of freedom are equal.\footnote{For massless fields in 1 dimension or in Euclidean theories, the fact that $P^2=0$ implies $P=0$, hence translations are not invertible in that case.} For the field theories that we will study, the defining relation~(\ref{Q^2=P}) is not realized exactly on the fields since in the right-hand side there appears a functional $\Gamma$ of the fields that is second order in derivatives for bosonic or first order for fermionic fields. Hence, the fields only form a representation when the constraint $\Gamma=0$ is imposed. Due to the fact that $\Gamma$ contains the required number of derivatives, the constraint can be interpreted as an equation of motion. Hence, the fields form a so-called on-shell multiplet as the equations of motion have to be satisfied in order to form a representation of the supersymmetry algebra (in contrary to the off-shell realizations where such a constraint does not appear in the right-hand side of the defining relation for the supersymmetry). As a consequence, the number of fermionic and bosonic modes of an on-shell multiplet only matches after elimination of degrees of freedom using the equations of motion. These on-shell realizations will be studied extensively in the following chapters.

In the case at hand, three multiplets exist for rigid supersymmetric field theories, namely the hypermultiplet, the vector multiplet and its dual, the tensor multiplet. Their field content is given in Table~\ref{tbl:multiplets}. Therein, $w$ denotes the Weyl weight, which we will explain in Chapter~\ref{suptencalc}. As can be seen from this Table, the fields in the vector and tensor multiplet bear another index $I$ or $M$. These multiplets not only form a representation of the $\mathcal{N}=2$ super-Poincar\'e algebra, but also transform under a scalar gauge algebra. Another point to make is that this vector multiplet is off-shell, in contrary to the others as can be deduced from Table~\ref{tbl:multiplets}. However, it is also possible to consider on-shell vector multiplets, which we will in Chapter~\ref{onshellsusy}.\footnote{The hypermultiplet can also be taken off-shell, but this requires an infinite number of auxiliary fields~\cite{Galperin:1984av}.}
\begin{table}[tb]
\begin{center}
\begin{tabular}{|c|ccc||c|ccc||c|ccc|}
\hline\rule[.2cm]{0cm}{.2cm}
Field&$\sharp$&$\su(2)$&$w$&Field&$\sharp$&$\su(2)$&$w$&Field&$\sharp$&$\su(2)$&$w$\\
\hline\hline\rule[.2cm]{0cm}{.2cm}
$A_\mu^I$&$4$&$1$&$0$&$B^M_{\mu\nu}$&$3$&$1$&$0$&$q^X$&$4$&  $2$&$-$\\\rule[.2cm]{0cm}{.2cm}
$Y^{ijI}$&$3$&$3$&$2$&$\phi^M$&$1$&$1$&$1$&&&&\\\rule[.2cm]{0cm}{.2cm}
$\sigma^I$ &$1$&$1$&$1$&&&&&&&&\\\hline\rule[.2cm]{0cm}{.2cm}
$\psi^{iI}$&$8$&$2$&$\frac32$&$\lambda^{iM}$&$4$&$2$&$\frac32$&$ \zeta^A$&$4$&$1$&$2$\\\hline
\end{tabular}
\end{center}
\caption{\it The $\mathcal N=2$, $d=5$ vector-, tensor- and hypermultiplets. Indicated are
the number of degrees of freedom, the $\su(2)$ representation and the Weyl weight.}\label{tbl:multiplets}
\end{table}
\subsection{Target space geometry}\label{targetspacegeometry}
To give a flavour of the intimate connection between supersymmetry and differential geometry, let us summarize the results of the classical paper~\cite{Alvarez-Gaume:1981hm}. As already mentioned, a nonlinear sigma model is a field theory where the scalar fields take values in a manifold, called the target space $M_t$. The simplest example is\footnote{Note that from now on, we will use capital letters to denote target space indices.}
\begin{eqnarray}
\mathcal{S}&=&-\frac12 \int d^dx \;g_{XY}(\varphi)\partial_a \varphi^X\partial^a \varphi^Y \, \Rightarrow\, \Box \varphi^X=\partial_a\partial^a \varphi^X+\Gamma_{YZ}^X(\varphi)\partial_a\varphi^Y\partial^a\varphi^Z=0\,.\nonumber\\\label{simplesigmamodel}
\end{eqnarray}
Thus, the kinetic terms are multiplied by a field-dependent object $g_{XY}$. First of all, note that $g_{XY}$ can be taken symmetric, as its antisymmetric part would not contribute to the action. If we moreover ask that the model is invariant under target space diffeomorphisms $\varphi'=\varphi'(\varphi)$, $g_{XY}$ has to transforms as a (2,0) tensor on $M_t$. This implies that $g_{XY}$ is actually the local expression for a metric on the target space. The corresponding equation of motion either is covariant under these diffeomorphisms since the d'Alembertian contains a connection term. It can easily be checked that this connection is the Levi-Civita connection with respect to the target space metric.

In~\cite{Alvarez-Gaume:1981hm}, the supersymmetry of such a nonlinear sigma model defined on a two-dimensional Minkowski space-time is studied. To repeat that discussion, let us introduce the spinor field $\psi^X$ which transforms as the direct product of a two-dimensional space-time Grassmann-valued Majorana spinor and a target space vector. It now turns out that since $M_t$ is a Riemannian manifold, it is always possible to extend~(\ref{simplesigmamodel}) to a theory invariant under the $\mathcal{N}=1$ two-dimensional super-Poincar\'e group. The corresponding action reads
\begin{equation}\label{minimalsigmamodel}
\mathcal{L}=-\frac12 \left(g_{XY}\partial_a \varphi^X\partial^a \varphi^Y+\rmi g_{XY}\bar{\psi}^X \slashed{\mathfrak{D}}\psi^Y+\frac16R_{WXYZ}\bar{\psi}^W\psi^Y\bar{\psi}^X\psi^Z\right)\,,
\end{equation}
where $R_{WXYZ}=R_{WXY}{}^Vg_{VZ}$ is the curvature tensor corresponding to the Levi-Civita connection on $TM_t$ and $\mathfrak{D}_\mu\psi^X=\partial_\mu\psi^X+\Gamma_{YZ}{}^X\psi^Z\partial_\mu \varphi^Y$. This action is invariant under the following supersymmetry transformation rules.
\begin{eqnarray}
\delta \varphi^X&=&\bar{\epsilon}\psi^X\,,\nonumber\\
\delta \psi^X&=&-\rmi \slashed{\partial}\varphi^X\epsilon-\Gamma_{YZ}{}^X\bar{\epsilon}\psi^Y\psi^Z\,,
\end{eqnarray}
where $\epsilon$ is the parameter for this supersymmetry transformation.
If we want to look for other supersymmetries leaving~(\ref{minimalsigmamodel}) invariant, we have to consider the most general transformation rules, consistent with dimensional arguments, Lorentz-invariance and the defining relation of the supersymmetry algebra~(\ref{Q^2=P}). It was found~\cite{Alvarez-Gaume:1981hm} that this new supersymmetry should be of the following form.
\begin{eqnarray}
\delta \varphi^X&=&J_Y{}^X\bar \epsilon'\psi^Y\,,\nonumber\\
\delta(J_Y{}^X\psi^Y)&=&-\rmi \slashed{\partial}\varphi^X\epsilon'+\frac12 \Gamma_{YZ}{}^XJ_V{}^YJ_W{}^Z\bar{\psi}^V\psi^W\epsilon'\,,\label{transfrulessigmamodel}
\end{eqnarray}
where $\epsilon'$ parametrizes this new supersymmetry transformation.
Moreover, $J$ has to be a traceless (1,1) tensor on $M_t$ that is covariantly constant with respect to the Levi-Civita connection and $J$ should square to minus the identity. The metric should moreover be hermitian with respect to $J$. As explained in Section~\ref{sss:complexmanifolds}, this implies that an $\mathcal N=2$ theory admits a target space manifold that is K\"ahler.

It might be possible to extend~(\ref{transfrulessigmamodel}) such that we have a set of three complex structures, generating three extra supersymmetry transformations of the above form. These complex structures should each obey all requirements of a complex structure, together with~(\ref{defhcstr}). In that case, the target manifold satisfies all demands of Section~\ref{sss:quatlikestruct} to be a hyperk\"ahler manifold.
\section{The Batalin-Vilkovisky formalism}\label{s:BV}
We will now give a brief technical account of the classical Batalin-Vilkovisky (BV) formalism~\cite{Batalin:1983jr}. We refer the reader to~\cite{Gomis:1995he} for conventions and a more extended expos\'e. This technique, which was originally developed in the context of the quantization of gauge theories, can also be used to study the algebraic structure of classical field theories. The key-role is played by the extended BV action, which generates all algebraic relations from the vanishing of a Poisson bracket.

In the BV formalism, we introduce a ghost field for every symmetry of the action,\footnote{In case of a rigid symmetry, the ghost field is taken to be constant.} and an antifield for every field, including for the ghosts. A ghost corresponding to a bosonic symmetry is anticommuting, and vice versa in the fermionic case. Similarly, antifields have the opposite statistics of the corresponding fields. A ghost number $g$ is assigned to every generation of ghosts, and an antifield number to every (anti)field. The ghost number of a field and its antifield should always sum to $-1$, i.e. 
\begin{equation}
g(\Phi^A)+g(\Phi^*_A)=-1\,,
\end{equation}
where $\Phi^A$ can denote any field and $\Phi_A^*$ is its antifield.
We then proceed with the introduction of the antibracket
\begin{equation}
(G,H)=G\frac{\overleftarrow{\delta}}{\delta \Phi^A}\frac{\overrightarrow{\delta}}{\delta \Phi^*_A}H-G\frac{\overleftarrow{\delta}}{\delta \Phi_A^*}\frac{\overrightarrow{\delta}}{\delta \Phi^A}H\,,
\end{equation}
where the arrows show the direction in which the derivative is taken, i.e.
\begin{equation}
\Phi^B\frac{\overleftarrow{\delta}}{\delta \Phi^A}=\delta^B_A\,,\qquad\frac{\overrightarrow{\delta}}{\delta \Phi^A}\Phi^B=\delta^B_A\,.
\end{equation}
We now add extra terms with strictly positive antifield number to the classical action, in such a way that the classical master equation,
\begin{equation}\label{master}
(\mathcal{S}_{BV},\mathcal{S}_{BV})=0\,,
\end{equation}
holds. The above relation can then be expanded in the antifield number and the terms of definite antifield number have to sum up to zero. 

We will now explain in more detail how the BV formalism encodes the algebraic relations by discussing the following simple example. Consider a field theory with classical action functional $\mathcal{S}_0$ containing scalar fields $\varphi^X$ labelled by $X$, which is invariant under bosonic symmetries $\delta_I(\epsilon^I)\varphi^X=R_I{}^X[\varphi]\epsilon^I$, with infinitesimal parameters $\epsilon^I$ labelled by $I$. In other words,
\begin{equation}\label{symmrel}
\int d^d x \frac{\delta \mathcal{S}_0}{\delta \varphi^X} R_I{}^X \epsilon^I=0\,.
\end{equation}
The ghosts corresponding to these symmetries will be called $c^I$. Suppose moreover that the commutator of two symmetries yields
\begin{equation}\label{commsymm}
[\delta_K(\epsilon_1^K),\delta_J(\epsilon_2^J)]\varphi^X=\delta_I(\epsilon_2^K\epsilon_1^Jf_{JK}{}^I)\varphi^X\,.
\end{equation}
The corresponding BV action that encodes all algebraic relations then reads
\begin{equation}
\mathcal{S}_{BV}=\mathcal{S}_0+\varphi_X^* R_I{}^X c^I+\frac12c^*_If_{JK}{}^Ic^Kc^J\,,
\end{equation}
where $\varphi_X^*$ and $c_I^*$ are respectively the antifield and the ghost antifield.
Looking at the antifield number in Table~\ref{ghostandaf}, we can see that the second and third term in this extended BV action have respectively antifield number one and two.
\begin{table}[tb]
\begin{center}
\begin{tabular}{|c|cc||c|cc|}
\hline
field&g&afn&field&g&afn\\
\hline
$\varphi^X$&$0$&$0$&$\varphi^*_X$&$-1$&$1$\\
$c^I$&$1$&$0$&$c^*_I$&$-2$&$2$\\
\hline
\end{tabular}
\caption{\it The ghost number (g) and antifield number (afn) of the fields.}\label{ghostandaf}
\end{center}
\end{table}
The validity of the master equation then yields the relation~(\ref{symmrel}) for a symmetry, the commutator~(\ref{commsymm}) and the Bianchi identity
\begin{equation}
f_{M[I}{}^Lf_{JK]}{}^M=0\,.
\end{equation}
Finally note that we did not have to use the explicit form of the action in order to be able the retrieve the properties of the symmetry algebra. In the following chapter, we will discuss on-shell theories that are defined via their equations of motion. When these field equations have to be derivable from an action, this will restrict the possible theories. We will use the BV formalism to systematically study these restrictions. Since we will not have to refer to the explicit form of the classical action, we will be able to hold the discussion in its full generality.
\chapter{Rigid on-shell supersymmetry}\label{onshellsusy}
The purpose of this work is the study of on-shell realized supersymmetric theories. More specifically, we want to look if such theories can be generalized by considering the possibility of having equations of motion that are not derivable from a Poincar\'e-invariant and supersymmetric action. In the present chapter, we will take a first step in that direction by considering $\mathcal N=2$ rigid supersymmetric theories. 

The content of this chapter is as follows. In the introductory Section~\ref{appearances}, we will list in what circumstances we can encounter theories without actions. In Section~\ref{ss:algebra}, we will enlighten the special properties of the symmetry algebra of the theories under study and in the final Section~\ref{ss:rigidsusy}, we will apply the ideas in the context of rigid supersymmetric vector- and hypermultiplets.
\section{Introduction}\label{appearances}
As we want to study field equations rather than actions, it is helpful to find some organizing principle. Our approach is to consider supersymmetric theories where the superalgebra is realized on-shell. Hence, as was already explained in the previous chapter, the defining relation for supersymmetry~(\ref{Q^2=P}) will be satisfied modulo nonclosure functionals that can be interpreted as equations of motion. As an introduction, we will therefore list the various circumstances in which both rigid and local supersymmetric theories without action can be encountered.

One of the most prominent examples of an on-shell realization of the superalgebra is the unique Poincar\'e supergravity
theory in eleven dimensions, also called classical M-theory~\cite{Cremmer:1978km}. The bosonic fields comprise a metric $g$ and a three-form potential $A_{(3)}$, gauging an Abelian symmetry. The only fermionic field is
the gravitino, which transforms as the direct product of a one-form and a Majorana spinor. It can easily be checked that with such a field content, the number of bosonic and fermionic modes only equals on the mass-shell. Moreover, the anticommutator of the supersymmetries calculated on the gravitino yields its equation of motion $\Gamma$. The field equations for the metric and the three-form can subsequently be found by applying a super-transformation on $\Gamma$. However, there is no obstruction to the construction of an action. 

Such obstructions do exist in other theories. First of all, a self-dual tensor field
(e.g. like in type IIB supergravity) does prohibit a standard action formulation. Secondly, if a theory is
dimensionally reduced using the generalized Scherk-Schwarz formalism~\cite{Scherk:1979zr}, it might be possible that (at least) the potential for e.g. the Kaluza-Klein scalar prohibits the existence of an action. A well-known example of this is the generalized Scherk-Schwarz reduction of classical M-theory to ten dimensions using the scaling symmetry of the equations of motion~\cite{Howe:1998qt,Lavrinenko:1998qa}. The resulting massive IIA theory does not admit an action.  The use of Kaluza-Klein ideas in the construction of theories without actions will be the subject of Chapter~\ref{dimred}. 

The above examples were supergravity theories. However, the final possibility that we consider can also be realized for theories with global supersymmetric invariance. We will namely study nonlinear sigma models on target space manifolds (together with a connection on the tangent bundle) that do not admit a metric that is preserved under parallel transport. Since the target space metric does not appear in the equations of motion, see e.g.~(\ref{simplesigmamodel}), while it does in the standard kinetic term of a nonlinear sigma-model action, this is still another possibility to prohibit the existence of an action. The present chapter together with Chapter~\ref{suptencalc} will be devoted to the study of theories allowing for such target spaces.
\section{Algebraic structure}\label{ss:algebra}
When looking for a representation of a symmetry algebra on the fields of a theory, it might happen that the structure constants of the algebra actually become functionals of the fields, that are also called structure functions. This has nothing to do with the fact that a symmetry is realized off or on the mass-shell, but rather is a typical feature of classical field theory.

We can illustrate the appearance of these so-called soft algebras in a supersymmetric gauge theory. In general, the supersymmetry transformation rules have to be covariant with respect to the gauge symmetry. As a consequence, the defining relation for supersymmetry~(\ref{Q^2=P}) gets changed, as it should transform covariantly under the action of the gauge group. To be more concrete, let us suppose that the theory contains a scalar $\varphi$ that couples to the gauge group with a charge $q$, and suppose $\epsilon_1$ and $\epsilon_2$ are two parameters for supersymmetry transformations. The translation operator in the right-hand side of~(\ref{Q^2=P}) should then be covariantized. Introducing the covariant derivative $\mathfrak D \varphi=(\partial-\rmi q A_\mu) \varphi$, we thus have 
\begin{equation}
[\delta_Q(\epsilon_1),\delta_Q(\epsilon_2)]\varphi=\bar{\epsilon}_2\gamma^\mu \epsilon_1 \mathfrak{D}_\mu \varphi=\delta_P(\bar{\epsilon}_2\gamma^\mu \epsilon_1)\varphi-\delta_G(\bar{\epsilon}_2\gamma^\mu \epsilon_1A_\mu)\varphi\,.
\end{equation}
Hence, the anticommutator of two supersymmetries does not only contain a translation but also a field-dependent gauge transformation.\footnote{From now on however, the complete right-hand side of the above relation will be called a covariant translation, which we will also denote by $P$.} The Jacobi-identities might get changed in the same way. 

Something more specific to the case at hand is the appearance of open algebras~\cite{deWit:1978cd}, which can be discussed starting from the remark that any action $\mathcal{S}$ is invariant under a set of trivial symmetries. Suppose the theory contains a set of scalars $\varphi^X$. A trivial transformation then reads
\begin{equation}
\delta_{triv}\varphi^X=\Omega^{XY}(\varphi)\frac{\delta \mathcal{S}}{\delta \varphi^Y}\,,
\end{equation}
where $\Omega^{XY}$ is any antisymmetric matrix (which can be field-dependent). The invariance of the action is then proven easily,
\begin{equation}
\delta_{triv} \mathcal{S}=\Omega^{XY}\frac{\delta \mathcal{S}}{\delta \varphi^X}\frac{\delta \mathcal{S}}{\delta \varphi^Y}=0\,.
\end{equation}
In the open algebras we will encounter, these transformations are part of the symmetry algebra. More concretely, the commutator of two supersymmetries~(\ref{Q^2=P}) will not only yield a translation but will also give rise to such a trivial symmetry, 
\begin{equation}\label{Q^2=Pbis}
[\delta_Q(\epsilon_1),\delta_Q(\epsilon_2)]\varphi^X=\delta_P(\bar{\epsilon}_2\gamma^\mu \epsilon_1)\varphi^X+\Omega^{XY}\frac{\delta \mathcal{S}}{\delta \varphi^Y}\,.
\end{equation}
Hence, an on-shell algebra is a symmetry algebra that contains trivial transformations. We can rephrase this by stating that the algebra is realized on the fields in such a way that the nontrivial symmetries are only represented correctly under the constraint that the fields satisfy the equations of motion. 

Such an on-shell multiplet can sometimes be taken off the mass-shell by introducing auxiliary fields. This means that in such a case, the trivial transformations in e.g.~(\ref{Q^2=Pbis}) can be made to vanish. In that case, the algebra does not depend any more on the specific theory we are looking at.\footnote{This will become clear in Chapter~\ref{suptencalc}, where we will couple such an off-shell vector multiplet to hypermultiplets without changing the transformation properties of the former.} Note that we can always return to the previous case by eliminating the auxiliary fields by their algebraic equations of motion.

We can now generalize this construction in the following way. Given the supersymmetry transformation rules of an on-shell multiplet, we are thus able to find the equations of motion prior to the introduction of an action due to the fact that trivial transformations enter the symmetry algebra. In that case,~(\ref{Q^2=Pbis}) should be rewritten slightly to
\begin{equation}\label{Q^2=Ptris}
[\delta_Q(\epsilon_1),\delta_Q(\epsilon_2)]\varphi^X=\delta_P(\bar{\epsilon}_2\gamma^\mu \epsilon_1) \varphi^X+f(\varphi,\epsilon)\Gamma^X\,,
\end{equation}
where $\Gamma^X$ is an equation of motion for $\varphi^X$ and $f(\varphi,\epsilon)$ is a functional of the fields and the supersymmetry parameters. It is obvious that introducing an action requires the existence of a new object, since we then ask that the equation of motion $\Gamma^X\equiv 0$ is proportional to the extremization conditions for the action $\delta \mathcal S / \delta \varphi^X\equiv 0$,
\begin{equation}
\frac{\delta \mathcal{S}}{\delta \varphi^X}=g_{XY}\Gamma^Y\equiv 0\,.
\end{equation}
Below, we will show that obstructions to the existence of such an object $g_{XY}$ can appear. More specifically, it will turn out that this object is (related to) the target space metric, and it can happen that such a tensor cannot be introduced in a consistent way.
\section{$\mathcal{N}=2$, $d=5$ supersymmetry}\label{ss:rigidsusy}
We will now elucidate some of the features of on-shell supersymmetry in the context of rigid $\mathcal{N}=2$ super-Poincar\'e invariance in five dimensions. The algebra of this group has already been given in~(\ref{N2d5poinc}).
\subsection{Vector multiplets}\label{5d}
Similar to the off-shell vector multiplet mentioned in Table~\ref{tbl:multiplets}, the set of on-shell fields comprises a gauge potential $A_\mu$, a real scalar field $\sigma$ and two symplectic-Majorana gauginos $\psi^i$, transforming as a doublet under the R-symmetry. A summary is given in Table~\ref{fieldsdof}.
\begin{table}[tb]
\begin{center}
\begin{tabular}{|c|c|c|}
\hline
Field&$\mathfrak{su}(2)$&on-shell dof\\
\hline \hline
$\sigma$&1&1\\
$A_\mu$&1&3\\
$\psi^{i}$&2&4\\
\hline
\end{tabular}
\end{center}
\caption {\it Fields in the $\mathcal{N}=2, d=5$ vector multiplet.}\label{fieldsdof}
\end{table}
The vectors in the multiplet gauge a group $G$. The fields will therefore also carry a gauge index $I$ and the corresponding gauge algebra will be able to rotate different multiplets into each other. 

We will show that on-shell vector multiplets generalize the off-shell case in two ways. In the latter, the kinetic terms of the theory are completely fixed by a constant gauge-invariant symmetric three-tensor $C_{IJK}$~\cite{Gunaydin:1984bi,Bergshoeff:2002qk}. To be more specific, the bosonic part of the action for the physical fields contains
\begin{equation}\label{offshellaction}
\mathcal S=\int d^5x \; C_{IJK}\sigma^K(-\frac12\mathfrak{D}_\mu \sigma^I\mathfrak{D}^\mu \sigma^J-\frac14F^I_{\mu \nu}F^{J\mu \nu})-g^2C_{IJK}\sigma^If^J_{(ij)}f^{(ij)K}\,. 
\end{equation}
From this, it can be seen that the target space of the nonlinear sigma model for the fields $\sigma^I$ carries the metric $C_{IJK}\sigma^K$. Secondly, the objects $f_{ij}^I$ appearing in the potential, are constants (in the $\mathbf 3$ of the $\symp(1)$ R-symmetry) that can only be present in the Abelian sectors of the gauge theory and that are called Fayet-Iliopoulos (FI) terms~\cite{deWit:1981tn,Breitenlohner:1981sm}. It will turn out that in the on-shell case, the target space can be more general and the Fayet-Iliopoulos terms may be present in the non-Abelian case as well. Moreover, they will not need to be constants.

Before embarking, let us first discuss the transformation under the gauge algebra $\mathfrak g$ gauged by the vectors. They satisfy the commutation
relations ($I=1, \ldots ,n$)
\begin{equation}
\left[\d_G(\beta_1), \d_G(\beta_2)\right] = \d_G(\beta_3 ) \,, \qquad \beta_3^K
= g \beta_1^I \beta_2^J f_{IJ}{}^K \,, \label{Lambda3}
\end{equation}
where $\beta^I$ denote the parameters for the gauge transformations and $g$ is the coupling constant of the algebra $\mathfrak g$. 
The gauge fields $A_\mu^I$\ ($\mu = 0,1, \ldots ,4$) and general matter
fields of the vector multiplet $Z^I$ transform under gauge
transformations according to
\begin{equation}
\d_G(\beta^J) A_\mu^I = \partial_\mu \beta^I + g A_\mu^J f_{JK}{}^I \beta^K \,,
\qquad \d_G(\beta^J) Z^I = -g \beta^J f_{JK}{}^I Z^K \,. \label{eq:gaugecov}
\end{equation}
The expressions for the
gauge covariant derivative and d'Alembertian of $Z^I$  and the field strength are given by
\begin{eqnarray}
\mathfrak D_\m Z^I &=& \partial_\m Z^I + g A_\m^J f_{JK}{}^I Z^K \,, \nonumber\\
\Box Z^I&=&\partial_\mu \mathfrak{D}^\mu Z^I+gf_{JK}{}^IA_\mu^J\mathfrak{D}^\mu Z^K\,,\nonumber\\
F_{\mu\nu}^I &=& 2 \partial_{[\mu} A_{\nu]}^I + g f_{JK}{}^I A_\mu^J
A_\nu^K\,. \label{eq:F}
\end{eqnarray}
The field strength satisfies the Bianchi identity
\begin{equation}
{\mathfrak D}_{[\m} F_{\n\l]}^I = 0 \,. \label{eq:BIvec}
\end{equation}

To be able to construct the most general multiplet, we have to start from supersymmetry transformation rules that contain unknown functionals of the fields, similar to the discussion in Section~\ref{targetspacegeometry}. Demanding that these symmetries form a closed algebra, modulo trivial transformations, then fixes the form of the supersymmetries completely. Concentrating first on the bosonic fields, it turns out~\cite{Gheerardyn:2003rf} that the most general transformation rules realizing the defining condition \`a la~(\ref{Q^2=P}) are ($\epsilon$ parametrizes the supersymmetry) 
\begin{eqnarray}
\delta \sigma^I&=&\frac12 \rmi \bar{\epsilon}\psi^I-gf_{JK}{}^I\beta^J\sigma^K\; ,\label{transfsigma2} \\
\delta A_\mu^I&=&\frac12 \bar{\epsilon}\gamma_\mu \psi^I+\partial_\mu\beta^I+gf_{JK}{}^IA_\mu^J\beta^K\; , \label{transfA2}\\
\delta \psi^{iI}&=&-\frac12 \rmi \slashed{\mathfrak{D}}\sigma^I\epsilon^i-\frac14\slashed{F}^I\epsilon^i+A^{(ij)I}\epsilon_j -gf_{JK}{}^I\beta^J\psi^{iK}\; , \label{transfpsi2} 
\end{eqnarray}
where the only remaining unknown functional is $A^{(ij)I}$, which transforms in the $\mathbf 3$ of $\symp(1)$. 
These rules imply that the following superalgebra commutator similar to~(\ref{Q^2=P}) is realized on the bosons,
\begin{equation}\label{algebra}
[\delta_Q(\epsilon_1),\delta_Q(\epsilon_2)]=\delta_P(\frac12\bar{\epsilon}_2\gamma^\mu\epsilon_1)+\delta_G(-\frac12\rmi \sigma^I \bar{\epsilon}_2\epsilon_1)\,.
\end{equation}
If we compare our result with the literature~\cite{Gunaydin:1984bi,Bergshoeff:2002qk}, where the off-shell case is described, it turns out that~(\ref{algebra}) is the same. Moreover, paralleling the discussion of the previous section, the authors of~\cite{Gunaydin:1984bi,Bergshoeff:2002qk} have taken the vector multiplet off the mass-shell by introducing the auxiliary field $Y^{(ij)I}$, listed in Table~\ref{tbl:multiplets}. Hence, $A^{(ij)I}$ in~(\ref{transfpsi2}) stands in place of that auxiliary field. It will however turn out that our discussion of the on-shell case yields more general theories~\cite{Gheerardyn:2003rf}. 

Trying to realize~(\ref{algebra}) on the gaugino will fix the remaining functional $A^{(ij)I}$ as we will need an Ansatz for its transformation. The most general possibility is 
\begin{equation}\label{transfAijt}
\delta_Q A^{(ij)I}=-\bar{\epsilon}_k\zeta^{k,(ij)I}\; .
\end{equation}
where $\zeta^{k,(ij)I}$ is a field-dependent spinor in the $\mathbf{2}\times \mathbf{3}$ of $\mathfrak{su}(2)$. Expanding in irreducible representations $\zeta^{k,(ij)I}=\varepsilon^{k(i}\zeta^{j)I}+\zeta^{(ijk)I}$, we can see that $\zeta^{(ijk)I}$ should be zero in order to be able to close the algebra~(\ref{algebra}) on the fermions. In conclusion, the transformation rule~(\ref{transfAijt}) becomes
\begin{equation}\label{transfAij}
\delta_Q A^{(ij)I}=\bar{\epsilon}^{(i}\zeta^{j)I}\; .
\end{equation}
The algebra on the fermions $\psi^{iI}$ then yields
\begin{eqnarray}
[\delta_Q(\epsilon_1),\delta_Q(\epsilon_2)]\psi^{iI}&=&\delta_P(\frac12\bar{\epsilon}_2\gamma^\mu\epsilon_1)\psi^{iI}+\delta_G(-\frac12 \sigma^I \bar{\epsilon}_2\epsilon_1)\psi^{iI}-\frac3{16}\bar{\epsilon}_2\epsilon_1\Gamma^{iI}-\frac3{16}\bar{\epsilon}_2\gamma^\mu \epsilon_1\gamma_\mu\Gamma^{iI}\nonumber\\&& -\frac1{16}\bar{\epsilon}_2^{(i}\gamma^{\mu \nu}\epsilon_1^{j)}\gamma_{\mu \nu}\Gamma_j^I \qquad {\rm with}  \qquad\Gamma^{iI}=\slashed{\mathfrak{D}}\psi^{iI}+\rmi gf_{JK}{}^I\sigma^J\psi^{iK}-2\zeta^{iI}\; ,\nonumber\\
\end{eqnarray}
where the nonclosure functional $\Gamma^{iI}$ can be interpreted as the equation of motion for the fermions.

To know the explicit expression for $\zeta^{iI}$ we need the field-dependence of the object $A^{(ij)I}$, which can be inferred from its transformation rule~(\ref{transfAij}) together with dimensional considerations. The latter yields as most general form for $A^{(ij)I}$, 
\begin{equation}\label{AnsatzA}
A^{(ij)I}=gf^{(ij)I}(\sigma)-\frac12 \rmi \gamma^I_{JK}(\sigma)\bar{\psi}^{iJ}\psi^{jK}\; ,
\end{equation}
where $\gamma^I_{JK}$ is symmetric in its lower indices. The first term will turn out to be the on-shell counterpart of a Fayet-Iliopoulos term. By using the rules~(\ref{transfsigma2}-\ref{transfpsi2}) we can calculate the transformation of $A^{(ij)I}$, which now only is compatible with~(\ref{transfAij}) if the following equations hold,
\begin{eqnarray}\label{FI1}
&&\partial_Jf^{(ij)I}+2\gamma^I_{JK}f^{(ij)K}=0\; ,\\
&&\gamma^I_{LM}\gamma^M_{JK}=-\frac12 \partial_L \gamma^I_{JK}\; .\label{gamma1}
\end{eqnarray}
Note that the requirement~(\ref{gamma1}) follows from the terms in $\delta_Q A^{(ij)I}$ that are cubic in the gaugino.

From the study of the symmetry algebra, we can infer two more conditions which should hold in the non-Abelian sectors of the gauge theory. As can be checked on the bosons, the commutator of a supersymmetry and a gauge transformation should vanish. On the fermions, this condition implies that $A^{(ij)I}$ transforms in the adjoint representation of the gauge group. This leads to the other defining conditions for the geometry:
\begin{eqnarray}\label{FI2}
&&f_{JL}{}^K\sigma^L\partial_Kf^{ijI}-f_{JK}{}^If^{Kij}=0\; ,\\
&&2f_{J(L}{}^M\gamma_{K)M}^I-f_{JM}{}^I\gamma_{KL}^M+f_{JM}{}^N\sigma^M\partial_N\gamma_{KL}^I=0\; .\label{gamma2}
\end{eqnarray}

We can now completely determine all equations of motion as the nonclosure functional $\Gamma^{iI}$ transforms under supersymmetry as
\begin{eqnarray}
\delta_Q\Gamma^{iI}&=&-\frac38\rmi\gamma^I_{JK}\bar{\psi}^{iJ}\Gamma^{jK}\epsilon_j-\frac38\rmi\gamma^I_{JK}\bar{\psi}^{iJ}\gamma^\mu \Gamma^{jK}\gamma_\mu \epsilon_j+\frac1{16}\rmi\gamma^I_{JK}\bar{\psi}^{iJ}\gamma^{\mu \nu}\Gamma^{jK}\gamma_{\mu \nu}\epsilon_j\nonumber\\&&
-\frac12\rmi \Delta^I\epsilon^i-\frac12\Xi_\mu^I\gamma^\mu \epsilon^i\; ,
\end{eqnarray}
where $\Delta^I$ is the equation of motion for the real scalar $\sigma^I$ and $\Xi^I_\mu$ the one for the real vector $A_\mu^I$. All dynamical constraints thus read
\begin{eqnarray}
\Gamma^{iI}&=&\slashed{\mathfrak{D}}\psi^{iI}+\gamma^I_{JK}\slashed{\mathfrak{D}}\sigma^J\psi^{iK} +\frac12\rmi\gamma^I_{JK}\slashed{F}^J\psi^{iK}-\frac12 \partial_K\gamma_{JL}^I \bar{\psi}^{iJ}\psi^{jL}\psi^K_j +2\rmi g\gamma^I_{JK}f^{ijI}\psi_j^K\nonumber\\&&+\rmi gf_{JK}{}^I\sigma^J\psi^{iK}\equiv0\; ,\label{psiEOM}\\
\Delta^I&=&\Box \sigma^I+ \gamma_{JK}^I\mathfrak{D}_\mu \sigma^J\mathfrak{D}^\mu \sigma^K-\frac12 \gamma^I_{JK}F_{\mu \nu}^JF^{\mu \nu K}-\frac12 \partial_L\gamma^I_{JK}\bar{\psi}^L\slashed{\mathfrak{D}}\sigma^J\psi^K-\frac14\rmi\partial_K\gamma^I_{JL}\bar{\psi}^J\slashed{F}^L\psi^K\nonumber\\&&-\frac5{32} \partial_M\partial_K\gamma^I_{JL}\bar{\psi}^{Lj}\psi^{Jk}\bar{\psi}^M_k\psi_j^K
-\frac18\partial_K\gamma^I_{JL}\gamma^J_{MN}\bar{\psi}^{Kj}\psi^{Lk}\bar{\psi}^M_k\psi^N_j\nonumber\\&&+\frac14\partial_K\gamma^I_{JL}\gamma^K_{MN}\bar{\psi}^{Lj}\psi^{Jk}\bar{\psi}^N_k\psi^M_j
+\frac12\rmi g f_{JK}{}^I\bar{\psi}^J\psi^K+\rmi g\gamma^I_{JK}f^J_{LM}\sigma^M\bar{\psi}^L\psi^K
\nonumber\\&&+\rmi g\partial_J\gamma^I_{LM}f^{ijL}\bar{\psi}_i^J\psi_j^M+\rmi g \partial_J\gamma^I_{LM}f^{ijJ}\bar{\psi}_i^L\psi_j^M
+2g^2\gamma^I_{JK}f^{ijJ}f_{ij}^K\equiv0\; ,\label{sigmaEOM}\\
\Xi_\mu^I&=&\mathfrak{D}^\nu F_{\nu \mu}^I-\frac14\gamma^I_{JK}\varepsilon_{\mu \nu \rho \sigma \tau}F^{\nu\rho J} F^{\sigma \tau K} +2 \gamma^I_{JK}\mathfrak{D}^\nu \sigma^KF_{\nu \mu}^J+\rmi\gamma^I_{JK}\bar{\psi}^J\mathfrak{D}_\mu \psi^K\nonumber\\&&+\frac14 \partial_M\gamma_{JK}^I\bar{\psi}^M\gamma_\mu \slashed{F}^J\psi^K
-\frac12\rmi\partial_L\gamma_{JK}^I\bar{\psi}^L\gamma_\mu \slashed{\mathfrak{D}}\sigma^J\psi^K
-\frac5{32}\rmi\partial_M\partial_K\gamma^I_{JL}\bar{\psi}^{Lj}\psi^{Jk}\bar{\psi}^M_k\gamma_\mu \psi_j^K
\nonumber\\&&-\frac18 \rmi \partial_{(K}\gamma_{L)J}^I\gamma^J_{MN}\bar{\psi}^{Mk}\psi^{Nj}\bar{\psi}^L_k\gamma_\mu \psi_j^K-gf_{JK}{}^I\sigma^J\mathfrak{D}_\mu \sigma^K+\frac12gf_{JK}{}^I\bar{\psi}^J\gamma_\mu \psi^K\nonumber\\&&-g\partial_J\gamma^I_{LM}f^{ijL}\bar{\psi}^J_i\gamma_\mu\psi_j^M\equiv0\; .\nonumber\\\label{AEOM}
\end{eqnarray}
These results can now be compared to the known off-shell case. There, the theory is completely determined by the constant symmetric gauge-invariant three-tensor $C_{IJK}$ as we have already mentioned before. Similarly, the dynamical constraints are determined by a new object $\gamma^I_{JK}$ which is symmetric in its lower indices. In Section~\ref{os}, we will prove that in the off-shell case, the object $\gamma_{JK}^I$ is fixed in terms of $C_{IJK}$. As this is not necessarily the case at present, our discussion is already more general at this level. In the simplest case where the gauge theory is Abelian and no FI terms are present,~(\ref{gamma1}) is the only condition and we will show in the next Section~\ref{os} that it is the counterpart of the fact that $C_{IJK}$ is constant. When we consider a non-Abelian gauge group, the transformation of the object $\gamma_{JK}^I$ should be compatible with~(\ref{gamma2}), which is to be compared to the demand that $C_{IJK}$ be gauge-invariant.

In the Abelian off-shell case, we can add a constant term (the Fayet-Iliopoulos term) to the equation of motion of the auxiliary field, which yields a potential in the action for the physical fields, as shown in~(\ref{offshellaction}). 
However, these FI terms can only be present in Abelian theories.\footnote{In the case of local (off-shell) supersymmetry, $\mathfrak{su}(2)$ FI terms exist.} This is to be contrasted with the on-shell multiplets where in the Abelian case, $f^{(ij)I}$ should only satisfy~(\ref{FI1}), which does not imply that it is constant. Moreover, in the non-Abelian case these terms should simultaneously obey~(\ref{FI1}) and~(\ref{FI2}) but it is clear that they are not eliminated. This is a major generalization as in the off-shell case, non-Abelian FI terms are not possible.
\subsection{Hypermultiplets}\label{rigidhypers}
We will now repeat the above discussion for hypermultiplets. This will lead us to the conclusion that the allowed target spaces can be generalized from hyperk\"ahler to hypercomplex manifolds~\cite{Bergshoeff:2002qk}.

A single hypermultiplet contains (see Table~\ref{tbl:multiplets}) four real scalars and two spinors, called the hyperinos, subjected
to a symplectic Majorana reality condition, as explained in Appendix~\ref{spinors}. For $n$ hypermultiplets, we
introduce real scalars $q^X(x)$, with $X=1,\dots ,4n$, and spinors
$\zeta^A(x)$ with $A=1,\dots ,2n$. To formulate the symplectic Majorana
condition, we introduce two matrices $\rho_A {}^B$ and $E_i{}^j$, with $i,j=1,2$, satisfying~(\ref{defrho}).
These define symplectic Majorana conditions for the fermions and
supersymmetry transformation parameters~\cite{VanProeyen:2001ng}:
\begin{equation}
  \alpha {\cal C}\gamma _0\zeta ^B \rho_B{}^A=\left(\zeta ^A\right)^*, \qquad
\alpha {\cal C}\gamma _0\epsilon ^j E_j{}^i=\left(\epsilon ^i\right)^*,
\end{equation}
where ${\cal C}$ is the charge conjugation matrix, and $\alpha $ is an
irrelevant number of modulus 1. We can always adopt the basis where
$E_i{}^j=\varepsilon_{ij}$, and will further restrict to that.

Since the scalar fields will be described in terms of a nonlinear sigma model, they are interpreted as coordinates on some target space,
and requiring the on-shell closure of the supersymmetry algebra imposes
certain conditions on the target space, which we derive below.
We will show how the closure of the supersymmetry transformation laws
leads to a hypercomplex manifold. More precisely, the closure of the algebra on the
bosons will yield the defining equations for this geometry, whereas the
closure of the algebra on the fermions and its further consistency will lead
again to equations of motion, independent of an action.

The supersymmetry transformations (with $\epsilon^i$ constant parameters)
of the bosons $q^X\!(x)$, are parametrized by arbitrary functions
$f^X_{iA}(q)$. For the transformation rules of the fermions we write
the general form compatible with the supersymmetry algebra. This
introduces other general functions $f_X^{iA}(q)$ and
$\omega_{XB}{}^A(q)$:\footnote{In fact, we can write down a more general
supersymmetry transformation rule for the fermions than
in~(\ref{SUzeta}), but using Fierz relations and simple considerations
about the supersymmetry algebra, we can always bring its form into the one
written above.}
\begin{eqnarray}
\delta_Q (\epsilon) q^X
&=& - \rmi \bar\epsilon^i \zeta^A f_{iA}^X \,,\nonumber\\
\delta_Q (\epsilon) \zeta^A &=& \frac 12 \rmi \slashed{\partial} q^X
f_X^{iA} \epsilon_i -\zeta^B \omega_{XB}{}^A \big( \delta_Q (\epsilon) q^X
\big)\,. \label{SUzeta}
\end{eqnarray}
The functions satisfy reality properties~(\ref{complconjViel}) and 
\begin{equation}
 \left( \omega_{XA}{}^B\right)^*= \left(\rho ^{-1}\omega _X\rho
 \right)_A{}^B\,.
 \label{realfunctions}
\end{equation}
in order to be consistent with the reality of $q^X$ and the symplectic Majorana condition.

A priori, the functions $f_{iA}^X$ and $f^{iA}_X$ are independent, but the
commutator of two supersymmetries on the scalars only gives a translation
if we impose
\begin{eqnarray}\label{cov_const}
f^{iA}_Y f^X_{iA} &=& \delta_Y^X \,,\qquad f^{iA}_Xf^X_{jB}=\delta^i_j
\delta^A_B\,,\nonumber\\
\covder_Y f_{iB}^X &\equiv&
\partial_Y f_{iB}^X - \omega_{YB}^{\quad A} f_{iA}^X + \Gamma _{ZY}^{\quad
X} f_{iB}^Z = 0\,,
\end{eqnarray}
where $\Gamma _{XY} {}^Z(q)$ is some object, symmetric in the lower indices.

We ask that the supersymmetry transformation rules are covariant with respect to two
kinds of reparametrizations. The first ones are the target space
diffeomorphisms, $q^X\rightarrow {\widetilde q}^X(q)$, under which
$f^X_{iA}$ transforms as a vector, $\omega_{XA}{}^B$ as a one-form, and
$\Gamma_{XY}{}^Z$ as a connection. The second set are the
reparametrizations which act on the tangent space indices $A,B,\ldots$ On
the fermions, they act as
\begin{equation}
\zeta^A \rightarrow {\widetilde \zeta}^A(q)=\zeta^B
U_B{}^A(q)\,,\label{ferm-equiv}
\end{equation}
where $U(q)_A{}^B$ is any invertible matrix compatible with the reality conditions, i.e. any $\Gl(n,\mathbb{H})$ transformation, written as a complex $2n\times 2n$ matrix. In general, such a
transformation brings us into a basis where the fermions depend on the
scalars $q^X$. In this sense, the hypermultiplet is written in a special
basis where $q^X$ and $\zeta^A$ are independent fields. The supersymmetry
transformation rules~(\ref{SUzeta}) are covariant
under~(\ref{ferm-equiv}) if we transform $f^{iA}_X(q)$ in the fundamental of $\Gl(n,\mathbb{H})$ and
$\omega_{XA}{}^B$ as a connection,
\begin{equation}
\omega_{XA}{}^B\rightarrow {\widetilde \omega}_{XA}{}^B=[(\partial_X
U^{-1})U+ U^{-1}\omega_X U]_A{}^B\,.
\end{equation}
Hence, $\{f_X{}^{iA}\}$ are covariantly constant Vielbeine on the target space $M_t$ and define frames. Due to their transformation properties and reality conditions, they can be used to reduce the structure group of the frame bundle $LM_t$ to $\Gl(n,\mathbb H)$. This makes us conclude that the target manifold is almost hypercomplex.

As a consequence, we can build the $\Gl(n,\mathbb H)$ invariant tensor fields, i.e. the almost hypercomplex structures~(\ref{defhcstr}), defining the corresponding G-structure using~(\ref{defJf}).
As the Vielbeine are covariantly constant~(\ref{cov_const}), the almost hypercomplex structures either are with respect to the torsionless connection $\Gamma$. Hence, the G-structure is 1-flat and the target space is hypercomplex.

Using the integrability condition for the Vielbeine, we arrive at~(\ref{RdecompJ}) with vanishing $\symp(1)$ curvature. We can now introduce a new object $W$ which is symmetric in its three lower indices. 
\begin{eqnarray}
  f_{Ci}^X f_{jD}^Y{R}_{XYB} {}^A&=& - \frac 12 \varepsilon_{ij}
W_{CDB}{}^A\,,\quad W_{CDB}{}^A \equiv  f^{iX}_C f^Y_{iD} {R}_{XY}{}_B {}^A = \frac
12f^{iX}_C f^Y_{iD} f_{jB}^Z f_V^{Aj} R_{XYZ}{}^V\,, \label{def-W}\nonumber\\
\end{eqnarray}
where the curvatures were defined in~(\ref{defcurvs}). This tensor $W$ will appear in the equations of motion.

Let us now summarize what are the independent objects and relations. We have found that $q^X$ are coordinates on the target space $M_t$ and that $\{f_X^{iA}\}$ are Vielbeine, covariantly constant with respect to the torsionless affine connection $\Gamma$ on $TM_t$ and the induced $\Gl(n,\mathbb H)$ connection $\omega$ on $aLM_t$.\footnote{Since the $\Symp(1)$ factor of the structure group of $LM_t$ is trivial, we work in a gauge in which the corresponding connection vanishes.} With these Vielbeine, we could construct a field of hypercomplex structures, that is preserved by $\Gamma$. Moreover, $\Gamma$ equals the Obata connection and the manifold $M_t$ is hypercomplex (see Section~\ref{sss:quatlikestruct}). Moreover, the fermions $\zeta^A$ are sections of the $\Gl(n,\mathbb{H})$ subbundle of $aLM_t$.

All this considerations lead us to define the covariant variation of the
fermions:
\begin{equation}
{\widehat \delta} \zeta^A\equiv \delta \zeta^A+\zeta^B\omega_{XB}{}^A
\delta q^X\,, \label{cov-var}
\end{equation}
for any transformation $\delta$. Two models related by either target space
diffeomorphisms or fermion reparametrizations of the
form~(\ref{ferm-equiv}) are equivalent; they are different coordinate
descriptions of the same system. Thus, in  a covariant formalism, the
fermions can be functions of the scalars. However, the expression
$\partial _X\zeta ^A$ makes only sense if we compare different bases and similarly for $\zeta^B \omega_{XB}{}^A$  as the connection has no
absolute value. The only covariant object is therefore the covariant derivative
\begin{equation}
  \covder_X \zeta ^A\equiv \partial_X \zeta ^A+\zeta ^B\omega
  _{XB}{}^A\,,
\end{equation}
which is a special case of~(\ref{cov-var}).
The covariant transformations are also a useful tool to calculate any
transformation on e.g.\ a quantity $W_A(q)\zeta ^A$:
\begin{eqnarray}
  \delta \left(W_A(q)\zeta ^A\right)&=& \partial _X \left(W_A\zeta ^A\right) \delta q^X +
   W_A\left.\delta \zeta ^A\right|_{q} \nonumber\\
   &=& \covder_X \left(W_A\zeta ^A\right)\delta q^X+ W_A\left(\widehat \delta \zeta
   ^A - \covder_X\zeta ^A\delta q^X\right) \nonumber\\
   &=& \left( \covder_XW_A\right)\delta q^X \zeta ^A + W_A\,\widehat \delta \zeta
   ^A \,.
\end{eqnarray}
We will frequently use the covariant transformations~(\ref{cov-var}). It
can similarly be used on target space vectors or tensors. For instance, for a
target space vector field $\Delta^X$:
\begin{equation}
  \widehat \delta \Delta ^X= \delta \Delta ^X + \Delta ^Y \Gamma _{ZY}{}^X\,\delta
  q^Z\,.
\end{equation}

Now we consider the commutator of supersymmetry on the fermions, which
will determine the equations of motion for the hypermultiplets.
Using~(\ref{cov_const}), (\ref{RdecompJ}) and~(\ref{def-W}), we find\footnote{To obtain this result,
we use Fierz identities expressing that only the cubic fermion
combinations of~\cite[(A.11)]{Bergshoeff:2001hc} are independent:
\[ \zeta^{(B}\bar{\zeta}^C \gamma_a \zeta^{D)} =- \gamma_a\zeta^{(B} \bar{\zeta}^C \zeta^{D)}\,.
\]}
\begin{equation}
[\delta_Q ( \epsilon_1),\delta_Q ( \epsilon_2)] \zeta^A =\frac 12\partial_a
\zeta^A \bar{\epsilon}_2 \gamma^a \epsilon_1 + \frac 14 \Gamma ^A
\bar{\epsilon}_2 \epsilon_1 -\frac 14 \gamma_a \Gamma ^A \bar{\epsilon}_2
\gamma^a \epsilon_1 \,. 
\end{equation}
The $\Gamma ^A$ are the nonclosure functions, and define the equations
of motion for the fermions,
\begin{equation}
\Gamma^A = \slashed{\covder} \zeta^A + \frac 12 W_{CDB} {}^A
 \zeta^B\bar{\zeta}^D \zeta^C\equiv 0\,, \label{eqmozeta}
\end{equation}
where we have introduced the covariant derivative with respect to the
transformations~(\ref{cov-var}),
\begin{equation}
  \covder_\mu  \zeta^A \equiv \partial_\mu  \zeta^A + (\partial_\mu  q^X)
\zeta^B\omega_{XB} {}^A\,.
 \label{defDzeta}
\end{equation}
By varying the equations of motion under supersymmetry, we derive the
corresponding equations of motion for the scalar fields:
\begin{equation}
\widehat \delta_Q(\epsilon ) \Gamma^A =  \frac 12 \rmi f_{X}^{iA}
\epsilon_i \Delta ^X \,,\label{delQGamma3}
\end{equation}
where
\begin{eqnarray}
 \Delta^X&=&\Box q^X -\frac 12 \bar{\zeta}^B \gamma_a
\zeta^D
\partial^a q^Y f_Y^{iC}f_{iA}^X W_{BCD} {}^A-\frac{1}{4}\covder_Y W_{BCD}{}^A \bar{\zeta}^E \zeta^D
\bar{\zeta}^C \zeta^B f_E^{iY}f_{iA}^X\equiv 0\,,\nonumber\\ \label{covscal}
\end{eqnarray}
and the covariant d'Alembertian is given by
\begin{equation}
\Box q^X = \partial_a \partial^a q^X+ \left( \partial_a q^Y\right)
\left(\partial^a q^Z \right) \Gamma_{YZ} {}^X\,.
\end{equation}

The supersymmetry algebra thus imposes the
constraints~(\ref{cov_const}) and the equations of
motion~(\ref{eqmozeta}) and~(\ref{covscal}). These form a multiplet, as acting with a supercharge on $\Delta^X$ again yields the bosonic and fermionic equations of motion.

The most important conclusion to draw from the above is the fact that hypermultiplets can parametrize a hypercomplex manifold. This is a generalization of the old result~\cite{Alvarez-Gaume:1981hm} that the scalars in the corresponding nonlinear sigma model Lagrangian are coordinates on a hyperk\"ahler manifold. 
\subsection{Taking everything off-shell}\label{os}
We will now show that the introduction of an action is only possible if the target spaces for the vector and hypermultiplets are respectively very special real and hyperk\"ahler. To be as general as possible, we will use the BV formalism to prove this, since in that case, we only have to assume the existence of an action without referring to its explicit form.
\subsubsection{Vector multiplets}
We will show that the existence of an action reduces the set of equations in Section~\ref{5d} to the well-known ones of very special geometry~\cite{Gunaydin:1984bi,deWit:1992cr}. More specifically, we will show that we can construct a standard action only if the object $\gamma^I_{JK}$ can be linked to the symmetric, gauge-invariant three-tensor $C_{IJK}$.

In a general nonlinear sigma model, the equations of motion transform covariantly under coordinate transformations and as a consequence, there appears a connection in the kinetic term for the scalars  
\begin{equation}
\Box \varphi^X=\partial_\mu\partial^\mu \varphi^X+\Gamma^X_{YZ}\partial_\mu\varphi^Y\partial^\mu\varphi^Z\; .
\end{equation}
In general, this (torsionless) connection is affine. If we demand that the equation of motion is to be derivable from an action with a standard kinetic term, i.e.
\begin{equation}
\mathcal{L}=-\frac12g_{XY}\partial_\mu \varphi^X\partial^\mu \varphi^Y+\dots\; ,
\end{equation}
the connection moreover has to be the Levi-Civita connection. Thus, since the existence of an action requires the introduction of a new object, namely the metric on the target space, the target space geometry becomes Riemannian.   

This is a rather general observation for nonlinear sigma models, but in the case of vector multiplets, more care is needed. The reason is mainly that we are working in special coordinates in which target space diffeomorphisms are not manifest. Therefore, the equations of motion~(\ref{psiEOM})-(\ref{AEOM}) are not covariant under general coordinate transformations $\sigma^I\to \sigma'{}^I(\sigma)$ if $\gamma$ would transform as a bona fide affine connection. Hence, the object $\gamma^I_{JK}$ should rather be seen as the on-shell counterpart of $C_{IJK}$ than to be considered a connection. This cautionary remark aside, we will prove that in the off-shell case, $\gamma^I_{JK}$ can be related to the Levi-Civita connection corresponding to the special real metric (in special coordinates) $g_{IJ}=C_{IJK}\sigma^K$. 

Another way to encounter the need to introduce one more object to characterize the theory if the equations of motion~(\ref{psiEOM})--(\ref{AEOM}) are to be derivable from an action functional $\mathcal S$, is to note that we want the nonclosure functional $\Gamma^I$ to be proportional to the equation of motion computed from the action
\begin{equation}\label{EOMpsi}
\frac{\delta \mathcal S}{\delta \bar{\psi}_i^I}=g_{IJ}\Gamma^{iJ}\equiv 0\, ,
\end{equation}
and similarly for the other field equations.
We will suppose that the object $g_{IJ}$ is symmetric, depends on the scalars $\sigma^I$ only and that it is invertible, $g_{IJ}g^{JK}=\delta_I^K$.  

Paralleling the discussion in~\cite{Stelle:2003rr}, we will use the Batalin-Vilkovisky formalism~\cite{Batalin:1983jr,Henneaux:1990jq,Gomis:1995he} to retrieve the conditions for the existence of an action.
We will therefore introduce translational ghosts $c^\mu$, supersymmetry ghosts $c^i$, gauge symmetry ghosts $\alpha^I$ and antifields (denoted by ${}^*$) and expand the BV action up to antifield number 2.
\begin{eqnarray}
\mathcal S_{BV}&=&\int d^5x\left( \mathcal{L}_0+\mathcal{L}_1+\mathcal{L}_2+\dots\right)\, ,\nonumber\\
\mathcal{L}_1&=&\sigma_I^*c^\mu\partial_\mu\sigma^I+A_I^{*\mu}c^\nu\partial_\nu A_\mu^I +\bar{\psi}_{iI}^*c^\mu\partial_\mu\psi^{iI}+\frac12\rmi\sigma_I^*\bar{\psi}^Ic +\frac12A_I^{*\mu}\bar{\psi}^I\gamma_\mu c\nonumber\\&& 
+\bar{\psi}^*_{iI}\left(-\frac12 \rmi \slashed{\mathfrak{D}}\sigma^Ic^i-\frac14\slashed{F}^Ic^i-\frac12\rmi\gamma^I_{JK}\bar{\psi}^{iJ}\psi^{jK}c_j +f^{(ij)I}c_j\right)\nonumber\\&&-g\sigma_I^*f_{JK}{}^I\alpha^J\sigma^K
+A_I^{\mu *}\mathfrak{D}_\mu \alpha^I-g\bar{\psi}^*_{iI}f_{JK}{}^I\alpha^J\psi^{iK}\; ,\\
\mathcal{L}_2&=& -\frac14 c_\mu^*\bar{c}\gamma^\mu c+\bar{c}^{(i}\psi_I^{*j)}\bar{c}_i\psi_{jJ}^*g^{IJ}
-\frac12g\alpha_I^*f_{JK}{}^I\alpha^J\alpha^K+\frac14\rmi \alpha_I^*\sigma^I\bar{c}c+\frac14\alpha_I^*A_\mu^I\bar{c}\gamma^\mu c\;, \nonumber
\end{eqnarray}
where $\mathcal{L}_0$ denotes the (unknown) classical Lagrangian. Note that the inverse of the new object $g_{IJ}$ introduced in~(\ref{EOMpsi}) is now appearing in $\mathcal L_2$, in order to encode correctly the supersymmetry algebra. Of course, the BV action can have terms with higher antifield number and to consider the form of these terms, we need to know the dimension, ghost number and antifield number of all fields (see Table~\ref{dimtable}). For consistency, the dimension of each field together with its antifield should add up to the same number, which we take to be equal to one. As a consequence, none of the fields have negative dimension. Using the fact that each term in the Lagrangian should have dimension two and vanishing ghost number, we can construct terms with antifield number higher than two, but none of these terms would spoil the arguments below. 

\begin{table}[!tb]
\begin{center}
\begin{tabular}{|c|c|c|c||c|c|c|c|}
\hline
field&dim&g&afn&antifield&dim&g&afn\\
\hline \hline
$\sigma^I$&$0$&$0$&$0$&$\sigma_I^*$&$1$&$-1$&$1$\\
$A_\mu^I$&$0$&$0$&$0$&$A_I^{*\mu}$&$1$&$-1$&$1$\\
$\psi^{iI}$&$1/2$&$0$&$0$&$\psi_{iI}^*$&$1/2$&$-1$&$1$\\
$c^\mu$&$0$&$1$&$0$&$c_\mu^*$&$1$&$-2$&$2$\\
$c^i$&$1/2$&$1$&$0$&$c_i^*$&$1/2$&$-2$&$2$\\
$\alpha^I$&$0$&$1$&$0$&$\alpha_I^*$&$1$&$-2$&$2$\\
$\mathcal{L}$&$2$&$0$&$-$&&&&\\
\hline
\end{tabular}
\end{center}
\caption{\it Dimension (dim), ghost number (g) and antifield number (afn) of all fields.}\label{dimtable}
\end{table}

To find the conditions imposed on $g_{IJ}$, we now concentrate on terms in the master equation~(\ref{master}) that are cubic in the supersymmetry ghost $c^I$ and quadratic in the gaugino antifield $\psi_{iI}^*$ and find
\begin{eqnarray}
&&\frac12 \rmi \partial_Ig^{JK}\bar{\psi}^Ic \bar{c}^{(i}\psi^{*j)}_J\bar{c}_i\psi_{jK}^*-2\rmi g^{IJ}\gamma^K_{IL}\bar{\psi}^*_{(iK}c_{k)}\bar{\psi}^{kL}c_j\bar{c}^{(i}\psi^{*j)}_J\\&&=
\frac12\rmi\bar{\psi}^Ic \bar{c}^{(i}\psi^{* j)}_J\bar{c}_i\psi^*_{jK}(\partial_Ig^{JK}+2\gamma^{(J}_{IL}g^{K)L})
+2\rmi\gamma_{IL}^{[J}g^{K]L}\bar{c}_{(i}\psi_{j)J}^*\bar{c}^{(i}\psi^{*k)}_K\bar{\psi}^{jI}c_k=0\; .\nonumber
\end{eqnarray}
This is equivalent with the following conditions. 
\begin{eqnarray}
&&\partial_Ig^{JK}+2g^{L(J}\gamma^{K)}_{IL}=0\; ,\nonumber\\
&&\gamma_{I,JK}=g_{IL}\gamma^L_{JK}=\gamma_{(I,JK)}\; .
\end{eqnarray}
The first condition means that the object $\gamma$ introduced in~(\ref{AnsatzA}) can now be seen as a Levi-Civita connection with respect to $g_{IJ}$, which we will call the metric from now on. The second is the complete symmetry of the first Christoffel connection coefficients. Applying those conditions to~(\ref{gamma1}), we find that these first Christoffel connection coefficients should be constant, and we will take
\begin{equation}
\gamma_{I,JK}=\frac12 C_{IJK}\; .
\end{equation}
This relation implies that $g_{IJ}=C_{IJK}\sigma^K+a_{IJ}$ where the last term is an integration constant. 
In the Abelian case, the only other condition we need to consider is~(\ref{FI1}), which implies that $f^{(ij)}_I=g_{IJ}f^{(ij)J}$ now is constant.

The non-Abelian case is slightly more involved as there are two more conditions to be solved. Using the above expressions for the metric and the connection,~(\ref{gamma2}) can first of all be solved trivially if the metric is constant, as the corresponding $\gamma^I_{JK}$ is zero. If the three-tensor $C_{IJK}$ is not vanishing, the same condition~(\ref{gamma2}) is solved if $a_{IJ}=0$ and the three-tensor is gauge-invariant, i.e.
\begin{equation}
f_{I(J}{}^MC_{KL)M}=0\; .
\end{equation}
In both cases, the condition on the FI terms~(\ref{FI2}) implies that they are vanishing.
Hence, by imposing the existence of an action, we have recovered the old results of very special geometry~\cite{Gunaydin:1984bi,deWit:1992cr}. This means first of all that the real scalars parametrize a manifold with metric $\sigma^K C_{IJK}$ where $C_{IJK}$ is a symmetric gauge-invariant three-tensor. Secondly, the FI terms can only be present in the Abelian sectors of the theory and are forced to be constants.

To conclude this section, we give the action (for $a_{IJ}=0$) together with the field equations.  
\begin{eqnarray}
S\!\!\!\!&=&\!\!\!\!\int d^5x \Bigg[-\frac12 g_{IJ}(\mathfrak{D}_\mu \sigma^I\mathfrak{D}^\mu \sigma^J+\frac12F^I_{\mu \nu}F^{J\mu \nu}+\bar{\psi}^I\slashed{\mathfrak{D}}\psi^J)-g^2g_{IJ}f^I_{ij}f^{ijJ}
-\frac18\rmi C_{IJK}\bar{\psi}^I\slashed{F}^J\psi^K\nonumber\\&&-\frac14\rmi g f_{IJ}{}^Kg_{KL}\sigma^L\bar{\psi}^I\psi^J
+\frac12\rmi gC_{IJK}f^{ijI}\bar{\psi}_i^J\psi_j^K
 +\frac1{16} C_{IJM}C_{KLN}g^{MN}\bar{\psi}^{iI}\psi^{jJ}\bar{\psi}_i^K\psi_j^L\nonumber\\&&
-\frac1{24}C_{IJK}\varepsilon^{\mu \nu \rho \sigma \tau}A_\mu^I(F^J_{\nu \rho}F^K_{\sigma \tau} +gf_{LM}{}^JA_\nu^LA_\rho^MF^K_{\sigma \tau} +\frac25g^2f_{LM}^Jf_{NP}{}^KA_\nu^LA_\rho^MA_\sigma^NA_\tau^P)\Bigg]\; ,\nonumber\\
&&\!\!\!\!\!\!\!\!\!\!\!\!\!\!\!\frac{\delta S}{\delta \bar{\psi}_i^I}=g_{IJ}\Gamma^{iJ}\; ,\quad 
\frac{\delta S}{\delta \sigma^I}=g_{IJ}\Delta^J-\frac12 C_{IJK}\bar{\psi}^J\Gamma^K\; ,\quad\frac{\delta S}{\delta A^{\mu I}}=g_{IJ}\Xi_\mu^J-\frac12 \rmi C_{IJK}\bar{\psi}^J\gamma^\mu \Gamma^K\; .\nonumber\\
\end{eqnarray}
The action without FI terms equals the one in~\cite{Bergshoeff:2002qk} after elimination of the auxiliary field $Y^{ijI}$ by its algebraic equation of motion. 
\subsubsection{Hypermultiplet}\label{sss:hypers}
We can repeat the above construction for the hypermultiplets. In that case, the equation of motion for the fermions is deduced from an action in a way similar to~(\ref{EOMpsi}),
\begin{equation}
\frac{\delta \mathcal{S}}{\delta \zeta^A}=\Gamma^BC_{BA}\,.
\end{equation}
Using again the BV formalism, the existence of this action then implies that $C_{AB}$ should be covariantly constant. 
\begin{equation}\label{Cq}
\mathfrak{D}_X C_{AB}=0\,.
\end{equation}
Otherwise stated, the requirement of an action implies the existence of an invariant fibrewise symplectic form on the associated bundle $aLM_t$ of the target manifold. As a consequence, we can construct a covariantly constant metric using the Vielbeine
\begin{equation}
g_{XY}= f_X^{iA}f_Y^{jB}C_{AB}\varepsilon_{ij}\,.
\end{equation}
This is equivalent with the fact that the target space is hyperk\"ahler, as $(g,\vec J)$ defines a 1-flat almost hypercomplex hermitian structure on the target space $M_t$. Note that due to this G-structure, the structure group of the frame bundle $LM_t$ can be reduced to $\Sp(n)$ (or a noncompact version thereof, see below). Consequently, the holonomy group should also be included in that group. 

We may moreover choose $C_{AB}$ to be constant. To prove this, we look at the
integrability condition for~(\ref{Cq})
\begin{equation}
\left[\covder_X,\covder_Y\right]C_{AB}=0=-2{R}_{XY[A}{}^CC_{B]C}\,.
\end{equation}
This implies that the antisymmetric part of the connection
$\omega_{XAB}\equiv \omega _{XA}{}^CC_{CB}$ is in pure gauge, and can be
chosen to be zero. If we do so, the covariant constancy condition for
$C_{AB}$ reduces to the equation that $C_{AB}$ is just constant. For this
choice, the connection $\omega_{XAB}$ is symmetric, so the structure
group $\Gl (r,\mathbb{H})$ breaks to $\USp(2r-2p,2p)$. The signature is
the signature of $d_{CB}$, which is defined as $C_{AB}=\rho_A {}^C
d_{CB}$ where $\rho_A {}^C$ was given in~(\ref{defrho}). However, we will
allow $C_{AB}$ also to be nonconstant, but covariantly constant.

Thus, similar to the case of vector multiplets, we have recovered the old result~\cite{Alvarez-Gaume:1981hm} that the target space of hypermultiplet scalars has to be hyperk\"ahler by requiring the existence of an action, which then reads
\begin{equation}
\mathcal S= \int d^5x [-\frac 12 g_{XY} \partial_a q^X
\partial^a q^Y+ \bar{\zeta}_A \slashed{\covder} \zeta^A-\frac 14 W_{ABCD}
\bar{\zeta}^A \zeta^B \bar{\zeta}^C \zeta^D]\,.
\label{rigidhyperaction}
\end{equation}

\chapter{Local on-shell supersymmetry}\label{suptencalc}
In this chapter, we will discuss on-shell realized supersymmetry in theories coupled to gravity. Following closely the ideas in the previous chapter, we will focus on one interesting example, namely that of local $\mathcal N=2$ five-dimensional super-Poincar\'e vector- and hypermultiplets. However, as the construction of local supersymmetric theories requires rather tedious calculations, we will rely on the so-called superconformal tensor calculus. 

During the past decades, the conformal approach to the construction of various supergravities has proven to be very powerful. The method starts from a theory with superconformal invariance, while the final super-Poincar\'e theory results from breaking the extra symmetries. Since the conformal theory admits a larger symmetry algebra, it has a simpler although nontrivial form. Breaking the conformal symmetries yields drastic changes in the target space, such that this approach often gives clear insights in the geometry of the models under study. In the present case, the construction will lead us to conclude that the different quaternionic-like manifolds discussed in Chapter~\ref{prelim} are pairwise related in a natural way.

The content of this chapter is as follows. We will first list the ingredients of the conformal construction and subsequently apply it on a simple example. Then, we will take the different steps of the conformal approach in the context of $\mathcal N=2$, $d=5$ theories. Firstly, we will add extra structure to the rigid super-Poincar\'e off-shell vector- and on-shell hypermultiplet to make them invariant under superconformal symmetry and we will subsequently gauge that group. Then we will discuss the geometry of the hypermultiplet target space of the models and show how it decomposes if we fix the extra symmetries. The fixing of these symmetries in the physical theory is the final step, which we will however only perform explicitly in the context of quaternionic-K\"ahler target spaces, i.e. in theories with actions. 

The superconformal approach was discussed in e.g.~\cite{VanProeyen:1983wk,Bergshoeff:1986mz} and was applied in the context of $\mathcal N=2$ four-dimensional theories described by an action functional in~\cite{Andrianopoli:1997cm,DeWit:1999fp,DeWit:2001dj}. It was reviewed recently in a few Ph.D. theses~\cite{vanhoof,claus,derix,halbersma,dewit,Cucu:2003kj}. Five-dimensional $\mathcal N=2$ supergravities that have an action prescription were discussed in~\cite{Cremmer:1980gs,Gunaydin:1984bi,Gunaydin:1985ak,Gunaydin:1999zx,Gunaydin:2000xk,Ceresole:2000jd,Kugo:2000af}. The conformal programme in five dimensions was followed in~\cite{Bergshoeff:2001hc,Bergshoeff:2002qk,Bergshoeff:inprep1,Bergshoeff:inprep2} yielding more complete results in the context of $\mathcal N=2$ theories with actions, and entirely new geometrical insights.  
\section{Introduction}
The importance of the conformal group, although in a completely different setting, was first stressed in the seminal paper~\cite{Coleman:1967ad}. There, the maximal bosonic space-time symmetry group that can act on the S-matrix elements, without rendering them trivial, was studied. It was found that when the symmetries act
in finite dimensional representations and the field theory is defined on four-dimensional Minkowski space-time, the maximal space-time symmetry group is the conformal group, if all particles are massless. A similar result followed in the context of supersymmetry~\cite{Haag:1975qh}, yielding the superconformal groups.
\subsection{Algebraic point of view}
The conformal group is the set of transformations of flat space that preserves angles, i.e. if $\eta$ is a Minkowski metric with signature $(1,d-1)$, if $V,W \in \mathbb R^d$ and if $|V|=\sqrt{\eta(V,V)}$, then the conformal group preserves $\eta(V,W)/(|V||W|)$. These transformations are generated by so-called conformal Killing vectors, defined in~(\ref{confkilvect}). We will start our discussion of the conformal approach by listing some peculiarities about the (super)conformal algebra.
\subsubsection{The conformal algebra}
In Cartesian coordinates and with the standard Minkowski metric, the conformal Killing equation~(\ref{confkilvect}) can be written as
\begin{equation}\label{confkilvecteq}
\partial_{(\mu}\xi_{\nu)}-\frac1d\eta_{\mu \nu}\partial_\rho \xi^\rho=0\,,
\end{equation}
where $\xi$ is the conformal Killing vector. In the case that $d>2$, this equation admits a finite number of solutions, parametrized by $(\lambda_P,\lambda_M,\lambda_D,\lambda_K)$. A general conformal Killing vector of flat space thus reads
\begin{equation}\label{intrxi}
\xi^\mu=\lambda_P^\mu +\lambda_M^{\mu \nu}x_\nu+\lambda_Dx^\mu+(\lambda_K^\mu x^2-2 \lambda_K \cdot x x^\mu )\,.
\end{equation}
The separate terms in this expansion generate different symmetries, and referring to the original definition~(\ref{confkilvect}), these generators can correspond to different functions $f$. Calculating their Lie-brackets using~(\ref{compLieder}), we arrive at the conformal Lie-algebra.
Symbolically, if we  denote the infinitesimal transformations induced by $\xi$ as
\begin{equation}
\xi=\lambda_P^\mu P_\mu+\lambda_M^{\mu \nu}M_{\mu\nu}+\lambda_DD+\lambda_K^\mu K_\mu\,.
\end{equation}
we arrive at the following algebra
\begin{equation}
\begin{array}{ll}
{\lbrack }M_{\mu \nu},M^{\rho \sigma}\rbrack =-2\delta_{\lbrack \mu}^{\lbrack \rho}M_{\nu\rbrack }{}^{\sigma\rbrack }\,,\qquad&\lbrack P_\mu,M_{\nu \rho}\rbrack =\eta_{\mu\lbrack \nu}P_{\rho\rbrack }\,,\\
\lbrack P_\mu,K_\nu\rbrack =2(\eta_{\mu \nu}D+2M_{\mu \nu})\,,&\lbrack K_\mu,M_{\nu \rho}\rbrack =\eta_{\mu\lbrack \nu}K_{\rho\rbrack }\,,\\
\lbrack D,P_\mu\rbrack =P_\mu\,,&\lbrack D,K_\mu\rbrack =-K_\mu\,.
\end{array}\label{confalg}
\end{equation}
It is standard knowledge that this algebra is isomorphic to $\so(2,d)$. More specifically, the map between the different generators reads 
\begin{equation}
\hat{M}^{\hat \mu \hat \nu}=\left( \begin{array}{ccc} M^{\mu \nu}&\frac14(P^\mu-K^\mu)&\frac14(P^\mu+K^\nu)\\
-\frac14(P^\mu-K^\mu)&0&-\frac12D\\
\frac14(P^\mu+K^\nu)&\frac12D&0\end{array}\right) \,.
\end{equation}
Note that this algebra preserves the scalar product defined by a metric $\hat \eta$, 
\begin{equation}
\hat{\eta}=\mbox{diag}(-,\underbrace{+,\dots,+}_{d\;\;\mbox{times}},-)\,.
\end{equation}
An important fact is that the $d$-dimensional Poincar\'e algebra is a subalgebra of~(\ref{confalg}). This is very important in the conformal construction of (super)gravity theories, since in the end, only this subalgebra will be represented nontrivially on the physical fields.
\subsubsection{Induced representations}
Another subalgebra of~(\ref{confalg}) is the one spanned by the generators $M$, $D$ and $K$, which enjoys the special property that the transformations leave fixed the space-time point $x=0$. Therefore it is called the stability subalgebra. Since fields are functions on space-time, they change under the action of the conformal algebra because space-time itself transforms. In general however, the fields themselves can also transform nontrivially under the action of the space-time symmetry algebra. For instance, a field can have different components that transform as a vector under rotations. This in general makes the transformation properties rather complicated. However, it turns out that the transformations at $x=0$ under the action of the stability subalgebra are sufficient to determine the transformation under the full conformal group at a general point in space-time. 

To make this more clear, we will concentrate on a theory with rigid conformal invariance, defined on flat Minkowski space-time. Focusing on the fields at $x=0$, the stability subalgebra is generated by $\Sigma$, $\Delta$ and $\kappa$ corresponding to $M$, $D$ and $K$ respectively. From the theory of induced representations, it then follows that the full transformation of the fields can be reconstructed. Indeed, for a general field $\phi$ the transformations read
\begin{eqnarray}
\delta_P \phi (x)&=&\xi^\mu \partial_\mu \phi(x)\, ,\nonumber\\
 \delta_M \phi
(x)&=&\frac 12 \lambda_M^{\mu\nu}(x_\nu\partial_\mu - x_\mu
\partial_\nu) \phi(x) + \delta_\Sigma(\lambda_M)\phi(x)\, ,\nonumber\\
 \delta_D
\phi(x)&=&\lambda_D x^\lambda \partial_\lambda \phi (x) + \d_\Delta
(\lambda_D)\phi(x)\, , \nonumber\\ 
\delta_K
\phi(x)&=&\lambda_K^\mu(x^2\partial_\mu - 2 x_\mu
x^\lambda\partial_\lambda)\phi(x)\ + \nonumber\\ &&{}+\big[\delta_\D (-2
x\cdot \lambda_K) + \d_\Sigma (-4x_{[\mu}\lambda_{K\nu]}) + \d_\kappa
(\lambda_K)\big] \phi(x)\, , \label{indreps}
\end{eqnarray}
with $\xi$ given in~(\ref{intrxi}).
Conversely, the transformations of a field under the conformal group at an arbitrary point in Minkowski space-time, can always be written in such a form. Hence, a general transformation on the field reads
\begin{equation}\label{conftrans}
\delta_C \phi=\big[\xi^\mu\partial_\mu+\delta_\Sigma(\partial_{[\nu}\xi_{\mu]})+\delta_\Delta(\frac1d \partial_\mu \xi^\mu)+\delta_\kappa(\lambda_K)\big]\phi\,.
\end{equation}
The Lagrangians (and the equations of motion) we will encounter do not explicitly depend on the space-time coordinates, hence they are translational invariant in a trivial way. Therefore, conformal invariance can be checked at $x=0$ and we can make use of the simpler transformations $\delta_\Sigma$, $\delta_\Delta$ and $\delta_\kappa$. Thus, if we write $\delta \phi$, we will always mean $\delta \phi\vert_{x=0}$. Note however that when computing the transformation properties of derivatives of fields, we have to take into account that some parameters in~(\ref{conftrans}) are position-dependent.
\subsubsection{The superconformal algebra}
The conformal algebra can be extended to a superalgebra in several ways. We will discuss its extension to the $\mathcal{N}=2$ case in five dimensions. As in the case of the super-Poincar\'e algebra~(\ref{N2d5poinc}), an extra bosonic $\su(2)$ R-symmetry has to be added, generated by $U_i{}^j$ acting on the supersymmetries in the fundamental representation. In addition to the super-Poincar\'e case, the algebra contains another $\su(2)$ doublet of fermionic symmetries, called special superconformal transformations and denoted by $S_\alpha^i$. 
The nontrivial relations of the complete superconformal algebra now read
\begin{equation}\label{supconfalg}
\begin{array}{ll}
{[}M_{ab},M_{cd}{]}=\eta_{a{[}c}M_{d{]}b}-\eta_{b{[}c}M_{d{]}a}\, ,\qquad&{[}P_a,M_{bc}{]}=\eta_{a{[}b}P_{c{]}}\,,\\[.1cm]
{[}K_a,M_{bc}{]}=\eta_{a{[}b}K_{c{]}}\,,&{[}U_i{}^j,U_k{}^l{]}=\delta_i^lU_k{}^j-\delta_k^jU_i{}^l\,,\\[.1cm]
{[}D,P_a{]}=P_a\,,&{[}D,K_a{]}=-K_a\,,\\[.1cm]
{[}P_a,K_b{]}=2(\eta_{ab}+2M_{ab})\,,&\\[.1cm]
{[}M_{ab},Q^i_\alpha{]}=-\frac14 (\gamma_{ab}Q^i)_\alpha\,,& {[}M_{ab},S_\alpha^i{]}=-\frac14(\gamma_{ab}S^i)_\alpha\,,\\[.1cm]
{[}D,Q_\alpha^i{]}=\frac12 Q_\alpha^i\,,&{[}D,S_\alpha^i{]}=-\frac12 S^i_\alpha\,,\\[.1cm]
{[}U_i{}^j,Q_\alpha^k{]}=\delta_i^kQ_\alpha^j-\frac12 \delta_i^jQ_\alpha^k\,,& {[}U_i{}^j,S_\alpha^k{]}=\delta_i^kS_\alpha^j-\frac12 \delta_i^jS_\alpha^k\,,\\[.1cm]
{[}K_a,Q_\alpha^i{]}=\rmi(\gamma_aS^i)_\alpha\,,&{[}P_a,S_\alpha^i{]}=-\rmi(\gamma_aQ^i)_\alpha\,,\\[.1cm]
\{Q_{i\alpha},Q^{j \beta}\}=-\frac12 \delta_i^j(\gamma^a)_\alpha{}^\beta P_a\,,& \{S_{i\alpha},S^{j\beta}\}=-\frac12 \delta_i^j (\gamma_a)_\alpha{}^\beta K_a\,,\\[.1cm]
\multicolumn{2}{l}{\{Q_{i\alpha},S^{j\beta}\}=-\frac12 \rmi (\delta_i^j\delta_\alpha^\beta D+\delta_i^j(\gamma^{ab})_\alpha{}^\beta M_{ab}+3\delta_\alpha^\beta U_i{}^j)\,,}
\end{array}
\end{equation}
with $a,\dots,d=0,\dots,4$, $\alpha,\beta=1,\dots,4$ and $i,j=1,2$.
In the classification of the superalgebras by Nahm~\cite{Nahm:1978tg},~(\ref{supconfalg}) is known as $F^2(4)$. 
It is important to note that the $\mathcal{N}=2$ super-Poincar\'e algebra~(\ref{N2d5poinc}) is a sub-superalgebra of the superconformal algebra. This is again very important in the conformal approach.

A general transformation of a field will now be given by a similar relation as~(\ref{conftrans}) and reads
\begin{equation}
\delta_C \phi=\big(\xi^\mu\partial_\mu+\delta_\Sigma(\partial_{[\nu}\xi_{\mu]})+\delta_\Delta(\frac1d \partial_\mu \xi^\mu)+\delta_\kappa(\lambda_K) +\delta_R(\lambda_R^{ij})+\delta_Q(\epsilon)+\delta_S(\eta)\big)\phi\,,
\end{equation}
where $\lambda_R$ parametrizes the R-symmetry, $\epsilon$ the supersymmetry and $\eta$ the special superconformal transformations. Closure of the algebra is now realized on the field $\phi$ if the parameter of supersymmetry is space-dependent,
\begin{equation}
\epsilon^i=\epsilon^i_0+\rmi x^\mu \gamma_\mu \eta^i\,,
\end{equation}
where both $\epsilon_0^i$ and $\eta^i$ are constant. This fact of a space-time dependent parameter is similar to~(\ref{conftrans}). 
\subsection{Scalar field coupled to general relativity}\label{scalfield}
To give a flavor of the way the conformal approach works and to introduce some more concepts, we will now discuss how the theory of relativity can be constructed using the conformal approach. This means that we will first of all develop a gauge theory of the conformal group. Subsequently, we will have to construct conformal matter fields and build conformally invariant actions with these fields. Note that in these actions, the conformal gauge fields will act as background fields, meaning that they are nondynamical. Finally, we will have to break the extra (conformal) symmetries, yielding a Poincar\'e-invariant theory. Let us therefore start by briefly reviewing how we can gauge the conformal group.
\subsubsection{Gauge theory}
The algebra of conformal symmetries~(\ref{confalg}), with generators $T_A$, can symbolically be written as
\begin{equation}
[\delta_A(\epsilon_1^A),\delta_B(\epsilon_2^B)]=\delta_C(\epsilon_2^B\epsilon_1^Af_{AB}{}^C)\,,
\end{equation}
where the structure constants determine the commutators in the usual way ($[T_A,T_B]=f_{AB}{}^CT_C$).\footnote{In case of fermionic symmetries in five dimensions, we have $\{T_A,T_B\}=-f_{AB}{}^CT_C$.} If we want to gauge the algebra, we have to introduce gauge fields $h_\mu^A$ for every generator, which is done in Table~\ref{confgauge}.
\begin{table}[tb]
\begin{center}
\begin{tabular}{|c|c|c|c|}
\hline
$P_a$&$M_{ab}$&$D$&$K_a$\\
\hline
$e_\mu^a$&$\hat{\omega}_\mu{}^{ab}$&$b_\mu$&$f_\mu^a$\\
\hline
\end{tabular}
\caption{\it The gauge fields of the conformal group.}\label{confgauge}
\end{center}
\end{table}
These gauge fields transform as
\begin{equation}\label{transfgauge}
\delta(\epsilon)h_\mu^A=\partial_\mu \epsilon^A+\epsilon^Ch_\mu^Bf_{BC}{}^A\,.
\end{equation}
We then proceed by introducing the covariant derivative $\mathfrak{D}_\mu=\partial_\mu-\delta_A(h_\mu^A)$. The commutator of two covariant derivatives now gives rise to curvatures
\begin{eqnarray}
[\mathfrak{D}_\mu , \mathfrak{D}_\nu]&=&-\delta_A(R_{\mu \nu}{}^A)\,,\nonumber\\
R_{\mu \nu}{}^A&=&2\partial_{[\mu}h_{\nu]}^A+h_\mu^Bh_\nu^Cf_{BC}{}^A\,.\label{defcurvgauge}
\end{eqnarray}
These curvatures satisfy the usual Bianchi identity $\mathfrak{D}_{[\mu}R_{\nu \rho]}{}^A=0$. 
\paragraph{Covariant general coordinate transformations}
Having introduced the gauge fields, we have come to face with two problems. To make this clear, it is necessary to keep in mind that in the end, we want to recover the theory of relativity. Hence, some of the properties of that theory should already be reflected in the conformal gauge theory. First of all, in general relativity, the $\hat{\omega}_\mu{}^{ab}$ is the spin-connection and can be written in terms of the Vielbeine. In the case at hand, we would similarly like $\hat{\omega}_\mu{}^{ab}$ to depend on the other gauge fields. Secondly, we want to substitute the local translations for arbitrary coordinate transformations, again similar to general relativity which in its canonical formulation, is invariant under coordinate transformations, rather than being a gauge theory of translations.

Because of the special role of translations, we will split the general index $A$ in translations $a$ and the rest, denoted by $I$. We then replace the translations by a linear combination of transformations, which we will call covariant general coordinate transformations (cgct), defined as follows
\begin{equation}\label{defcgct}
\delta_{\rm cgct}\equiv\delta_{\rm gct}(\xi)-\delta_I(\xi^\mu h_\mu^I)\,,
\end{equation}
where the subscript gct denotes the usual general coordinate transformations for which $\xi^\mu=\xi^ae_a{}^\mu$ is the parameter. The latter transformations are defined via the following action on all gauge fields
\begin{equation}
\delta_{\rm gct}h_\mu^A\equiv\xi^\nu \partial_\nu h_\mu^A+\partial_\mu \xi^\nu h_\nu^A=\delta_B(\xi^\nu h_\nu^B)h_\mu^A-\xi^\nu R_{\mu \nu}{}^A\,.
\end{equation}
Now that we have split the transformations into coordinate and gauge transformations, we can refine the concept of a covariant object, that was already introduced in Chapter~\ref{prelim}. This is a field or combination of fields that do not transform to a derivative on $\epsilon^I$ under a finite number of transformations $\delta_I$. The notion of a covariant object will be encountered frequently.

We can now study the properties of all fields under cgct and $I$-trans\-for\-ma\-tions. Starting with the Vielbeine, the cgct's are
\begin{equation}\label{cgctViel}
\delta_{\rm cgct}(\xi)e_\mu{}^a=\partial_\mu\xi^a+\xi^bh_\mu^Af_{Ab}{}^a-\xi^\nu R_{\mu \nu}(P^a)\,.
\end{equation}
We are now ready to make $\hat \omega$ a field-dependent object, by introducing the constraint that 
\begin{equation}\label{convconstr1}
R_{\mu \nu}{}^a(P)=2\partial_{[\mu}e_{\nu]}{}^a+2b_{[\mu}e_{\nu]}{}^a+2\hat{\omega}_{[\mu}{}^{ab}e_{\nu]b}=0\,,
\end{equation}
such that we can now solve $\hat \omega$ as
\begin{equation}
\hat{\omega}_{\mu}{}^{ab}=2e^{\nu[a}\partial_{[\mu}e_{\nu]}-e^{\nu[a}e^{b]\rho}e_{\mu c}\partial_\nu e_\rho{}^c+2e_{\mu}{}^{[a} b^{b]}\,.
\end{equation}
All transformations of the Vielbein, including the cgct of~(\ref{cgctViel}), therefore read
\begin{equation}
\delta e_\mu{}^a=\underbrace{\partial_\mu\xi^a+b_\mu\xi^a+\hat{\omega}_{\mu}{}^{ab}\xi_b}_{\delta_{\rm cgct}(\xi)}-\lambda_De_\mu{}^a -\lambda_M^{ab}e_{\mu b}\,.
\end{equation}
We can now repeat this for the other gauge fields. According to~(\ref{transfgauge}), they transform under $I$-transformations as follows
\begin{equation}
\delta_J(\epsilon^J)h_\mu^I=\partial_\mu \epsilon^I+\epsilon^J h_\mu^K f_{KJ}{}^I+\epsilon^J\mathcal{M}_{\mu J}{}^I\,.
\end{equation}
The object $\mathcal{M}$ is the covariant part in the transformation of the gauge field. In the case at hand, 
it only contains terms as a consequence of our use of splitting the gauge indices. In other words, 
\begin{equation}
\mathcal{M}_{\mu J}{}^I=e_\mu{}^af_{aJ}{}^I\,.
\end{equation}
In general however, this $\mathcal M$ can also contain auxiliary fields, present in some multiplets.
Under covariant general coordinate transformations, we find a transformation rule for the gauge fields similar to the one for Vielbeine~(\ref{cgctViel}),
\begin{eqnarray}
\delta_{cgct}(\xi)h_\mu^I&=&-\xi^\nu \hat{R}_{\mu \nu}{}^I-\xi^a h_\mu^J\mathcal{M}_{aJ}{}^I\,,\nonumber\\ \hat{R}_{\mu \nu}{}^I&=&2\partial_{[\mu}h_{\nu]}^I+h_\mu^J h_\nu^K f_{JK}{}^I-2h_{[\mu}^J \mathcal{M}_{\nu]J}{}^I\,.\label{defcovcurvgauge}
\end{eqnarray}
Here we have introduced the covariant curvature, denoted by a hat.
\paragraph{Conventional constraints}
The constraint~(\ref{convconstr1}) permitted us to solve for the spin connection algebraically and therefore it is called conventional. Hence, in such constraints, the field that should be solved for, has to appear linearly and without derivatives. In the case under consideration, the gauge field for special conformal transformations can be solved in a similar way by constraining the $\hat{R}(M)$ curvature. Imposing
\begin{equation}\label{convconstr2}
e^\nu{}_b\hat{R}_{\mu \nu}{}^{ab}(M)=0\,,
\end{equation}
the expression for that field reads
\begin{equation}\label{exprf}
f_\mu{}^a=\frac16\mathcal{R}_\mu{}^a-\frac1{48}e_\mu{}^ae^\nu{}_b\mathcal{R}_\nu{}^b \,,\qquad \mathcal{R}_\mu{}^a=e^\nu{}_b R_{\mu \nu}{}^{ba}\,,
\end{equation} 
where $R_{\mu \nu a}{}^b$ was defined in~(\ref{Rspinconnection}) but is computed in the above using $\hat \omega$. 
Note that we could also have solved $f_\mu{}^a$ from the D-curvature.\footnote{With `D-curvature', we mean the curvature computed with the gauge field for dilatations, i.e. containing $\partial b$ as in~(\ref{defcurvgauge}). We will use similar names to denote the other curvatures.} However, the Bianchi identity of the P-curvature links these two possibilities as it leads to
\begin{equation}
e_a{}^\mu e_b{}^\nu e_c{}^\rho \mathfrak D_{[\mu}R_{\nu \rho]}(P^d)= -\delta_{[a}^d \hat{R}_{bc]}(D)+\hat{R}_{[ab}(M_{c]}{}^d)=0\,.
\end{equation}
In conclusion, we were able to impose two constraints~(\ref{convconstr1}) and~(\ref{convconstr2}) which has led to the fact that the gauge fields $\hat{\omega}$ and $f$ became dependent on the Vielbeine and on the gauge field for dilatations $b$. Note that constraining these gauge fields is a way to eliminate degrees of freedom from the theory in order to build a gauge theory of minimal size.
\subsubsection{Constructing Poincar\'e gravity} \label{easyexample}
We will now use the above formalism to reconstruct the Einstein-Hilbert action. The starting point is a real scalar field $\varphi$ (called the compensator) that has Weyl weight $w$, meaning that it transforms under dilatations as $\delta_D \varphi=w\lambda_D \varphi$. Hence, the covariant derivative and d'Alembertian read
\begin{eqnarray}
\mathfrak{D}_\mu \varphi&=&\partial_\mu \varphi -w b_\mu \varphi\,,\nonumber\\
\Box \varphi&=&\eta^{ab}\mathfrak{D}_a\mathfrak{D}_b\varphi\nonumber\\
&=&e^{\mu a}(\partial_\mu \mathfrak{D}_a\varphi-(w+1)b_\mu \mathfrak{D}_a\varphi+\hat{\omega}_{\mu a}{}^b\mathfrak{D}_b\varphi +2w f_{\mu a } \varphi)\,.
\end{eqnarray}
We now consider the action 
\begin{equation}\label{easyconfapp}
\mathcal{S}=\int d^5x\; e (-1) \varphi\Box \varphi\,.
\end{equation}
Note that the action has actually the wrong sign for the kinetic energy, a feature that will be clarified below. To be the starting point for the conformal construction, the above action needs to admit full conformal invariance. Imposing $K$-invariance therefore fixes the Weyl weight to $w=3/2$ (in five space-time dimensions). 

The second step is to substitute all dependent fields for their expressions. For the gauge field $f_\mu{}^a$, it can be seen that only its contraction with a Vielbein appears. In that case, we have from~(\ref{exprf}) that
\begin{equation}
f_a{}^a\equiv f_\mu{}^a e_a{}^\mu=-\frac1{16}R(\hat{\omega})\,,
\end{equation}
where the right-hand side denotes the Ricci scalar computed with $\hat \omega$ (see Appendix~\ref{appconventions}). For the latter field, we will use the following decomposition 
\begin{equation}
\hat{\omega}_{\mu}{}^{ab}=\omega_\mu{}^{ab}+2e_{\mu}{}^{[a} b^{b]}\,,
\end{equation}
where $\omega$ is the usual expression for the spin connection in terms of the Vielbeine~(\ref{expromega}).

The next step is to break the symmetries that are not present in a pure Poincar\'e-invariant theory. First of all, note that when we substitute above expressions for the dependent fields in the action~(\ref{easyconfapp}), it cannot depend on the gauge field $b_\mu$ any more since this would be the only field transforming under $K$ symmetry. Therefore, we can break the special conformal transformations by choosing an arbitrary value for $b_\mu$, which we will take to be zero for convenience, and this does not change anything in the action. Next we have to deal with the dilatations, which we can fix by freezing the scalar field to a constant. More precisely, as the action can be written as
\begin{equation}
\mathcal{S}=-\int d^5x\;e \varphi \big[\partial_\mu g^{\mu \nu}\partial_\nu -\frac3{16}R(\omega)\big]\varphi\,,
\end{equation}
to retrieve the canonical normalized Einstein-Hilbert action, we have to put the scalar to the value 
\begin{equation}
\varphi=\sqrt{\frac 23}\frac{2}{\kappa}\,.
\end{equation}
Note that this clarifies why we had to start from an action with the wrong sign for the kinetic term. If we had not, we should have had to freeze the scalar to an imaginary value.
In conclusion, the action~(\ref{easyconfapp}) results in the standard Einstein-Hilbert action
\begin{equation}
\mathcal{S}=\frac1{2\kappa^2}\int d^5x\; \sqrt{g}R(\omega)\,.
\end{equation} 
The strength of this construction method lies in the fact that we are able to build every Poincar\'e action from the conformal approach. Conversely, every theory of gravity coupled to arbitrary matter must be extendable to a theory of matter fields in a conformal background (i.e. with local conformal symmetry). This is indeed the case. To make that statement clear, let us consider a simple example of gravity coupled to a real scalar $\lambda$ with a potential $V$,
\begin{equation}\label{HEsc}
\mathcal S=\int d^5x\;e\big[\frac1{2\kappa^2}R-\frac12\partial_\mu \lambda \partial^\mu \lambda + V(\lambda)\big]\,.
\end{equation}
To find the conformal origin of this theory, we would again have to introduce the compensator field $\varphi$. We would then write a term of the form~(\ref{easyconfapp}) in the conformal action, and this would correspond to the Einstein-Hilbert term in~(\ref{HEsc}). The other terms in~(\ref{HEsc}) can also be lifted to a conformal action. Taking $\lambda$ to have zero Weyl weight, we would have to multiply the kinetic term for $\lambda$ and every term in its potential with an appropriate factor of the compensator. In this way, we can construct the conformal lift of~(\ref{HEsc}) and similarly for any Poincar\'e-invariant theory. 

Aficionados of the conformal tensor calculus approach believe\footnote{By now, no proof exists.} that every supersymmetric theory coupled to gravity can be constructed using this approach such that in every step, the theory remains fully supersymmetric. This conjecture thus implies that in a theory with local superconformal invariance, we can always find appropriate compensating fields to reduce the theory to one that is invariant under some super-Poincar\'e group. Moreover, since this is believed to be an entirely supersymmetric construction, these compensating fields should fall in (superconformal) multiplets.
\section{$\mathcal{N}=2$, $d=5$ superconformal fields}\label{locsupconfie}
We now want to embark on the superconformal construction of local $\mathcal N=2$ five-dimensional super-Poincar\'e vector- and hypermultiplets. In the present section, we will therefore review some of the $\mathcal{N}=2$ superconformal multiplets. First of all, we will have to discuss the Weyl multiplet, which contains the superconformal gauge fields. Next, we will discuss the off-shell superconformal vector multiplet and study the superconformal extension of the hypermultiplet, discussed in Chapter~\ref{onshellsusy}. The vector multiplets will then be of use, as they will allow us to gauge symmetries of the hypermultiplet target space. Note that tensor multiplets and on-shell vector multiplets will not be discussed.  
\subsection{Weyl multiplet}
As in the purely bosonic case, the first step is a discussion of the gauge theory of the superconformal algebra, which can be found in~\cite{Bergshoeff:2001hc}. We again start by introducing a gauge field for every symmetry, as shown in Table~\ref{tbl:superconformal5}. 
\begin{table}[tb]
\begin{center}
\begin{tabular}{|c|c|c|c|c|c|c|c|}
\hline \rule[-1mm]{0mm}{6mm}
Generators & $P_a$ & $M_{ab} $& $D$ & $K_a$ & $U_{ij}$ & $Q_{\a i}$ &
$S_{\a i}$\\
\hline \rule[-1mm]{0mm}{6mm}
Fields & $e_\m{}^a$ & $\hat\o_\m^{ab}$ & $b_\m$ & $f_\m{}^a$ & $V_\m^{ij}$ &
$\p_\m^i$ & $\phi_\m^i$\\
 \hline
\rule[-1mm]{0mm}{6mm}
Parameters & $\xi^a$ & $\l^{ab}$ & $\l_D$ & $\l_K^a$ & $\l_R^{ij}$ & ${\e}^i$ &
${\eta}^i$\\
\hline
\end{tabular}
\caption{\it  The gauge fields  and parameters of the superconformal
algebra $F^2(4)$. \label{tbl:superconformal5}}
\end{center}
\end{table}
From the algebra~(\ref{supconfalg}), we can now read off the transformation rules for the gauge fields, using~(\ref{transfgauge}). Similarly, we can construct the different curvatures from~(\ref{defcurvgauge}).  We again trade the local translations for the covariant general coordinate transformations and split the gauge indices in an $a$ for translations and an $I$ for the rest. We then also introduce the covariant curvatures $\hat R$ using~(\ref{defcovcurvgauge}), for which we have not yet a complete expression, since we will later on be forced to introduce auxiliary fields that will contribute to covariant terms in the transformations of the gauge fields. We can impose the following three conventional constraints
\begin{eqnarray}
\label{constraints}
R_{\mu\nu}{}^a(P) &=& 0 \nonumber\,, \\
e^{\nu}{}_b \hat R_{\mu\nu}{}^{ab} (M) &=& 0 \nonumber\,,\\
\gamma^\mu \hat R_{\mu\nu}{}^i (Q)& =& 0 \,.
\end{eqnarray}
These can be solved for the dependent gauge fields. Introducing first
\begin{equation}
  \hat R'_{\mu \nu }{}^I= \hat R_{\mu \nu }{}^I+ 2h_{[\mu }^J
e_{\nu ]}^a
  f_{aJ}{}^I\,,
\end{equation}
we can write
\begin{eqnarray}
 \hat \o^{ab}_\m
&=& 2 e^{\n[a} \partial_{[\m} e_{\n]}^{~b]} - e^{\n[a} e^{b]\s} e_{\m c}
\partial_\n e^{~c}_\s
 + 2 e_\m^{~~[a} b^{b]} - \frac12 \bar{\p}^{[b} \g^{a]} \p_\m - \frac14
\bar{\p}^b \g_\m \p^a \,,\nonumber\\
\f^i_\m &=& \frac13\rmi \g^a \hat{R}^\prime _{\m a}{}^i(Q) - \frac1{24}\rmi
\g_\m \g^{ab} \hat{R}^\prime _{ab}{}^i(Q)\,, \label{transfDepF} \\
f^a_\m &=& -\frac16{\cal R}_\mu {}^a +\frac1{48}e_\mu {}^a {\cal R}\,,\qquad
{\cal R}_{\mu \nu }\equiv \hat{R}_{\rho \mu }^{\prime~~ba}(M) e_b{}^\rho
e_{\nu a}\,,\qquad {\cal R}\equiv {\cal R}_\mu {}^\mu \,.\nonumber
 \end{eqnarray}
Counting the number of bosonic and fermionic (off-shell) degrees of freedom, taking into account the constraints~(\ref{constraints}), it turns out that these do not match. The standard cure for this is to introduce auxiliary (matter) fields. These are respectively a real scalar $D$, an antisymmetric real two-form $T_{ab}$ and an $\su(2)$ doublet of fermions $\chi^i$. The introduction of these fields deforms the ($Q$- and $S$-) transformation rules of the different gauge fields. As a consequence, these fields contribute to the object $\mathcal{M}_{aI}{}^J$, introduced previously.
To be precise, the $Q$- and $S$-transformation rules for the independent fields in the Weyl multiplet are~\cite{Bergshoeff:2001hc}
\begin{eqnarray}
\delta e_\mu{}^a   &=&  \frac 12\bar\epsilon \gamma^a \psi_\m  \nonumber\, ,\\
\delta \psi_\mu^i   &=& {\mathfrak  D}_\mu \epsilon^i + \rmi \g\cdot T \gamma_\mu \epsilon^i -\rmi
\gamma_\mu \eta^i  \nonumber\, ,\\
\delta V_\mu{}^{ij} &=&  -\frac32\rmi \bar\epsilon^{(i} \phi_\mu^{j)} +4
\bar\epsilon^{(i}\gamma_\mu\chi^{j)}
  + \rmi \bar\epsilon^{(i} \g\cdot T \psi_\mu^{j)} + \frac32\rmi
\bar\eta^{(i}\psi_\mu^{j)} \nonumber\, ,\\
\delta T_{ab}     &=&  \frac12\rmi \bar\epsilon \gamma_{ab} \chi -\frac
{3}{32}\rmi \bar\epsilon\hat R_{ab}(Q)    \nonumber\, ,\\
\delta \chi^i     &=&  \frac 14 \epsilon^i D -\frac{1}{64} \gamma\cdot \hat R^{ij}(V)
\epsilon_j+ \frac18\rmi \gamma^{ab} \slashed{\mathfrak{D}} T_{ab} \epsilon^i
                   - \frac18\rmi \gamma^a \mathfrak{D}^b T_{ab} \epsilon^i \nn\\
               &&  -\frac 14 \gamma^{abcd} T_{ab}T_{cd} \epsilon^i + \frac 16 T^2 \epsilon^i
             +\frac 14 \g\cdot T \eta^i  \nonumber\, ,\\
\delta D         &=&  \bar\epsilon \slashed{\mathfrak D} \chi - \frac {5}{3} \rmi
\bar\epsilon\gamma\cdot T
\chi - \rmi  \bar\eta\chi \nonumber\, ,\\
\delta b_\mu       &=& \frac 12 \rmi\bar\epsilon\phi_\mu -2 \bar\epsilon\gamma_\mu \chi +
\frac12\rmi \bar\eta\psi_\mu  \,. \label{modifiedtransf}
\end{eqnarray}
Here, the covariant derivative on the supersymmetry parameter is given by
\begin{equation}
\mathfrak{D}_\mu \epsilon^i=\partial_\mu \epsilon^i+\frac12 b_\mu \epsilon^i+\frac14 \hat{\slashed{\omega}}_\mu\epsilon^i-V_\mu^{ij}\epsilon_j\,.
\end{equation}
The different Weyl and $\su(2)$ weights are shown in Table~\ref{tbl:fieldsWeyls}.
\begin{table}[bt]
\begin{center}
\begin{tabular}{|c|ccc||c|cc||c|ccc|}\hline\rule[.2cm]{0cm}{.2cm}
Field&$\sharp$&$\su(2)$&$w$&Field&$\su(2)$&$w$&Field&$\sharp$&$\su(2)$&$w$\\\hline\hline\rule[.2cm]{0cm}{.2cm}
$e_\mu{}^a$&$9$&$1$&$-1$&$\hat\omega_\mu{}^{ab}$&$1$&$0$&$T_{[ab]}$&$10$&$1$&$1$\\\rule[.2cm]{0cm}{.2cm}
$b_\mu$&$0$&$1$&$0$&$f_\mu{}^a$&$1$&$1$&$D$&$1$&$1$&$2$\\\rule[.2cm]{0cm}{.2cm}
$V_\mu{}^{(ij)}$&12&3&0&&&&&&&\\\hline\rule[.2cm]{0cm}{.2cm}
$\psi_\mu^i$&$24$&$2$&$-\frac12$&$\phi_\mu^i$&$2$&$\frac12$&$\chi^i$&$8$&$2$&$\frac32$\\\hline
\end{tabular}
\end{center}
\caption{\it Elementary gauge fields, dependent gauge fields and matter fields of the Weyl multiplet.}\label{tbl:fieldsWeyls}
\end{table}

Given the transformation rules~(\ref{modifiedtransf}) of the fields, we can calculate how the supersymmetry algebra is realized on them. This is nontrivial since the rigid structure of~(\ref{supconfalg}) gets changed into a soft algebra, as the structure constants have become field-dependent structure functionals. Below, we list the relevant commutators.

The full commutator of two supersymmetry transformations is
\begin{eqnarray}
 \left[\d_Q(\e_1),\d_Q(\e_2)\right] &=&  \d_{cgct}(\xi_3^\m)+
\d_M(\l^{ab}_3) + \d_S(\eta_3)  + \d_R(\l^{ij}_3) 
+\d_K(\l^a_{K3})\,.\nonumber\\
 \label{algebraQQ}
\end{eqnarray}
The covariant general coordinate transformations have been defined in
(\ref{defcgct}). The parameters appearing in~(\ref{algebraQQ})
are
\begin{eqnarray}
\xi^\m_3       &=& \frac 12 \bar\e_2 \g_\m \e_1 \,,\nn\\
\l^{ab}_3      &=& - \rmi \bar \epsilon _2\gamma ^{[a}\gamma \cdot T
\gamma ^{b]}\epsilon _1 \,, \nonumber\\
\lambda^{ij}_3 &=& \rmi \bar\e^{(i}_2 \g\cdot T \e^{j)}_1\,, \nn\\
\eta^i_3       &=& - \frac {9}{4}\rmi\, \bar \e_2 \e_1 \chi^i
                   +\frac {7}{4}\rmi\, \bar \e_2 \g_c \e_1 \g^c \chi ^i  + \frac1{4}\rmi\,  \bar \e_2^{(i} \g_{cd} \e_1^{j)}
\left( \g^{cd} \chi_j
                   + \frac 14\, \hat R^{cd}{}_j(Q) \right) \, ,\nonumber\\
\lambda^a_{K3} &=& -\frac 12 \bar\e_2\g^a\e_1 D + \frac{1}{96}
\bar\e^i_2\g^{abc}\e^j_1 \hat R_{bcij}(V) + \frac1{12}\rmi\bar\e_2\left(-5\g^{abcd} \mathfrak D_b T_{cd} +
9 \mathfrak D_b T^{ba} \right)\e_1  \nn\\
               && + \bar\e_2\left(  \g^{abcde}T_{bc}T_{de}
                  - 4 \g^c T_{cd} T^{ad} +  \frac 23  \g^a T^2
                  \right)\e_1 \,.
\end{eqnarray}
For the other commutators containing $Q$ and $S$, we find the following algebra:
\begin{eqnarray}
 \left[\d_S(\eta),
\d_Q(\e)\right] &=& \d_D( \frac12\rmi \bar\e\eta ) + \d_M( \frac12\rmi \bar\e
\g^{ab} \eta) +      \d_R(  -\frac32\rmi \bar\e^{(i} \eta^{j)} )  +
\delta_K(\tilde \lambda_{3K}^a ) \,,\nn\\
\left[ \d_S(\eta_1), \d_S(\eta_2) \right] &=& \d_K( \frac 12 \bar\eta_2 \g^a
\eta_1 ) \,,
\end{eqnarray}
where
\begin{equation}
 \tilde \lambda_{3K}^a= \frac16 \bar{\e} \left(\g \cdot T \g_a - \frac12 \g_a \g \cdot T
\right) \eta \,. 
\end{equation}

To find the modifications to the transformation rules for the dependent gauge fields, we can use their expressions~(\ref{transfDepF}), which can be computed completely as the expressions for the covariant curvatures are completely known by now. Another way to derive these modifications is to note that asking the constraints~(\ref{constraints}) to be invariant, modifies the transformation properties of the dependent gauge fields. 

To conclude this brief review of the Weyl multiplet, we give the complete expressions for the dependent gauge fields.
Introducing for notational ease
\begin{equation}\label{Lor-covder}
\mathcal{D}_{\mu}\psi^i_{\nu}=\partial_{\mu}\psi^i_{\nu} +\frac14
\hat{\omega}_{\mu}{}^{ab} \gamma_{ab}\psi^i_{\n}\,,
\end{equation}
we have
\begin{eqnarray}
\phi_\mu^i&=&\frac12\rmi \g^a \mathcal{D}_{[\mu}\psi^i_{\nu]}e_a^\nu-\frac1{12}\rmi
\g_\m{}^{ab} \mathcal{D}_{\kappa}\psi^i_{\nu}e_a^\kappa e_b^\nu
-\frac12\rmi V_{[\m}{}^{ij} \g^a \p_{a]\,j}+\frac1{12}\rmi V_{a}{}^{ij}
\g_\m{}^{ab} \p_{b\,j}\nonumber\\&& -T^a{}_\mu
\psi_a^i-\frac13T^{ab}\gamma_{b\mu}\psi_a^i-\frac23T_{b\mu}\gamma^{ab}\psi_a^i-\frac13T_{bc}\gamma^{abc}{}_\mu\psi_a\,,\nonumber\\
f_a{}^a&=&f_\mu{}^ae_a{}^\mu\nonumber\\&=&-\frac1{16}(R(\hat\o)+\frac13\bar{\psi}_\kappa\gamma^{\kappa \mu
\nu}\mathfrak{D}_\mu \psi_\nu -\frac13
\bar{\psi}_a^i\gamma^{abc}\psi_b^jV_{cij}-16\bar{\psi}_a\gamma^a\chi+4\rmi\bar{\psi}^a\psi^bT_{ab}\nonumber\\&&-\frac43\rmi\bar{\psi}^b\gamma_{abcd}\psi^aT^{cd})\,,\nonumber\\
\hat{\omega}_{\mu}{}^{ab}
&=& 2 e^{\n[a} \partial_{[\m} e_{\n]}^{~b]} - e^{\n[a} e^{b]\s} e_{\m c}
\partial_\n e^{~c}_\s
 + 2 e_\m^{~~[a} b^{b]} - \frac12 \bar{\p}^{[b} \g^{a]} \p_\m - \frac14
\bar{\p}^b \g_\m \p^a \nonumber\\
&=& \omega_{\mu}{}^{ab}
 + 2 e_\m^{~~[a} b^{b]} - \frac12 \bar{\p}^{[b} \g^{a]} \p_\m - \frac14
\bar{\p}^b \g_\m \p^a\,.\label{depgaugefields}
\end{eqnarray}
\subsection{Vector multiplet}
We have already discussed the rigid on-shell $\mathcal{N}=2$ super-Poincar\'e vector multiplet in Section~\ref{5d}. Much of this discussion will carry over to the present case. However, in case we would use an on-shell vector multiplet, the transformations would get changed during the coupling to the hypermultiplets. This problem is circumvented if we take the realization of the symmetry algebra off the mass-shell, which is done by introducing an auxiliary bosonic field $Y^{(ij)}$ in the $\mathbf{3}$ of $\su(2)$. 
\subsubsection{Rigid case}
We will start by giving the transformation rules for a vector multiplet
in the adjoint representation~\cite{Fujita:2001kv}. An off-shell vector
multiplet has $8+8$ real degrees of freedom whose $\SU(2)$ labels and
Weyl weights we have already indicated in Table~\ref{tbl:multiplets}. 

The transformation properties under the gauge algebra are completely similar to the ones of the on-shell multiplet~(\ref{Lambda3})-(\ref{eq:BIvec}).
The rigid $Q$- and $S$-supersymmetry transformation rules for the
off-shell Yang-Mills multiplet are given by~\cite{Fujita:2001kv}
\begin{eqnarray}
\d A_\m^I
&=& \frac 12 \bar{\e} \g_\m \p^I \,, \nonumber\\
\d Y^{ij I}
&=& -\frac 12 \bar{\e}^{(i} \slashed{\mathfrak D} \p^{j) I} -  \frac {1}{2} \rmi g \bar \e^{(i} f_{JK}{}^I \sigma^J \psi^{j) K} + \frac 12 \rmi \bar{\eta}^{(i} \p^{j)I} \,, \nonumber \\
\d \p^{i I}
&=& - \frac 14 \g \cdot F^I \e^i -\frac 12\rmi \slashed{\mathfrak D} \s^I \e^i - Y^{ij I} \e_j + \s^I \eta^i \,, \nonumber\\
\d \s^I &=& \frac 12 \rmi \bar{\e} \p^I \,. \label{ymflat}
\end{eqnarray}
The commutator of two $Q$-supersymmetry transformations equals the on-shell case~(\ref{algebra}), without the possibility of having a nonclosure functional.
\subsubsection{Local case}
Following the conformal recipe, we now need to place the vector multiplet in the background of the Weyl multiplet, i.e. we have to gauge all superconformal symmetries. This is done in the following way.

The local supersymmetry rules are given by
\begin{eqnarray}
\d A_\m^I
&=& \frac 12 \bar{\e} \g_\m \p^I -\frac 12 \rmi \s^I \bar\e \p_\mu \,, \nonumber\\
\d Y^{ij I}
&=& -\frac 12 \bar{\e}^{(i} \slashed{\mathfrak D} \p^{j) I}
+\frac 12 \rmi \bar\e^{(i} \gamma \cdot T \p^{j)  I}
- 4 \rmi \s^{ I} \bar\e^{(i} \chi^{j)}  -\frac 12 \rmi g \bar \e^{(i} f_{J K}{}^{ I} \sigma^{ J} \psi^{j) { K}} + \frac 12 \rmi \bar{\eta}^{(i} \p^{j)  I} \,, \nonumber \\
\d \p^{i  I} &=& - \frac 14 \g \cdot \hat{F}{}^{ I} \e^i -\frac 12\rmi \slashed{\mathfrak D} \s^{ I}
\e^i - Y^{ij  I} \e_j + \s^{ I} \gamma \cdot T \e^i
 + \s^{ I} \eta^i \,, \nonumber\\
\d \s^{ I} &=& \frac 12 \rmi \bar{\e} \p^{ I} \,.
\label{tensorlocal}
\end{eqnarray}
The covariant derivatives read
\begin{eqnarray} \label{localderiv-tensor}
\mathfrak D_\mu\, \sigma^{ I} &=& {\cal D}_\m \s^{ I} - \frac12\, \rmi \bar{\psi}_\mu \psi^{ I}
\,, \nn \\[2pt]
{\cal D}_\m \s^{ I} &=& (\partial_\mu - b_\mu)
\sigma^{ I}
+ g f_{J K}{}^{ I} A_\mu^J \sigma^{ K} \,, \nn \\[2pt]
\mathfrak D_\mu \psi^{i  I} &=& {\cal D}_\m \psi^{i  I} + \frac
14 \g \cdot \hat{{F}}^{ I} \p_\mu^i + \frac 12\rmi
\slashed{\mathfrak D} \s^{ I} \p_\mu^i - Y^{i  I}_j \p_{\mu}^j
 - \s^{ I} \g \cdot T \p_\mu^i - \s^{ I} \phi_\mu^i  \,, \nn \\
{\cal D}_\m \psi^{i  I} &=& (\partial_\mu - \frac 32 \, b_\mu +
\frac 14\, \g_{ab}\, \o_\mu {}^{ab}) \p^{i  I} - V_\mu^{ij}
\p_j^{ I} + g f_{J K}{}^{ I} A_\mu^J
\psi^{i  K} \,.
\end{eqnarray}
The superconformal field strength is defined as
\begin{equation}
\hat{F}_{\m\n}^I = 2 \partial_{[\mu } A_{\nu ]}^I + g f_{JK}{}^I
A_\mu^J A_\nu^K - \bar{\p}_{[\m} \g_{\n]} \p^I + \frac 12 \rmi \s^I
\bar{\p}_{[\m} \p_{\n]}
 \,,
\end{equation}
while the superconformal d'Alembertian reads
\begin{eqnarray}
&& \Box^{\rm c}
\s^{ I}
= \mathfrak D^a \mathfrak D_a \s^{ I} \nn \\
&&= \left( \partial^a  - 2 b^a + \o_b^{~ba} \right) \mathfrak D_a \s^{ I}
+ g t_{J  K}{}^{ I} A_a^J \mathfrak D^a \sigma^{ K} -
\frac{1}2\rmi \bar{\p}_\m \mathfrak D^\m \p^{ I}  - 2 \s^{ I}
\bar{\p}_\m \g^\m \chi \nonumber\\
& &
 + \frac12 \bar{\p}_\m \g^\m \g \cdot T \p^{ I} + \frac12
\bar{\f}_\m \g^\m \p^{ I} + 2 f_\m{}^\m \s^{ I}
-\frac 12 g \bar\p_\m \g^\m t_{ J  K}{}^{ I}\p^J
\s^{ K}\,.
\end{eqnarray}
The above transformation rules leave the following action invariant.
\begin{eqnarray}\label{conf-VTaction}
e^{-1} {\cal L}_{v}&=&\biggl[ \big(-
\frac14 {\hat F}_{\m\n}^{I} {\hat F}^{\m\n
{ {J}}} - \frac 12 \bar{\p}^{ {I}} \slashed{\mathfrak D} \p^{
{ J}} + \frac 13 \s^{I} \Box^c \s^{J} +
\frac 16 \mathfrak D_a \s^{ {I}} \mathfrak D^a \s^{ {J}}
+  Y_{ij}^{ {I}} Y^{ij { {J}}}  \big) \s^{ {K}}\nn \\
& &\hphantom{\! \biggl[}  {-}\frac 43 \s^{I} \s^{J}
\s^{K} \left(D + \frac{26}{3} T_{ab} T^{ab} \right)
+ 4 \s^{I} \s^{J} {\hat  F}_{ab}^{K} T^{ab}  -\frac 18 \rmi \bar{\p}^{ {I}} \g
\cdot {\hat F}^{ {J}} \p^{ {K}} \nn \\
& &\hphantom{\! \biggl[} -\frac 12
\rmi \bar{\p}^{i { {I}}} \p^{j { {J}}} Y_{ij}^{
{K}}  + \rmi \s^{I} \bar\p^{J}
\g \cdot T \p^{K} - 8 \rmi \s^{I} \s^{J} \bar\p^{K} 
{+} \frac 16 \s^{I} \bar \p_\m \g^\m \big(\rmi \s^{J} \slashed{\mathfrak D} \p^{K}\chi\nn \\
& &\hphantom{\! \biggl[} + \frac 12 \rmi \slashed{\mathfrak D} \s^{J} \p^{K} - \frac 14 \g
{\cdot} {\hat  F}^{J} \p^{K} 
+ 2 \s^{J} \g {\cdot} T \p^{K} - 8 \s^{J} \s^{K} \chi \big)\nn\\
&&\hphantom{\! \biggl[}
{-} \frac 16 \bar \p_a \g_b \p^{I} \left(\s^{J} {\hat  F}^{ab {K}} -8 \s^{J} \s^{K} T^{ab} \right) -\frac 1{12} \s^{I} \bar \p_\l \g^{\m\n\l} \p^{J} {\hat  F}_{\m\n}^{K}   \nn\\
&&\hphantom{\! \biggl[}{+}\frac 1{12} \rmi \s^{I} \bar \p_a
\p_b  \left(\s^{J} {\hat  F}^{ab {K}} -8
\s^{J} \s^{K} T^{ab} \right) +\frac 1{48} \rmi
\s^{I} \s^{J} \bar \p_\l \g^{\m\n\l\r} \p_\r
{\hat  F}_{\m\n}^{K}  \nn \\&& \hphantom{\!
\biggl[} {-} \frac 12 \s^{I} \bar \p_\m^i \g^\m \p^{j
J} Y_{ij}^{K} +\frac 1{6} \rmi \s^{I}
\s^{J} \bar\p_\m^i \g^{\m\n} \p_\n^j Y_{ij}^{K}
-\frac{1}{24} \rmi \bar \p_\m \g_\n \p^{I} \bar \p^{J} \g^{\m\n} \p^{K} \nn \\&& \hphantom{\! \biggl[}
{+}\frac{1}{12} \rmi \bar \p_\mu^i \g^\mu \p^{j {I}} \bar
\p_i^{J} \p_j^{K} -\frac{1}{48} \s^{I}
\bar \p_\m \p_\n \bar \p^{J} \g^{\m\n} \p^{K}
+\frac {1}{24} \s^{I} \bar\p_\m^i \g^{\m\n} \p_\n^j
\bar\p_i^{J} \p_j^{K}  \nn\\&& \hphantom{\!
\biggl[}
{-}\frac 1{12} \s^{I} \bar \p_\l \g^{\m\n\l} \p^{J} \bar{\p}_{\m} \g_{\n} \p^{K}{+} \frac 1{48} \rmi \s^{I} \s^{J} \bar \p_\l \g^{\m\n\l\r} \p_\r \bar{\p}_{\m} \g_{\n} \p^{K} \nn \\
& &\hphantom{\! \biggl[} + \frac 1{24} \rmi
\s^{I} \s^{J} \bar \p_\l \g^{\m\n\l} \p^{K} \bar{\p}_{\m} \p_{\n}
  + \frac1{96}\s^{I} \s^{J} \s^{K} \bar \p_\l \g^{\m\n\l\r} \p_\r \bar{\p}_{\m} \p_{\n}\biggr]
  C_{IJK}  \nn\\
& &  -\frac 1{24} e^{-1} \ve^{\m\n\l\r\s} C_{IJK} A_\m^I\!
 \Big(\!F_{\n\l}^J F_{\r\s}^K + f_{FG}{}^J \!A_\n^F A_\l^G \!  \big(\!{-} \frac 12 g \, F_{\r\s}^K\nn \\
 && + \frac1{10} g^2 f_{HL}{}^K A_\r^H A_\s^L  \!\big)\!\Big)\!\! 
+ \frac14 \rmi g \bar{\p}^{I}\psi
^{{J}}\sigma ^{{K}}\sigma ^{{L}} f_{IJ}{}^{M}C_{MKL}\,.
\end{eqnarray}
\subsection{Hypermultiplet}
In Chapter~\ref{onshellsusy}, we have already introduced the hypermultiplet. In the superconformal case, the target space still is hypercomplex or hyperk\"ahler for the same reasons as was explained there. However, since more symmetries can now act nontrivially on the fields, one more object (a homothetic symmetry vector field) has to be introduced on the manifold. 
\subsubsection{Rigid case}\label{rigidsuconfhypers}
\paragraph{Algebra}
The symmetry algebra that is nontrivially realized on the hypermultiplet fields is an extension of the invariances discussed in Section~\ref{rigidhypers}. As a consequence, the supersymmetry transformation rules are the same as in~(\ref{SUzeta}).
However, we should now also consider the transformation properties under dilatations, special conformal transformations, $\su(2)$-transformations and the special superconformal transformations.
The scalars do not transform under special
conformal transformations and special superconformal symmetry, but under
dilatations and $\su(2)$-transformations, we parametrize
\begin{eqnarray}
\delta_D(\lambda_D) q^X&=& \lambda_D k^X(q) \,, \nonumber\\
\delta_{R}(\vec \lambda) q^X&=& 2\vec\lambda\cdot\vec k^X(q) \,,
\label{SUzetacov}
\end{eqnarray}
for some unknown target space vector fields $k^X(q)$ and $\vec k^X(q)$.

To derive the appropriate transformation rules for the fermions, we first
note that the hyperinos should be invariant under special conformal
symmetry. This is due to the fact that this symmetry changes the Weyl
weight with one. If we realize the $[K,Q]$-commutator on the
fermions $\zeta^A$,\footnote{Note that $\partial q$ does transform under $K$-symmetry, due to~(\ref{indreps}).} we read off the special supersymmetry transformation
\begin{equation}
\delta_S(\eta)\zeta^A=-k^X f_{X}^{iA} \eta_i\,.
\end{equation}
To proceed, we consider the commutator of regular and special
supersymmetry. Realizing this on the scalars, we determine
the expression for the generator of $\su(2)$-transformations in terms of
the dilatations and complex structures,
\begin{equation}\label{dil-su2}
\vec k^X= \frac 13 k^Y \vec J_Y {}^X  \,.
\end{equation}
Realizing the same commutator on the hyperinos, we determine the covariant
variations (see Section~\ref{rigidhypers})
\begin{equation}
\hat \delta_D \zeta^A = 2 \lambda_D \zeta^A \,, \qquad \hat
\delta_{R} \zeta^A =0\,,
\end{equation}
and furthermore this commutator only closes if we impose
\begin{equation}\label{conf-constr}
\covder_Y k^X 
= \frac 32 \delta_Y{}^X\,,
\end{equation}
which also implies
\begin{equation}
 \covder_Y \vec k^X=\frac 12\vec J_Y{}^X \,.
 \label{DkSU2}
\end{equation}
Due to the resemblance of~(\ref{conf-constr}) to~(\ref{homkilvect}), we will call the vector $k^X$ a homothetic symmetry vector field. Similarly, we will call $\vec k$ the $\su(2)$ symmetry vector fields. Note that~(\ref{conf-constr}) is imposed by the algebra. In a more
usual derivation, where one considers symmetries of the Lagrangian, we
would find this constraint by imposing dilatation invariance of the
action. Our result, though, does not require the
existence of an action. The relations~(\ref{conf-constr})
and~(\ref{dil-su2}) further restrict the geometry of the target space,
and it is easy to derive that the Riemann tensor has four zero
eigenvectors,
\begin{equation}
k^XR_{XYZ}{}^W=0\,,\qquad  \vec k^X\,R_{XYZ}{}^W=0\,. \label{konRis0}
\end{equation}
Also, under dilatations and $\su(2)$-transformations, the hypercomplex
structure is scale invariant and rotated into itself,
\begin{eqnarray}
\lambda_D \left( k^Z \partial_Z \vec J_X{}^Y -\partial_Z k^Y \vec J_X{}^Z+\partial_X k^Z \vec J_Z{}^Y\right) &=&0\,, \nonumber\\
\vec \lambda \cdot k^{Z}  \partial_Z \vec J_X{}^Y -\vec \lambda\cdot \partial_Z
\vec k^Y \vec J_X{}^Z+\vec \lambda \cdot \partial_X \vec k^Z \vec J_Z{}^Y
&=&-\vec \lambda\times  \vec J_X{}^Y\,.\label{normofcomplstr}
\end{eqnarray}

All these properties are similar to those derived from superconformal
hypermultiplets in four space-time
dimensions~\cite{DeWit:1980gt,DeWit:1999fp}. There, the $\symp(1) \times
\gl(r,\mathbb{H})$ sections, or simply, hypercomplex sections, were
introduced
\begin{equation}\label{defsections}
A^{iB}(q)\equiv k^Xf_X^{iB}\,, \qquad
(A^{iB})^*=A^{jC}E_j{}^i\rho_C{}^B\,,
\end{equation}
with $\rho$ and $E$ defined in~(\ref{defrho}). These sections were used to rewrite all equations and transformation rules without the occurrence of the $q^X$ fields. For example,
the hypercomplex sections are zero eigenvectors of the
$\Gl(r,\mathbb{H})$ curvature~(\ref{def-W}),
\begin{equation}
A^{iB}W_{BCD}{}^E=0\,,
\end{equation}
and have the following supersymmetry, dilatation and $\su(2)$-transformation laws.
\begin{equation}
\hat \delta A^{iB} = \frac 32f^{iB}_X\delta q^X = -\frac 32 \rmi{\bar
\epsilon}^i\zeta^B + \frac 32 \Lambda_D A_i{}^B - \Lambda^i{}_jA^{jB}\,,
\label{deltaA}
\end{equation}
where $\hat \delta$ is understood as a covariant variation, in the
sense of~(\ref{cov-var}).

We now assume the action of a symmetry algebra on the hypermultiplet, which must commute with the
full superconformal algebra. Hence, we will construct hypermultiplet couplings to vector multiplets with a non-Abelian gauge group $G$.
The symmetries are parametrized by
\begin{eqnarray}
\delta_G q^X&=& -g \beta^I k_I^X(q)\,, \nonumber\\
\hat \delta_G \zeta^A &=& -g \beta^I t_{IB}{}^A(q) \zeta^B\,.
\label{gauge-tr}
\end{eqnarray}
The vectors $k_I^X$ depend on the scalars and generate the algebra $\mathfrak g$ of $G$
with structure constants $f_{IJ}{}^K$,
\begin{equation} \label{G-alg}
k_{[I|}^Y \partial_Y k^X_{|J]}=-\frac 12 f_{IJ}{}^K k_K^X \,.
\end{equation}
The commutator of two gauge transformations~(\ref{Lambda3}) on the
fermions requires the following constraint on the field-dependent matrices
$t_I(q)$,
\begin{equation} \label{tt-comm}
[t_I,t_J]_B{}^A=-f_{IJ}{}^Kt_{KB}{}^A -2k^X_{[I|}\covder_Xt_{|J]B}{}^A
+k^X_Ik^Y_J{R}_{XYB}{}^A\,,
\end{equation}
where the curvature is defined as in~(\ref{defcurvs}).

Requiring that the gauge transformations commute with supersymmetry leads
to further relations between the quantities $k_I^X$ and $t_{IB}{}^A$.
Vanishing of the commutator on the scalars yields
\begin{equation} \label{GG-q}
t_{IB}{}^Af^X_{iA}=\covder_Yk^X_If^{Y}_{iB}\,.
\end{equation}
These constraints determine $t_I(q)$ in terms of the Vielbeine $f^{iA}_X$
and the vectors $k^X_I$,
\begin{equation} \label{def-t}
t_{IA}{}^B=\frac 12f^Y_{iA}\covder_Yk^X_If^{iB}_X\,,
\end{equation}
and furthermore
\begin{equation} \label{symm-ff}
f^{Y(i}_Af^{j)B}_X \covder_Y k^X_I=0\,.
\end{equation}
The relations~(\ref{symm-ff}) and~(\ref{def-t}) are equivalent
to~(\ref{GG-q}). We interpret~(\ref{def-t}) as the definition for
$t_{IA}{}^B$. The vanishing of an $(ij)$-symmetric part in an equation
as~(\ref{symm-ff}) can be expressed as the vanishing of the commutator of
$\covder_Y k^X_I$ with the complex structures:
\begin{equation}
  \left(\covder_X k^Y_I\right)\vec J {}_Y{}^Z= \vec J {}_X{}^Y\left(\covder_Y
  k^Z_I\right)\,.
 \label{commDkJ}
\end{equation}
Extracting connections from this equation, it can be written as
\begin{equation}
\label{intJ} \left( {\cal L}_{k_I} \vec J\right) _X{}^Y\equiv  k^Z_I
\partial_Z \vec J_X{}^Y -\partial_Z k^Y_I \vec J_X{}^Z+\partial_X
k^Z_I \vec J_Z{}^Y=0\,.
\end{equation}
The left-hand side is the Lie-derivative of the complex structure in the
direction of the vector $k_I$. The relation~(\ref{intJ}) is a special case of the
statement that the vector $k_I$ normalizes the quaternionic-like structure, which means that
\begin{equation}\label{kInormalJ}
\mathcal{L}_{k_I} J^{\bar \alpha}{}_X{}^Y=b^{\bar \alpha \bar \beta} J^{\bar \beta}{}_X{}^Y\,,
\end{equation}
with $\bar \alpha, \bar \beta=1,2,3$ and $b \in \su(2)$ in the adjoint and possibly depending on $\{q^X\}$.
The latter would allow that this Lie-derivative is proportional to a
complex structure. Killing vectors which normalize the hypercomplex
structure can be decomposed in an $\su(2)$ part and a $\gl(r,\mathbb{H})$
part. The vanishing of this Lie-derivative, or~(\ref{symm-ff}), is
expressed by saying that the gauge transformations act
\emph{triholomorphically}. Thus, it says that all the symmetries are embedded
in $\gl(r,\mathbb{H})$.

Vanishing of the gauge-supersymmetry commutator on the fermions requires
\begin{equation}
\covder_Y t_{IA}{}^B= k^X_I {R}_{YXA} {}^B\,. \label{t-id}
\end{equation}
Using~(\ref{GG-q}) this implies a new constraint,
\begin{equation}
\label{DDk} \covder_X\covder_Y k^Z_I=R_{XWY}{}^Zk_I^W\,.
\end{equation}
Note that this equation is in general valid for any Killing vector of a
metric. As we want our discussion to hold as well in the absence of an action, we could not rely on this fact, but
here the algebra imposes this equation. It turns out that~(\ref{symm-ff})
and~(\ref{DDk}) are sufficient for the full commutator algebra to hold.
In particular,~(\ref{t-id}) follows from~(\ref{DDk}), using the
definition of $t$ as in~(\ref{def-t}).

A further identity can be derived: substituting~(\ref{t-id})
into~(\ref{tt-comm}) one gets
\begin{equation}
[t_I,t_J]_B{}^A=-f_{IJ}{}^Kt_{KB}{}^A-k^X_Ik^Y_J{R}_{XYB}{}^A\,.
\end{equation}
This identity can also be obtained from substituting~(\ref{def-t}) in the
commutator on the left-hand side, and then
using~(\ref{G-alg}),~(\ref{symm-ff}), and~(\ref{DDk}).

The group of gauge symmetries should also commute with the superconformal
algebra, in particular with dilatations and $\su(2)$-transformations.
This leads to
\begin{equation}
\label{comDG} k^Y \covder_Y k_I^X=\frac {3}{2} k_I^X\,, \qquad \vec k^Y\covder_Yk_I^X=\frac 12 k_I^Y\vec J_Y{}^X\,,
\end{equation}
coming from the scalars, and there are no new constraints from the
fermions and neither from other commutators. Since $\covder_Yk_I^X$ commutes with
$\vec J_Y{}^X$, the second equation in~(\ref{comDG}) is a consequence of
the first one.

In the above analysis, we have taken the parameters $\beta^I$ to be
constants. In the following, we also allow for local gauge
transformations. The gauge coupling is done by introducing vector
multiplets (which can be nondynamical) in the definition of the covariant derivatives
\begin{eqnarray}
\covder_{\mu} q^X&\equiv& \partial_{\mu} q^X +g A_{\mu}^I k_I^X \,, \nonumber\\
\covder_{\mu} \zeta^A &\equiv& \partial_\mu \zeta^A + \partial_\mu q^X
\omega_{XB}{}^A \zeta^B + g A_{\mu}^I t_{IB} {}^A \zeta^B\,.
\label{covderg}
\end{eqnarray}
The commutator of two supersymmetries should now also contain a local
gauge transformation, as in~(\ref{algebra}). This requires an extra term in
the supersymmetry transformation law of the fermion,
\begin{equation}
\hat \delta (\epsilon) \zeta^A = \frac 12 \rmi \slashed{\covder} q^X
f_X^{iA} \epsilon_i +\frac {1}{2}g \sigma^I k_I^X f_{iX}^A \epsilon^i\,.
\end{equation}
With this additional term, the commutator on the scalars closes, whereas
on the fermions, it determines the equations of motion
\begin{equation}
\Gamma^A\equiv \slashed{\covder} \zeta^A +\frac 12 W_{BCD} {}^A
\bar{\zeta}^C \zeta^D \zeta^B-g(\rmi k_I^X f_{iX}^A \psi^{iI}+ \rmi
\zeta^B\sigma^I t_{IB} {}^A )=0\,.
\end{equation}

Acting on $\Gamma^A$ with supersymmetry determines the equation of motion
for the scalars
\begin{eqnarray}
\Delta^X&=&\Box q^X -\frac 12 \bar{\zeta}^B \gamma_a \zeta^D \covder^a
q^Y f_Y^{iC}f_{iA}^X W_{BCD} {}^A-\frac {1}{4} \covder_Y W_{BCD}{}^A
\bar{\zeta}^E \zeta^D \bar{\zeta}^C \zeta^B f_E^{iY}f_{iA}^X \nonumber\\
&&-\,g \left(  2\rmi \bar{\psi}^{iI} \zeta^B t_{IB}{}^A f_{iA}^X- k_I^Y
J_Y{}^X {}_{ij} Y^{ijI} \right) +g^2 \sigma^I \sigma^J \covder_Y k_I^X
k_J^Y=0\,.
\end{eqnarray}
The first line is the same as in~(\ref{covscal}), the second line
contains the corrections due to the gauging. The gauge-covariant
d'Alembertian is here given by
\begin{equation}
\Box q^X=\partial_a \covder^a q^X +g \covder_a q^Y \partial_Y k_I^X
A^{aI}+\covder_a q^Y \covder^a q^Z \Gamma_{YZ}^X\,.
\end{equation}
\paragraph{Action}
As discussed in Section~\ref{sss:hypers}, the introduction of an action requires a metric. 
Hence, a first consequence is that the vector field $k^X$ that is generating dilatations becomes a homothetic Killing vector field~(\ref{homkilvect}), while the $\su(2)$ vectors $\vec k^X$ become Killing vector fields.
Secondly, the symmetries of before become isometries, i.e. the vector fields $k_I^X$ either become Killing vector fields
\begin{equation}
  \covder_X k_{YI} + \covder_Y k_{XI}=0\,.
 \label{Killingeq}
\end{equation}
This makes the requirement~(\ref{DDk}) superfluous, but we still have to
impose the tri\-holo\-morphicity expressed by either~(\ref{symm-ff})
or~(\ref{commDkJ}) or~(\ref{intJ}).
In order to integrate the equations of motion to an action, we have to
define (locally) triples of `moment maps', according to
\begin{equation}
\partial_X \vec P_I=-\frac 12\vec J_{XY} k_I^Y\,. \label{momentmap}
\end{equation}
The integrability condition that makes this possible is the
triholomorphic condition~(\ref{intJ}).

In the kinetic terms of the action, the derivatives should be
covariantized with respect to the new transformations. We are also forced
to include some new terms proportional to $g$ and $g^2$ as compared to~(\ref{rigidhyperaction})
\begin{eqnarray}
\mathcal L&=& -\frac 12 g_{XY} \covder_a q^X
\covder^a q^Y+ \bar{\zeta}_A \slashed{\covder} \zeta^A-\frac 14 W_{ABCD}
\bar{\zeta}^A \zeta^B \bar{\zeta}^C \zeta^D
\label{Sna}\\
&& -g\left( 2 \rmi k_I^X f_{iX}^A
\bar{\zeta}_A \psi^{iI}+\rmi \sigma^I t_{IB} {}^A \bar{\zeta}_A \zeta^B-2
P_{Iij} Y^{Iij}\right)-g^2\frac 12 \sigma^I \sigma^J k_I^X k_{JX}\!
\,, \nonumber
\end{eqnarray}
[where the covariant derivatives $\covder$ now also include
gauge-covariantization proportional to $g$ as in~(\ref{covderg})]. Supersymmetry
of the action imposes
\begin{equation}
  k_I^X \vec J_{XY} k_J^Y= 2f_{IJ}{}^K \vec P_K\,.
 \label{constraintfP}
\end{equation}
As only the derivative of $P$ appears in the defining
equation~(\ref{momentmap}), one may add an arbitrary constant to $P$. But
that changes the right-hand side of~(\ref{constraintfP}). One should then
consider whether there is a choice of these coefficients such
that~(\ref{constraintfP}) is satisfied. This is the question about the
center of the algebra, which is discussed
in~\cite{D'Auria:1991fj,Andrianopoli:1997cm}. For simple groups there is
always a solution. For Abelian theories the constant remains undetermined.
This free constant is the so-called Fayet--Iliopoulos term. 

In Section~\ref{5d}, we have discussed FI terms in the context of vector multiplets. The reason why both terms are given the same name is that if we couple vector to hypermultiplets, the hypermultiplet FI term will amount to adding a constant term to the algebraic equation of motion for the auxiliary $Y^{ij}$.

In a conformal invariant theory, the Fayet--Iliopoulos term is not
possible. Indeed, dilatation invariance of the action needs
\begin{equation}
3\vec P_I=k^X \partial_X\vec P_I\,.
\end{equation}
Thus, $\vec P_I$ is completely determined [using~(\ref{momentmap})
or~(\ref{comDG})] as (see also~\cite{Wit:2001bk})
\begin{equation}\label{fixmomentmaps}
  -6\vec P_I=k^X\vec J_{XY}  k_I^Y= -\frac 23k^Xk^Z\vec J_Z{}^Y \covder_Y k_{IX}\,.
\end{equation}
The proof of the invariance of the action under the complete
superconformal group, uses the equation obtained from~(\ref{comDG})
and~(\ref{momentmap}):
\begin{equation}
\vec k^{X} \covder_X k_I^Y= \partial^Y \vec P_I\,.
\end{equation}
If the moment map $\vec P_I$ has the value that it takes in the
conformal theory, then~(\ref{constraintfP}) is satisfied due
to~(\ref{G-alg}), as can be seen by acting on that equation with
$k_Xk^Z\vec J_Z{}^W\covder_W$ and using~(\ref{commDkJ}),~(\ref{DDk})
and~(\ref{konRis0}). Thus, in the superconformal theory, the moment maps
are determined and there is no further relation to be obeyed, i.e.\ the
Fayet-Iliopoulos terms of the rigid theories are absent in this case.

To conclude, isometries of the scalar manifold that commute with
dilatations, see~(\ref{comDG}), can be gauged. The resulting theory has an
extra symmetry group $G$, its algebra is generated by the corresponding
Killing vectors.
\subsubsection{Local case}
We should now formulate the hypermultiplet in a background of the Weyl multiplet, i.e. make the superconformal symmetries local. This is a straightforward computation and leads to the following.

Imposing the local superconformal algebra we find the
supersymmetry rules:
\begin{eqnarray}
\delta q^X
&=& - \rmi \bar\epsilon^i \zeta^A f_{iA}^X \,,\\
\hat \delta \zeta^A &=& \frac 12 \rmi \slashed{\mathfrak D} q^X f_X^{iA}
\epsilon_i
  - \frac13 \gamma \cdot T k^X f^A_{iX} \epsilon^i - \frac 12 g \sigma^I k_I^X f_{iX}^A \epsilon^i + k^X f^A_{iX} \eta^i\, .\nonumber
\end{eqnarray}
The covariant derivatives are given by
\begin{eqnarray}
\mathfrak D_\mu q^X
&=& {\mathcal D}_\mu q^X + \rmi \bar{\psi}_\mu^i \zeta^A f_{iA}^X\, , \nonumber\\
{\mathcal D}_{\mu} q^X &=& \partial_\mu  q^X - b_\mu  k^X - V_\mu^{jk}
k_{jk}^X +
 g A_{\mu}^I k_I^X \,, \nn\\
\mathfrak D_\mu \zeta^A &=& {\mathcal D}_\mu \zeta^A - k^X f_{iX}^A \phi_\mu^i +
\frac 12 \rmi \slashed{\mathfrak D} q^X f_{iX}^A \psi_\mu^i + \frac13 \gamma \cdot
T k^X f_{iX}^A \psi_\mu^i \nonumber\\&&+ g \frac 12 \sigma^I k_I^X f_{iX}^A \psi^i_{\mu}\\
{\mathcal D}_\mu \zeta^A &=& \partial_\mu  \zeta^A + \partial_\mu  q^X
\omega_{XB} {}^A \zeta^B + \frac14 \omega_\mu {}^{bc} \gamma_{bc} \zeta^A
- 2 b_\mu  \zeta^A + g A_\mu^I t_{IB}{}^A \zeta^B\, .\nonumber
\end{eqnarray}
Similarly to the rigid case, requiring closure of the
supersymmetry algebra using these transformation rules, yields the equation of
motion for the fermions
\begin{eqnarray}
\Gamma_{\rm conf}^A &=& \slashed{\mathfrak D} \zeta^A + \frac 12 W_{CDB} {}^A
\zeta^B \bar{\zeta}^D \zeta^C - \frac 83 \rmi k^X f_{iX}^A \chi^i + 2
\rmi \gamma \cdot T \zeta^A 
\nonumber\\&& -g\left(\rmi k_I^X f_{iX}^A \psi^{iI}+ \rmi \sigma^I t_{IB} {}^A
\zeta^B\right). \label{fermion_eom_local}
\end{eqnarray}
The scalar equation of motion can be obtained from
varying~(\ref{fermion_eom_local}):
\begin{equation}
\hat \delta_Q \Gamma^A = \frac 12 \rmi f_X^{iA} \Delta^X \epsilon_i +
\frac14 \gamma^\mu \Gamma^A \bar{\epsilon} \psi_\mu - \frac14 \gamma^\mu
\gamma^\nu \Gamma^A \bar{\epsilon} \gamma_\nu \psi_\mu \,,
\end{equation}
where
\begin{eqnarray}
\Delta^X_{\rm conf}
&=&\Box^c q^X + \frac89 T^2 k^X +\frac43 D k^X + 8 \rmi \bar{\chi}^i \zeta^A f_{iA}^X  + \frac12 \bar{\zeta}^B \gamma^a
\zeta^C \mathfrak D_a q^Y {\cal R}_{YZC} {}^A f_B^{iZ} f_{iA}^X \nn\\
&& - \frac{1}{4} \mathfrak{D}_ZW_{CDB} {}^A \bar{\zeta}^D \zeta^C
\bar{\zeta}^E \zeta^B f_E^{iZ} f_{iA}^X -\frac14 W_{CDB}{}^A \bar{\zeta}^C
\gamma^a \zeta^B \mathfrak D_a q^Z f_Z^{iD} f_{iA}^X \nn\\
&&  -g \big( 2\rmi \bar{\psi}^{iI} \zeta^B t_{IB}{}^A f_{iA}^X - k_I^Y J_Y{}^X {}_{ij} Y^{Iij} \big)   + g^2 \sigma^I \sigma^J \mathfrak{D}_Y k_I^X k_J^Y\, ,
\end{eqnarray}
with
\begin{eqnarray}
\mathfrak D_X W_{ABC}{}^D&=&\partial_X W_{ABC}{}^D-3\omega_{X(A}{}^EW_{BC)E}{}^D+\omega_{XE}{}^DW_{ABC}{}^E\,,\nonumber\\
\mathfrak D_X k_I^Y&=&\partial_X k_I^Y+\Gamma_{XZ}{}^Y k_I^Z\,.
\end{eqnarray}
The superconformal d'Alembertian is given by
\begin{eqnarray}
\Box^{\rm c} q^X
&\equiv& \mathfrak D_a \mathfrak D^a q^X \nn \\
&=& \partial_a \mathfrak D^a q^X - \partial_Y k^X b_a \mathfrak D^a q^Y - \partial_Y k^{Xjk} V_a^{jk}
 \mathfrak D^a q^Y + \rmi \bar{\psi}_a^i \mathfrak D^a \zeta^A f_{iA}^X \nn\\
&& + 2 f_a{}^a k^X - 2\bar{\psi}_a \gamma^a \chi k^X + 4
\bar{\psi}_a^{(j} \gamma^a \chi^{k)} k_{jk}^X - \bar{\psi}_a^i \gamma^a
\gamma \cdot T \zeta^A f_{iA}^X \nn \\
&&- \bar{\phi}_a^i \gamma^a \zeta^A f_{iA}^X + \omega_a {}^{ab} \mathfrak D_b q^X
-\frac 12 g \bar{\psi}^a \gamma_a \psi^I k_I^X-\mathfrak D_a q^Y \partial_Y k_I^X
A^{aI} \nonumber\\&&+\mathfrak D_a q^Y \mathfrak D^a q^Z \Gamma_{YZ}{}^X\,.
\end{eqnarray}
Note that until this point, we managed to put the hypermultiplet in a background of the Weyl and vector multiplet, without having to introduce an action. Hence, the local superconformal hypermultiplets can also parametrize a hypercomplex target space.
Introducing a metric, we can construct the locally conformal supersymmetric action, which is then given by
\begin{eqnarray}\label{conf-hyperaction}
\mathcal{L}_h
&=& - \frac 12 g_{XY} \mathcal{D}_a q^X {\mathcal{D}^a} q^Y +\bar{\zeta}_A \slashed{\mathfrak D} \zeta^A + \frac 49 Dk^2 + \frac {8}{27} T^2 k^2
 - \frac{16}{3} \rmi \bar{\zeta}_A \chi^i k^X f_{iX}^A\nonumber\\&&
+ 2 \rmi \bar{\zeta}_A \gamma \cdot T \zeta^A
-\frac14 W_{ABCD} \bar{\zeta}^A \zeta^B \bar{\zeta}^C \zeta^D - \frac29 \bar{\psi}_a \gamma^a \chi k^2 \nonumber\\&&+ \frac 13 \bar{\zeta}_A \gamma^a \gamma \cdot T \psi_a^i k^X f_{iX}^A
+ \frac 12 \rmi \bar \zeta_A \gamma^a \gamma^b \psi_a^i \mathcal{D}_b q^X f_{iX}^A  + \frac23 f_a {}^a k^2 \nonumber\\&&- \frac16 \rmi \bar{\psi}_a \gamma^{ab} \phi_b k^2
 - \bar{\zeta}_A \gamma^a \phi_a^i k^X f_{iX}^A  + \frac{1}{12} \bar{\psi}_a^i \gamma^{abc} \psi_b^j \mathcal{D}_c q^Y J_Y {}^X {}_{ij} k_X\nn\\&& - \frac 19 \rmi \bar{\psi}^a \psi^b T_{ab} k^2 
 + \frac {1}{18} \rmi \bar{\psi}_a \gamma^{abcd} \psi_b T_{cd} k^2
 + g\Big( -\rmi \sigma^I t_{IB} {}^A  \bar{\zeta}_A \zeta^B \nn\\
&&- 2 \rmi k_I^X f_{iX}^A \bar{\zeta}_A \psi^{iI} -\frac 12 \sigma^I k_I^X f_{iX}^A \bar{\zeta}_A \gamma^a \psi_a^i
 - \bar{\psi}_a^i \gamma^a \psi^{jI}
P_{Iij}\nonumber\\&&+\frac{1}{2} \rmi \bar{\psi}_a^i \gamma^{ab} \psi^j_b \sigma^I
P_{Iij}+2 Y^{ij}_I P_{ij}^I\Big)  -\frac 12 g^2 \sigma^I \sigma^J k_I^X
k_{JX}\,.
\end{eqnarray}
\section{The map between quaternionic-like spaces}\label{geomchange}
Having completed the first step in the conformal construction, we now have to break the extra symmetries. In the easy example of Section~\ref{easyexample}, we have put the compensator to a constant value in order to fix dilatations. For the superconformal hypermultiplet, this will correspond to gauge fixing the scalar field that parametrizes the flow of the homothetic symmetry vector field $k$. However, as we want to keep the construction supersymmetric, this compensator should actually be part of a compensating multiplet. Hence, we have to find three other scalars that can be integrated out. It is obvious that these scalars should correspond to the flows generated by the $\su(2)$ symmetry vector fields $\vec k$, as we either have to break this local symmetry. Since the hypermultiplet scalars take values in a manifold, the fixing of these extra symmetries will have severe consequences on the geometry of that target space, as we will explain in the present section. It will turn out that this gauge fixing process geometrically is a projection from a hypercomplex to a quaternionic and from a hyperk\"ahler to a quaternionic-K\"ahler manifold~\cite{Bergshoeff:inprep1}. Hence, in the resulting $\mathcal N=2$ super-Poincar\'e theory, the hypermultiplet target space will be quaternionic if the theory does not have an action, or quaternionic-K\"ahler if it does. The gauge fixing process on the target space of the vectors will briefly be discussed in the next Section~\ref{respoinc}.
\subsection{Notation}
Before we start the construction, let us first comment on the notation. We will start from an $4(n+1)$-dimensional hypercomplex/hyperk\"ahler manifold and construct a projection to a $4n$-dimensional quaternionic/quaternionic-K\"ahler manifold. The higher-dimensional space will be called the \emph{large space} in contrast with the $4n$-dimensional \emph{small space}. The objects defined on the large space will be identified by `hats'. Indices on the large space are $\hat X$ running from $1$ to $4(n+1)$ and $\hat A=1,\dots , 2(n+1)$, while on the small space we will use respectively $X$ and $A$. The large space will be called $M_t^L$ and the small $M_t^S$.
\subsection{Coordinate choice and the Ansatz for the hypercomplex structure}
We will start with a manifold $M_t^L$ on which there is a 1-flat hypercomplex structure and a homothetic symmetry vector field $k$~(\ref{conf-constr}). Due to this vector field, we can build $\vec k$ following~(\ref{dil-su2}). To keep the discussion as general as possible, we will always explicitly mention when we are restricting to the case with metric. Note that we will call a 1-flat hypercomplex (hyperk\"ahler) structure together with $k$ a conformal hypercomplex (hyperk\"ahler) structure, and the corresponding manifold a conformal hypercomplex (hyperk\"ahler) manifold.

As the construction will turn out to be a projection on the space orthogonal to the flows generated by $k$ and $\vec k$, the first step is to find coordinates adapted to our setting. In other words, we want to find a set of coordinate functions that explicitly singles out these four directions. A similar discussion can be found in~\cite{Pons:2003ka}.

We start by considering the homothetic symmetry vector field $k^{\hat X}$ and we will choose one coordinate such that this vector has a convenient form. For $k\neq 0$, which we will suppose, it is always possible to
find coordinates $q^{\hat X}=\{z^0,y^p\}$, where $p=1,...,4n+3$,
such that the components of the homothetic symmetry vector field are
\begin{equation}\label{dilat-coord}
k^{\hat{X}}=3z^0 \delta _0^{\hat{X}}\,.
\end{equation}
Having singled out the direction of the flow of $k$, we now proceed in a similar way by considering the $\su(2)$ vector fields $\vec k$. Frobenius' theorem~\cite{Choquet} tells
that the three-dimensional hypersurface spanned by the three $\su(2)$ vectors can be parametrized by coordinates
$\{z^\alpha\}$ with ${\alpha= 1,2,3}$, such that $\vec k^{\hat X}$ is
nonzero only for $\hat X$ being one of the indices $\alpha $. Note that since the
vectors $k$ and $\vec k$ commute, this coordinate choice is possible while keeping $k$ in its fixed form~(\ref{dilat-coord}). The other $4n$ coordinates will be indicated by $q^X$, and will be said to parametrize the small space $M_t^S$. We thus have at this point
\begin{eqnarray}
&&q^{\hat X}=\left\{z^0 ,\,y^p\right\} = \left\{ {z^0,\, z^\alpha},\,
{q^X}\right\}\,, \qquad \alpha =1,2,3\,,\quad
X=1,\ldots ,4n\,,\nonumber\\
&&  k^{\hat X}=3z^0 \delta _0^{\hat{X}}\,, \qquad \vec k^0=\vec k^X=0\,.
 \label{splitcoord}
\end{eqnarray}
In this coordinates, the fact that $[k,\vec k]=0$ moreover yields
\begin{equation}
\partial_0 {\vec k}^\alpha =0\,.
\end{equation}
In conclusion, the Lie-algebra spanned by the vectors $\vec k$ can be written as
\begin{equation}\label{bundle}
\vec k^\beta \times \partial_\beta \vec k^\alpha=\vec k^\alpha\,,
\end{equation}
where $\vec k=\vec k(z^\alpha,q^X)$. From now on, we will assume ${\vec k}^\beta=(k^{\bar\alpha})^\beta$ to be
invertible as a three by three matrix, such that we can define ${\vec
k}_\alpha$ as $(-z^0)$ times the inverse of ${\vec k}^\alpha$  (the
normalization will be motivated below),
\begin{equation}\label{kinv}
(k^{\bar \alpha })_\alpha (k^{\bar \beta })^\alpha=-z^0\delta^{\bar
\alpha \bar \beta}\,,\qquad \mbox{or}\qquad \vec k^\alpha \cdot \vec
k_\beta =-z^0\delta ^\alpha _\beta \,.
\end{equation}


Having determined the set of coordinates in which we want to perform the projection, we can find an expression for the hypercomplex structures as they are listed below. The first column has been found by using the definition of the $\su(2)$ symmetry vector fields~(\ref{dil-su2}). The form of the other components has been determined by demanding that the quaternionic algebra~(\ref{defhcstr}) is satisfied. We have introduced the following object
\begin{equation}
  A_X^\alpha \equiv -\frac{1}{z^0}{\vec k}^\alpha \cdot {\vec J}_X{}^0\,,
 \label{defAXalpha}
\end{equation}
which will turn out to be proportional to the nontrivial $\su(2)$ connection on the small space.
\begin{equation}
   \begin{array}{lll}
    \hat{\vec J}_0{}^0=0 \,,\quad \quad & \hat{\vec J}_\alpha{}^0=
    \vec k_\alpha  \,,
     & \hat{\vec J}_X{}^0=A_X^\alpha \vec k_\alpha  \,,\\
    \hat{\vec J}_0{}^\beta=\frac{1}{z^0}\vec k^\beta   \,,& \hat{\vec J}_\alpha{}^\beta=\frac{1}{z^0}\vec k_\alpha \times \vec k^\beta  \,,\quad&
    \hat{\vec J}_X{}^\beta=\frac{1}{z^0}A_X^\gamma \vec k_\gamma \times \vec k^\beta
                               -\vec J_X{}^ZA_Z^\beta  \,,    \\
    \hat{\vec J}_0{}^Y=0 \,,& \hat{\vec J}_\alpha{}^Y=0 \,,& \hat{\vec J}_X{}^Y=\vec
    J_X{}^Y\,.
  \end{array}
 \label{allhatJ}
\end{equation}
The last equation in~(\ref{allhatJ}) means that the components of the hypercomplex
structure $\hat{\vec J}$ that lie along the small space space satisfy the defining relation~(\ref{defhcstr}) of an admissible base of an almost quaternionic structure on the small space. As a hypercomplex manifold is completely specified by a set of 1-flat hypercomplex structures, the above list completely specifies our Ansatz in which we want to perform the projection. In other words, starting from~(\ref{allhatJ}), we can in principle compute all relevant quantities (connections, curvatures,...).

Having introduced coordinates adapted to our setting, we can use the action of $k$ and $\vec k$ to determine the dependence on $z^0$ and $z^\alpha$ of the relevant objects. Hence, the first line of~(\ref{normofcomplstr}) determines the $z^0$ dependence of all quantities. We find:
\begin{equation}
  \partial _0\vec k^\alpha=0\,,\qquad\partial _0 \vec k_\alpha
=\frac{1}{z^0}\vec k_\alpha\,,\qquad\partial _0 A_X^\alpha =0\,,\qquad
\partial _0\vec J_X{}^Y=0\,.
 \label{partial0}
\end{equation}
The second line of~(\ref{normofcomplstr}) determines the $z^\alpha $ dependence
of the objects appearing in the complex structures. Replacing $\vec
\lambda $ by $-\frac{1}{z^0}\vec k_\alpha$, we obtain
\begin{equation}
  \left( \partial _\alpha -\frac1{z^0} \vec k_\alpha \times \right)
  (A_X^\beta \vec k_\beta )=\partial_X \vec k_\alpha\,,\qquad
  \left( \partial _\alpha -\frac{1}{z^0}\vec k_\alpha \times \right) \vec
  J_X{}^Y=0\,.
 \label{partialalpha}
\end{equation}
Using the above, we can compute that the transformation of $A_X^\alpha$ along the flow surfaces generated by $\vec k$ is determined in terms of the $\{q^X\}$ dependence of $\vec k$,
\begin{equation}
{\cal L}_{\vec k} A^\alpha _X \equiv \vec k^\beta
\partial_\beta  A^\alpha _X -  A^\beta _X \partial_\beta  \vec k^\alpha
=-\partial_X \vec k^\alpha\,.
 \label{LkAalphaX}
\end{equation}
\subsection{Admissible frames}
As~(\ref{allhatJ}) determines a G-structure, the structure group
of the frame bundle can be reduced from $\Gl(n+1,\mathbb{H})$ to
$\SU(2)\cdot\Gl(n,\mathbb{H})$, which is the group of fibrewise automorphisms that leaves~(\ref{allhatJ}) in its specified form. It is therefore useful to construct explicitly a set of admissible frames $\hat f_{\hat X}^{i\hat A}$ for this new G-structure, which will play the role of a set of Vielbeine. To do so, we split the index $\hat A$
into $(i,A)$ with $i=1,2$ and $A=1,\dots,2n$.
To be more precise, we split any section of $aLM_t^L$, i.e. any $\Gl(n+1,\mathbb{H})$-vector $\zeta
^{\hat{A}}(q^{\hat{X}})$ in $(\zeta ^i,\zeta ^A)$, which is a vector of
$\SU(2)\cdot\Gl(n,\mathbb{H})$, and where $\zeta^i$ is defined by
\begin{equation}
\sqrt{\frac12z^0 } \, \zeta ^i\equiv -\rmi \varepsilon ^{ij}\hat f^0
_{j\hat{A}} \zeta ^{\hat{A}}\,.
 \label{defzetai}
\end{equation}
This implies
\begin{equation}
  \hat f^0_{ij} = -\rmi\varepsilon_{ij}\sqrt{\frac12z^0}\,,\qquad \hat f^0_{iA}=0\,.
 \label{fchi}
\end{equation}
The factors $\rmi$ have been introduced to have proper reality conditions~(\ref{realfunctions}), such that
$\rho$ in~(\ref{defrho}) has components $\rho
_i{}^j= -E_i{}^j=-\varepsilon _{ij}$, and  has no off-di\-a\-go\-nal
elements like $\rho_i{}^A$. This is convenient for the formulation of quaternionic-like
manifolds that appear in supergravity, as discussed in Section~\ref{respoinc}.
Writing the first of~(\ref{allhatJ}) in terms of Vielbeine then yields
\begin{equation}
{\hat f}_0^{ij}=\rmi \varepsilon^{ij}\sqrt{\frac{1}{2z^0}}\,,
\label{f0ij}
\end{equation}
where we have made use of~(\ref{defJf}).
Furthermore, we can always construct frames for which
\begin{equation}
  \hat f^X_{ij}=0\,,
 \label{fXij0}
\end{equation}
by redefining $\zeta^A$ according to
\begin{equation}
  \zeta^{\prime A} = \zeta^A - \rmi\sqrt{2z^0}\hat f^{iA}_0 \zeta^j\varepsilon
  _{ji}\,.
 \label{Sinvzeta}
\end{equation}
Using $\hat{\vec J}_0{}^Y=0$ and~(\ref{f0ij}) this implies
\begin{equation}
  f^X_{i{\hat A}}\zeta ^{\hat A}=f^X_{iA}\zeta ^{\prime A}\,.
 \label{prooffXij}
\end{equation}
The active counterpart of the above passive transformation amounts to choosing a frame for which (\ref{fXij0}) holds. As a consequence, we can drop the primes.

From~(\ref{defJf}) and using~(\ref{allhatJ}),
the Vielbeine on the large space can now be determined in terms of $z^0$, $\vec k^\alpha $, $A_X^\alpha $ and the frames defined on the small space $\{f_X^{iA}\}$:
\begin{equation}
  \begin{array}{lll}
\hat f^0_{ij} = -\rmi\varepsilon _{ij}\sqrt{\frac12z^0}\,, \quad &
  \hat f^\alpha_{ij} =\sqrt{\frac{1}{2z^0}}\vec k^\alpha \cdot \vec
\sigma_{ij}\,,\quad &
 \hat f^X_{ij}=0\,, \\
\hat f^0_{iA}=0\,,\quad & \hat{f}^\alpha _{iA}=-f^X_{iA}A_X^\alpha\,,\quad & \hat f_{iA}^X=f_{iA}^X\,,\\
 \hat f_0^{ij}=\rmi\varepsilon^{ij}\sqrt{\frac{1}{2z^0}}\,,\quad & \hat f_\alpha^{ij}
  =  \sqrt{\frac{1}{2z^0}} \vec k_\alpha\cdot
\vec\sigma^{ij}
 \,,\quad &\hat f_X^{ij}= \sqrt{\frac{1}{2z^0}} \vec k_\alpha\cdot
\vec\sigma^{ij}A_X^\alpha \,,
 \\
\hat f_0^{iA}=0\,,\quad  & \hat f_\alpha^{iA}=0\,,\quad &\hat
f_X^{iA}=f_X^{iA}\,.
  \end{array} \label{allf}
\end{equation}
Remember that the index raising and lowering conventions imply $\vec
\sigma ^{ij}=\varepsilon ^{ik}\vec \sigma _k{}^j$ and
$\vec \sigma _{ij}=\vec \sigma _i{}^k\varepsilon _{kj}=\vec \sigma
_{ji}$.
\subsection{Mapping hypercomplex to quaternionic}
\label{ss:maphcquat}
In the previous sections, we have specified our coordinates and have listed the form of the hypercomplex structures in~(\ref{allhatJ}) and the form of the Vielbeine in~(\ref{allf}). Using this result, we will now compute the Obata connection coefficients on $TM_t^L$ in these particular coordinates, and use this to prove that the manifold parametrized by $\{q^X\}$ is quaternionic. We will then choose a particular set of connections on the quaternionic space, using the `$\xi$ transformation~(\ref{xitf}), that will prove to be very useful in the study of the case with a metric and in the physical applications of Section~\ref{respoinc}. We will then briefly discuss how the nontrivial $\su(2)$ connection on the quaternionic manifold is embedded in the higher space and will conclude by commenting on the inverse process of lifting a quaternionic into a hypercomplex space.
\subsubsection{Connections on the hypercomplex space}
\label{ss:connhc}
As the large space is hypercomplex, we can determine the Obata connection once the 1-flat hypercomplex structure is given. Hence, we can in principle compute $\hat \Gamma$ in our Ansatz~(\ref{allhatJ}) by using the definition~(\ref{defobata}). Subsequently, our choice of frames explained in the previous section and the fact that the Vielbeine are covariantly constant enables us to determine $\hat \omega$ on $aLM_t^L$.

There is however an easier way to compute the relevant connection coefficients in our Ansatz. 
From~(\ref{conf-constr}), we find that 
\begin{equation}\label{Gammanometric}
\hat\Gamma_{\hat{X}0}{}^{\hat{Y}}=\frac 1{z^0}(\frac12
\delta_{\hat{X}}^{\hat{Y}}-\delta_{\hat{X}}^0\delta_0^{\hat{Y}})\,.
\end{equation}
The covariant derivative on the $\su(2)$ symmetry vector fields~(\ref{DkSU2}) yields some other components of
the connection $\hat \Gamma$. Using the covariant constancy of the Vielbeine, we can
already compute most of the components of $\hat{\omega}$. Hence, only the connection
components $\hat{\Gamma}_{XY}{}^Z$ and $\hat{\omega}_{XA}{}^B$ should be
obtained from their definition since the remaining components
$\hat{\Gamma}_{XY}^0$ and $\hat{\Gamma}_{XY}^\alpha$ can be found by
considering the covariant constancy of $\hat{f}_X^{ij}$.

To simplify the notation, we introduce the following objects
\begin{equation}\label{defhatg}
\hat{g}_{\hat{X}\hat{Y}}\equiv
2\hat{\Gamma}_{\hat{X}\hat{Y}}{}^0\,,\qquad h_{XY}\equiv
\frac{1}{z^0}\left( \hat{g}_{XY}- A_X^\alpha \hat g_{\alpha \beta
  }A_Y^\beta\right) \,.
\end{equation}
We will show in Section~\ref{ss:proposalmetric} that if the large
manifold is hyperk\"ahler, then the metric will coincide with this
definition for $\hat g$ in the coordinates we are using. 
A specific component of $\hat g$ can be found by taking the $\hat{X}=0$ and $\hat{Y}=\alpha $ components of~(\ref{DkSU2})
in the basis~(\ref{splitcoord}). Using~(\ref{allhatJ}), we obtain
\begin{equation}\label{inversekg}
\vec{k}_\alpha=\hat{g}_{\alpha \beta}\vec{k}^\beta \Rightarrow
\hat{g}_{\alpha \beta}=-\frac1{z^0}\vec{k}_{\alpha}\cdot \vec{k}_\beta \,.
\end{equation}
Note that this is a nontrivial equation, as $\vec k_\alpha $ was so far
only defined as being proportional to the inverse of $\vec k^\alpha $,
see~(\ref{kinv}). This equation~(\ref{inversekg}) thus motivates the
normalization of $\hat{g}_{\alpha \beta}$. The different components of the Obata connection on the higher
dimensional space thus are
\begin{equation}
\begin{array}{ll}
\hat\Gamma_{00}{}^0 = -\frac{1}{2z^0}\,,&
\hat\Gamma_{00}{}^p = 0\,,\\
\hat\Gamma_{0p}{}^0 = 0\,,&\hat\Gamma_{0q}{}^p =
\frac{1}{2z^0} \delta^p_q \,, \\[2mm]
\hat\Gamma_{\alpha\beta}{}^0 = \frac12\hat{g}_{\alpha\beta}\,,&
\hat\Gamma_{\alpha\beta}{}^\gamma = \frac{1}{z^0} \vec{k}_{(\alpha}\cdot
\partial_{\beta)} \vec{k}^\gamma\,,\\
\hat\Gamma_{\alpha\beta}{}^X = 0 \,,&\hat{\Gamma}_{XY}{}^0=\frac12\hat{g}_{XY}\,,\\
\hat\Gamma_{X\alpha}{}^0 = \frac12 A^\beta _X \hat{g}_{\beta \alpha }=
-\frac{1}{2z^0} \hat{\vec J}_X{}^0 \cdot \vec k_\alpha \,,\quad&
\hat\Gamma_{X\alpha}{}^p = -\frac{1}{2z^0} \hat{\vec J}_X{}^p \cdot
\vec k_\alpha+\frac{1}{z^0}\vec{k}_\alpha \partial_X \vec{k}^p
\,,\\
\multicolumn{2}{l}{\hat{\Gamma}_{XY}{}^Z    = \Gamma^{\rm
Ob}{}_{XY}{}^Z-\frac{1}{3z^0} \hat{\vec J}
   {}_{(X}{}^\delta\cdot\left(  \vec k_\delta \delta _{Y)}^Z
    +\frac12 \vec k{}_\delta \times \vec J{}_{Y)}{}^Z\right)\,,}\\
\multicolumn{2}{l}{\hat\Gamma_{XY}{}^\alpha
 =\partial_{(X} A_{Y)}^\alpha - \hat\Gamma_{XY}{}^W A^\alpha_W-\frac12
 h_{Z(X}\vec J_{Y)}{}^Z\cdot \vec k^\alpha
 -\frac1{z^0}A_{(Y}^\beta\vec{k}^\alpha\cdot \partial_{X)}\vec{k}_\beta
\,.}
\end{array}
\label{summGammabl}
\end{equation}
Here $\Gamma^{\rm Ob}{}_{XY}{}^Z$ is the Obata connection defined
by~(\ref{defobata}) using the $\vec J$ complex structures, while
$\hat{\Gamma }$ are the Obata connection coefficients using $\hat{\vec J}$.
The induced connection has the following component expression
\begin{equation}
   \begin{array}{ll}
  \hat{\omega }_{0 i}{}^j=\hat{\omega }_{0 i}{}^A=\hat{\omega }_{0
  A}{}^j=0\,,\qquad &\hat{\omega }_{0 A}{}^B=\frac12f_Y^{iB}\partial _0 f_{iA}^Y+ \frac{1}{2z^0}
  \delta_A^B\,,\\
\hat{\omega }_{\alpha i}{}^A=0\,,\qquad &\hat{\omega }_{\alpha  A}{}^j=
0\,,\\
\hat{\omega }_{\alpha i}{}^j=-\rmi\frac{1}{2z^0}\vec
k_\alpha \cdot\vec \sigma _i{}^j\,,\qquad   & \hat{\omega }_{\alpha A}{}^B=\frac12 f_Y^{iB}\partial _\alpha f_{iA}^Y\,, \\
\hat\omega_X{}_i{}^A = \rmi\sqrt{\frac{1}{2z^0}} \varepsilon_{ik} \hat
f_X^{kA} \,,\quad &  \hat{\omega}_{XA}{}^i= - \rmi\sqrt{\frac{z^0
}{2}}\varepsilon ^{ij}f_{jA}^Y h_{YX}\,,\\
\hat{\omega }_{Xi}{}^j=A_X^\alpha \hat{\omega }_{\alpha i}{}^j\,,&
\\ \multicolumn{2}{l}{
\hat{\omega}_{XA}{}^B=\frac12f_Y^{iB}\partial_Xf_{iA}^Y+\frac12f_Y^{iB}f_{iA}^Z\left(
\hat{\Gamma}_{XZ}{}^Y +\frac1{2z^0}A_Z^\alpha\vec{k}_\alpha\cdot
\vec{J}_X{}^Y \right)\,.}
  \end{array}
 \label{allOmega}
\end{equation}
The last expression is just a rewriting of the covariant constancy of the Vielbeine, explicit for the
hatted quantities (remember that we work in a gauge for which there is no $\su(2)$
connection on the large space).
\subsubsection{Proof that the small space is quaternionic} \label{ss:proofQuat}
As already explained in Chapter~\ref{prelim}, the $\SU(2)$ bundle is trivial on
hypercomplex (hyperk{\"a}hler) manifolds as
it admits globally defined, everywhere nonzero sections (the complex structures). On
quaternionic(-K{\"a}hler) manifolds however, there is a nontrivial $\su(2)$ connection.
As a consequence, the integrability condition for the quaternionic
structure differs from the hypercomplex (hyperk{\"a}hler) case, as the
diagonal Nijenhuis tensor now is proportional to the $\su(2)$ Oproiu
connection~(\ref{defOproiu}).

Starting with the fact that the Nijenhuis tensor vanishes on the large
space, the corresponding tensor on the small space reads
\begin{equation}
  N^{\rm d}{}_{XY}{}^Z= -\frac 16\hat{\vec J}{}_{[X}{}^\alpha  \cdot
  \partial _\alpha \hat{\vec J}{}_{Y]}{}^Z=\frac{1}{6z^0} \vec k {}_\alpha \cdot
  \hat{\vec J}{}_{[X}{}^\alpha  \times \hat{\vec J}_{Y]}{}^Z\,,
 \label{Nij1}
\end{equation}
for which it can be checked that it satisfies~(\ref{condNquat}). Hence, the lower-di\-men\-sio\-nal manifold pa\-ra\-meterized by coordinates $\{q^X\}$ is quaternionic. Using~(\ref{defOproiu}), the Nijenhuis tensor determines the $\su(2)$ Oproiu connection
\begin{equation}
  \vec {\tilde \omega} _X =-\frac{1}{6z^0} \hat{\vec J}{}_X{}^\alpha  \times
  \vec k{}_\alpha  \,.
 \label{tilomega}
\end{equation}
\subsubsection{Connections on the quaternionic space}
\label{ss:connectionsQuat}
The purpose of this section is to obtain
an expression for the connection components on the lower-dimensional
quaternionic space. We have already found an $\su(2)$ connection~(\ref{tilomega}) and will start by computing the corresponding affine connection $\Gamma^{\rm Op}$. Thereafter, we will discuss the $\xi$-transformation~(\ref{xitf}).

Using~(\ref{defOproiu}), we find
\begin{eqnarray}
\Gamma^{\rm Op}{}_{XY}{}^Z&=&\Gamma^{\rm Ob}{}_{XY}{}^Z
-\vec{J}_{(X}{}^Z\cdot\vec{\tilde{\omega}}_{Y)}\label{exprOp}\\
&=&\hat{\Gamma}{}_{XY}{}^Z-\frac1{3z^0} \vec k{}_\alpha\cdot \vec J
{}_{(X}{}^V\delta_{Y)}^ZA_V^\alpha +\frac2{3z^0} A_{(X}^\alpha \vec
k{}_\alpha \cdot \vec J{}_{Y)}{}^Z+\frac1{3z^0} A_V^\alpha \vec
k{}_\alpha\cdot \vec J{}_{(X}{}^V\times \vec J{}_{Y)}{}^Z\,,\nonumber
\end{eqnarray}
where we have used the last equation of~(\ref{summGammabl}). Finally using
(\ref{allOmega}), we can write the undetermined component of the $\Gl(n+1,\mathbb{H})$
connection on $aLM_t^L$ as
\begin{eqnarray}
\omega^{\rm Op}{}_{XA}{}^B&=&\hat
\omega_{XA}{}^B-\frac1{6z^0}\hat{f}_{iA}^Z\hat{f}_{(X}^{iB}\vec
J{}_{Z)}{}^V A_V^\alpha \cdot \vec k{}_\alpha
\\
&&-\frac1{6z^0}\hat{f}_{iA}^Z\hat{f}_{Y}^{iB}A_{(Z}^\alpha \vec
k{}_\alpha\cdot \vec
J{}_{X)}{}^Y+\frac1{6z^0}\hat{f}_{iA}^Z\hat{f}_{Y}^{iB}A_V^\alpha \vec
k{}_\alpha\cdot  J{}_{(Z}{}^V\times \vec J{}_{X)}{}^Y \,,\nonumber
\end{eqnarray}
where ${\omega}^{\rm Op}{}_{XA}{}^B$ is the $\Gl(n,\mathbb{H})$ connection on $aLM_t^S$
corresponding to the Oproiu connections (\ref{tilomega}) and~(\ref{exprOp}).

As already explained in Chapter~\ref{prelim}, there is a family
of possible connections on a quaternionic manifold, related to each other
by a transformation parametrized by a one-form $\xi$
(\ref{xitf}). We can drastically simplify the expressions for the connections on the quaternionic space by performing such a transformation with the following parametrization,
\begin{equation}
\xi_X=-\frac1{6z^0}\vec J_X{}^Y\cdot \vec k_\alpha  A_Y^\alpha  \,.
\label{xigauge}
\end{equation}
This leads to
\begin{eqnarray}
\omega_{XA}{}^B&=& \hat{\omega}_{XA}{}^B \,,\qquad
\vec \omega_X = -\frac1{2z^0}A_X^\alpha \vec k_\alpha \,,\nonumber\\
\Gamma_{XY}{}^Z&=&\hat{\Gamma}_{XY}{}^Z + \frac1{z^0} A_{(X}^\alpha \vec
k{}_\alpha \cdot \vec J{}_{Y)}{}^Z
 = \hat{\Gamma}_{XY}{}^Z - 2 \vec \omega_{(X}\cdot \vec J{}_{Y)}{}^Z\,.
\label{eq:su2-connection}
 \end{eqnarray}
While in quaternionic manifolds this is just one possible choice, we will
later show that in qua\-ter\-ni\-o\-nic-K{\"a}hler manifolds the connection in this
gauge coincides with the Levi-Civita connection. Moreover, for this choice of $\xi$, the $\un(1)$ part of the curvature on the small space equals the one on the large space, as we will show in~(\ref{RisR}).

The values of the complex structures~(\ref{allhatJ}) can be re-expressed in terms of this new $\su(2)$ connection
as
\begin{equation}
   \begin{array}{lll}
    \hat{\vec J}_0{}^0=0 \,,\quad & \hat{\vec J}_\alpha{}^0={\vec k_\alpha}
      \,, & \hat{\vec J}_X{}^0=-2{z^0} {\vec \omega _X}  \,,\\
    \hat{\vec J}_0{}^\beta=\frac{1}{{z^0}}{\vec k^\beta }  \,,&
    \hat{\vec J}_\alpha{}^\beta=
    \frac{1}{{z^0}}{\vec k_\alpha} \times {\vec k^\beta }
    \,,\quad&
    \hat{\vec J}_X{}^\beta= {\vec k^\beta}  \times {\vec \omega _X}
                               -{\vec J_X{}^Z}\left({\vec \omega _Z}\cdot {\vec k^\beta}
                                \right)   \,,    \\
    \hat{\vec J}_0{}^Y=0 \,,& \hat{\vec J}_\alpha{}^Y=0 \,,& \hat{\vec J}_X{}^Y={\vec
    J_X{}^Y}\,.
  \end{array}
 \label{Jinomega}
\end{equation}
Note that we can of course perform any $\xi$ transformation on the connections of the quaternionic manifold as this does not change our Ansatz. We have however chosen to work with the above but the formulas can easily be translated to any other choice of $\xi$ by using~(\ref{transcurvxi}).
\subsection{The local $\SU(2)$-invariance}\label{ss:localinv}
All quaternionic-like manifolds admit a local $\SU(2)$-invariance, in the sense that we can rotate the complex structures in a position-dependent way. However, on the large space, we have been working in a gauge for which the corresponding $\su(2)$ connection was vanishing, as this connection is trivial. The fact that the $\SU(2)$ subbundle on $LM_t^L$ is trivial in contrary to the $\SU(2)$ subbundle of $LM_t^S$ can be understood by noting that the latter is embedded in the $\Gl(n+1,\mathbb H)$ subbundle of $LM_t^L$, which is in general nontrivial. In the present section, we will clarify this statement.

Suppose we perform a local $\SU(2)$ rotation on the quaternionic structures of the small space. This implies that the $\SU(2)$ connection $\vec \omega$ gets a gauge transformation. If we apply these transformations to the Ansatz (\ref{Jinomega}), one can check that the resulting hypercomplex structure transforms in a new hypercomplex structure if the $\SU(2)$ vectors $\vec k$ and their inverses transform as well. More concretely, we have to apply the following transformation:
\begin{eqnarray}
    \delta_{\rm SU(2)} \vec \omega_X &=& -\frac12 \partial_X \vec{\ell}   + \vec{\ell} \times \vec \omega_X, \qquad \delta_{\rm SU(2)}\vec J_X{}^Y=\vec{\ell} \times \vec J_X{}^Y,\nonumber\\
   \delta_{\rm SU(2)}\vec k^\alpha&=&\vec{\ell} \times \vec k^\alpha,
   \qquad\qquad\qquad \delta_{\rm SU(2)} \vec k_\alpha =\vec{\ell} \times \vec k_\alpha, \label{delSU2A}
\end{eqnarray}
where the $\SU(2)$ parameter $\vec{\ell}(q^X)$ cannot depend on $z^0$ or
$z^\alpha$. 

The $\SU(2)$ connection on the hypercomplex manifold is always trivial while on the quaternionic manifold, it is not. This can be explained by noting that the quaternionic $\SU(2)$ is embedded in the hypercomplex $\Gl(n+1,\mathbb H)$ as can be deduced from~(\ref{allOmega}) using~(\ref{eq:su2-connection}),
\begin{equation}\label{indconn}
\hat \omega_{Xi}{}^j=\rmi \vec \omega_X \cdot \vec \sigma_i{}^j\,.
\end{equation}
This connection is in general nontrivial, meaning that its curvature does not vanish.\footnote{Note that due the Ansatz~(\ref{allhatJ}), this $\su(2)$ part of the $\gl(n+1,\mathbb H)$ connection does not mix with its other components.} 
\subsubsection{Inverse mapping} \label{ss:invqhc}
In the above, we have shown that by carefully projecting out four specific directions on a conformal hypercomplex manifold, we can construct a family of quaternionic spaces related by a $\xi$ transformation~(\ref{xitf}). A natural question to ask then is if the inverse is also possible. Hence, we start from a manifold with a 1-flat quaternionic structure $\vec J$. Suppose we calculate the Oproiu connection, using~(\ref{defOproiu}) and perform a random $\xi$-transformation~(\ref{xitf}). Let us write this as 
\begin{eqnarray}
\Gamma(\xi)_{XY}{}^Z&=&\Gamma^{\rm Op}{}_{XY}{}^Z+S_{XY}^{ZV}\xi_V\,,\nonumber\\
\vec \omega(\xi)_X&=&\vec \omega^{\rm Op}{}_X+\vec J_X{}^Y \xi_Y\,.
\end{eqnarray}
Consider now the left-invariant vector fields $\vec k^\alpha$ on the $\SU(2)$ group manifold, with coordinates $z^\alpha$, and consider their inverse $\vec k_\alpha$ via~(\ref{kinv}), in which $z^0$ is a parameter. We can now give the quaternionic structure $\vec J$ a dependence on $z^\alpha$ using~(\ref{partialalpha}), and similarly for the Oproiu connection using
\begin{equation}
(\partial_\alpha-\frac1{z^0} \vec k_\alpha \times)\vec \omega^{\rm Op}{}_X =0\,.
\end{equation}
Subsequently, we can use~(\ref{Jinomega}), with $\vec \omega$ substituted by this $\vec \omega(\xi)$, to construct a set of $\hat {\vec J}(\xi)$. It is an easy exercise to show that the latter form an almost hypercomplex structure. However, the structure is not necessarily 1-flat. To check this, we should compute the diagonal Nijenhuis tensor or equivalently, check if the almost hypercomplex structure is preserved by the Obata connection calculated with this set of $\hat{\vec J}(\xi)$. It then turns out that $\xi$ should parametrize an honest quaternionic $\xi$ transformation~(\ref{xitf}), i.e.
\begin{equation}
\partial_0\xi_X=\partial_\alpha\xi_X=0\,.
\end{equation}
Moreover, in order for $\hat {\vec J}(\xi)$ to be 1-flat, we find that the following relation should hold
\begin{equation}\label{defxirel}
\vec R\big[\omega(\xi)\big]_{XY}=-\frac12\vec J_{[X}{}^Z h_{Y]Z}(\xi)\,,
\end{equation}
where the $\su(2)$ curvature~(\ref{defcurvs}) is calculated with respect to $\vec \omega (\xi)$ and $h(\xi)$ is defined similar to~(\ref{defhatg}),
\begin{eqnarray}
h(\xi)_{XY}&=&\frac1{z^0}(2\hat \Gamma(\xi)_{XY}{}^0+4z^0 \vec \omega(\xi)_X \cdot \vec \omega(\xi)_Y)\nonumber\\
&=&-\frac13\left[ 4\vec J_{(X}{}^Z\cdot \vec R_{Y)Z}(\xi )
  +(\vec J_X{}^U\times\vec
  J_Y{}^Z)\cdot \vec R _{UZ}(\xi )\right] \,, \label{hinvecR}
\end{eqnarray}
where $\hat \Gamma$ is the Obata connection computed with $\hat{\vec J}$. Note that the expression in the final line of~(\ref{hinvecR}) can also be found by solving $h$ from~(\ref{defxirel}) directly. Hence, comparing~(\ref{defxirel}) with~(\ref{R=JB}), we see that we can find a choice of quaternionic connections that is compatible with the Ansatz~(\ref{Jinomega}) by demanding that the antisymmetric part of $B$ vanishes. This part of $B$ is moreover proportional to the antisymmetric part of the quaternionic Ricci tensor as can be seen in~(\ref{defRB}). From the $\xi$ transformation of the antisymmetric part of the quaternionic Ricci tensor
\begin{equation}\label{transfxiu1}
R(\xi)_{[XY]}=R^{\rm Op}{}_{[XY]}-4(n+1)\partial_{[X}\xi_{Y]}=-\partial_{[X}\Gamma^{\rm Op}{}_{Y]Z}{}^Z-4(n+1)\partial_{[X}\xi_{Y]}\,,
\end{equation}
we can see that 
\begin{equation}
\xi_X=-\frac1{4(n+1)}\Gamma^{\rm Op}{}_{XY}{}^Y\,,
\end{equation}
is certainly a valid choice. Hence, we have found a value for the one-form $\xi$, for which $\hat{\vec J}$ in~(\ref{Jinomega}) is a 1-flat hypercomplex structure on a manifold with coordinate functions $\{z^0,z^\alpha,q^X\}$. Note that this choice of $\xi$ certainly is not unique as e.g. $\xi_X+\partial_X \lambda(q)$ would also yield a 1-flat hypercomplex structure. Actually, any connection for which the antisymmetric part of the Ricci tensor satisfies
\begin{equation}
R(\xi)_{[XY]}=\Pi_{XY}{}^{VW}u_{[VW]}\,,
\end{equation}
for any antisymmetric $u$ and
with $\Pi$ defined in~(\ref{defpi}) is such that the antisymmetric part of $B$ vanishes, implying that any such choice of $\xi$ represents a set of connections for which we can use our Ansatz~(\ref{Jinomega}) to build a 1-flat hypercomplex structure on the manifold with coordinates $\{z^0,z^\alpha,q^X\}$.
\subsection{Mapping hyperk{\"a}hler to quaternionic-K{\"a}hler}
\label{ss:maphkqk}
\subsubsection{The candidate metric}\label{ss:proposalmetric}
As we already mentioned in Chapter~\ref{prelim}, a hyperk{\"a}hler space is a manifold with a 1-flat hypercomplex hermitian structure. This means that the hyperk\"ahler metric $\hat{g}_{\hat{X}\hat{Y}}$ (for which we will discuss its relation to~(\ref{defhatg}) shortly) is hermitian with respect to the 1-flat hypercomplex structure and its Levi-Civita connection equals the Obata connection,
i.e.
\begin{eqnarray}
\hat{\vec J}_{\hat{X}}{}^{\hat{Z}}\hat{g}_{\hat{Z}\hat{Y}}
  +\hat{\vec J}_{\hat{Y}}{}^{\hat{Z}}\hat{g}_{\hat{Z}\hat{X}}&=&0\,,
 \label{hermitHatg}\\
  \partial _{\hat Z} \hat{g} _{\hat{X}\hat{Y}}
  -2\hat{\Gamma }_{\hat{Z}(\hat{X}}{}^{\hat{W}}\hat{g}
  _{\hat{Y})\hat{W}}&=&0\,.
 \label{condQuatK}
\end{eqnarray}
Because of the introduction of the metric, the vector $k$ becomes a homothetic Killing vector~(\ref{homkilvect}), while the vectors $\vec k$ are now Killing vector fields. The consequence of this is that we can find coordinates~(\ref{splitcoord}) in which the metric takes the following form 
\begin{eqnarray}\label{QK}
\rmd \hat s^2  & = & -\frac{(\rmd z^0)^2}{z^0}
 +\Big\{ z^0h_{XY} (q)\rmd q^X \rmd q^Y \\
&& + \hat{g}_{\alpha\beta} (z^0,z^\alpha)[\rmd
z^\alpha + 
A_X^\alpha(z^\alpha,q)\rmd q^X][\rmd z^\beta
+ 
A_Y^\beta (z^\alpha,q)\rmd q^Y]\Big\} \,.\nonumber
\end{eqnarray}
Note that we will take as signature for this metric $(-,-,-,-,+,\dots,+)$ where the minus signs cor\-respond to the coordinates $\{z^0,z^\alpha\}$. In the conformal approach, these coordinates are the bosonic part of the compensating hypermultiplet. Therefore, they have to have the wrong sign for the kinetic energy, which is reflected in the signature of this metric.
In~(\ref{QK}), $\hat{g}_{\alpha\beta}$ is defined in~(\ref{inversekg}), while $A_X^\alpha$ does not depend on $z^0$~(\ref{partial0}) and transforms along the flows of $\vec k$ as in~(\ref{LkAalphaX}). 

This form of the metric~(\ref{QK}) already reveals a lot of the underlying geometrical structure. First of all, substituting $r=2\sqrt{z^0}$ the metric becomes
\begin{equation}
\rmd\hat s^2=dr^2+r^2 h_{pq}^t(z^\alpha,q^X) \rmd y^p\rmd y^q\,.
\end{equation}
Thus, the hyperk\"ahler manifold is a cone over some $4n+3$-dimensional space with metric $h^t$. Moreover, this $4n+3$-dimensional space is a so-called 3-Sasakian manifold~\cite{Boyer:1998sf}. Such a manifold is a three-sphere fibration\footnote{Actually, the three-sphere might be divided by the action of a discrete group.} over a quaternionic-K\"ahler manifold, which in our case is the space parametrized by $\{q^X\}$, as we will show shortly.  

The $\hat{Z}=p$, $\hat{X}=q$, $\hat{Y}=0$ part of~(\ref{condQuatK}) now leads
to
\begin{equation}
  \hat{g}_{\hat{X}\hat{Y}}= 2\hat{\Gamma}_{\hat{X}\hat{Y}}{}^0\,.
 \label{hatg}
\end{equation}
Notice the difference with~(\ref{defhatg}), which was a definition for
arbitrary hypercomplex manifolds. Here we prove that any good metric on
hyperk\"ahler manifolds is of the form~(\ref{hatg}) after choosing suitable
coordinates.
\subsubsection{Levi-Civita Connection in quaternionic-K{\"a}hler manifolds} \label{ss:LCconnection}
For a hyperk\"ahler manifold, the Levi-Civita connection
necessarily equals the Obata connection. In the quaternionic-K{\"a}hler case,
there is a family of possible connections that preserves the quaternionic structure. The unique Levi-Civita connection selects one particular
member of this family. 

In the next section, we will show that $h$ is the correct metric to consider on the small space. A first sign of this is that when calculating the Levi-Civita connection on the hyperk{\"a}hler manifold using~(\ref{QK}), it is related to the one on the
quaternionic-K{\"a}hler manifold $\Gamma^{\rm LC}{}_{XY}{}^Z$ computed with $h$. More precisely, we find that
the quaternionic-K\"ahler connection equals the one mentioned in~(\ref{eq:su2-connection}).
\begin{equation}
\Gamma^{\rm LC}{}_{XY}{}^Z\equiv \frac12
h^{ZV}(2\partial_{(X}h_{Y)V}-\partial_Vh_{XY})=\hat{\Gamma}^{\rm
LC}{}_{XY}{}^Z
  +\frac{1}{z^0}A_{(X}^\alpha \vec J_{Y)}{}^Z \cdot \vec k_\alpha \,,
 \label{resultLCvsOb}
\end{equation}
where $\hat{\Gamma}^{\rm LC}{}_{XY}{}^Z$ is a component of the metric
connection with respect to $\hat{g}_{\hat{X}\hat{Y}}$.
\subsubsection{Conditions for a good metric}
Following the discussion in Section~\ref{ss:maphcquat}, we already know that the space parametrized by $\{q^X\}$ is quaternionic. We will now show that the manifold is actually quaternionic-K\"ahler when we start the projection from a hyperk\"ahler manifold. To be able to do so, we have to consider the components of the conditions~(\ref{hermitHatg}) and~(\ref{condQuatK}) along the small space. Using our reduction Ansatz, we find for the hermiticity condition
\begin{equation}
  \hat{\vec J}_{(\hat{X}}{}^{\hat{Z}}\hat{g}_{\hat{Y})\hat{Z}}=0\ \Leftrightarrow
  \vec J_{(X}{}^Z h_{Y)Z}=0\,,
 \label{hermgIsHermh}
\end{equation}
where $\hat{g}$ and $h$ are those defined in~(\ref{defhatg}). We can thus
state that the candidate metric $\hat{g}$ on the hypercomplex manifold is
hermitian if and only if the candidate metric on the quaternionic space
is.

The condition~(\ref{condQuatK}) finally leads to 
\begin{equation}
\partial _{\hat Z} \hat{g} _{\hat{X}\hat{Y}}
  -2\hat{\Gamma }_{\hat{Z}(\hat{X}}{}^{\hat{W}}\hat{g}
  _{\hat{Y})\hat{W}}=0\Leftrightarrow
\partial _Z h_{XY} -2\Gamma _{Z(X}{}^Wh_{Y)W}=0\,.
 \label{condhconst}
\end{equation}
In this expression, $\Gamma$ is the Levi-Civita connection calculated with $h$~(\ref{resultLCvsOb}).
We can thus conclude that the small space is quaternionic-K\"ahler if and only if the large space is hyperk\"ahler.

As explained in Chapter~\ref{prelim}, the introduction of a metric is equivalent with the introduction of an invariant scalar product on $aLM_t^L$, which in the case at hand boils down to the introduction of an invariant symplectic structure $\hat C$. Similarly to the metric $\hat g$, we can decompose it in our Ansatz, which yields
\begin{equation}
 \hat C_{AB} =  C_{AB} \,, \qquad \hat C_{ij} = \varepsilon _{ij}\,, \qquad \hat C_{iA} = 0 \,. \\[1mm]
 \label{mapCAB}
\end{equation}
\subsubsection{Inverse mapping}\label{ss:invmapqkhk}
Due to the above, the lifting of a quaternionic-K\"ahler manifold to a hyperk\"ahler manifold is much more simple than in the affine case. The reason is that we do not have to worry about finding the right `choice' of $\xi$ for use in~(\ref{Jinomega}). Hence, the lifting can be done as follows.

Starting from a quaternionic hermitian structure $(\vec J,h)$, we can compute the Levi-Civita connection $\Gamma$ and the corresponding $\su(2)$ connection $\vec \omega$. Similar to the quaternionic case, we construct $\vec k^\alpha$ and $\vec k_\alpha$ starting from the left-invariant vector fields on $\SU(2)$, and we use them to give a $z^\alpha$ dependence of the complex structures using~(\ref{partialalpha}) and similarly for the $\su(2)$ connection,
\begin{equation}
(\partial_\alpha-\frac1{z^0}\vec k_\alpha \times)\vec \omega_X=0\,.
\end{equation}
Using~(\ref{Jinomega}), we can directly construct the 1-flat hypercomplex structure $\hat{\vec J}$. To find the corresponding metric $\hat g$, note that $A_X^\alpha=2\vec \omega_X \cdot \vec k^\alpha$, which can be used in the Ansatz for the metric~(\ref{QK}). Hence, we have found a 1-flat hypercomplex hermitian structure, meaning that the manifold parametrized by $\{z^0,z^\alpha,q^X\}$ is hyperk\"ahler.
\subsection{Curvatures}\label{ss:curvatures}
We will now discuss how the curvature is reduced during our projection. We will keep everything as general as possible, by only specifying at the very end to the case where there exists a compatible metric. Note that we will perform all calculations for the connections on the small space corresponding to the specific choice of $\xi$ as in~(\ref{eq:su2-connection}). 

In the large space, some of the components of the curvature are zero
due to fact that the homothetic and $\su(2)$ vector fields are eigenvectors of the curvature with zero eigenvalue~(\ref{konRis0}). This (together with the first Bianchi identity of the curvature) shows
that the only possible nonzero components of the curvature are
$\hat{R}_{XYZ}{}^{\hat W}$. A first direct consequence of this is that the Ricci tensor on the large space is a two-form\footnote{The Ricci tensor on a hypercomplex manifold is antisymmetric.} that lies along the small space.

We start the reduction process by deriving the following relation. An expression for the
$\su(2)$ curvature can be found by considering $\hat{D}_{[X}
\hat{f}_{Y]}^{ij}=0$:
\begin{equation}\label{RisJ}
\vec{\mathcal{R}}_{XY}=-\frac12 \vec{J}_{[X}{}^Zh_{Y]Z}\, ,
\end{equation}
which is the well-known equivalence between the $\su(2)$ curvature and the
triple of fundamental two-forms in the quaternionic-K{\"a}hler case~\cite{AM1996}. However, the above is also valid on quaternionic manifolds, where $h$ then is defined via~(\ref{defhatg}). Note that we found exactly this condition~(\ref{defxirel}) in order to be able to lift a quaternionic manifold to a hypercomplex one.

We can now use~(\ref{RisJ}) in computing the reduction of the curvature, and find 
\begin{eqnarray}
  &&\!\!\!\!\!\!\!\hat R_{XYZ}{}^W =   R_{XYZ}{}^W+\frac12\delta _{[X}^Wh_{Y]Z}
  -\frac12\vec J_Z{}^{W}\cdot \vec J_{[X}{}^{V}h_{Y]V}
  -\frac12\vec J_Z{}^{V}\cdot \vec J_{[X}{}^{W}h_{Y]V}\,.\nonumber\\
 \label{hatRRRRicS}
\end{eqnarray}
Taking the trace of the above relation yields the reduction of the Ricci tensor,
\begin{equation}
  \hat R_{XY}=  R_{XY}+\frac{2n+1}2h_{XY}
  +\frac12\vec{J}_X{}^Z\cdot\vec{J}_Y{}^Wh_{ZW}\,.
 \label{decompricci}
\end{equation}

The left-hand side has only an antisymmetric part and determines the
$\un(1)$ part of the curvature, see~(\ref{defu1}). We thus find that the
antisymmetric part of the Ricci tensor is the same for the large and for
the small space:
\begin{equation}\label{RisR}
\hat R_{XY}= R_{[XY]}\,.
\end{equation}
This leads us to the important conclusion that for the present choice of $\xi$, the structure group of the frame bundle on the quaternionic space admits a $\U(1)$ part if and only if it does on the hypercomplex space. This also clarifies why~(\ref{RisJ}) and the final line in~(\ref{R=JB}) are compatible. First note that the tensor $B$ defined in~(\ref{defRB}) on the quaternionic manifold reads
\begin{equation}
  B_{XY}=\frac1{4(n+1)}\hat{R}_{XY}-\frac14h_{XY}\,.
 \label{BisRplush}
\end{equation}
Starting now from the general identity on a hypercomplex manifold
\begin{equation}
\hat \Pi_{\hat X\hat Y}{}^{\hat V \hat W}\hat R_{\hat V \hat W}=\hat R_{\hat X \hat Y}\,,
\end{equation}
with $\hat \Pi$ as in~(\ref{defpi}),
it can easily be checked that the antisymmetric part of $B$ cancels in~(\ref{R=JB}) yielding~(\ref{RisJ}). 

The symmetric part of~(\ref{decompricci}) determines the symmetric part
of the Ricci tensor for the quaternionic manifold
\begin{equation}
R_{(XY)}=-\frac{2n+1}2h_{XY}
  -\frac12\vec{J}_X{}^Z\cdot\vec{J}_Y{}^Wh_{ZW}\,.
\end{equation}
Hence, this part of the quaternionic Ricci tensor admits a universal expression in terms of the candidate metric $h$. This is an extension of the standard result that the Ricci tensor of a quaternionic-K\"ahler manifold formally equals the one calculated from the natural metric on $\mathbb HP^n=\Gl(n+1,\mathbb H)/\Gl(n,\mathbb H)$~\cite{Alekseevsky1968}. 

Using the expression for $B$~(\ref{BisRplush}), we are then able to identify the reduction of the curvature tensor~(\ref{hatRRRRicS}) as
\begin{eqnarray}
  \hat R_{XYZ}{}^W =   \left( R- R^{\rm Ric}_{\rm symm}\right) _{XYZ}{}^W
  = \left( R^{({\rm
  W})}+R^{\rm Ric}_{\rm antis}\right)_{XYZ}{}^W\,.
 \label{hatRRRic}
\end{eqnarray}
Otherwise stated, the part of the curvature on the hypercomplex manifold that lies completely in the small space equals the antisymmetric Ricci part plus the Weyl part of the quaternionic curvature. Moreover, the former is proportional to the antisymmetric part of the hypercomplex Ricci tensor,
\begin{equation}
  \hat B_{XY}=\frac{1}{4(n+2)}\hat R_{[XY]}\,\Rightarrow
   \hat{R}^{\rm Ric}_{\rm antis}{}_{XYZ}{}^W= \frac{n+1}{n+2}
  R^{\rm Ric}_{\rm antis}{}_{XYZ}{}^W\,.
 \label{BhatB}
\end{equation}
A similar consideration holds for the Weyl curvature of the hypercomplex manifold. The latter is
determined by a traceless tensor $\hat{\mathcal W}$, see~(\ref{WfromRW}). But for
hypercomplex manifolds there is also a tensor
$\hat{W}$~(\ref{def-W}) whose trace part
determines the antisymmetric part of the Ricci tensor, and whose
traceless part is $\hat{\mathcal{W}}$. The
vanishing of all the curvature components mentioned in the beginning of this section implies that the
nonvanishing parts of $\hat{W}$
are only $\hat{W}_{ABC}{}^{\hat{D}}$, i.e. $\hat{W}_{ABC}{}^D$
and $\hat{W}_{ABC}{}^i$. The latter will not be important for the
reduction to the small space. Note, that the traceless 
$\hat{\mathcal{W}}_{\hat{A}\hat{B}\hat{C}}{}^{\hat{D}}$ has as
nonzero components 
\begin{eqnarray}
&&\hat{\mathcal{W}}_{ABC}{}^D   =\hat{W}_{ABC}{}^D
-\frac{3}{2(n+2)}\delta ^D_{(A}\hat{W}_{BC)E}{}^E\,, \qquad
 \hat{\mathcal{W}}_{ABC}{}^i   =\hat{W}_{ABC}{}^i \,,\nonumber\\
&&\hat{\mathcal{W}}_{iAB}{}^j =
\hat{\mathcal{W}}_{AiB}{}^j=\hat{\mathcal{W}}_{ABi}{}^j=
-\frac{1}{2(n+2)}\delta ^j_i\hat{W}_{ABE}{}^E \,.
 \label{hatcalW}
\end{eqnarray}
This implies that the Weyl part on the hypercomplex space depends also on
the trace $\hat{W}_{ABE}{}^E$. Bearing this in mind, we can understand the reduction of $\hat W$,
\begin{eqnarray}\label{hatWiscW}
\hat{W}_{ABC}{}^D=\mathcal{W}_{ABC}{}^D+\frac3{2(n+1)}\delta^D_{(A}\hat{W}_{BC)E}{}^E\,.
\end{eqnarray}
Hence, $\mathcal{W}_{ABC}{}^D$ is the traceless part of
$\hat{W}_{ABC}{}^D$.

Symbolically, we can represent the dependence of parts of the curvature
tensors on basic tensors as follows:
\begin{equation}
  \begin{array}{ccccccc}
    \hat{R} & = &   &   & \hat{R}^{\rm Ric}_{\rm antis} & + & \hat{R}^{({\rm W})} \\
            &   &   &   & \uparrow & \nearrow & \uparrow \\
     &   & h_{XY} &   & \hat{W}_{ABC}{}^C &   & \mathcal{W}_{ABC}{}^D \\
      &   & \downarrow &   & \downarrow &   & \downarrow \\
    R & = & R^{\rm Ric}_{\rm symm}  & + & R^{\rm Ric}_{\rm antis}  & + & R^{(\rm W)} \\
  \end{array}
 \label{schemCurvDecomp}
\end{equation}
For the mapping between a hyperk{\"a}hler manifold and a quaternionic-K{\"a}hler
manifold, there is no antisymmetric part of the Ricci tensor, and hence
$\hat{W}_{ABC}{}^D$ is traceless. Therefore, the relation
(\ref{hatWiscW}) reduces to
\begin{equation}
  \hat{W}_{ABC}{}^D={\cal W}_{ABC}{}^D\,.
 \label{hatWisW}
\end{equation}
It implies that the hyperk{\"a}hler curvature components along the
quaternionic directions are the Weyl part of the quaternionic-K{\"a}hler
curvature. As already mentioned, the Ricci part of this quaternionic-K{\"a}hler curvature is the
same expression as for $\mathbb{H}P^{n}$. In fact, we find
\begin{equation}
  B_{XY}=-\frac14h_{XY}\,.
 \label{Bsymm}
\end{equation}
In quaternionic-K\"ahler manifolds, the following relation holds in general~\cite{AM1996},
\begin{equation}\label{BinqK}
B_{XY}=\frac14\nu g_{XY}\,,\qquad \nu=\frac1{4n(n+2)}R\,.
\end{equation}
Comparing this with~(\ref{Bsymm}), and
using the metric $\hat g_{XY}=z^0 h_{XY}=g_{XY}$ implies
\begin{equation}
  \nu =-\frac{1}{{z^0}} \,.
 \label{valuenu}
\end{equation}
In the context of supergravity, the value of ${z^0}$ will be fixed to
$\kappa ^{-2}$, where $\kappa $ is the gravitational coupling constant (see Section~\ref{respoinc}).

Finally note that for 1-dimensional case, $n=1$, we had restricted the
definition of quaternionic manifolds in
appendix~B.4 of~\cite{Bergshoeff:2002qk} by adding special requirements, as was also done in the
mathe\-ma\-tical literature~\cite{Alekseevsky1975}.  Here we find that these
relations are automatically fulfilled in the embedded quaternionic
manifolds. Hence, they are unavoidable in a supergravity context.
\subsection{Reduction of the symmetries}\label{ss:symmetries}
\subsubsection{Introduction}
In the case of manifolds on which there is no metric, the question of
defining symmetries needs some careful consideration. We will therefore call $\{k_I\}$ a set of symmetry vector fields generating $\delta(\beta) q^X=\beta^I k^X_I(q)$ if the following conditions are met,\footnote{These conditions for a symmetry have to be met for all quaternionic-like manifolds. We have therefore dropped the `hats' in this paragraph.} which we already encountered in the study of symmetries of rigid superconformal hypermultiplets in Section~\ref{rigidsuconfhypers}. 
\begin{enumerate}
\item We impose the condition~(\ref{DDk}) on the curvature. 
This relation can be understood by noting that it is exactly the condition in order that the harmonic equation
\begin{equation}
\Box q^X\equiv \partial^2q^X +\Gamma_{YZ}{}^X\partial q^Y\partial q^Z=0\,,
\end{equation}
is left invariant under $\delta(\beta)q^X$. 
\item We also demand that the symmetry generators normalize the complex structures~(\ref{kInormalJ}).
\end{enumerate}
In case of a compatible metric, we will ask that the Killing equation 
\begin{equation}
\mathfrak D_{(X}k_{IY)}=0\,,
\end{equation}
is obeyed instead of the first demand, which is then just the integrability condition. Below, we will explicitly check in what case the symmetry vector fields $\hat k_I$ on the large space reduce to symmetries on the small space. 

But before we do so, let us first comment briefly on the definition of a moment map, which is important in the context of Section~\ref{respoinc}. On a hyperk\"ahler manifold, the action of a Killing vector field can be encoded in a so-called moment map, introduced in~(\ref{momentmap}), which will be denoted by $\hat{\vec P}_I$ from now on. Moreover, for conformal hyperk\"ahler manifolds, the symmetries should also commute with the dilatational and $\su(2)$
symmetry vector fields, which was expressed~(\ref{comDG}).
Consistency then completely fixes the moment maps as~(\ref{fixmomentmaps}).                                  

We can similarly define a moment map on quaternionic and quaternionic-K\"ahler manifolds. The definition then reads
\begin{equation}\label{momentmapq}
4n\tilde \nu \vec P_I=-\vec J_X{}^Y\mathfrak D_Y k_I^X\,,
\end{equation}
where $k_I$ is a symmetry vector field and $\tilde \nu$ is some number that can be taken equal to $\nu$ in~(\ref{BinqK}) on a quaternionic-K\"ahler manifold.
This implies that the normalization of the quaternionic structure~(\ref{kInormalJ}) can be rewritten as        \begin{equation}\label{normal2}
\mathcal{L}_{k_I} \vec J_X{}^Y=2\vec J_X{}^Y\times \left(
{\vec \omega _Z} {k_I^Z} + \tilde \nu {\vec P_I}\right)\,,
\end{equation}                                                                                                 
leading to a decomposition
of the derivatives of the vector field $k_I$, as
\begin{equation}
 \mathfrak{D}_X{k^Y_I}=\tilde \nu \vec J_X{}^Y\cdot {\vec P_I} +
L_X{}^Y{}_A{}^B {t_{IB}{}^A} \,,
 \label{DXkIYdecomp}
\end{equation}
where $t_{IB}{}^A$ was introduced in~(\ref{def-t}) and $L_X{}^Y{}_A{}^B$ in~(\ref{RdecompJ}). The integrability condition of the above relation leads to projections of curvature
tensors along the symmetry vectors:
\begin{equation}
 {\vec {R}_{XY}} {k^Y_I}= -\tilde \nu \mathfrak{D}_X
{\vec P_I}\,, \qquad
  {{R}_{XYB}{}^A}{k^Y_I}= \mathfrak{D}_X {t_{IB}{}^A}\,.
 \label{Rkin2} 
\end{equation}
An important property of these moment maps is the `equivariance relation'
\begin{equation}
  -2\tilde \nu ^2 \vec P_I \times \vec P_J + \vec {R} _{YW} k_I^Y k_J^W-\tilde \nu f_{IJ}{}^K \vec P_K=0\,.
 \label{equivariance} 
\end{equation}
\subsubsection{Reduction}
In the present section, we start from a symmetry vector field $\hat k_I$ on the large space and we want to find the condition for which its components along the small space yield a symmetry vector field on the latter.
We will again start with a discussion in the hypercomplex/quaternionic case to keep everything as general as possible and perform our calculations for the choice of $\xi$ as in~(\ref{eq:su2-connection}).

First of all, the symmetry vector fields $\hat k_I$ commute with $k$ and $\vec k$~(\ref{comDG}) which implies in our Ansatz the following conditions,
\begin{equation}\label{chider}
\partial_0 \hat k_I^0 = \frac{1}{z^0} \hat k_I^0 \,, \qquad \partial_0 \hat k_I^{p} = \partial_\alpha \hat k_I^0 = \partial_\alpha \hat k_I^Y =0\,,\qquad\partial_\alpha \hat k_I^{\beta} =  \frac1{z^0} \hat k_I^\gamma \, {\vec k}^\beta \cdot \partial_\gamma  {\vec k}_\alpha \,.
\end{equation}
Secondly, the symmetries on the large space are tri-holomorphic~(\ref{intJ}). Hence, rewriting $\mathcal L_{\hat k_I} \hat{\vec J}=0$ in our Ansatz yields the normalization of the quaternionic structure~(\ref{normal2}) on the small space. In this way, we can solve for the moment maps $\vec P_I$,
\begin{equation}
\tilde \nu \vec P_I=-\hat k_I^X\vec \omega_X-\frac12 \vec Q_I\,,\qquad \vec Q_I=-\frac1{z^0}\hat k_I^\alpha \vec k_\alpha\,.
\end{equation}
Hence, the second condition~(\ref{normal2}) for $\hat k_I^X$ to be a symmetry on the quaternionic manifold is already satisfied. Another component of the tri-holomorphicity of $\hat k_I$ moreover leads to the fact that $\hat k_I^0/z^0$ is a constant. Therefore, with a suitable choice of the symmetry generators, this component can be made to vanish.
The other condition we have to check, is~(\ref{DDk}). Starting from this identity on the large space, we find after reduction that 
\begin{equation}
R_{XVY}{}^Z\hat k_I^V=\mathfrak{D}_X\mathfrak{D}_Y \hat
k_I^Z\, , \label{modifiedKillingeqn}
\end{equation}
Hence,$\hat k_I^X$ is a symmetry vector field on the quaternionic manifold. 

We now consider the dimensional reduction of the Lie-bracket on the large space
\begin{equation}
  2 \hat k_{[I}^{\hat Y} \partial_{\hat Y} \hat k_{J]}^{\hat X} = - f_{IJ}{}^K \hat k_K^{\hat X} \,.
\end{equation}
The $X$ component yields~(\ref{G-alg}) implying that $\hat k_I^X$ generate a Lie-algebra with the same structure constants $f$, the $\alpha$ component gives~(\ref{equivariance}) while the $0$ component is satisfied trivially.
We can conclude that the symmetry vector fields $\hat k_I$ reduce to symmetry vector fields on the small space.
In case there exists a compatible metric, all the above remains valid. Moreover, contracting the first relation of~(\ref{comDG}) with $k_{\hat X}$ implies immediately that $\hat k_I^0=0$ in our coordinates~(\ref{splitcoord}).

Hence, in the projection it turns out that there is a one-to-one correspondence of isometries on the large and on the small space. This can be seen as follows. First of all, $\hat k_I^X$ is a symmetry vector field on the quaternionic manifold as follows from (\ref{DDk}) on the large space. Conversely, given a symmetry vector field $k_I$ on the small space, we can construct a symmetry vector of the hypercomplex manifold that results from the lifting process of Section~\ref{ss:invmapqkhk}. To do so, in addition to that lifting process, we take $\hat k_I^X=k_I^X$ and $\hat k_I^0=0$. Subsequently defining 
\begin{equation}
\vec Q_I=-2(\tilde \nu \vec P_I+k_I^X \vec \omega_X)\,,
\end{equation}
and giving it a $z^\alpha$ dependence using
\begin{equation}
(\partial_\alpha-\frac1{z^0} \vec k_\alpha \times)\vec Q_I=0\,,
\end{equation}
we can construct the final components of the higher-dimensional symmetry vector field as  
\begin{equation}
\hat k_I^\alpha =\vec Q_I\cdot \vec k^\alpha\,.
\end{equation}
From the above, we can also proof that $\hat{\vec P}_I=\vec P_I$.
\begin{table}[tb]
\begin{center}
  \leavevmode \epsfxsize=110mm
 \epsfbox{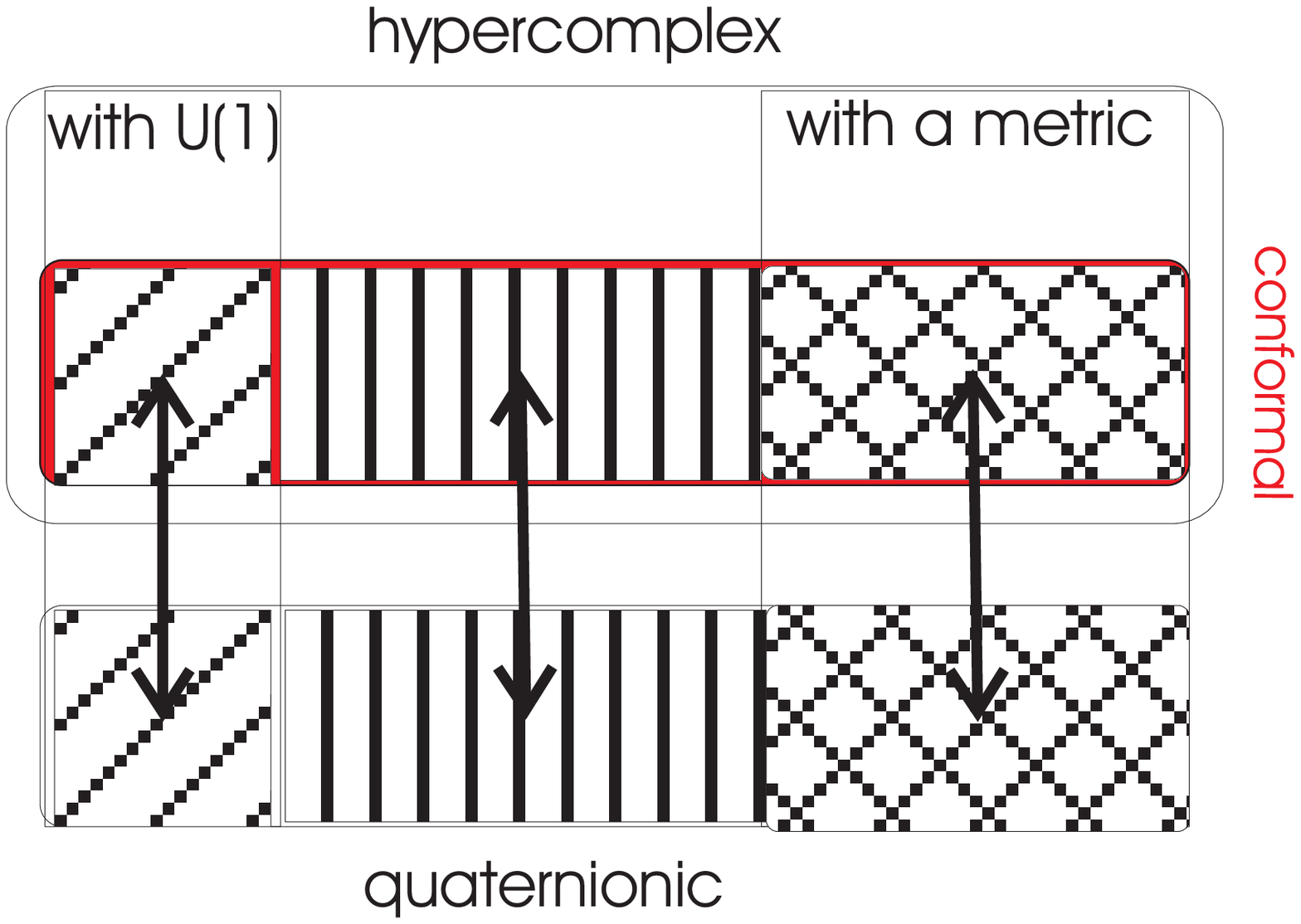}
\end{center}
\caption{\it The projection from a hypercomplex/hyperk\"ahler to a quaternionic/quaternionic-K\"ahler manifold for $k\neq 0$. }\label{tbl:resultmap}
\end{table}
\section{The Poincar\'e supergravity}\label{respoinc}
By now, we have given a description of coupled local superconformal vector- and hypermultiplets in Section~\ref{locsupconfie}. We then have developed the necessary mathematics in Section~\ref{geomchange} to be able to describe the target space effects induced by the gauge fixing process of the extra conformal symmetries. We thus are now in a position to perform the final step in the conformal approach. We will do this in the specific case that the theory has a description in terms of an action functional.
\subsection{Gauge fixing\label{s:gf}}
The actions given in~(\ref{conf-VTaction}) and~(\ref{conf-hyperaction}) are invariant under the full super-covariant group. In order to break the symmetries that are not present in the super-Poincar{\'e} algebra, we will impose the necessary gauge conditions in the following sections.

\begin{figure}[tb]
\centerline{\epsfig{file=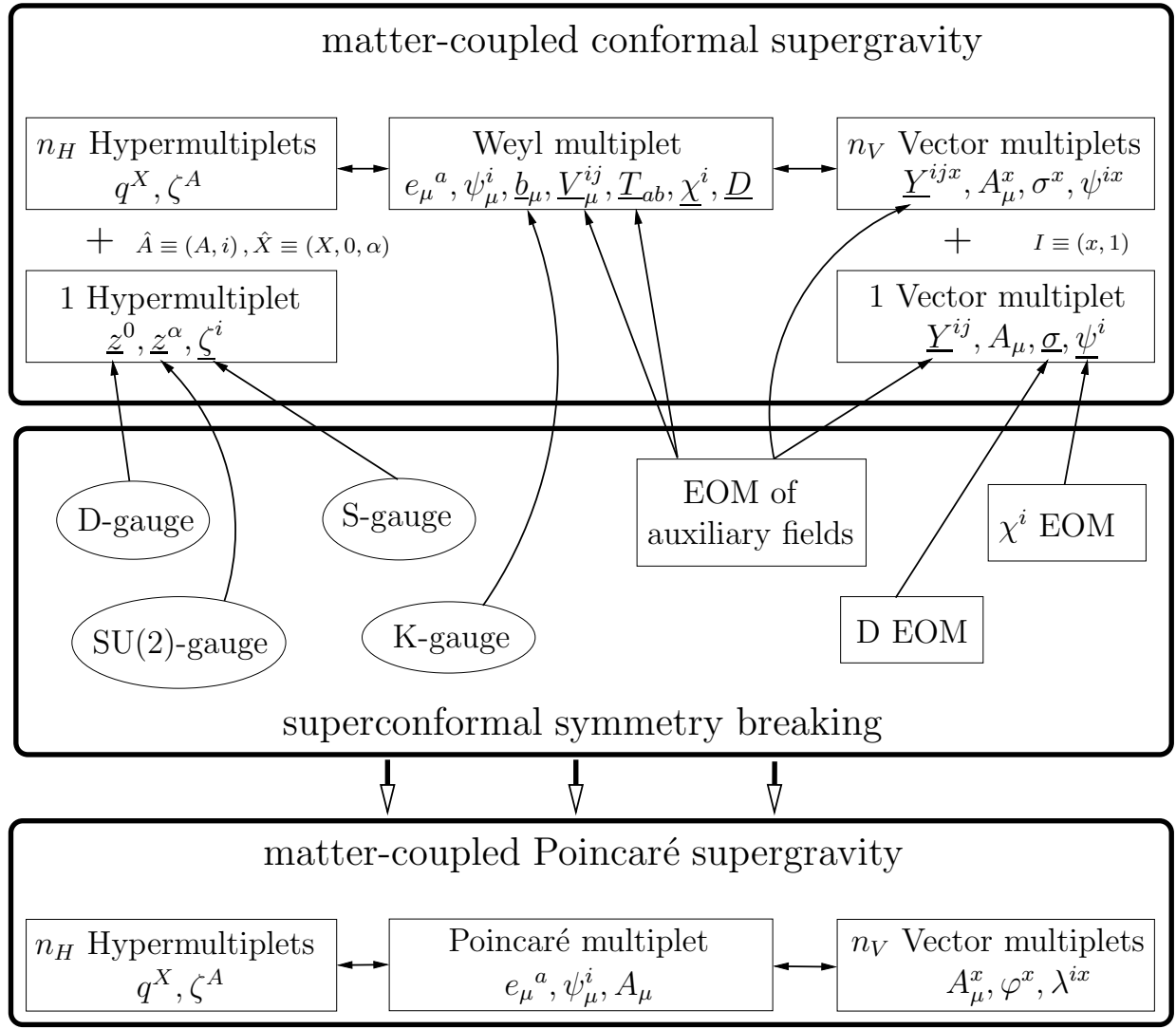,width=\textwidth}}
 \caption{\it The gauge fixing procedure: the underlined fields are
eliminated when passing to Poincar{\'e} supergravity. The  arrows indicate how these
fields are eliminated: by gauge fixing a symmetry or by applying an
equation of motion.}
 \label{fig:matter}
\end{figure}

Before carrying out all technicalities implied by the gauge fixing
procedure, it is instructive to outline the steps we are going to follow.
Just like in the example of Section~\ref{scalfield}, the extra
(superconformal) symmetries can be removed with the help of compensating
fields, which should fall in compensating multiplets, in order to keep everything supersymmetric. In our particular case, one hypermultiplet together with  one
vector multiplet will play the role of compensator. Our
strategy is illustrated in Figure~\ref{fig:matter}, where we
have summarized which fields are eliminated by gauge fixing and/or
solving the equations of motion.\footnote{The arrows show that one can
imagine another construction. Namely, we may obtain conformally invariant
actions with consistent field equations by only using a compensating
vector multiplet and no compensating hypermultiplet. The compensating
vector multiplet is necessary to obtain consistent field equations for
the auxiliary $D$ and $\chi _i$. However, the compensating hypermultiplet
is only needed to break conformal invariance.}

The field content of the matter-coupled conformal supergravity is given
by the Weyl multiplet, $n_H +1$ hypermultiplets and $n_V+1$ Yang-Mills
vector multiplets. Note that in Figure~\ref{fig:matter}, we
represented only the independent fields of the Weyl multiplet. The
dependent fields $f_\mu{}^a$, $\hat \omega_\mu{}^{ab}$ and $\phi_\mu^i$ are
expressed in terms of the former as is shown in~(\ref{depgaugefields}). Note that
the $b_\mu$ field does not enter the action; it can therefore be set to
zero as a $K$-gauge condition. There exist several auxiliary fields both
in the Standard Weyl multiplet ($V_\mu^{ij}$ and $T_{ab}$) as in the
vector multiplets ($Y^{ij  I}$) that can be eliminated by solving from the
corresponding (algebraic) field equations.

\landscape
\endlandscape

The other auxiliary fields $\chi$ and $D$ play the role of Lagrange multipliers.  Their equations of motion collaborate with
the $S$-and $D$-gauge-fix condition, respectively, to remove the fermionic
degrees of freedom of the compensating multiplets and two of their
scalars, i.~e.~$\sigma$ and $z^0$. We see from the Figure that all field
components of the compensating hypermultiplet are eliminated
by the
gauge fixing procedure. The only field component of the compensating
vector multiplet surviving the gauge fixing procedure is the gauge
potential $A_\mu$, which contributes to the graviphoton field in the
Poincar{\'e} multiplet. As we eliminated the auxiliary field $Y^{ij}$, the
vector multiplets are
realized on-shell in the Poincar{\'e} theory. We thus end up with a
matter-coupled Poincar{\'e} supergravity theory containing, besides the
Poincar{\'e} multiplet, $n_V$ vector- and $n_H$
hypermultiplets. The geometry described by the scalars of the latter
modifies during the gauge fixing according to our discussion in
Section~\ref{geomchange}. We will see below that the vector scalars also
parametrize  a particular type of manifold at the Poincar{\'e} level, namely
a very special real manifold (see Section~\ref{ss:hypersurfaces}).

One should keep in mind that during the gauge fixing procedure the
definition of the covariant derivatives changes. Indeed, when passing
from a superconformal invariant theory to a super-Poincar{\'e} theory, the
remaining fields are chosen such that they do not transform under the
broken symmetries, e.g.~the scale symmetries. These scale symmetries
generated terms in the superconformal covariant derivative that are
absent in the Poincar{\'e} covariant derivatives. Something else to keep in
mind has to do with the transformation rules. In the full conformally
invariant theory these transformation rules involve the parameters of the
conformal transformations. Due to the gauge fixing conditions, these
parameters become dependent and are expressed in terms of the parameters
of the Poincar{\'e} theory through the so-called decomposition rules. The
super-Poincar{\'e} transformation rules are therefore inferred from the
superconformal ones after eliminating auxiliary fields and employing the
decomposition rules. This finishes our overview of the gauge fixing
procedure. We now proceed with a more technical discussion of the same procedure.
\subsubsection{Preliminaries}
The first step in the gauge fixing process is the elimination of the
dependent gauge fields $\phi_\m^i$ and $f_\m{}^a$, associated to $S$- and
$K$-symmetry respectively, for which we already have listed their form in~(\ref{depgaugefields}). 

After writing out all covariant derivatives and dependent gauge fields,
the gauge field $b_\mu$ does not appear in the action anymore. This can be
understood from $K$-invariance of the action. We will choose the
conventional gauge choice for $K$-symmetry, namely
\begin{equation}\label{eq:K}
K\mbox{-gauge:}\quad b_\mu=0\,.
\end{equation}
At this point we are left with one more gauge field corresponding to a
non-Poincar{\'e} symmetry: the $\SU(2)$ gauge field $V_\m^{ij}$. Solving for
its equation of motion originating from the
Lagrangian~$\mathcal L_t=\mathcal L_{v}+\mathcal L_{h}$ given in~(\ref{conf-VTaction}) and~(\ref{conf-hyperaction}), yields the following expression
\begin{eqnarray}
&&\!\!\!\!\!\!V_\m^{ij}
= \frac{9}{2 k^2}\Big( \hat g_{{\hat X}{\hat Y}} (\partial_\m q^{\hat X} + g A_\m^I \hat k_I^{\hat X}) k^{ij{\hat Y}} + \frac12\rmi  k^{\hat X} {\hat f}_{{\hat X}}^{i{\hat A}}  \bar{\zeta}_{\hat A} \g_{\m\n} \p^{\n\,j}\\&&\!\!\!\!\!\!  - \rmi k^{ij{\hat X}} {\hat f}_{k{\hat X}}^{\hat A} \bar \zeta_{\hat A} \g_\n\g_\m\p^{k\n} - \frac 12 C_{ I J K} \s^{ K} \bar{\p}^{i I}\g_\m \p^{j J}+ \frac14 \rmi C_{ I J K}\s^{ K}\s^{ I} \bar{\psi}^{i J}\g_{\m\n} \p^{j\n} \Big)\,.\nonumber
\end{eqnarray}
The action further contains four auxiliary matter fields: $D, T_{ab}$ and
$\chi^i$ from the Weyl multiplet, and $Y^{ I}_{ij}$ from the
vector multiplet. Both $D$ and $\chi^i$ appear as Lagrange
multipliers in the action, leading to the following constraints,
respectively
\begin{eqnarray}
D &:&
C - \frac13 k^2 = 0 \,, \qquad \mbox{with} \quad C \equiv C_{ I J K} \s^{ I}\s^{ J}\s^{ K} \,, \label{eq:D-EOM} \\
\chi^i &:& -8 \rmi C_{ I J
K}\s^{ I}\s^{ J}\p^{ K}_i - \frac43 \left( C
-\frac13 k^2 \right) \g^\m \p_{\m i} + \frac{16}{3}\rmi A^{\hat A}_i
\zeta_{\hat A} = 0\, .\nonumber\\\label{eq:chi-EOM}
\end{eqnarray}
The sections $A_i^{\hat A}$ were defined in~(\ref{defsections}).
The equations of motion for $Y^{ I}_{ij}$ and $T_{ab}$ are given by
\begin{eqnarray}
 &&\!\!\!\!\!\!\!\!\!\!Y^{ij J} C_{ I J K} \s^{ K} = - g \, \delta_{ I}^L\, \hat P_L^{ij} + \frac14\rmi C_{ I J K} \bar\psi^{i J} \psi^{j K} \,, \label{eq:Y-EOM} \\
&&\!\!\!\!\!\!\!\!\!\!T_{ab}= \frac{9}{64 k^2} \Big( 4 \s^{ I}\s^{ J}\hat{F}^{ K}_{ab}C_{ I J K} + \s^{ I}\s^{ J}\bar{\psi}^{ K}\g_{[a}\psi_{b]}C_{ I J K} + \s^{ I}\s^{ J}\bar{\psi}^{ K}\g_{abc}\psi^c C_{ I J K}\nonumber\\
&&\!\!\!\!\!\!\!\!\!\!
 + \rmi\s^{ I}\bar{\psi}^{ J}\g_{ab}
\psi^{ K}C_{ I J K} + \frac23
k^{\hat X} {\hat f}_{i{\hat X}}^{\hat A} \bar{\zeta}_{\hat A}
\g_{[a}\psi_{b]}^i + \frac23k^{\hat X} {\hat f}_{i{\hat X}}^{\hat A}
\bar{\zeta}_{\hat A} \g_{abc}\psi^{ic} + 2\rmi \bar{\zeta}_{\hat
A}\g_{ab}  \zeta^{\hat A}\Big) \, .\nonumber\\\label{eq:T-EOM}
\end{eqnarray}
These equations have been simplified by using~(\ref{eq:D-EOM}).

\subsubsection{Gauge choices and decomposition rules}
\label{ss:gaugechoice}

Apart from the $K$-gauge~(\ref{eq:K}) that we have already introduced to
fix the special conformal symmetry, we now choose
gauges for the other non-Poincar{\'e} (super)symmetries as well.
\paragraph{D-gauge}
Demanding canonical factors for the Einstein-Hilbert and Rarita-Schwinger
kinetic terms in $\mathcal L_t$, we have to impose the
following $D$-gauge:
\begin{equation}
 \label{eq:D-gauge}
D\mbox{-gauge:}\quad  \frac{1}{24} \left( C + k^2 \right) = -\frac{1}{2
\kappa^2} \,,
\end{equation}
where $\kappa$ has dimensions of [length]$^{3/2}$. 
If we combine the D-gauge~(\ref{eq:D-gauge}) and the equation of motion for the auxiliary $D$~(\ref{eq:D-EOM}),
we obtain
\begin{equation}
\label{eq:hypersurfaces}
 k^2 = -\frac{9}{\kappa^2}\,, \qquad  C = -\frac{3}{\kappa^2}\,.
\end{equation}
Using the coordinates introduced in Section~\ref{geomchange}, the first constraint implies
that
\begin{equation}\label{eq:z0}
z^0=\kappa^{-2}\, ,
\end{equation}
whereas the second constraint effectively eliminates one of the
vector scalars.\footnote{The constraint~(\ref{eq:z0}) implies that
the parameter $\nu$ defined in~(\ref{BinqK}) is given by $\nu
= -\kappa^2$~(\ref{valuenu}). The parameter $\nu$ will therefore not appear any more in this section.}
\paragraph{S-gauge}
\label{ss:S-gauge}   
The action $\mathcal L_t$ contains terms where $\gamma ^\mu
\phi _\mu= \frac14\rmi\gamma ^{\mu \nu }\partial _\mu \psi _\nu +$
(nonderivative terms) is multiplied by hyperino and gaugino fields.
These terms imply a mixing of the kinetic terms of the gravitino with the
hyperino and gaugino fields. A suitable $S$-gauge can eliminate this
mixing, as we can use it to put the coefficient of the above expression equal to zero:
\begin{equation}
  S\mbox{-gauge:}\quad C_{ I J K}
  \sigma^{ I} \sigma^{ J} \psi_i^{ K} +
  2 A_i^{\hat{A}} \zeta _{\hat{A}} =0\,.
 \label{eq:S-gauge}
\end{equation}
Combining this with the $\chi $ field equation~(\ref{eq:chi-EOM}) leads
to
\begin{equation}
  C_{ I J K}
  \sigma^{ I} \sigma^{ J} \psi_i^{ K} =0\,,\qquad
   A_i^{\hat{A}} \zeta _{\hat{A}}=0\,.
 \label{eq:S-gaugesplit}
\end{equation}
In the frame we have chosen in~(\ref{allf}), we obtain the following expression for the
sections $A^{i{\hat A}}$
\begin{equation}
\label{eq:genSU2gauge}
 A^i_{\hat A} \equiv \varepsilon^{ij}k_{\hat X} \hat f^{\hat X}_{j\hat{A}} = -3\varepsilon^{ij} \hat{f}^0_{j\hat{A}}
 = -\rmi 3\sqrt\frac{z^0}{2} \delta_{\hat A}^i \,.
\end{equation}
Therefore, our choice of frame on the hyperk{\"a}hler manifold is
consistent with the fact that the hyperinos of the compensating
hypermultiplet carry no physical degree of freedom:
\begin{equation}
\label{eq:zetai} \zeta^i = 0\,.
\end{equation}
\paragraph{$\SU(2)$-gauge}
The gauge for dilatations was chosen such that $z^0=\kappa^{-2}$.
Similarly we may also choose a gauge for $\SU(2)$. Such a gauge would select
a specific point in the 3-dimensional space of the coordinates $\{z^\alpha\}$. In
principle we could choose $z^\alpha =z_0^\alpha (q)$ for any
function $z_0^\alpha (q)$, but we will restrict ourselves here to
constants $z_0^\alpha$:
\begin{equation}
  \mbox{SU(2)-gauge:} \qquad z^\alpha = z_0^\alpha \,.
 \label{SU2gauge}
\end{equation}
\paragraph{Decomposition rules}
As a consequence of the gauge choices, the corresponding transformation
parameters can be expressed in terms of the others by so-called
decomposition rules. For example, the requirement that the $K$-gauge
(\ref{eq:K}) should be invariant under the most general superconformal
transformation, i.e. $\d b_\m=0$, leads to the following decomposition
rule for $\lambda_K^a$:
\begin{equation}
\lambda_K^a = -\frac12 e^{\mu a} \left( \partial_\mu \lambda_D + \frac12\rmi
\bar\e\phi_\mu - 2 \bar\e\g_\mu\chi + \frac{1}{2}\rmi \bar\eta\psi_\mu
\right)\,.
\end{equation}
Similarly, demanding $\delta z^0=0$ yields
\begin{equation}
  \lambda_D = 0\,.
\end{equation}
The decomposition rule for $\eta^i$ can be found by varying the S-gauge and demanding that
\begin{equation}
\d\,\left( C_{ I J K}\s^{ I}\s^{ J}\psi^{i K} \right) = 0\,.
\label{DemanddelC0}
\end{equation}
We find
\begin{eqnarray}
\label{eq:eta} \kappa ^{-2}\eta^i
 &=& - \frac{1}{12} C_{ I J K} \s^{ I}
  \s^{ J} \g \cdot {\hat{F}}^{ K} \e^i
  + \frac{1}{3} g \s^I P_I^{ij} \e_j
 + \frac{1}{32 } \rmi \g^{ab} \e^i \bar\zeta_A \g_{ab} \zeta^A \nn\\
&& + \frac{1}{16} \rmi C_{ I J K}
\s^{ I} \Big( \g^a \e_j \bar\psi^{i J}
\g_a\psi^{j K} - \frac{1}{16} \g^{ab} \e^i \bar\psi^{
J} \g_{ab}\psi^{ K}\Big) \,.
\end{eqnarray}
The $\SU(2)$ decomposition rule can be found by requiring that $\delta
z^{\alpha} = 0$:
\begin{eqnarray}
\vec \lambda_{\SU(2)} &=& \vec \omega_X (\d_Q + \d_G) q^X + g \lambda_G^I
\vec P_I\,,
\end{eqnarray}
where $\vec \omega_X$ was defined in~(\ref{indconn}).
\subsubsection{Hypersurfaces\label{ss:hypersurfaces}}
We now discuss the geometry for the vector multiplets after gauge fixing. In order to get a standard normalization we rescale the
$C_{ I J K}$ symbol and the vector
multiplet scalars as follows:
\begin{eqnarray}
\s^{ I} &\equiv& \sqrt{\frac{3}{2\k^2}} h^{ I}\,, \nonumber\\
C_{ I J K} &\equiv& - 2\sqrt{\frac{2\k^2}{3}}
{\cal C}_{ I J K} \,,\qquad   {\cal
C}_{ I J K} h^{ I} h^{
J} h^{ K} = 1\,. \label{eq:hypersurface_vector}
\end{eqnarray}
The constraint~(\ref{eq:hypersurface_vector}) then defines a $n_V$-dimensional hypersurface pa\-ra\-me\-tri\-zed by scalars $\phi^x$ called a `very special real' manifold, which is embedded into a $n_V +1$-dimensional
space spanned by the scalars $h^{ I}(\phi)$.

The metric on the embedding $h^{ I}$-manifold can be determined by substituting the equation of motion for $T_{ab}$~(\ref{eq:T-EOM}) back into the action, and defining the kinetic term for the vectors as
\begin{equation}
 {\mathcal L}_{\rm kin, vec-ten} = - \frac14 a_{ I J} \hat{F}_{\m\n}^{ I} \hat{F}^{\m\n J}\,.
\end{equation}
We find
\begin{eqnarray}
a_{ I J} &=& -2 {\cal C}_{ I
J K} h^{ K} + 3 h_{ I} h_{ J} \,,
\label{eq:a_IJ-identity}
\end{eqnarray}
where
\begin{eqnarray} \label{eq:normalized_hypersurface_vector}
  h_{ I} &\equiv& a_{ I J} h^{ J} = {\cal C}_{ I J K} h^{ J} h^{ K} \qquad \Rightarrow \qquad h_{ I} h^{ I} = 1 \,.
\end{eqnarray}
In the following we will assume that $a_{ I J}$ is
invertible, which will enable us to solve~(\ref{eq:Y-EOM}) for $Y^{
I ij}$. For convenience we introduce
\begin{equation}
 h_x^{ I} \equiv -\sqrt{\frac{3}{2\kappa^2}} h^{ I}_{,x}(\phi)\,, \quad \rightarrow \quad h_{ Ix} \equiv a_{ I J} h^{ J}_x(\phi) = \sqrt{\frac{3}{2\kappa^2}} h_{ I,x}(\phi) \,.
\end{equation}
The metric on the hypersurface is then defined as the pull-back of the metric $a_{IJ}$,
\begin{equation}
  g_{xy} = h^{ I}_x h^{ J}_y a_{ I J}\,.
\end{equation}
We will use this metric to raise and lower indices $x$,
\begin{equation}
h^x_{ I} \equiv g^{xy} h_{ I y}\, .
\end{equation}
It follows from~(\ref{eq:normalized_hypersurface_vector}) that
\begin{equation}
  h_{ I} h^{ I}_x = h_{ I}^x h^{ I} =
  0\, .
\end{equation}
Another useful identity that can be deduced is
\begin{equation}
h_{ I}h_{ J} + h_{ I}^x h_{ Jx}=a_{ I J}\,.
\label{eq:a=hh+hh}
\end{equation}
The gauginos $\psi^{ I}$ are constrained fields, due to the
S-gauge. In order to translate these to $n_V$ unconstrained
gauginos, we introduce the fields $\lambda^{i\,x}$ that are (proportional to) the pull-back of $\psi^I$ to the tangent bundle on the hypersurface. As we will see later, a
convenient choice is given by (for agreement with the
literature~\cite{Ceresole:2000jd})\footnote{We avoid here the
introduction of a local basis for the fermions indicated by indices
$\tilde a$ in~\cite{Ceresole:2000jd}.}
\begin{equation}
\label{eq:lambda} \lambda^{i\,x} \equiv - h^x_{ I}\,
\psi^{i I}\,, \qquad \psi^{i I} = - h^{
I}_x \lambda^{i\,x} \,.
\end{equation}
Note that this choice for $\psi^{i I}$ indeed solves the
S-gauge~(\ref{eq:S-gaugesplit}).
\subsection{Results}\label{s:results}
After applying the steps outlined in the previous section, i.e.~using a
special coordinate basis, substituting the expressions for the dependent
gauge fields and matter fields, and `reducing' the objects on the
hyperk{\"a}hler manifold to the quaternionic-K{\"a}hler manifold, we obtain the
$N=2$ super-Poincar{\'e} action.

We give in this section the full action for a number of vector multiplets
(indices $I$) and hypermultiplets (indices $X$ for the scalars and $A$ for
the spinors). The couplings of the vector multiplets are
determined by the constants ${\cal C}_{ I J K}$ and the structure constants
$f_{I J}{}^{ K}$. The related quantities are defined in
section~\ref{ss:hypersurfaces}.

We define the supercovariant field strengths $\widehat F^I_{ab}$ and a
tensor field $ B^M_{ab}$ such that
\begin{equation}
\hat F^{I}_{ab}=F_{ab}^I-\bar{\psi}_{[a}\gamma_{b]}\psi^I
+\frac{\sqrt{6}}{4\kappa}\rmi \bar{\psi}_a\psi_b
h^I\,,\qquad
F_{\mu\nu}^I \equiv  2 \partial_{[\mu} A_{\nu]}^I + g f_{JK}{}^I A_\mu^J
A_\nu^K\,.
\end{equation}
The $\hat F_{ab}$ transforms covariantly, while
the action gets a simpler form using $F_{ab}$.

The hypermultiplets are determined by the Vielbeine $f_X^{iA}$, that
determine complex structures, $\symp(n_H)$ and $\su(2)$ connections. They
transform in general under the gauge group of the vector multiplets. The
Killing vectors $k_I^X$ determine $t_{IA}{}^B$ and are restricted
by~(\ref{def-t}) and~(\ref{symm-ff}). They determine the moment maps
by~(\ref{momentmap}). As mentioned in Section~\ref{rigidsuconfhypers}, the moment
maps can also exist without a quaternionic-K{\"a}hler manifold ($n_H=0$), in
which case they are the constant `Fayet--Iliopoulos (FI) terms'. These
are possible for two cases. First, in case the gauge group contains an
$\SU(2)$ factor, we can have
\begin{equation}
  \vec P_I= \vec e_I \,\xi \,,
 \label{SU2FI}
\end{equation}
where $\xi$ is an arbitrary constant, and $\vec e_I$ are constants that
are nonzero only for $I$ in the range of the $\SU(2)$ factor and satisfy
\begin{equation}
  \vec e_I\times \vec e_J= f_{IJ}{}^K \vec e_K\,,
 \label{vece}
\end{equation}
in order that~(\ref{equivariance}) is satisfied.

The second case is $\U(1)$ FI terms. In this case
\begin{equation}
   \vec P_I= \vec e\, \xi_I \,,
 \label{U1FI}
\end{equation}
where $\vec e$ is an arbitrary vector in $\SU(2)$ space and $\xi_I$ are
constants for the $I$ corresponding to $\U(1)$ factors in the gauge
group.

To be able to write down the potential and the supersymmetry
transformation rules in an elegant fashion, we define
\begin{equation}
\begin{array}{ll} 
\vec P \equiv \kappa ^2 h^I \vec P_I \,, \qquad & \vec P_x \equiv \kappa
^2 h_x^I \vec P_I \,, \\ {\mathcal N}^{iA} \equiv \frac{\sqrt6}{4}\kappa
h^I k_I^X f_X^{iA}\,,&   K^x_I \equiv -\frac1\kappa \sqrt\frac32 f_{I{
J}}{}^{ K} h^{ J} h^x_{ K} =-\frac1\kappa
\sqrt\frac32 f_{I{ J}}{}^{ K} h^{{ J}x}
h_{ K}
\,,  \\
T_{xyz} \equiv
\mathcal{C}_{{I}{J}{K}}h^{{I}}_xh^{{J}}_yh^{{K}}_z\,,
& \Gamma_{xy}^w
= h^w_{ I} h^{ I}_{x,y} + \kappa \sqrt{\frac23} T_{xyz}
g^{zw}\,,
\end{array}
\end{equation}
the latter being the Levi-Civita connection of $g_{xy}$.

The covariant derivatives now read
\begin{eqnarray}
\mathfrak{D}_\mu\, h^{ I}
&=& \partial_\mu h^{ I} + g f_{J K}{}^{ I} A_\mu^J h^{ K}
 \\&=&-\sqrt{\frac23}\kappa h_x^{{I}}\left( \partial_\mu \phi^x+gK_J^xA_\mu^J\right) =-\sqrt{\frac23}\kappa h_x^{{I}}\mathfrak{D}_\mu \phi^x  \,, \nn \\
{\mathfrak D}_\m q^{X}
&=& \partial_\m q^{X}+ g A_\m^I k_I^{X} \,, \nn \\
\mathfrak{D}_\mu \lambda^{xi}&=&\partial_\mu\lambda^{xi}+\partial_\mu
\phi^y \Gamma_{yz}^x\lambda^{zi}+\frac14 \omega_\mu{}^{ab}\gamma
_{ab}\lambda^{xi}\nonumber\\&& +\partial_\mu q^X
\omega_{Xj}{}^i\lambda^{xj}
- g\kappa ^2 A^I_\mu P_{I}{}^{ij}\lambda^{x}_j +g A_\mu^I K_I^{x;y} \lambda_y^i\,,\nonumber\\
\mathfrak{D}_\m \zeta^{A}
&=& \partial_\m \zeta^{A} + \partial_\m q^{X} \omega_{{X}{B}}{}^{A} \zeta^{B} + \frac14 \omega_\m{}^{bc} \g_{bc} \zeta^{ A} + g A^I_\m t_{IB}{}^{A} \zeta^{B} \, , \nonumber\\
\mathfrak{D}_\mu \psi_{\nu }^i&=&(\partial_\mu +\frac14
\omega_\mu{}^{ab}\gamma _{ab})\psi_{\nu }^i-\partial_\mu
q^X\omega_{X}{}^{ij}\psi_{\nu j}-g\kappa ^2 A_\mu^IP_{I}{}^{ij}\psi_{\nu
j} \,.\nonumber
\end{eqnarray}
Here $K_I^{x;y}$ stands for the covariant derivative, where the index is
raised with the inverse metric $g^{xy}$. We choose to extract the
fermionic terms from the spin connection as in the final line of~(\ref{depgaugefields}), using $\omega_{\mu a}{}^b$
instead of $\hat{\omega}_{\mu a}{}^b$ in the covariant derivatives and
the Ricci scalar, unless otherwise mentioned.

Performing all the steps of the conformal programme we find the following
action:
\begin{eqnarray}\label{final}
&&\!\!\!\!\!e^{-1}\mathcal{L}=\nonumber\\&&\!\!\!\!\!
\frac1{2\kappa^2} R(\omega) -\frac14a_{{I}{J}}
\hat{F}^{{I}}_{\mu
\nu}\hat{F}^{{J}\mu \nu} -\frac12
g_{xy}\mathfrak{D}_\mu \phi^x\mathfrak{D}^\mu \phi^y
-\frac1{2}g_{XY}\mathfrak{D}_\mu q^{X}\mathfrak{D}^\mu q^{Y}\nonumber\\&&\!\!\!\!\!
-\frac1{2\kappa ^2}\bar{\psi}_\rho \gamma^{\rho \mu \nu}\mathfrak{D}_\mu
\psi_\nu -\frac12 \bar \lambda _x\slashed{\mathfrak{D}}\lambda ^x
-\bar{\zeta}^A\slashed{\mathfrak{D}}\zeta_A
+\frac{g^2}{\kappa^4}\big( 4\vec{P}\cdot \vec{P}
-2\vec{P}^x\cdot\vec{P}_x\nonumber\\&&\!\!\!\!\!
-2\mathcal{N}_{iA}\mathcal{N}^{iA}\big)
+\frac{\kappa}{6\sqrt{6}} e^{-1} \ve^{\m\n\l\r\s} {\cal C}_{IJK} A_\m^I \Big[ F_{\n\l}^J F_{\r\s}^K + f_{FG}{}^J A_\n^F A_\l^G \big(- \frac12 g \, F_{\r\s}^K \nonumber\\&&\!\!\!\!\!+ \frac1{10} g^2 f_{HL}{}^K A_\r^H A_\s^L \big)\Big] 
-\frac14 h_{{I}x} F_{bc}^{{I}}
\bar{\psi}_a\gamma^{abc}\lambda^x-\frac{\sqrt{6}}{16\kappa }\rmi
h_{{I}}
F^{cd{I}}\bar{\psi}^a\gamma_{abcd}\psi^b\nonumber\\&&\!\!\!\!\!+\frac14\sqrt{\frac23}
\kappa \rmi \big(\frac14g_{xy}h_{{I}}+ T_{xyz}
h_{{I}}^z \big) \lambda^x \gamma \cdot
F^{{I}}\lambda^y +\frac18 \sqrt{6}\rmi \kappa
h_{{I}}\bar{\zeta}_A\gamma \cdot F^{{I}}\zeta^A\nonumber\\&&\!\!\!\!\!
+\frac12 \rmi \bar{\psi}_a\slashed{\mathfrak{D}} \phi^x \gamma^a\lambda_x
+\rmi \bar{\zeta}_A \gamma^a\slashed{\mathfrak{D}}q^X\psi_a^if_{iX}^A
 +g\Big[-\sqrt{\frac32}\frac1{\kappa}\rmi h^It_{IB}{}^A\bar{\zeta}_A\zeta^B \nonumber\\&&\!\!\!\!\!+2\rmi
k_I^Xf_{iX}^Ah^I_x \bar{\zeta}_A\lambda^{ix} -\sqrt{\frac23}\frac1{\kappa} \rmi \big( \frac14g_{xy} P_{ij}+T_{xyz}
P_{ij}^z\big) \bar{\lambda}^{ix}\lambda^{jy}\nonumber\\&&\!\!\!\!\!
+\frac1{\kappa}\frac12\rmi\bar{\lambda}^x\lambda^y h^I_x K_{Iy} 
 \,\, -\frac2{\kappa^2}{\cal
N}_i^A\bar{\zeta}_A\gamma^a\psi_a^i +\frac1{\kappa
^2}\bar{\psi}_a^i\gamma^a\lambda^{jx} P_{xij} \nonumber\\&&\!\!\!\!\!+\sqrt{\frac38}\frac1{\kappa^3}\rmi
P_{ij}\bar{\psi}_a^i\gamma^{ab}\psi_b^j \Big] 
-\frac{1}{32}\bar{\psi}_a^i \psi^{ja}\bar{\lambda}_i^x\lambda_{jx}
-\frac{1}{32}\bar{\psi}_a^i
\gamma_b\psi^{ja}\bar{\lambda}_i^x\gamma^b\lambda_{jx}\nonumber\\&&\!\!\!\!\!
-\frac{1}{128}\bar{\psi}_a \gamma_{bc}\psi^{a}
\bar{\lambda}^x\gamma^{bc}\lambda_x 
-\frac{1}{16}\bar{\psi}_a^i\gamma^{ab}
\psi_b^j\bar{\lambda}^x_i\lambda_{jx}
-\frac{1}{32}\bar{\psi}^{ai}\gamma^{bc}
\psi^{dj}\bar{\lambda}^x_i\gamma_{abcd}\lambda_{jx} \nonumber\\&&\!\!\!\!\!+\frac{1}{8\kappa
^2}\bar{\psi}_a\gamma_b\psi^b\bar{\psi}^a\gamma_c\psi^c
-\frac{1}{16\kappa^2}\bar{\psi}_a\gamma_b\psi_c\bar{\psi}^a\gamma^c\psi^b
 -\frac{1}{32\kappa^2}\bar{\psi}_a\gamma_b\psi_c\bar{\psi}^a\gamma^b\psi^c\nonumber\\
&&\!\!\!\!\!
+\frac{1}{32\kappa^2}\bar{\psi}_a\psi_b\bar{\psi}_c\gamma^{abcd}\psi_d
 -\frac{1}{16}\bar{\psi}^a\gamma^b\psi^c\bar{\zeta}_A\gamma_{abc}\zeta^A
 +\frac1{16}\bar{\psi}_a\gamma^{bc}\psi^a\bar{\zeta}_A\gamma_{bc}\zeta^A\nonumber\\&&\!\!\!\!\!
-\frac1{16}\bar{\psi}^a\psi^b\bar{\zeta}_A\gamma_{ab}\zeta^A 
+\frac{\kappa}{6}\sqrt{\frac23}\rmi T_{xyz}\big(
\bar{\psi}_a\gamma_b\lambda^x\bar{\lambda}^y\gamma^{ab}\lambda^z
 +\bar{\psi}_a^i\gamma^a\lambda^{jx}\bar{\lambda}_i^y\lambda_j^z \big)
\nonumber\\&&\!\!\!\!\! +\frac{\kappa^2}{32}\bar{\lambda}^x\gamma_{ab}\lambda_x
\bar{\zeta}_A\gamma^{ab}\zeta^A
  -\frac{\kappa^2}{16}\bar{\lambda}^{ix}\gamma_a \lambda^j_x\bar{\lambda}^y_i
  \gamma^a\lambda_{jy}\nonumber\\&&\!\!\!\!\!
+\frac{\kappa^2}{128}\bar{\lambda}^x\gamma_{ab}\lambda_x\bar{\lambda}^y\gamma^{ab}\lambda_y
+\frac{\kappa^2}6g^{zt}T_{xyz}T_{tvw}
\bar{\lambda}^{ix}\lambda^{jy}\bar{\lambda}_i^v\lambda^w_j\nonumber\\&&\!\!\!\!\!
-\frac{\kappa^2}{48}\bar{\lambda}^{ix}\lambda^j_x\bar{\lambda}_i^y\lambda_{jy}
+\frac{\kappa^2}{32}\bar{\zeta}_A\gamma_{ab}\zeta^A\bar{\zeta}_B\gamma^{ab}\zeta^B
-\frac1{4}\mathcal{W}_{ABCD}\bar{\zeta}^A\zeta^B\bar{\zeta}^C\zeta^D\,.\nonumber\\
\end{eqnarray}

This action admits the following $\mathcal N=2$ supersymmetry:
\begin{eqnarray}
\d e_\mu{}^a &=& \frac12 \bar\e \gamma^a \psi_\mu \, ,\nonumber\\
\delta \psi_\mu^i&=&\mathfrak D_\m(\hat\omega)\e^i + \frac{\rmi\kappa }{4\sqrt6}
h_{ I} \hat{F}^{ I\nu \rho }
 ( \g_{\mu\nu \rho } - 4 g_{\mu \nu } \g_\rho  ) \e^i + \delta q^X \omega_X^{ij} \psi_{\mu j}\nonumber\\
&& - \frac{1}{\kappa\sqrt6} \rmi g P^{ij}\gamma _\mu  \e_j -\frac{\kappa^2}{6} \e_j \bar\lambda^{ix} \gamma_\mu \lambda^j_x
+\frac{\kappa^2}{12}\gamma_{\mu \nu }\e_j \bar\lambda^{ix} \gamma^\nu
\lambda^j_x\nonumber\\
&&
 -\frac{\kappa^2}{48}\gamma_{\mu \nu \rho } \e_j \bar\lambda^{ix} \gamma^{\nu \rho } \lambda^j_x
 +\frac{\kappa^2}{12} \gamma^\nu  \e_j\bar\lambda^{ix} \gamma_{\mu \nu } \lambda^j_x + \frac{\kappa ^2}{16} \g_{\mu \nu \rho } \e^i\bar\zeta_A \g^{\nu \rho } \zeta^A  \, ,\nonumber\\
\d h^{ I}&=& -\frac{\kappa}{\sqrt6} \rmi\bar\e\lambda^x
h_x^{ I}\quad ,\qquad
\d\phi^x= \frac{1}{2} \rmi\bar\e\lambda^x\, ,\nonumber\\
\d A_\m^I&=&\vartheta_\m^I\,,\nonumber\\
\d \lambda^{xi}
&=& - \frac{\rmi}{2} \hat{\slashed{\mathfrak D}} \phi^x \e^i - \d\phi^y \Gamma_{yz}^x\lambda^{zi} + \delta q^X \omega_X{}^{ij} \lambda_j^x + \frac14 \gamma\cdot\hat{F}^{ I} h^x_{ I} \e^i \nn\\
&& \frac{\kappa }{4\sqrt6} T^{xyz} \left[ 3 \epsilon_j\bar\lambda_y^i
\lambda_z^j -  \g^\m\epsilon_j\bar\lambda_y^i \g_\m \lambda_z^j
- \frac12 \g^{\m\n}\epsilon_j\bar\lambda_y^i \g_{\m\n} \lambda_z^j  \right] - \frac{1}{\kappa^2} g P^{x\, ij} \e_j \,,\nonumber\\
\d q^X &=& -\rmi\bar\e^i\zeta^A f^X_{iA} \, ,\nonumber\\
\d \zeta^A&=& \frac12 \rmi \gamma^\m \hat{\mathfrak D}_\m q^X f_X^{iA}
\e_i - \d q^X \omega_{XB}{}^A \zeta^B + \frac{1}{\kappa^2} g {\mathcal
N}_i^A \e^i  \,.\label{tfrules}
\end{eqnarray}
We denoted
\begin{eqnarray}
 \vartheta_\m^{ I} &=& -\frac12 \bar{\e} \gamma_\m \lambda^x h_x^{
I} - \frac{\sqrt6}{4\kappa } \rmi h^{ I} \bar\e \psi_\m \, , \nonumber\\
\hat{\mathfrak D}_\m q^X &=& \partial_\m q^X + g A_{\mu}^I k_I^X
+ \rmi \bar\psi_\mu^j \zeta^B f_{jB}^X\nonumber\\
\hat{\mathfrak D}_\m \phi^x &=& \partial_\m \phi^x + g A_\m^I K^x_I -
\frac{1}{2}\rmi
\bar\psi_\m \lambda^x\,,\nonumber\\
\mathfrak D_\m(\hat\omega) \e^i &=& {\mathcal D}_\m(\hat\omega) \e^i - \partial_\m
q^X \omega_X^{ij}\e_j - g\kappa ^2 A_\mu^I P_I^{ij}\e_j \,,
\end{eqnarray}
where ${\mathcal D}_\m(\hat\omega)$ defined as in~(\ref{Lor-covder}).

In conclusion, by starting from a superconformal Lagrangian $\mathcal L_t$, we have been able to construct the $\mathcal N=2$ super-Poincar\'e theory with an arbitrary number of vector- and hypermultiplets and arbitrary gaugings. In~\cite{Bergshoeff:inprep2}, we have given a generalization of this result, by including also an arbitrary number of tensor multiplets. Hence, the conformal approach has enabled us to construct the most general $\mathcal N=2$ super-Poincar\'e gravity, yielding more complete results than already known in the literature~\cite{Gunaydin:1999zx,Ceresole:2000jd,Cremmer:1980gs,Gunaydin:1984bi,Gunaydin:1985ak,Gunaydin:2000xk}. As a final note, let us point out that if we would study the superalgebra~(\ref{tfrules}) on the hypermultiplets directly, we would again find dynamical constraints, leading to the conclusion that the target space can also be quaternionic, if these equations of motion are not integrable into a Poincar\'e-invariant and supersymmetric action.

\chapter{Kaluza-Klein Theory}\label{dimred}
By now, we have listed some peculiarities of on-shell supersymmetric field theories. We have used supersymmetry as an organizing principle to compute the equations of motion. This direct construction does however not answer the question what the string theory origin of such theories might be. In the present chapter, we will show that the process of dimensional reduction can yield theories that do not have a Lagrangian description.

The content of this chapter is as follows. After a brief introduction, we will discuss the meaning of consistent truncation in Section~\ref{construnc}, after which we will explain the theory of the Scherk-Schwarz dimensional reduction in Section~\ref{schsch}. In the final Section~\ref{massive}, we will apply these ideas to the construction of a massive IIA supergravity that does not admit a least action principle and present a method to find supersymmetric solutions.
\section{Introduction}
\subsection{Circle reduction}
The technique of dimensional reduction was initiated in the old papers~\cite{Kaluza:1921tu,Klein:1926tv} in which Kaluza and Klein constructed the four-dimensional theory of general relativity coupled to Maxwell theory (with gauge potential $A$) and a real scalar field $\varphi$ starting from the five-dimensional Einstein-Hilbert action. In their construction, they assumed that the extra dimension was a circle (with radius $R$) and supposed that the five-dimensional metric $\hat g$ was constant along that circle. They then decomposed this metric in terms of four-dimensional fields 
\begin{equation}\label{standardred}
\hat g=e^{\frac{\sqrt 3}{3} \varphi} g_{\mu \nu}dx^\mu dx^\nu+e^{-\frac{2\sqrt{3}}{3}\varphi}(dz+A_\mu dx^\mu)^2\qquad\mbox{with}\qquad \mu=0,\dots,3\,,
\end{equation} 
where $z$ parametrizes the compact direction and none of the fields $\{g,A,\varphi\}$ depends on $z$. Note that the decomposition of a five-dimensional metric in a four-dimensional one, together with a real scalar and a four-dimensional vector, is compatible with the following branching rule
\begin{equation}
\so(1,4)\to\so(1,3)\qquad:\qquad \mathbf{14}\to \mathbf{1}+\mathbf{4}+\mathbf{9}\,.
\end{equation}
The starting point for a dimensional reduction thus is a form for the higher-dimensional fields in which the dependence on the compact directions is fixed. This is called an Ansatz. Using the Ansatz~(\ref{standardred}) in the five-dimensional Einstein-Hilbert action then yields the four-dimensional theory
\begin{eqnarray}
\hat{\mathcal S}&=&\frac1{16 \pi G_N}\int d^5x \sqrt{-\hat g}\hat R\,\Rightarrow\nonumber\\ \mathcal S&=&\frac{R}{8G_N}\int d^4 x \sqrt{-g}\big[R-\frac12\partial_\mu \varphi \partial^\mu \varphi -\frac14 e^{-\sqrt 3\varphi}F_{\mu \nu}F^{\mu \nu}\big]\,,\label{effaction}
\end{eqnarray}
where $G_N$ is  the five-dimensional Newton's constant and the field strength is defined in the usual way, $F_{\mu\nu}=2\partial_{[\mu}A_{\nu]}$.
Note that the higher-dimensional origin of the $\un(1)$ gauge transformation is a coordinate transformation on $z$ that preserves the Ansatz~(\ref{standardred}),
\begin{equation}\label{coordu1}
z\to z'=z+\lambda(x^\mu)\;\Rightarrow \qquad A_\mu \to A_\mu'=A_\mu-\partial_\mu \lambda\,.
\end{equation}
  
The physical reason why we can accept the prescribed form~(\ref{standardred}) for the metric is that we want the resulting four-dimensional action to capture the effective four-dimensional physics in a space-time in which the fifth direction is a circle. To understand this, consider a real, massless scalar field $\hat \phi$ in four-dimensional Minkowski space-time times a circle. As a consequence, the field has to be periodic around this circle. Expanding $\hat\phi$ in Fourier modes
\begin{equation}
\hat \phi(x^\mu,z)= \sum_{n=-\infty}^{\infty} \phi_{|n|}(x^\mu)e^{\rmi \frac{n}{R}z}\,,
\end{equation}
we can use this Ansatz in the wave equation and find
\begin{equation}
\Box_5 \hat \phi=0 \;\Rightarrow \Box \phi_n=\left(\frac{n}{R}\right)^2 \phi_n\,,
\end{equation}
where $\Box_5$ is the five-dimensional d'Alembertian and $\Box$ is the four-dimensional one.
Hence, the nonzero Fourier modes give rise to four-dimensional fields with an effective mass inversely proportional to $R$. Choosing therefore $R$ to be sufficiently small, we can forget about the massive modes, and the theory becomes effectively four-dimensional. If we want to describe processes with energies higher that $1/R$, the Kaluza-Klein tower becomes visible and space-time appears to be five-dimensional again.\footnote{In the dimensional reduction of gravity, the definition of massive modes of the metric is problematic. However, expanding the metric as $\hat g_{\hat \mu\hat\nu}=\hat \eta_{\hat \mu\hat\nu}+\delta \hat g_{\hat \mu\hat \nu}$ would yield a five-dimensional theory of a massless spin-two particle that would give rise to massive particles in four dimensions after dimensional reduction.}

Notice that in the four-dimensional action~(\ref{effaction}), there is no potential for $\varphi$. This means that $\varphi$ can be e.g. any constant, labelling an infinite number of degenerate configurations corresponding to a choice of the physical radius of the compact direction. Therefore, $\varphi$ is called a modulus.
\subsection{Generalizations}
Since the invention of this technique to `curl up' dimensions, numerous generalizations have been investigated. First of all, Yang-Mills theory with gauge group $G$, was constructed by dimensionally reducing pure gravity on the group manifold $G$~\cite{Dewitt}. Similar reductions on coset manifolds turned out to be much harder, though some examples exist, e.g. $\mathcal N=8$ four-dimensional gauged $\SO(8)$ supergravity was constructed by reducing classical M-theory on a seven sphere ($S^7\equiv\SO(8)/\SO(7)$)~\cite{deWit:1987iy}.
Of course, more general compact manifolds can be used as internal spaces. In the context of superstring theory, compactifications on Calabi-Yau three-folds are often encountered~\cite{Polchinski}, leading to entirely new mathematical concepts such as mirror symmetry (see e.g.~\cite{Hosono:1994av}). Note that in all these generalized compactifications, the fields do depend on the compact coordinates, but only in a very specific way.

Another generalization was the discovery of a natural mechanism to introduce masses (and gaugings) in the compactified theory~\cite{Scherk:1979zr}. If the higher-dimensional theory admits a global symmetry, it is possible to generalize the Ansatz in the sense that the fields can have a more general dependence on the compact coordinates, leading to the introduction of masses (and gaugings) in the effective theory. An example of such a reduction can be found in the context of the $\Sl(2,\mathbb R)$ symmetry of IIB supergravity. This global symmetry can be used to perform a Scherk-Schwarz reduction on a circle, yielding the most general massive, gauged $\mathcal N=2$ nine-dimensional supergravity~\cite{Cowdall:2000sq,Gheerardyn:2001jj}. 

In compactifications on complicated manifolds, like Calabi-Yau three-folds (i.e. six-di\-men\-sio\-nal Calabi-Yau manifolds), the metric Ansatz is not completely known. However, we can still partially deduce the field content of the effective theory. First of all, it is a general feature of compactifications that deformations of the internal metric, preserving the structures defined on the manifold, give rise to moduli (like $\varphi$ in the circle reduction). Secondly, the effective fields coming from the reduction of the other (bosonic) fields follow from general (cohomology) arguments. The simplest example of this is the reduction of a scalar field. Suppose that space-time is a direct product of four-dimensional Minkowski space and some internal manifold $X$. The equation of motion for a scalar field $\hat \phi$ then becomes
\begin{equation}\label{eqsc}
\Box_t\hat \phi=\Box \hat\phi+\Box_i\hat\phi=0\,,
\end{equation}
where $\Box_t$ is the d'Alembertian on the higher-dimensional space while $\Box_i$ denotes the one on the internal space $X$. Supposing that there exist $n$ independent harmonic functions $f_i$, $i=1,\dots,n$ on $X$, the dimensional reduction of $\hat\phi$ gives rise to $n$ massless fields $\phi_i$ (that do not depend on the internal space any more), using the Ansatz
\begin{equation}
\hat \phi= \sum_{i=1}^n  f_i\phi_i\,,
\end{equation}
in the equation of motion~(\ref{eqsc}).
Using these general remarks, one can often deduce some qualitative features of general compactifications, see e.g.~\cite{Ferrara:1989qu} for compactifications on Calabi-Yau three-folds. Note however that, starting with some solution of e.g. ten-dimensional supergravity, a full compactification Ansatz should capture all fully nonlinear interactions of the massless\footnote{Seen from the lower-dimensional viewpoint.} fluctuations around that background geometry. Finding such an Ansatz and retrieving the corresponding effective theory is a nontrivial calculation. This can be illustrated with the historical remark that the spectrum of massless deformations of the $AdS_5\times S^5$ solution of IIB supergravity was already computed in~\cite{Kim:1985ez} while the five-dimensional theory governing the full nonlinear interactions of these deformations was only found recently~\cite{Cvetic:2000nc}.
\section{Consistent truncations}\label{construnc}
The process of dimensional reduction always amounts to a certain truncation of the field content of the original theory. As it might be possible that the truncation is in conflict with the original equations of motion, the issue of consistency of a truncation requires some careful examination. 

We will call a reduction consistent if and only if every solution of the equations of motion of the lower-dimensional theory yields a solution to the equations of motion of the higher-dimensional theory. To make this more precise, let us suppose that we start with a field theory with field content denoted by $\hat \Phi (x,y)$, where $x$ are the coordinates on the noncompact and $y$ on the compact space. A compactification Ansatz now is a demand that $\hat \Phi$ depends in a specific way on the compact coordinates $y$, or more symbolically,
\begin{equation}
\hat \Phi(x,y)=\phi(x) \star f(y)\,.
\end{equation} 
Note that this dependence is not always a product, hence our $\star$ notation. The next step is to use this Ansatz in the equations of motion for $\hat \Phi$. If the result is a set of equations of motion for $\phi$ in which all dependence on $y$ vanishes, the reduction is called consistent. Note that the resulting effective theory may or may not admit an action functional. If it does, it again may or may not be the action that would be found by using the Ansatz in the Lagrangian of the higher-dimensional theory and integrating out the compact coordinates. 
\section{Scherk-Schwarz reduction}\label{schsch}
By now, we have discussed some elementary properties of dimensional reduction on a circle. It has been shown that coordinate transformations on the compact direction preserving the Ansatz, induce gauge transformations in the effective theory. However, none of the other fields were charged under this $\un(1)$ symmetry. If a theory admits a global symmetry, the reduction Ansatz can however be generalized, leading to a theory with charged fields~\cite{Scherk:1979zr}.

To introduce this concept of Scherk-Schwarz reduction, let us consider the following elementary example~\cite{Meessen:1998qm}. Our starting point is the five-dimensional theory of gravity coupled to a real scalar,
\begin{equation}\label{simpleth}
\hat{\mathcal S}=\frac{1}{16\pi G_N}\int d^5x \sqrt{-\hat g}\big[\hat R-\frac12 \partial_{\hat\mu}\hat \phi\partial^{\hat\mu}\hat \phi\big]\,.
\end{equation}
This theory obviously is invariant under $\delta \hat \phi=a$ with $a$ a constant. Taking one direction to be compact, i.e. $z\equiv z+2\pi R$, we now suppose that the field $\hat \phi$ is multi-valued, i.e. $\hat \phi\equiv \hat \phi +2\pi m$. Taking again the standard Ansatz for the metric~(\ref{standardred}), the following form for the scalar 
\begin{equation}\label{ssansatzphi}
\hat \phi=\frac{mN}R z +\phi(x^\mu)\,,
\end{equation}
yields a consistent reduction. Due to the fact that the theory is invariant under $\delta \hat \phi$, all $z$ dependence will vanish. Moreover, as the $\un(1)$ symmetry is generated by coordinate transformations~(\ref{coordu1}), we can see from~(\ref{ssansatzphi}) that under such a coordinate transformation, $\phi$ has to transform as well in order to remain in the same Ansatz~(\ref{ssansatzphi}). The dimensional reduction of the derivative on the scalar then yields
\begin{equation}
\hat \partial_a \hat \phi\equiv \hat e_a{}^{\hat \mu}\partial_{\hat \mu}\hat \phi=e^{-\frac{\sqrt{3}}6\varphi}\left(\partial_a \phi - \frac{mN}{R}A_a\right)\equiv e^{-\frac{\sqrt{3}}6\varphi}\mathfrak D_a \phi\,,
\end{equation}
where we used the standard Ansatz~(\ref{standardred}). The right-hand side transforms covariantly under the combined transformation
\begin{equation}
\phi \to \phi+ \frac{mN}R \lambda\,,\qquad A_\mu\to A_\mu-\partial_\mu \lambda\,.
\end{equation}
The resulting effective theory reads 
\begin{eqnarray}
\mathcal S&=&\frac{R}{8G_N}\int d^4 x \sqrt{-g}\big[R-\frac12\partial_\mu \varphi \partial^\mu \varphi -\frac14 e^{-\sqrt 3\varphi}F_{\mu \nu}F^{\mu \nu}-\frac12\mathfrak D_\mu \phi\mathfrak D^\mu \phi\nonumber\\&&-\frac12 \left(\frac{mN}{R}\right)^2e^{\sqrt 3\varphi}\big]\,.
\end{eqnarray}

Similarly, the circle reduction of the Einstein-Hilbert action can be made more general as well. To come as close as possible to the next section, we will start from pure gravity in eleven dimensions, i.e.
\begin{equation}\label{newEH}
\hat{\mathcal S}=\frac1{16\pi G_N}\int d^{11}x \sqrt{-\hat g}\hat R\,\Rightarrow\qquad \hat R_{\hat \mu\hat \nu}=0\,,
\end{equation}
where $G_N$ is now the eleven-dimensional Newton's constant.
It can easily be checked that under a rigid Weyl transformation $\hat g \to \exp(2\lambda)\hat g$ with $\lambda$ constant, the equation of motion is invariant, while the action is not. It now turns out that we can use this scaling symmetry in a Scherk-Schwarz reduction. Therefore, we generalize the Ansatz \`a la~(\ref{standardred}) to
\begin{equation}\label{ssansatz}
\hat g=e^{2mz}\left(e^{\frac16 \varphi}g_{\mu\nu}dx^\mu dx^\nu+e^{-\frac43 \varphi}(dz+A_\mu dx^\mu)^2\right)\,,
\end{equation}
where $\mu ,\nu=0,\dots,9$. 
This yields the following equations of motion
\begin{eqnarray}
\Box\varphi &=& -\frac38 e^{-\frac32\varphi}\, F^2 +\frac{27}2 m^2 A_\mu A^\mu + 9m A^\mu\partial_\mu\varphi -\frac32 m 
\mathfrak D_\mu{A}^\mu\,,\nonumber\\
\mathfrak D_\nu(e^{-\frac32\varphi}\, F_\mu{}^\nu) &=& 12m\partial_\mu\varphi
+18m^2\, A_\mu + 9m\, e^{-\frac32\phi}\, A^\nu\, F_{\mu\nu}\,, \nonumber\\
R_{\mu\nu} -\frac12 R\, g_{\mu\nu}&=& \frac12(\partial_\mu\varphi\,
\partial_\nu\varphi -\frac12 (\partial\varphi)^2\, g_{\mu\nu}) - 36m^2 e^{\frac32\varphi} g_{\mu\nu}\label{sseom}\\
&&
-9m^2 (A_\mu A_\nu + 4A_\rho A^\rho g_{\mu\nu}) +\frac12 e^{-\frac32\phi} (F_{\mu\rho} F_\nu{}^\rho 
-\frac14F^2\, g_{\mu\nu}) \nonumber\\
&&-\frac92 m (\mathfrak D_\mu A_\nu +\mathfrak D_\nu A_\mu
-2\mathfrak D_\rho A^\rho\, g_{\mu\nu})\nonumber\\&&+\frac34m\, (A_\mu\partial_\nu\varphi + A_\nu\partial_\mu\varphi -A^\rho\partial_\rho \varphi\, g_{\mu\nu})\,.\nonumber
\end{eqnarray}
The first issue to raise is that these equations of motion cannot be derived from an action. One can easily see this by noting that the Einstein equation contains a cosmological constant term $-36m^2e^{\frac32\varphi}g_{\mu \nu}$, which would have a counterpart in the equation of motion for $\varphi$ if these field equations were compatible with an action. Note that when we would use the Ansatz~(\ref{ssansatz}) in the action~(\ref{newEH}), we would be able to integrate out the $z$ dependence, but the resulting action would not have~(\ref{sseom}) as its Euler-Lagrange equations. However, since we have performed the dimensional reduction on the level of the equations of motion, the consistency of the truncation is automatic.

A second important issue is that all fields are charged under the $\mathbb R$ symmetry that is gauged by $A$. This can be deduced from the Ansatz~(\ref{ssansatz}). Transforming $z\to z+\lambda(x^\mu)$, the fields transform as
\begin{equation}\label{sttr}
\varphi\to\varphi+\frac{3m}2\lambda \,,\qquad A_\mu\to A_\mu-\partial_\mu \lambda\,,\qquad g_{\mu\nu}\to e^{-\frac{9m}4 \lambda}g_{\mu\nu}\,,
\end{equation}
in order to remain in the same Ansatz~(\ref{ssansatz}). The way the gauge potential appears in the equations~(\ref{sseom}) might look a bit unfamiliar, but can be understood from the above transformation properties. Another consequence of~(\ref{sttr}) is that we can make $\varphi$ to vanish using a gauge transformation with parameter given by
\begin{equation}
\lambda=-\frac{2}{3m}\varphi\,.
\end{equation}
In this way, we break gauge invariance, giving mass to the gauge potential. This is the so-called St\"uckelberg mechanism. 

Finally, note that if we do not want to suppose that the metric $\hat g$ is multi-valued, we have to accept that we have performed a dimensional reduction on a noncompact manifold. Therefore, the interpretation of the resulting `effective' theory still is unclear~\cite{Chamblin:2001jj}. Our reason to consider this reduction is that it is a toy model to show that dimensional reduction can yield theories without action. Moreover, this construction can be generalized to the supersymmetric context, yielding a massive ten-dimensional supergravity, discussed in the following section.
\section{Massive HLW IIA theory}\label{massive}
\subsection{Motivation}
In regular supergravities, the fields in the gauge multiplet are always massless. Massive theories are continuous deformations of the regular ones in which the gauge and supersymmetry transformations get extra dependence via a mass parameter $m$. Consequently, the equations of motion get extra terms linear and quadratic in this parameter. Some of the fields (the St\"uckelberg fields) can subsequently be gauged away, giving mass to other fields.

The prime example of such a massive theory is Romans' IIA supergravity~\cite{Romans:1986tz}. Its role in string theory was clarified after the discovery of D-branes~\cite{Polchinski:1995mt}. It then became clear that this supergravity is actually the low energy limit of string theory in the background of a D8-brane. Such a brane solution was found in Romans' theory~\cite{Polchinski:1996df,Bergshoeff:1996ui}, and the mass parameter was seen to be proportional to the charge of the D8.
    
A less known massive IIA theory is the one constructed by Howe, Lambert and West (HLW) \cite{Howe:1998qt}. Although they constructed the theory by introducing a conformal spin connection, the theory can also be built by performing a generalized Scherk-Schwarz reduction of the eleven-dimensional equations of motion, using their scaling symmetry~\cite{Lavrinenko:1998qa}. The resulting massive theory does not have an action and has got no fundamental strings as the two-form potential (together with the dilaton) are St\"uckelberg fields. It was noted that this theory admits a de Sitter solution~\cite{Lavrinenko:1998qa}. As only few other solutions are known~\cite{Howe:1998qt,Chamblin:1999ea,Bergshoeff:2002nv,Chamblin:2001jj}, we will present a method of constructing (supersymmetric) solutions to that theory, using cones of special holonomy~\cite{Gheerardyn:2002wp}. Contrary to a claim in~\cite{Howe:1998qt}, the resulting field configurations were the first supersymmetric solutions to this theory~\cite{Lambert}. Moreover, these solutions were the first field configurations that solve any supersymmetric theory that does not have an action principle.

It is a general property of massive supergravities that they admit domain wall solutions. Therefore, the study of this HLW IIA supergravity is relevant to e.g. the domain wall/CFT correspondence~\cite{Boonstra:1998mp,Behrndt:1999mk} and the Randall-Sundrum scenario~\cite{Randall:1999ee,Randall:1999vf}. Moreover, it is a toy model of how dimensional reduction can yield theories without action.\footnote{There is however the subtle issue of boundary conditions that have to be supplemented to the eleven-dimensional action, which we will not discuss in this text.} 

Finally, it is notable that the resulting IIA theory is again realized on the mass-shell. This implies that computing the defining relation for supersymmetry \`a la~(\ref{Q^2=P}) would yield the equations of motion. In this sense, the theory is closely related to the ones presented in the preceding chapters.
\subsection{Construction}
As already mentioned in Chapter~\ref{onshellsusy}, the unique supergravity in eleven dimensions contains a metric $\hat g$ (or equivalently a Vielbein $\hat e$), a three-form potential $\hat A$ and a gravitino $\hat \psi$~\cite{Cremmer:1978km}. The equations of motion of that theory admit a rigid scaling symmetry
\begin{equation}\label{scaling}
e_\mu{}^a \rightarrow \lambda e_\mu{}^a \; , \; A_{\mu \nu \kappa} \rightarrow \lambda^3 A_{\mu \nu \kappa}\; , \;
\psi_\mu \rightarrow \lambda^{\frac{1}{2}} \psi_\mu \; ,
\end{equation}
which can be used to construct the massive HLW IIA supergravity~\cite{Lavrinenko:1998qa}. Below we will suppose that all fermionic fields in the IIA theory are zero. Hence, we will only need the part if the equations of motion that contains the bosons and we will only have to consider the supersymmetry transformations for the fermions that contain solely the bosonic fields. Since it is impossible that an equation of motion for a fermion contains a purely bosonic term, the truncation by setting all fermions to zero is a consistent one.
\subsubsection{Bosonic Fields}
The Ansatz for the metric is~(\ref{ssansatz}) and can equivalently be written in terms of the Vielbein
\begin{equation} \label{ansatze}
\hat{e}_{\hat \mu}{}^{\hat a}=e^{mz}\left(
 \begin{array}{cc}
 e^{\frac{1}{12}\varphi}e_\mu{}^a & e^{-\frac{2}{3} \varphi}A_\mu \\
 0& e^{-\frac{2}{3} \varphi}
 \end{array} \right)
 \; , 
\end{equation}
with $\mu,a=0,\dots,9$ and $\hat \mu,\hat a=0,\dots,10$. Note that the coordinate along the compactifying direction is called $z$, while the other coordinates are $\{x^\mu\}$. The three-form must depend on $z$ in the following way
\begin{equation}
\hat{A}_{\mu\nu\rho}=e^{3mz}A_{(3)\mu\nu\rho}\,,\qquad \hat A_{\mu\nu z}=e^{3mz}A_{(2)\mu\nu}\, .
\end{equation}
With this three-form, we define the ten-dimensional field strengths to be
\begin{eqnarray}
F_{(2)ab}&=& 2\partial_{[a}A_{b]}=2e_{[a}{}^\mu e_{b]}{}^\nu\partial_\mu A_\nu\; ,\nonumber\\
F_{(3)abc}&=& 3\partial_{[a}A_{(2)bc]}-3mA_{(3)abc}\; ,\nonumber\\
F_{(4)abcd}&=&4\partial_{[a}A_{(3)bcd]}-3\partial_{[a}A_{(2)bc} A_{d]}-3mA_{(3)[abc}A_{d]}\; .
\end{eqnarray}
This leads to the following equations of motion that are a generalization of~(\ref{sseom}),
\begin{eqnarray}
&&\Box\varphi =-\frac38 e^{-\frac32\varphi}\, F_{(2)}^2 +\frac1{12}
e^\varphi\, F_{(3)}^2 -\frac1{96} e^{-\frac12\varphi}\, F_{(4)}^2 +\frac{27}2 m^2\,
A_\mu\, A^\mu \nonumber\\&&\qquad+ 9m\, A^\mu\partial_\mu\varphi -\frac32 m\,
\mathfrak D_\mu{A}^\mu\,,\nonumber\\
&&\mathfrak D_\nu(e^{-\frac32\varphi}\, F_{(2)\mu}{}^\nu) =12m\partial_\mu\varphi
+18m^2\, A_\mu + 9m\, e^{-\frac32\varphi}\, A^\nu\, F_{(2)\mu\nu} \nonumber\\&&\qquad-\frac16 e^{-\frac12\varphi}\, F_{(4)\mu\nu\rho\sigma}
F_{(3)}^{\nu\rho\sigma}\,,\nonumber\\
&&\mathfrak D^{\sigma}(e^{\varphi}F_{(3)\mu\nu\sigma})=6m\,e^{\varphi}\, A^\sigma \, 
F_{(3)\mu\nu\sigma}+\frac12\, e^{-\frac12\varphi}\, F_{(4)\mu\nu\sigma\rho}\, 
F_{(2)}^{\sigma\rho} \nonumber\\&&\qquad+ \frac1{1152}
\epsilon_{\mu\nu\rho_1\dots\rho_8}\, F_{(4)}^{\rho_1\dots\rho_4}
F_{(4)}^{\rho_5\dots\rho_8}\,,\nonumber\\ 
&&\mathfrak D^{\sigma}(e^{-\frac12\varphi}F_{(4)\mu\nu\rho\sigma})= -6m\,e^{\varphi}\,
F_{(3)\mu\nu\rho}+6m\, e^{-\frac12\varphi}\, A^{\sigma}F_{(4)\mu\nu\rho\sigma}
\nonumber\\&&\qquad-\frac1{144}\epsilon_{\mu\nu\rho\sigma_1\dots\sigma_7}\,
F_{(4)}^{\sigma_1\dots\sigma_4}F_{(3)}^{\sigma_5\sigma_6\sigma_7}\,,\nonumber\\
&&R_{\mu\nu} -\frac12 R\, g_{\mu\nu}= - 36m^2\, e^{\frac32\varphi}\, g_{\mu\nu}+\frac12(\partial_\mu\varphi\,
\partial_\nu\varphi -\frac12 (\partial\varphi)^2\, g_{\mu\nu}) 
\nonumber\\
&&\qquad-9m^2\, (A_\mu\, A_\nu + 4A_\rho\, A^\rho\, g_{\mu\nu})  -\frac92 m\, (\mathfrak D_\mu A_\nu +\mathfrak D_\nu A_\mu
-2\mathfrak D_\rho A^\rho\, g_{\mu\nu})\nonumber\\
&&\qquad+\frac34m\, (A_\mu\partial_\nu\varphi + A_\nu\partial_\mu\varphi -A^\rho\partial_\rho \varphi\, g_{\mu\nu})+\frac12 e^{-\frac32\varphi}\, (F_{(2)\mu\rho}\, F_{(2)\nu}{}^\rho 
\nonumber\\&&\qquad-\frac14F_{(2)}^2\, g_{\mu\nu})+\frac1{4} e^\varphi\, (F_{(3)\mu\rho\sigma}\,
F_{(3)\nu}{}^{\rho\sigma} -\frac16 F_{(3)}^2\, g_{\mu\nu}) \nonumber\\&&\qquad
+ \frac1{12} e^{-\frac12\varphi}\, (F_{(4)\mu\rho\sigma\lambda}\,
F_{(4)\nu}{}^{\rho\sigma\lambda} -\frac18 F_{(4)}^2\, g_{\mu\nu})\,.
\end{eqnarray}
\subsubsection{Fermionic Fields}
As usual, the eleven-dimensional spinors are Majorana. The reduction yields two Majorana-Weyl spinors in ten dimensions of both chiralities. We will combine them in a single $32$-dimensional Majorana spinor. The gamma matrices in eleven-dimensional space-time are denoted by $\Gamma^{\hat \mu}$ while the ones in ten dimensions read $\gamma^\mu$. We will call the ten-dimensional chirality matrix $\gamma_*$

The Ansatz for the gravitino is taken in such a way that the supersymmetry transformation of the lower-dimensional Vielbein has the usual form
\begin{equation}\label{viel}
\delta e_\mu {}^a= \bar{\epsilon} \gamma^a \psi_\mu\; .
\end{equation}
Taking into account the scaling property~(\ref{scaling}), the Ansatz reads
\begin{equation}
\hat{\psi}_a=e^{-\frac{1}{24} \varphi -\frac{1}{2}mz}(\psi_a-\frac18 \gamma_a \lambda)\; , \;
\hat{\psi}_{i}=e^{-\frac{1}{24} \varphi -\frac{1}{2}mz}\gamma_{11} \lambda\; ,
\end{equation}
while the parameter for supersymmetry has to satisfy
\begin{equation}\label{ansatzsusy}
\hat{\epsilon}=e^{\frac{1}{24} \varphi +\frac{1}{2}mz}\epsilon\; .
\end{equation}
Using this Ansatz, the supersymmetry transformation rules of the massive IIA theory become
\begin{eqnarray}
\delta \lambda&=& \frac12 m e^{\frac{3}{4} \varphi} \gamma_* \epsilon-\frac18 e^{-\frac{3}{4} \varphi} \slashed{F}_{(2)}
\gamma_* \epsilon-\frac13 \slashed{\mathfrak{D}} \varphi
\epsilon-\frac{1}{144} e^{-\frac{1}{4} \varphi} \slashed{F}_{(4)} \epsilon+\frac{1}{18} e^{\frac{1}{2} \varphi}
\slashed{F}_{(3)} \gamma_* \epsilon \nonumber\\
\delta \psi_a&=&D_a \epsilon+\frac{9}{16}m e^{\frac{3}{4} \varphi} \gamma_a \gamma_*
\epsilon-m\frac{9}{16} \gamma_a \slashed{A}\epsilon-\frac{1}{64} e^{-\frac{3}{4} \varphi}({}_a \slashed{F}_{(2)}-14
\slashed{F}_{(2)a})\gamma_* \epsilon \nonumber\\&&+\frac{1}{48} e^{\frac{1}{2} \varphi}(9\slashed{F}_{(3)a}-{}_a
\slashed{F}_{(3)}) \gamma_* \epsilon+\frac{1}{128} e^{-\frac{1}{4} \varphi}(\frac{20}{3} \slashed{F}_{(4)a}-{}_a
\slashed{F}_{(4)})\epsilon
\end{eqnarray}
with
\begin{equation}
\mathfrak{D}_a \varphi=\partial_a \varphi+\frac32 m A_a
\end{equation}
Note that all ten-dimensional fields now only depend on the ten coordinates $\{x^\mu\}$.
\subsection{Solutions}
\subsubsection{Homothetic Killing Vectors}\label{homothetic}
In the present section, we will derive the conditions for a purely gravitational solution of M-theory to reduce to a (supersymmetric) solution of the massive IIA theory. For an eleven-dimensional solution with only the metric being nontrivial, the only equation of motion is the demand that the metric $\hat g$ is Ricci flat~(\ref{newEH}).
If such a solution admits a homothetic Killing vector $\hat k$~(\ref{homkilvect}), there will exist a coordinate system in which the metric will satisfy the reduction Ansatz~(\ref{ssansatz}). This can easily be seen by choosing coordinates adapted to $\hat k$ ($\hat k=\partial_z$), as the condition for a homothetic Killing vector field~(\ref{homkilvect}) then reduces to
\begin{equation}\label{readyforansatz}
\mathcal{L}_{\hat k} \hat{g}_{\hat \mu \hat \nu}\equiv\partial_z \hat{g}_{\hat \mu \hat \nu}=2m\hat{g}_{\hat \mu \hat \nu}\; ,
\end{equation}
which is solved by
\begin{equation}\label{solg}
\hat{g}_{\hat \mu\hat \nu}(x^\mu,z)=e^{2m z} \hat{h}_{\hat \mu\hat \nu}(x^\mu)\; ,
\end{equation}
implying that $\hat g$ satisfies the Ansatz~(\ref{ssansatz}).

Such an eleven-dimensional solution is supersymmetric if the supersymmetry transformation of the gravitino, which reduces to
\begin{equation}
\delta_Q \hat \psi_{\hat \mu}=\hat{\mathfrak D}_{\hat \mu}\hat \epsilon\,,
\end{equation}
has solutions for the parameter $\hat \epsilon$. Equivalently, the eleven-dimensional solution preserves as many supersymmetries as there are parallel spinors $\hat \epsilon$.
Looking at the Ansatz~(\ref{ansatzsusy}) for the parameters of supersymmetry, we see that these parallel spinors would satisfy the Ansatz if they depend on $z$ as
\begin{equation}
\partial_z \hat{\epsilon}=\frac m2 \hat{\epsilon}\; .
\end{equation}
Using the results of Section~\ref{spinnen}, we can rewrite this in a coordinate invariant way and conclude that the dimensionally reduced solution will preserve as many supersymmetries as there are eleven-dimensional parallel spinors satisfying
\begin{equation}
\mathcal{L}_{\hat k} \hat{\epsilon}=\hat k^{\hat \mu} \mathfrak D_{\hat \mu} \hat{\epsilon}+\frac14 \partial_{\hat \mu} \hat k_{\hat\nu} \Gamma^{\hat\mu\hat\nu} \hat{\epsilon}=\frac m2 \hat{\epsilon} \; .
\end{equation}
This condition can even be simplified to
\begin{equation}\label{pressusy}
\partial_{\hat\mu} \hat k_{\hat\nu} \Gamma^{\hat\mu\hat\nu} \hat{\epsilon}=2m \hat{\epsilon}\; ,
\end{equation} 
since the spinors are parallel.

If we would also consider solutions with a three form differing from zero, it would satisfy the Ansatz if
\begin{equation}
\mathcal{L}_{\hat k} \hat{A}_3=3m\hat{A}_3 \; .
\end{equation}
In conclusion, a gravitational solution of classical M theory yields a solution of the massive theory if it admits a homothetic Killing vector. The number of preserved supersymmetries of the ten-dimensional field configuration equals the number of eleven-dimensional parallel spinors satisfying~(\ref{pressusy}).
\subsubsection{Ricci-flat Cones}\label{ricciflat}
If a manifold admits a homothetic Killing vector $\hat k$ that is hypersurface orthogonal, i.e. if there exists a function $\hat f$ such that
\begin{equation}
\hat k_{\hat \mu}=\hat g_{\hat \mu \hat \nu} \hat k^{\hat \nu}=\partial_{\hat \mu} \hat f \; ,
\end{equation}
then we can find a set of coordinates in which the metric can be written as a cone~\cite{Gibbons:1998xa}. The vector field $\hat k$ is then called the Euler vector. We will now set the stage by considering Ricci-flat cones and dimensionally reduce on the Euler vector. This will yield solutions of the IIA theory that do not preserve any supersymmetry.

We start the construction by taking a solution of eleven-dimensional supergravity which is $(11-d)$-dimensional Minkowski space times a $d$-dimensional Ricci-flat Euclidean cone. We then write Minkowski space as a cone over de Sitter space.\footnote{De Sitter space can be viewed as the $D$-dimensional hypersurface in $D+1$-dimensional Minkowski space (with Cartesian coordinates $x^a$, $a=0,\dots,D$ and metric $\eta$) defined as $x^a x^b\eta_{ab}=\rho^2$ for a certain real value of $\rho$.} The metric $\hat g$ therefore reads
\begin{equation}\label{g2}
\hat{g}=dR^2+R^2ds^2_{dS}+dr^2+r^2 ds^2_{d-1}\; ,
\end{equation}
where the first two terms denote the metric on Minkowski space, while the last two are the metric on the Ricci-flat cone.
Now, $\hat k=R\partial_R+r\partial_r$ is a homothetic Killing vector that is hypersurface orthogonal.
We can now change to coordinates adapted to $\hat k$,
\begin{equation}
r=e^z \cos \rho\, , \, \, R=e^z \sin \rho\; , \;\; \rho \in [0,\frac{\pi}2] \; .
\end{equation}
In this coordinates, the metric reads
\begin{equation}
\hat{g}=e^{2z}\left(dz^2+d\rho^2+\sin^2 (\rho) ds_{dS}^2+\cos^2 (\rho) ds^2_{d-1} \right) \; .
\end{equation}
From the reduction Ansatz~(\ref{ssansatz}), we can directly read of that 
\begin{equation}
g=d\rho^2+\sin^2 \rho ds_{dS}^2+\cos^2 \rho ds^2_{d-1}\; , \; \; m=1 \; ,
\end{equation}
together with all other fields zero, yields a solution of the massive supergravity. We can find solutions for any value of the mass parameter by reducing on $m\hat k$. Note that this solution does not preserve any supersymmetry, since
\begin{equation}
\mathcal L_{\hat k}\hat \epsilon=0\,,
\end{equation}
for any eleven-dimensional parallel spinor $\hat \epsilon$.

A large set of Ricci-flat Euclidean cones are cones of special holonomy, discussed extensively in~\cite{Acharya:1998db}. 
\subsubsection{Supersymmetric Reductions}\label{susy}
In this section, we will construct solutions of the massive IIA theory that are reductions on special holonomy cones (supplied with some extra flat directions) and that preserve some fraction of the supersymmetry. The vector field we will use to perform the dimensional reduction will now be the Euler vector from the previous section plus a boost in the flat directions. Therefore, the vector field will not be hypersurface orthogonal anymore and we will be able to preserve supersymmetry during the reduction process. 
\paragraph{Reduction of Flat Space}
We will start with the easiest case, which is the supersymmetric reduction of Minkowski space. The resulting solution will preserve half of the supersymmetries of the IIA theory.

As previously mentioned, Minkowski space admits a homothetic hypersurface orthogonal Kil\-ling vector $\hat K$ that can be written in Cartesian coordinates as 
\begin{equation}
\hat K=x^{\hat \mu} \partial_{\hat \mu} \; , \; \; \mathcal{L}_{\hat K}\hat \eta_{\hat \mu \hat \nu}=2\hat\eta_{\hat \mu\hat \nu}\; ,
\end{equation}
implying that Minkowski space is a cone over de Sitter space with $\hat K$ being the Euler vector.
Noting that all $\so(1,10)$ rotations induce Killing vectors $\hat l$ of Minkowski space, the vector field 
 $\hat k=\hat K+\hat l$ still is homothetic, but not hypersurface orthogonal anymore. The important observation to be able to construct supersymmetric solutions is that  $\so(1,10)$ rotations exist such that
\begin{equation}
\mathcal{L}_{\hat k}\hat{\epsilon}=\mathcal{L}_{\hat l} \hat{\epsilon}=\frac12 \hat{\epsilon}\; ,
\end{equation}
for 16 (out of the 32) parallel spinors $\hat{\epsilon}$ of eleven-dimensional Minkowski space. The spinors satisfying this condition reduce to supersymmetries of the ten-dimensional solution.

We will now show that $\hat l$ has to be a boost. We therefore start by writing
\begin{equation}
\hat l=B_{\hat \mu}{}^{\hat \nu} x^{\hat \mu} \partial_{\hat \nu} \; .
\end{equation}
Keeping in mind that $m=1$, the condition for supersymmetry~(\ref{pressusy}) now reads
\begin{equation}
\slashed{B}\hat{\epsilon}=2 \hat{\epsilon}\,.
\end{equation}
As $\slashed{B}$ has to have real eigenvalues, $\hat l$ has to be a boost, and it is always possible to take it in the $(x^0,x^1)$ plane. Because $\Gamma^{01}$ squares to one and is traceless, half of its eigenvalues are $1$, and the other half are $-1$. Therefore, we take $\hat l$ to be 
\begin{equation}
\hat l=x^1\partial_0+x^0\partial_1 \; .
\end{equation}
As a consequence, the homothetic Killing vector we want to reduce on, reads
\begin{equation}
\hat k=(x^0+x^1)\partial_0+(x^0+x^1)\partial_1+r\partial_{r}\; ,
\end{equation}
where $r=\sqrt{x^ax^a}$ and $a=2,\dots,10$. If we write the flat metric as
\begin{equation} \label{redmink}
\hat{g}=-(dx^0)^2+(dx^1)^2+\Big(dr^2+r^2d\Omega_8^2\Big)\; ,
\end{equation}
where $d\Omega_8^2$ is the natural metric on the 8-sphere, we can choose new coordinates\footnote{Note that the Jacobian for this coordinate transformation becomes singular at $y_2=0$, restricting the range of $y_2$ to $0<y_2<\infty$.} that are adapted to the vector field such that $k=\partial_z$.
\begin{eqnarray}
x^0&=&\frac12 y_2\left( e^{2z}+e^{-2y_1} \right)\; ,\nonumber \\
x^1&=&\frac12y_2 \left( e^{2z}-e^{-2y_1} \right)\; ,\nonumber\\
r&=&e^{z-y_1}\; .
\end{eqnarray}
In these new coordinates, the eleven-dimensional metric reads
\begin{eqnarray}
\hat{g}&=&e^{2(z-y_1)} \Big( (dz-(1-2y_2^2)dy_1-y_2dy_2)^2\nonumber\\&&+4y_2^2(1-y_2^2)dy_1^2-(1+y_2^2)dy_2^2+4y_2^3dy_1dy_2+d\Omega_8^2 \Big)\; .
\end{eqnarray}
From the reduction Ansatz~(\ref{ssansatz}), we can read of the ten-dimensional field configuration. 
The ten-dimensional solution is ($m=1$)
\begin{eqnarray}
g&=&e^{-\frac94y_1}\left( 4y_2^2(1-y_2^2)dy_1^2-(1+y_2^2)dy_2^2+4y_2^3dy_1dy_2+d\Omega_8^2 \right) ,\nonumber\\
A_{y_1}&=&-(1-2y_2^2) \; \;  , \; \; \;
A_{y_2}=-y_2\; ,\nonumber\\
\varphi&=&\frac32 y_1\; ,\label{flatsol}
\end{eqnarray}
and preserves $1/2$ of the supersymmetry. 
\paragraph{Reductions on Special Holonomy Cones}
\begin{table}
\begin{center}
\begin{tabular}[t]{|c|c|c|}
\hline
Dimension&Manifold& $\mathcal{N}$\\
\hline \hline
8&Spin(7)&1\\
8&$CY_4$&2\\
8&$HK_2$&3\\
7&$G_2$&2\\
6&$CY_3$&4\\
4&$CY_2= HK_1$&8\\
\hline
\end{tabular}
\caption{\it Number of preserved supersymmetries $\mathcal N$ of a solution of the HLW theory that is constructed from an eleven-dimensional solution which is the direct product of an Euclidean special holonomy cone (with dimension listed in the first column) and Minkowski space.}\label{numbsusies}
\end{center}
\end{table}
To find solutions with less supersymmetry, we can start from an eleven-dimensional configuration which is the product of an Euclidean simply connected special holonomy cone and flat Minkowski space. 

The eleven-dimensional solution we start with thus reads
\begin{equation}
\hat{g}=-(dx^0)^2+(dx^1)^2+dr_1^2+r_1^2d\Omega_{n-2}^2+dr_2^2+r_2^2 ds_{9-n}^2\; ,
\end{equation}
where the last two terms are the metric on the $10-n$-dimensional special holonomy cone and $d\Omega_{n-2}^2$ is the natural metric on the $n-2$-dimensional sphere. We first perform a coordinate transformation
\begin{equation}
r_1=r \cos \alpha \; \; , \; \; \; r_2=r\sin \alpha \; ; \; \; \alpha \in [0,\frac{\pi}2] \; , \; n>2 \; | \;\alpha \in [-\frac{\pi}2,\frac{\pi}2] \; , \; n=2 \; .
\end{equation}
The metric in these new coordinates reads
\begin{equation}
\hat{g}=-(dx^0)^2+(dx^1)^2+dr^2+r^2 \Big(d\alpha^2 + \cos^2 \alpha d\Omega_{n-2}^2+\sin^2 \alpha ds_{9-n}^2\Big)\; ,
\end{equation}
while the homothetic Killing vector we will use for the reduction equals
\begin{equation}
k=(x^0+x^1)\partial_0+(x^0+x^1)\partial_1+r\partial_r\; .
\end{equation}
Comparing with~(\ref{redmink}), we see that we only have to substitute the metric on the 8-sphere by
\begin{equation}
d\Omega_8^2 \to  d\alpha^2 + \cos^2 \alpha d\Omega_{n-2}^2+\sin^2 \alpha ds_{9-n}^2\, ,
\end{equation}
in the solution~(\ref{flatsol}).
The number of preserved supersymmetries is half of the number preserved by the eleven-dimensional solution.
Starting from the number of parallel spinors on special holonomy manifolds, as given in Section~\ref{berger}, we list the number of preserved supersymmetries $\mathcal{N}$ in Table~\ref{numbsusies}.
\chapter{Conclusion}\label{concel}
In this thesis, we have discussed some aspects of on-shell supersymmetry. First of all, we have shown that we can use supersymmetry as an organizing principle to construct the equations of motion of theories that do not admit a Poincar\'e-invariant and supersymmetric least action principle. In the context of rigid $\mathcal N=2$ Poincar\'e supersymmetry, this has led us in Chapter~\ref{onshellsusy} to generalizations of theories containing vector- and hypermultiplets.  For the former, we have generalized the target space and we have discovered new potentials. For the latter, we have unravelled the possibility of having hypercomplex target spaces. In the following Chapter~\ref{suptencalc}, we have performed the same study in the case the $\mathcal N=2$ super-Poincar\'e algebra is realized locally. This has led us to the conclusion that such a supergravity allows for quaternionic target spaces in the nonlinear sigma models of the hypermultiplets. As we used the superconformal tensor calculus approach, we have studied the target spaces during the different steps of the construction. This enabled us to show that we can project a hypercomplex manifold with a homothetic symmetry vector field onto a quaternionic manifold and a hyperk\"ahler space onto a quaternionic-K\"ahler one. In our final Chapter~\ref{dimred}, we have discussed a simple example of a dimensional reduction that yields a ten-dimensional theory without action. By carefully examining the reduction Ansatz, we have succeeded in constructing supersymmetric solutions to this massive type IIA supergravity. 

It is obvious that these theories without actions deserve further study. We therefore list some possible directions for future research. First of all, the vector multiplets discussed in Chapter~\ref{onshellsusy} can also realize the superalgebra locally. Hence, one could try to repeat the construction in this context. Moreover, if off-shell vector multiplets are dimensionally reduced to three dimensions, this results in a theory with hypermultiplets after dualizing the vector into a scalar (this process yields the so-called combined r- and c-map~\cite{Cecotti:1989qn,deWit:1992nm}). It would be interesting to know what are the geometries of the target spaces of these three-dimensional theories (e.g. quaternionic or hypercomplex). One could similarly try to complete the superconformal tensor calculus approach of Chapter~\ref{suptencalc} in a context without actions. 

These new $\mathcal N=2$ theories we have found, could yield new and interesting solutions that also require further study. Moreover, it is standard knowledge that five-dimensional $\mathcal N=2$ theories with action can be constructed by compactifying classical M-theory on a Calabi-Yau three-fold. It would be interesting to discover what type of compactification would yield our new theories. It might be possible that they can be retrieved by turning on a flux on the Calabi-Yau manifold. 

Mirror symmetry in its simplest form is the fact that the effective four-dimensional theory stemming from the compactification of type IIA supergravity on a certain Calabi-Yau manifold can also be constructed by compactifying type IIB supergravity on another Calabi-Yau, called the mirror manifold. Superstring theory predicts that such pairs of manifolds exist. Once we know how to construct the $\mathcal N=2$ theories without action from compactification, it would therefore be very interesting if this notion of mirror symmetry can be extended to the effective theories without actions.

Finally, we have often referred to the unique eleven-dimensional supergravity. One could try to generalize this theory by only considering its superalgebra and its equations of motion. More precisely, one could try to fit in a cosmological constant if the eleven-dimensional equations of motion do not have to be integrable into an action.

Let us conclude with a remark about supersymmetry. Although this fermi\-o\-nic invariance has been widely studied during more than thirty years, supersymmetric theories still reveal new and intriguing mathematical structures, even at the classical level. Whether we will one time discover or exclude the existence of supersymmetry in Nature, its impact on the insights in the structure of field theories will never lose importance. 
\appendix
\chapter{Spinors}\label{spinors}
\section{The spin group and the Clifford algebra}
By $V^d_{(t)}$ we will denote a $d$-dimensional vector space over $\mathbb{R}$ with metric 
\begin{equation}
\eta=\rm{diag}(\underbrace{-1,\dots,-1}_{\rm{t \; times}},\underbrace{+1,\dots,+1}_{\rm{d-t \; times}})\,.
\end{equation}
We take $\{e_i\}$ to be a set of $d$ orthonormal basis vectors of $V_{(t)}^d$ and let $v,w\in V^d_{(t)}$. Let $vw$ be an associative product of vectors that is distributive with respect to the usual addition and that satisfies
\begin{equation}
uv+vu=2<u,v>\; .
\end{equation}
The corresponding algebra is called the Clifford algebra $C(V^d_{(t)})$.
Obviously, a basis can be constructed starting with the $\{e_i\}$ as 
\begin{equation}
(1,e_i,e_ie_j,\dots,e_1\dots e_d)\; .
\end{equation}
As such, the Clifford algebra $C(V^d_{(t)})$ is a linear space of dimension $2^d$. The linear subspace $C_p(V_{(t)}^d)$ is spanned by the elements $\{e_{i_1}\dots e_{i_p}\}$. 

The mirror image $w$ of a vector $v$ with respect to a plane orthogonal to the vector $u$ reads
\begin{equation}
w=v-2\frac{<u,v>}{<u,u>}u\; .
\end{equation}
Using Clifford multiplication, we can also write such a reflection as $w=-uvu^{-1}$ with $u^{-1}={u/<u,u>}$. Any element of  $\SO(t,d-t)$ can be constructed from an even number of reflections. As a consequence, a general rotation can be written as $w=u_k\dots u_1vu_1^{-1}\dots u_k^{-1}$ with $k\in 2\mathbb{N}$. Therefore, with $\Lambda=u_1\dots u_k$, $\Lambda^{-1}=u_k^{-1}\dots u_1^{-1}$ and $\tilde{\Lambda}=u_k\dots u_1$, we can now introduce the spin group.
\begin{defi}
The group $\Spin(t,d-t)$ is defined to be the group of all elements $\Lambda \in C(V_{(t)}^d)$ such that
\begin{eqnarray}
\Lambda v \Lambda^{-1}\in C_1(V^d_{(t)}) &,& \forall v\in C_1(V^d_{(t)})\; ,\nonumber\\
\Lambda \tilde{\Lambda}=1\; . &&\label{defspin}
\end{eqnarray}
\end{defi}
The mapping $\mathcal{H}:\Spin(t,d-t)\to\SO(d,t-d):\Lambda e_i\Lambda^{-1}\mapsto a_i{}^j e_j$ is a 2-1 homomorphism.
This can easily be seen as follows. Obviously, the map is a homomorphism 
which is onto, as any element of $\SO(t,d-t)$ is an even number of reflections. Moreover, an element of the kernel of $\mathcal H$ has to lie in the center of $C(V_{(t)}^d)$, as it has to commute with any of the basis elements $e_i$. Hence, the kernel only contains $\Lambda=\pm 1$. Therefore, the homomorphism is 2-1. Note that the infinitesimal transformations of both groups are isomorphic.
\section{Clifford algebra representation}
In physics, we represent the basis vectors of the Clifford algebra by so-called Clifford (or gamma) matrices $\gamma_a$ of dimension $2^{d/2}$ for even and $2^{(d-1)/2}$ for odd $d$. Consider a space-time with $t$ timelike and $d-t$ spacelike directions. Consequently, we have a Minkowski metric $\eta_{ab}$ with $a,b=0,\dots,d-1$. In terms of these gamma matrices, the Clifford algebra reads
\begin{equation}\label{clifalggamma}
\gamma_a\gamma_b+\gamma_b\gamma_a=2\eta_{ab}\,.
\end{equation}
These matrices are taken to be hermitian for spacelike index $a$, and antihermitian in the timelike case. Therefore, they satisfy the hermiticity property 
\begin{equation}\label{hermgamma}
\gamma^\dagger_a=(-1)^tA\gamma_aA^{-1}\,,\qquad A=\gamma_0\dots\gamma_{t-1}\,,
\end{equation}
where $\dagger$ denotes hermitian conjugation.
Conditions~(\ref{clifalggamma}) and~(\ref{hermgamma}) are preserved by any unitary transformation of the Clifford matrices. In even dimensions, we can introduce a chirality matrix $\gamma_*$ that anticommutes with all gamma matrices and squares to one.
\begin{equation}
\gamma_*=(-\rmi)^{d/2+t}\gamma_0\dots\gamma_{d-1}\,.
\end{equation}
Due to this property, we can construct a representation of the Clifford algebra in $2m+1$ dimensions by taking the $2m$ matrices of the algebra in $2m$ dimensions and taking as the final $\gamma_{2m}$ the chirality matrix of the representation in $2m$ dimensions. antisymmetric combinations of gamma matrices are related to each other through Hodge duality,
\begin{eqnarray}
\gamma_{a_0\dots a_i}&=&\frac1{(d-i)!}\varepsilon_{a_0\dots a_{d-1}}\rmi^{d/2+t}\gamma_*\gamma^{a_{d-1}\dots a_{i+1}}\,,\qquad \mbox{with $d$ even}\,,\\
\gamma_{a_0\dots a_i}&=&\frac1{(d-i)!}\varepsilon_{a_0\dots a_{d-1}}\rmi^{(d-1)/2+t}\gamma^{a_{d-1}\dots a_{i+1}}\,,\qquad \mbox{with $d$ odd}\,.\nonumber
\end{eqnarray}
It is always possible to introduce a unitary charge conjugation matrix $\mathcal{C}$ satisfying
\begin{equation}\label{chargeconj}
\mathcal{C}^T=-\epsilon\mathcal{C}\,,\qquad \gamma_a^T=-\eta \mathcal{C}\gamma_a\mathcal{C}^\dagger\,,
\end{equation}
where $T$ denotes the operation of transposition.
By counting the number of symmetric and antisymmetric matrices $\mathcal{C}\gamma^{(i)}$ for all values of $i$, one can deduce the allowed values for the signs $\eta$ and $\epsilon$~\cite{VanProeyen:1999ni}.

Naturally, the gamma matrices $\gamma^a=\eta^{ab}\gamma_b$ transform as vectors of $V^d_{(t)}$. If $M$ is a space such that $\forall p\in M: T_pM\equiv V^d_{(t)}$, we can construct a field of gamma matrices that are sections of $TM$ by multiplying with an inverse Vielbein, $\gamma^\mu=e_a{}^\mu \gamma^a$.
\section{Irreducible spinors}
A spinor is a representation of $\Spin(t,d-t)$ that is not a representation of $\SO(t,d-t)$. If we restrict our attention to infinitesimal transformations, the spinor should transform under the action of $\spin(t,d-t)\equiv \so(t,d-t)$. This representation might be reducible. In even dimensions, we can always project a spinor $\lambda$ on its left- or right-handed part.
\begin{equation}
\lambda_{L,R}=\frac12 (1\pm\gamma_*)\lambda\,,
\end{equation}
which is compatible with Lorentz-transformations, meaning that transformations under $\spin(t,d-t)$ do not mix the handedness of the spinor. Such a spinor of definite handedness (chirality) is called a Weyl-spinor. 

\begin{table}[t]
\begin{center}\begin{tabular}{|c|lr|lr|lr|lr|}\hline
d $\backslash$ t &\multicolumn{2}{c|}{0} &\multicolumn{2}{c|}{ 1}
& \multicolumn{2}{c|}{2} & \multicolumn{2}{c|}{3} \\ \hline
1 & M     & 1 & M    & 1 &       &   & &   \\
2 & M$^-$ & 2 & MW   & 1 & M$^+$ & 2 & &   \\
3 &       & 4 & M    & 2 & M     & 2 & & 4 \\
4 & SW   & 4 & M$^+$& 4 & MW    & 2 & M$^-$ & 4 \\
5 &       & 8 &      & 8 & M     & 4 & M     & 4 \\
6 & M$^+$ & 8 & SW  & 8 &  M$^-$& 8 & MW    & 4 \\
7 & M     & 8 &      & 16 &      & 16 & M    & 8 \\
8 & MW    & 8 & M$^-$& 16 & SW  & 16 & M$^+$& 16 \\
9 & M     & 16 & M   & 16 &      & 32 &      & 32 \\
10& M$^-$ & 32& MW   & 16 & M$^+$ &32 & SW  & 32 \\
11&       & 64& M    & 32 & M     & 32 & & 64 \\
12& SW   & 64 & M$^+$& 64 & MW    & 32 & M$^-$ & 64\\ \hline
\end{tabular}\end{center}\caption{\it Possible spinors in various dimensions,
and for various number of timelike directions (modulo 4).}\label{tbl:MWSspinors}
\end{table}

It might also be possible to impose a reality condition on the spinor, i.e.
\begin{equation}\label{majcond}
\lambda^*=\tilde B \lambda\,,
\end{equation}
where $*$ denotes complex conjugation.
This must again be compatible with Lorentz-trans\-for\-ma\-tions, $\tilde B$ must be unitary and $\lambda^{**}$ must equal $\lambda$. If this is possible, a spinor satisfying~(\ref{majcond}) is called a Majorana spinor.
Finally, whenever the Majorana condition is not possible, we can impose a twisted reality condition. 
\begin{equation}
(\lambda_i)^*=\tilde B \Omega_{ij}\lambda^j\,,
\end{equation}
which should satisfy the same consistency demands of Majorana spinors (i.e. compatibility with Lorentz transformations, unitarity of $\tilde B$ and $**\equiv \unity$), and where $\Omega$ is some antisymmetric matrix with the property that $\Omega \Omega^*=-1$. Such a spinor is called symplectic Majorana.

In Table~\ref{tbl:MWSspinors}, taken from~\cite{VanProeyen:1999ni}, the minimal spinor representations are given together with the conditions imposed on it. As can be seen from there, it is sometimes possible to impose two requirements simultaneously. The Weyl-condition is denoted by `W', Majorana by `M' and symplectic Majorana by `S' while `$M^\pm$' indicates which sign of $\eta$~(\ref{chargeconj}) should be used. Note that the possibility of a Weyl spinor is not mentioned explicitly since this is always possible in even dimensions. The same holds for a symplectic Majorana spinor representation, which is always possible whenever a Majorana condition cannot be imposed.

\chapter{Conventions}\label{appconventions}
In this Appendix, we will explain our conventions.
\begin{enumerate}
\item The name of a \emph{Lie-group} always contains a capital, e.g. $G$, $\SO$, $\Gl$, etc. while the corresponding \emph{Lie-algebra} is denoted by small, curly letters, e.g. $\mathfrak g$, $\so$, $\gl$, etc.
\item \emph{Symmetrization} and \emph{antisymmetrization} is denoted by $(\dots)$ and $[\dots]$ respectively, and is done with unit weight. For instance
\begin{equation}
\xi_{[\mu \nu]}=\frac12 (\xi_{\mu\nu}-\xi_{\nu\mu})\,,\qquad \xi_{(\mu\nu\rho)}=\frac1{3!}(\xi_{\mu\nu\rho}+\rm{permutations\;\;in\;\;}\mu\nu\rho)\,.
\end{equation}
\item \emph{The Levi-Civita symbol} in $t$ timelike and $(d-t)$ spacelike dimensions is taken to be $\varepsilon_{01\dots (d-1)}$ $=+1$ and any interchange of two indices multiplies it with $-1$. Moreover, we take $\varepsilon^{01\dots (d-1)}$ $=(-1)^t$. A useful formula for the contraction yields
\begin{equation}
\varepsilon_{a_1\dots a_p b_1 \dots b_q}\varepsilon^{a_1\dots a_pc_1\dots c_q}=(-1)^t p!q!\delta_{[b_1}^{[c_1}\dots\delta_{b_q]}^{c_q]}\,.
\end{equation}
The above Levi-Civita symbol is actually a volume form on flat $d$-di\-men\-sio\-nal space-time with $t$ timelike directions. In curved space, we can extend its definition via
\begin{eqnarray}
\varepsilon_{\mu_1\dots\mu_d}&=&e^{-1}e_{\mu_1}{}^{a_1}\dots e_{\mu_n}{}^{a_n}\varepsilon_{a_1\dots a_n}\,, \nonumber\\ \varepsilon^{\mu_1\dots\mu_d}&=&e e_{a_1}{}^{\mu_1}\dots e_{a_n}{}^{\mu_n}\varepsilon^{a_1\dots a_n}\,,
\end{eqnarray}
where $e=\det(e_\mu{}^a)$ with $\{e_\mu{}^a\}$ the Vielbein where $\mu$ is a curved and $a$ a flat index.
\item \emph{Triplet} and \emph{doublet} notation for the adjoint of $\su(2)$ are connected via Pauli matrices. Consider a vector $T$ in the adjoint of $\su(2)$. In components, we may write $T^\alpha$ with $\alpha=1,2,3$ or $T^{ij}\equiv T^{(ij)}$ with $i,j=1,2$. We will almost always use a vector notation $\vec T$ for the former, and the connection with the latter is given by
\begin{equation}
\vec T=\frac12\rmi\vec \sigma_{i}{}^kT^{ij}\varepsilon_{kj}\,,
\end{equation}
where $\vec\sigma_i{}^j$ are the Pauli-matrices given by
\begin{equation}
\sigma^1\equiv\left(\begin{array}{cc}0&1\\1&0\end{array}\right)\,,\qquad\sigma^2\equiv\left(\begin{array}{cc}0&-\rmi\\\rmi&0\end{array}\right)\,,\qquad\sigma^3\equiv\left(\begin{array}{cc}1&0\\0&-1\end{array}\right)
\end{equation}
Given two such vectors $\vec S$ and $\vec T$, we can define scalar and vector product as
\begin{equation}
\vec S\cdot \vec T=S^\alpha T^\alpha\,,\qquad (\vec S\times \vec T)^\alpha=\varepsilon^{\alpha\beta\gamma}S^\beta T^\gamma\,.
\end{equation}
\item Indices on objects for which a \emph{symplectic product} is defined, are raised and lowered via the North-East South-West convention. For instance, in the study of hypermultiplets we encounter the tensor $C_{AB}$, which is used as follows
\begin{equation}
A_A = A^BC_{BA} \,,\qquad A^A = C^{AB} A_B\qquad \rm{with}\qquad C_{AC} C^{BC} =   \delta_A{}^B \,.
\end{equation}
Similar considerations hold for $\su(2)$ indices on e.g. symplectic Majorana spinors.
\item The space-time metric is always \emph{mostly-plus}.
\item The covariant constancy of the \emph{Vielbeine} on space-time read
\begin{equation}\label{covcteVielgr}
\mathfrak{D}_\mu e_\nu{}^a=\partial_\mu e_\nu{}^a+\omega_{\mu}{}^{ab}e_{\nu b}-\Gamma^\rho{}_{\mu \nu}e_\rho{}^a\,.
\end{equation}
The \emph{curvatures} are then defined as
\begin{eqnarray}
 R^\upsilon{}_{\rho\mu\nu} &\equiv & 2
\partial_{[\mu}\Gamma^\upsilon{}_{\nu]\rho} + 2 \Gamma^\upsilon{}_{\tau[\mu} \Gamma^\tau{}_{\nu]\rho}
\,,\nonumber\\
R_{\mu\nu b}{}^{a} &\equiv & 2\partial_{[\mu} \omega_{\nu]b}{}^{a}
+ 2\omega_{[\mu| c|}{}^{a} \omega_{\nu]b}{}^{c}\,, 
\end{eqnarray}
while the Ricci tensor and scalar read
\begin{equation}
R_{\mu\nu}\equiv R^\rho{}_{\mu\rho\nu}=R_{\rho\mu}{}^{ba} e_b{}^\rho e_{\nu a}\,,\qquad R\equiv R_{\mu\nu}g^{\mu\nu}\,.
\end{equation}
The target space Riemannian curvature on the contrary, is denoted by 
\begin{equation}
R_{XYZ}{}^V\equiv 2\partial_{[X}\Gamma_{Y]Z}{}^V+2\Gamma_{W[X}{}^V\Gamma_{Y]Z}{}^W\,.
\end{equation}
The target space Ricci tensor has the same sign as compared to the Ricci tensor on space-time,\footnote{Both definition have in common that a compact manifold corresponds to a positive Ricci scalar.} and the Ricci scalar has the same definition,
\begin{equation}
R_{XY}\equiv R_{ZXY}{}^Z\,,\qquad R\equiv R_{XY}g^{XY}\,.
\end{equation}
Finally, note that the Greek indices $\mu,\nu,\dots$ on space-time correspond to capitals $X,Y,\dots$ on target space and that the position of the upper index on the space-time connection differs from that on the target space connection, as we want to make clear that their corresponding curvatures are written in a different way.
\item We now list some basic relations for anticommuting \emph{spinors} in five dimensions. The fundamental Fierz relation reads
\begin{equation}\label{basicfierz}
\lambda \bar \zeta=-\frac14 (\bar \zeta\lambda+\bar \zeta\gamma^a\lambda\gamma_a-\frac12\bar \zeta\gamma^{ab}\lambda\gamma_{ab})\,.
\end{equation}
The signs $t_{(n)}$, defined as 
\begin{equation}
\bar \zeta \gamma_{(n)}\lambda=t_{(n)}\bar \lambda \gamma_{(n)}\zeta\,,
\end{equation}
with $\gamma_{(n)}=\gamma_{a_1\dots a_n}$, are listed in Table~\ref{signsconv}.
\begin{table}[t]
\begin{center}
\begin{tabular}{|c||c|c|c|c|}
\hline 
$n$&0,4&1,5&2&3\\
\hline
$t_{(n)}$&+&+&$-$&$-$\\
\hline
\end{tabular}
\caption{\it The signs $t_{(n)}$ for different values of $n$ in five space-time dimensions.}\label{signsconv}
\end{center}
\end{table}
In our discussion of $\mathcal N=2$ supersymmetry, the spinors carry an $\su(2)$ index. If the index is contracted, we do not write it, i.e.
\begin{equation}
\bar \zeta^i\gamma_{(n)}\lambda_i\equiv\bar \zeta\gamma_{(n)}\lambda\,.
\end{equation}
Note that when one or both of the spinors are commuting (which is often encountered in the Batalin-Vilkovisky formalism), the right-hand side of~(\ref{basicfierz}) together with the signs $t_{(n)}$ are multiplied with $-1$.
\item We use a notation with \emph{slashes} if we want to denote multiplication with gamma matrices. If $\Lambda$ is a $p$-form defined on a space $M$, we have
\begin{equation}
\slashed \Lambda\equiv \Lambda_{\mu_1\dots\mu_p}\gamma^{\mu_1\dots\mu_p}\,,\, \slashed{\Lambda}_\mu\equiv \Lambda_{\nu_1\dots \nu_p}\gamma^{\nu_1\dots \nu_p}{}_\mu\,,\, {}_\mu\slashed{\Lambda}\equiv \Lambda_{\mu\nu_1\dots \nu_{p-1}}\gamma^{\nu_1\dots \nu_{p-1}}\,.
\end{equation}
\item A \emph{set of elements} is denoted by $\{\dots\}$.
\item We denote the \emph{semi-direct product} of algebras by $\ltimes$. For instance, $\mathbb R^p\ltimes \so(p)$ is an algebra of $\so(p)$ rotations and $p$ translations that transform in the fundamental of $\so(p)$.
\item An \emph{infinitesimal symmetry transformation} of a field is denoted by $\delta$. If we want to be more specific, we sometimes write the parameter, like e.g. $\delta (\epsilon)$,  write a subscript, like $\delta_Q$, or combine both, as in $\delta_Q(\epsilon)$.
\item In the projection of Section~\ref{geomchange} and the dimensional reduction of Chapter~\ref{dimred}, \emph{hats} are used to denote objects on the higher-dimensional space.
\item With $\Sp(n)$ we mean $\Sp(n,\mathbb H)$ which is the associated compact group to $\Sp(2n,\mathbb R)$.
\end{enumerate}

\chapter{Samenvatting}\label{samenvatting}
\section{Inleiding}
In deze thesis lichten we de studie toe van klassieke veldentheorie\"en die geen Lagrangiaanse be\-schrij\-ving kennen, wat impliceert dat de formulering van deze theorie\"en volledig gebeurt in termen van bewegingsvergelijkingen. Uiteraard is het mogelijk om allerlei dynamische veldvergelijkingen op te schrijven, maar wij zullen ons beperken tot een klasse met een aantal restricties. Eerst en vooral moet het stelsel van bewegingsvergelijkingen invariant zijn onder de actie van de Poincar\'e groep. Ten tweede moet de theorie supersymmetrisch zijn, wat wil zeggen dat ze invariant is onder een zekere omwisseling van velden die overeenkomen met bosonische en fermionische deeltjes. De cor\-res\-pon\-de\-ren\-de algebra van infinitesimale symmetrie-transformaties wordt een superalgebra genoemd.  

In de theorie\"en die wij zullen bestuderen, vormen de velden slechts een re\-pre\-sen\-ta\-tie van de superalgebra wanneer zij voldoen aan dynamische be\-per\-king\-en\footnote{De onafhankelijke componenten van dergelijke velden voldoen steeds aan een golf\-ver\-ge\-lij\-king.} die we zullen interpreteren als bewegingsvergelijkingen. We zeggen daar\-om dat supersymmetrie in deze gevallen on-shell gerealiseerd wordt\footnote{Een representatie waar deze dynamische beperkingen niet hoeven te worden opgelegd, wordt een off-shell representatie genoemd.} en we zullen deze fermionische symmetrie dan ook kunnen gebruiken als het or\-ga\-ni\-se\-rend principe om de veldvergelijkingen te vinden. Onze studie zal leiden tot de vaststelling dat de superalgebra's die tot nu toe in de literatuur werden bestudeerd, meer algemene theorie\"en toelaten die echter geen be\-schrij\-ving kennen in termen van een supersymmetrische en Poincar\'e-invariante actie. Deze theorie\"en kunnen eerst en vooral andere potentialen bevatten, die verenigbaar blijven met de symmetrie-algebra. Bovendien kunnen sommige van deze veldentheorie\"en meer algemene geometrie\"en beschrijven. Dit is het geval voor niet-lineaire sigma modellen, waar de scalaire velden (ook kortweg scalairen genoemd) ge\"interpreteerd worden als co\"ordinaat functies op een bepaalde vari\"eteit (de doelruimte). Supersymmetrie beperkt de toegelaten doelruimtes, maar wij zullen aantonen dat theorie\"en zonder actie een volledig nieuwe klasse van vari\"eteiten kunnen beschrijven. Dit zal ons ertoe leiden om via fysische technieken nieuwe wiskundige resultaten te bekomen.

Een hedendaagse snarentheoreet is een theoretisch fysicus die veronderstelt dat snarentheorie het fundamentele beginsel is waar alle andere fysische wetten in principe afleidbaar van zijn. E\'en van de eerste problemen waar een dergelijk fysicus bij het ontdekken van deze klassieke veldentheorie\"en zonder actie mee wordt geconfronteerd, is daarom de vraag naar de oorsprong binnen snarentheorie. Om dit probleem te kunnen beantwoorden moeten we eerst enkele schalen introduceren. Wanneer we fysische processen willen bestuderen met een energieschaal $E$ die klein is ten opzichte van de Planck massa $M_p$ (i.e. voor energie\"en $E\ll M_p=10^{28}$ eV), dan kan snarentheorie met een tien-dimensionale supergravitatie\footnote{Dit is een veldentheorie die gravitatie bevat en waar supersymmetrie een lokale invariantie is.} worden beschreven. Willen we deze effectieve veldentheorie gebruiken om vier-dimensionale processen te beschrijven, dan kunnen we ze dimensionaal reduceren wat betekent dat we veronderstellen dat de tien-dimensionale ruimte bestaat uit zes compacte en vier niet-compacte richtingen. Wanneer de compacte ruimte een typische afstandsschaal $R$ bezit zodanig dat $M_p\gg R^{-1}\gg E$, dan blijkt namelijk dat de laag-energetische fluctuaties met een energie van de orde van $E$, constant zijn over deze compacte ruimte. We zullen aantonen dat het mogelijk is om via een veralgemening van deze dimensionale reductie theorie\"en zonder actie te construeren en deze zodoende een snarentheoretische oorsprong te geven. 

Tot slot van deze inleiding overlopen we nog kort de inhoud van de thesis. Hoofdstuk~\ref{intr} bevat de inleiding en de verantwoording, zoals hierboven werd samengevat. Vervolgens worden in Hoofdstuk~\ref{prelim} enkele elementaire wiskundige en fysische concepten ge\"introduceerd die in het vervolg van de tekst gebruikt worden. Daarna behandelt Hoofdstuk~\ref{onshellsusy} rigide supersymmetrische theorie\"en die geen Lagrangiaanse beschrijving hebben. In Hoofdstuk~\ref{suptencalc} wordt dit uitgebreid in de context van lokale supersymmetrie. In Hoofdstuk~\ref{dimred} wordt in een eenvoudig voorbeeld besproken hoe veralgemeende dimensionale reductie aanleiding kan geven tot theorie\"en zonder actie. Bovendien zullen we supersymmetrische oplossingen van de lager-dimensionale theorie construeren. We eindigen de hoofdtekst in Hoofdstuk~\ref{concel} met onze conclusies. Tot slot behandelt Appendix~\ref{spinors} enkele elementaire eigenschappen van spinor representaties terwijl in Appendix~\ref{appconventions} onze conventies verduidelijkt worden.
\section{Differentiaal geometrie}
Omdat een belangrijk deel van deze thesis handelt over de studie van klassieke niet-lineaire sigma modellen, worden in Hoofdstuk~\ref{prelim} enkele fundamentele concepten uit de differentiaal geometrie ge\"introduceerd. Deze tak van de wiskunde is immers de natuurlijke taal waarin de implicaties van supersymmetrie op de geometrie van de doelruimte geformuleerd kunnen worden. 

De kinetische term van een niet-lineair sigma model met re\"ele scalaire velden $\varphi^X$, ge\-pa\-ra\-me\-tri\-seerd door $X$, heeft als vorm
\begin{equation}\label{simplesigma}
\mathcal S=-\frac12\int d^d x\,g_{XY}(\varphi)\partial_\mu\varphi^X\partial^\mu\varphi^Y\,,\qquad \mu=0,\dots ,d-1\,.
\end{equation}
Merk eerst en vooral op dat we $g$ altijd symmetrisch kunnen kiezen in $XY$ vermits het antisymmetrische stuk niet voorkomt in de actie.
Zoals gezegd willen we in een dergelijk model de scalaire velden interpreteren als co\"ordinaat functies op de doelruimte. Eisen we dan dat deze actie invariant is onder co\"ordinaat diffeomorfismen $\varphi \to\varphi'(\varphi)$, dan moet het veldafhankelijke object $g$ transformeren als een metriek. Met andere woorden, de doelruimte van een dergelijk niet-lineair sigma model met een actie is altijd een Riemannse vari\"eteit. De bewegingsvergelijkingen afkomstig van bovenstaande actie lezen
\begin{equation}\label{bewvgl}
\Box \varphi^X\equiv\partial_\mu\partial^\mu \varphi^X+\Gamma_{YZ}{}^X\partial_\mu \varphi^Y\partial^\mu \varphi^Z=0\,.
\end{equation}
Willen we nu opnieuw dat deze vergelijkingen covariant transformeren onder co\"ordinaat diffeomorfismen van de doelruimte, dan heeft dit als implicatie dat $\Gamma$ een connectie is. Deze connectie is bovendien steeds symmetrisch in $YZ$, wat ze torsieloos maakt. Vermits verder~(\ref{bewvgl}) afkomstig is van de hoger vermelde actie, laat deze connectie de metriek $g$ invariant.\footnote{In dat geval wordt $\Gamma$ de Levi-Civita connectie met betrekking tot $g$ genoemd.} Willen we echter enkel de bewegingsvergelijkingen~(\ref{bewvgl}) bestuderen en eisen we niet meer dat deze afleidbaar zijn van een Lagrangiaan, dan kan $\Gamma$ meer algemeen elke torsieloze, affiene connectie zijn. Dit wil zeggen dat $\Gamma$ nog steeds het parallel transport in de raakbundel van de doelruimte beschrijft, maar dat deze ruimte niet noodzakelijk een invariante metriek moet dragen. Samenvattend kunnen we dus stellen dat het niet-lineair sigma model~(\ref{bewvgl}) als doelruimte een vari\"eteit moet bezitten met een affiene connectie op de raakbundel. Deze veldentheorie is dan enkel van een actie afleidbaar wanneer de doelruimte een metriek draagt die invariant is onder parallel transport. Dit is het meest eenvoudige voorbeeld van de veralgemeningen die wij in deze thesis willen bestuderen. Om echter supersymmetrische niet-lineaire sigma modellen te kunnen bespreken, hebben we nog enkele andere concepten uit de differentiaal geometrie nodig.

Een $2n$-dimensionale complexe vari\"eteit $M$ is een ruimte waarvoor het mogelijk is in elke kaart complexe co\"ordinaat functies te kiezen (i.e. $\{z^i,\bar z^{\bar i}\}:M\to \mathbb C^n$ met $i,\bar i=1,\dots,n$) zodanig dat de transitie functies holomorf zijn. Een complexe vari\"eteit draagt daarom een complexe structuur (i.e. een vezelachtig endomorfisme van de raakbundel) die in deze co\"ordinaten de volgende vorm aanneemt
\begin{equation}
J=\rmi \rmd z^i\otimes \partial_i-\rmi \rmd\bar z^{\bar i}\otimes \bar \partial_{\bar i}\,.
\end{equation}
$J$ is met andere woorden de veralgemening van de imaginaire eenheid `$\rmi$'.
In meer algemene co\"ordinaten is de complexe structuur een spoorloze (1,1)-tensor die kwadrateert tot `$-1$' en
waarvoor de corresponderende Nijenhuistensor $N$ overal verdwijnt,
\begin{equation}
N_{XY}{}^Z\equiv\frac16 J_X{}^V \partial_{[V}J_{Y ]}{}^Z-(X \leftrightarrow Y)=0\,.
\end{equation} 
Merk op dat alle objecten die hier besproken worden, gedefinieerd zijn op de doelruimte zodat we de corresponderende indices $X,Y,\dots$ gebruiken.
Deze conditie op de Nijenhuistensor is volgens het Newlander-Nirenberg theorema equivalent met het bestaan van een torsieloze connectie die de complexe structuur invariant laat. Wanneer de ruimte nu ook nog een metriek $g$ draagt die hermitisch is, i.e.
\begin{equation}
g_{XY}=J_X{}^Z J_Y{}^V g_{ZV}\,,
\end{equation}
en de Levi-Civita connectie bovendien de complexe structuur invariant laat, dan is de ruimte een K\"ahler vari\"eteit.

Een complexe ruimte wordt dus bepaald in termen van een complexe structuur $J$, die o\-ver\-een\-komt met de imaginaire eenheid in $\mathbb C$. Nu is dit concept te veralgemenen tot het geval waar drie complexe structuren bestaan die dan overeenkomen met de drie imaginaire eenheden van de quaternionen $\mathbb H$. Een $4n$-dimensionale quaternionisch-achtige ruimte bezit dus drie complexe structuren $\vec J$ die de algebra van de imaginaire quaternionen genereren,
\begin{equation}\label{structuren}
J^\alpha{}_X{}^Z J^\beta{}_Z{}^Y=-\delta^{\alpha\beta}\delta_X^Y+\varepsilon^{\alpha\beta\gamma}J^\gamma{}_X{}^Y\,,
\end{equation} 
met $\alpha,\beta=1,2,3$. Tijdens de studie van supersymmetrische niet-lineaire sigma modellen zullen we vier verschillende soorten quaternionisch-achtige ruimtes beschrijven. Zo zorgen de drie complexe structuren van een hypercomplexe vari\"eteit ervoor dat de diagonale Nijenhuistensor $N^d$,  
\begin{equation}
N^{\rm d}{}_{XY}{}^Z\equiv\frac16 \vec J_X{}^V \cdot \partial_{[V}\vec J_{Y ]}{}^Z-(X \leftrightarrow Y)=0\,,
\end{equation} 
overal verdwijnt. Analoog aan het theorema van Newlander en Nirenberg is het verdwijnen van $N^d$ dit keer equivalent met het bestaan van een unieke torsieloze connectie op de raakbundel van de doelruimte die elk van deze drie complexe structuren invariant laat. Dit is de Obata connectie. Wanneer de ruimte bovendien een metriek draagt die invariant is onder parallel transport met betrekking tot deze Obata connectie en deze metriek verder hermitisch is met betrekking tot elk van de complexe structuren, dan wordt de vari\"eteit hyperk\"ahler genoemd. Bij deze eerste twee quaternionisch-achtige ruimtes is elk van de complexe structuren invariant onder parallel transport. Wanneer deze echter onder parallel transport in elkaar worden geroteerd, spreken we van een quaternionische ruimte. In dit geval voldoet de diagonale Nijenhuistensor aan de volgende vergelijking:
\begin{equation}
(1-2n)N^{\rm d}{}_{XY}{}^Z=-\vec J_V{}^W\cdot\vec J_{[X}{}^Z N^{\rm d}{}_{Y]W}{}^V\,.
\end{equation}
Alle mogelijke lineaire combinaties van deze drie complexe structuren bepalen nu een zogenaamde quaternionische structuur. De bovenstaande conditie op de Nijenhuistensor is nu equivalent met het bestaan van een torsieloze connectie op de raakbundel die deze quaternionische structuur invariant laat. Vertrekken we echter van een vari\"eteit met daarop een quaternionische structuur, dan is deze torsieloze connectie niet uniek. Draagt de quaternionische ruimte ten slotte ook nog een metriek die hermitisch is ten opzichte van de drie complexe structuren, dan spreken we van een quaternionische K\"ahler vari\"eteit, wanneer de Levi-Civita connectie de quaternionische structuur invariant laat.  
\section{Rigide on-shell supersymmetrie}
In Hoofdstuk~\ref{onshellsusy} bespreken we rigide on-shell realisaties van de $\mathcal N=2$ super-Poincar\'e algebra.
Deze superalgebra bezit een $\mathbb Z_2$ gradatie, waarbij de Poincar\'e algebra samen met een hiermee commuterende $\su(2)$ algebra alle even ge\-ne\-ra\-toren opspannen. De oneven generatoren zijn gegroepeerd in een anti-commuterende symplectische Majorana spinor representatie\footnote{Dit is een acht-dimensionale irreducibele representatie van $\so(1,4)$, zie Appendix~\ref{spinors}.} van $\so(1,4)$. De\-ze generatoren die in de fysische theorie de bosonische en fermi\-o\-ni\-sche deeltjes met elkaar omwisselen, worden daarom ook superladingen genoemd. De $\su(2)$ algebra (ook wel gekend als de R-symmetrie) roteert de generatoren van de supersymmetrie transformaties in elkaar. Een verzameling van (bosonische en fermionische) velden die de superalgebra realiseren, wordt een multiplet genoemd. In het Hoofdstuk over rigide on-shell supersymmetrie construeren we twee soorten on-shell multiplets, namelijk vector-multiplets en hypermultiplets.

De fysische velden van de eerste zijn een re\"eel scalair veld $\sigma$, een re\"ele vector $A_\mu$ en een $\su(2)$ doublet van symplectische Majorana spinoren, de gaugini $\psi^i$. Wanneer we $n$ multiplets samen beschrijven, ijken de vectoren een bepaalde $n$-dimensionale algebra $\mathfrak g$ (met koppelingsconstante $g$) en transformeren de scalair en de gaugini in de toegevoegde representatie. 

Wordt deze verzameling van velden uitgebreid met een re\"eel hulpveld $Y$, dat in de toegevoegde representatie van zowel $\su(2)$ als van $\mathfrak g$ transformeert, dan realiseren deze vier velden de superalgebra off-shell. Bijgevolg start men de studie van de dynamica van dergelijke off-shell vector-multiplets met de constructie van de meest algemene actie die invariant is onder de superalgebra. Dergelijke actie blijkt nu volledig bepaalt te zijn in termen van twee objecten. De symmetrische constante drie-tensor $C_{IJK}$, waarin $I,J,K$ de multiplets doorlopen en dus indices zijn die waarden aannemen in $\mathfrak g$, bepaalt de kinetische termen van de velden en hun onderlinge koppelingen~\cite{Gunaydin:1984bi}. Deze tensor moet ijkinvariant zijn, wat betekent dat
\begin{equation}
f^M_{I(J}C_{KL)M}=0\,,
\end{equation}
waarbij $f$ de structuurconstanten zijn van $\mathfrak g$. Ten tweede kunnen de Abelse sectoren van de theorie nog een zogenaamde Fayet-Iliopoulos (FI) potentiaal bevatten (na eliminatie van het hulpveld) van de vorm~\cite{deWit:1981tn,Breitenlohner:1981sm}
\begin{equation}
-g^2 C_{IJK}\sigma^I f^J_{ij}f^{ijK}\,,
\end{equation}
waarbij $f^{ij}$ constanten zijn in de $\mathbf 3$ van $\su(2)$ en FI-termen worden genoemd.

In Hoofdstuk~\ref{onshellsusy} vertrekken we echter van een vector-multiplet dat enkel de fysische velden bevat en bijgevolg de superalgebra on-shell realiseert. Deze constructie start dan ook met de zoektocht naar de meest algemene supersymmetrie transformaties. Berekenen we vervolgens de anti-commutator van twee supersymmetrie\"en zoals deze gerealiseerd wordt op de verschillende velden, dan sluit de algebra slechts wanneer de velden voldoen aan dynamische be\-per\-king\-en, die we nu interpreteren als bewegingsvergelijkingen~\cite{Gheerardyn:2003rf}. Op deze manier hebben we dus de dynamische vergelijkingen van het vector-multiplet bekomen zonder te moeten refereren naar een actie. De vorm van de be\-we\-gings\-ver\-ge\-lij\-king\-en ligt nu niet vast in termen van de hoger vermelde drie-tensor maar wordt bepaald door een object $\gamma^I_{JK}$ dat veld-afhankelijk is en een meer algemene indexstructuur bezit. Bovendien zijn Fayet-Iliopoulos potentialen nu ook mogelijk in de niet-Abelse sectoren. Na het bepalen van de be\-we\-gings\-ver\-ge\-lij\-king\-en bewijzen we met behulp van het Batalin-Vilkovisky formalisme dat deze veralgemeningen enkel mogelijk zijn wanneer de veldvergelijkingen niet afleidbaar zijn van een supersymmetrische en Poincar\'e-invariante actie.

Een ander multiplet dat de vijf-dimensionale $\mathcal N=2$ super-Poincar\'e algebra kan realiseren, is het hypermultiplet. In tegenstelling tot het vector-multiplet kan dit set van velden enkel on-shell beschreven worden. Het multiplet bestaat uit vier re\"ele scalairen $q$ en twee symplectische Majorana spinoren $\zeta$ en de dynamica van het multiplet wordt beschreven aan de hand van een niet-lineair sigma model. Bezit dit model een Lagrangiaan, dan is de kinetische term van de vorm~(\ref{simplesigma}) waarbij de doelruimte omwille van supersymmetrie een hyperk\"ahler vari\"eteit is~\cite{Alvarez-Gaume:1981hm}. Vermits de superalgebra echter on-shell wordt gerealiseerd, volgen de bewegingsvergelijkingen al uit algebra\"ische o\-ver\-we\-ging\-en zodat we voor de scalairen~(\ref{bewvgl}) al kunnen vinden zonder het bestaan van een actie te veronderstellen. Bijgevolg kan de connectie die optreedt in deze dynamische beperkingen affien zijn. Omwille van supersymmetrie moet de doelruimte echter nog steeds drie passende complexe structuren~(\ref{structuren}) bezitten waaruit we concluderen dat deze doelruimte hypercomplex kan zijn, met de Obata connectie die parallel transport definieert op de raakbundel~\cite{Bergshoeff:2002qk}.
\section{Lokale on-shell supersymmetrie}
In Hoofdstuk~\ref{suptencalc} bestuderen we een analoge theorie waarbij de $\mathcal N=2$ super-Poincar\'e algebra lokale symmetrie\"en genereert. Omwille hiervan bevat de theorie een multiplet van ijkvelden. De translaties en de Lorentz-rotaties vertegenwoordigen het even deel van de geijkte generatoren zodat deze sector met als velden het Vielbein en de spinconnectie, correspondeert met algemene re\-la\-ti\-vi\-teits\-the\-o\-rie.\footnote{Het ijkmultiplet bevat ook nog een vector (het gravifoton) dat bij geijkte supergravitatie een $\un(1)$ subalgebra van de R-symmetrie algebra ijkt.} Het gravitino\footnote{In de $\mathbf 5 \times \mathbf 8$ van $\so(1,4)$.} is dan tenslotte het ijkveld dat van de superladingen in de $\mathcal N=2$ algebra lokale symmetrie-generatoren maakt. We zullen de meest algemene koppelingen bestuderen van deze ijktheorie aan een willekeurig aantal vector- en hypermultiplets en dit zal leiden tot een meer volledige kennis van de theorie met een actie. In het algemeen kan men ook nog een willekeurig aantal tensormultiplets toevoegen, waarvoor we echter verwijzen naar~\cite{Bergshoeff:inprep2}. Bovendien kunnen we in de hypermultiplet-sector dezelfde uitbreiding beschouwen als in vorig Hoofdstuk werd besproken~\cite{Bergshoeff:inprep1}. Dit multiplet wordt namelijk nog steeds beschreven in termen van een niet-lineair sigma model, waarbij de doelruimte van dit model nu een quaternionische K\"ahler vari\"eteit is in het geval de theorie een Lagrangiaanse beschrijving kent~\cite{Bagger:1983tt}. Maar vermits het multiplet de superalgebra on-shell realiseert, is het weer mogelijk om de bewegingsvergelijkingen uit de algebra af te leiden, zodat de connectie op de raakbundel van de doelruimte opnieuw affien kan zijn wat de vari\"eteit in dit geval quaternionisch maakt.

Omdat deze meest algemene $\mathcal N=2$ supergravitatie een vrij ingewikkelde structuur heeft, gebruiken we in Hoofdstuk~\ref{suptencalc} de superconforme tensorcalculus methode om deze veldentheorie te construeren. Deze werkwijze kan duidelijk gemaakt worden aan de hand van het elementaire voorbeeld waar we Einstein-Hilbert gravitatie uit een conforme\footnote{Conforme transformaties zijn transformaties die hoeken invariant laten.} actie van een re\"eel scalair veld reconstrueren. Beschouw daarom de volgende Lagrangiaan van een re\"ele scalair $\varphi$ gekoppeld aan een metriek $g$ in vijf dimensies:
\begin{equation}\label{confactie}
\mathcal L=\sqrt{-g}\big(\frac12 \partial_\mu \varphi \partial^\mu \varphi +\frac1{12} R \varphi^2\big)\,,
\end{equation}
met $R$ de Ricci scalair.
Deze actie is invariant onder een lokale dilatatie symmetrie die op de velden gerealiseerd wordt als $\delta \varphi=\lambda \varphi$ en $\delta g_{\mu\nu}=-2\lambda g_{\mu\nu}$. Breken we nu deze invariantie door het scalaire veld op een vaste waarde te houden, $\varphi=\sqrt{6}/\kappa$, dan bekomen we uit~(\ref{confactie}) de Einstein-Hilbert actie
\begin{equation}\label{EHactie}
\mathcal L=\frac1{2\kappa^2}\sqrt{-g}R\,.
\end{equation}
Deze twee acties~(\ref{confactie}) en~(\ref{EHactie}) zijn dus ijk-equivalent. Een alternatieve manier om tot dezelfde conclusie te komen, zou zijn om in~(\ref{confactie}) de metriek te herdefini\"eren als $\tilde g_{\mu\nu}=\kappa^2 \varphi^2 g_{\mu\nu}$. Op deze manier valt alle afhankelijkheid van $\varphi$ uit de actie weg zodat de resulterende theorie de vorm~(\ref{EHactie}) (met $\tilde g$ in plaats van $g$) aanneemt. Het veld $\varphi$ is dus niet-fysisch en wordt een compensator genoemd. Merk tenslotte op dat de kinetische term in~(\ref{confactie}) het verkeerde teken draagt, wat typisch is voor compenserende velden. 

Dezelfde werkwijze kan gebruikt worden om de vijf-dimensionale $\mathcal N=2$ super-Poincar\'e theorie te construeren. We vertrekken nu van een theorie die lokaal invariant is onder de $\mathcal N=2$ superconforme algebra en nadien breken we de `overbodige' symmetrie\"en. De $\mathcal N=2$ superconforme algebra bevat als even generatoren naast de Poincar\'e algebra en de R-symmetrie algebra ook nog infinitesimale dilataties en speciale conforme transformaties, terwijl de oneven generatoren naast de supersymmetrie transformaties ook nog de zogenaamde speciale superconforme transformaties genereren. Bovendien wordt in de conforme theorie ook de R-symmetrie geijkt, wat niet noodzakelijk het geval is in de Poincar\'e theorie. Omdat de superconforme veldentheorie waarvan we zullen vertrekken invariant is onder een grotere algebra van transformaties, is de structuur van deze theorie eenvoudiger, wat een eerste voordeel van deze constructiemethode is. 

In het eenvoudige voorbeeld~(\ref{confactie}) was de oorspronkelijke theorie invariant onder algemene co\"ordinaat transformaties en dilataties. In het supersymmetrische geval vertrekken we daarom van een lokaal superconforme theorie van vector- en hypermultiplets~\cite{Bergshoeff:2002qk}. Om dergelijke theorie te kunnen construeren, gebruiken we het ijkmultiplet van de superconforme algebra (het Weyl multiplet), zoals deze werd besproken in~\cite{Bergshoeff:2001hc}. Eerst en vooral bespreken we daarom in Hoofdstuk~\ref{suptencalc} de constructie van het superconforme vector-multiplet wat een vrij eenvoudige uitbreiding is van het on-shell vector multiplet dat we besproken hebben in Hoofdstuk~\ref{onshellsusy}. We gebruiken in Hoofdstuk~\ref{suptencalc} echter een off-shell beschrijving omdat de transformatie eigenschappen van deze vector-multiplets dan niet veranderen tijdens de koppeling aan hypermultiplets.

Voor de hypermultiplets geeft de conforme methode meer inzicht in de geometrie van qua\-ter\-ni\-o\-nisch-achtige ruimtes. Zoals we reeds hebben besproken in Hoofdstuk~\ref{onshellsusy} kunnen Poincar\'e hypermultiplets een hypercomplexe of hyperk\"ahler doelruimte beschrijven.
Wanneer we de symmetrie-algebra nu willen uitbreiden met dilatatie transformaties, die gerealiseerd worden als $\delta q^X=\lambda k(q)^X$, dan volgt uit algebra\"ische overwegingen dat 
\begin{equation}
\mathfrak D_X k^Y=\frac32 \delta_X{}^Y\,.
\end{equation}
Een superconform hypermultiplet zonder Lagrangiaanse beschrijving, bezit dus een hypercomplexe doelruimte die een zogenaamd homothetisch sym\-me\-trie-vec\-tor\-veld $k$ draagt. Wanneer deze constructie verenigbaar moet zijn met een metriek op de doelruimte, dan is dit vectorveld $k$ een homothetische Killing vector,
\begin{equation}\label{homKilvector}
\mathfrak D_{(X}k_{Y)}=\frac32 g_{XY}\,.
\end{equation}
Het bestaan van dit vectorveld impliceert dan onmiddellijk dat de R-symmetrie op de velden gerealiseerd wordt als 
\begin{equation}
\delta_R q^X\equiv 2 \vec \lambda_R \cdot \vec k=\frac23 \vec \lambda_R \cdot \vec J_Y{}^X k^Y\,.  
\end{equation}

Met deze kennis kunnen we nu analoog aan de actie~(\ref{confactie}) de conforme constructie starten met een theorie van $n_V+1$ vector-multiplets en $n_H+1$ hypermultiplets in een achtergrond van het Weyl-multiplet. De rol van het veld $\varphi$ dat in het eenvoudige voorbeeld gebruikt werd om de dilatatie symmetrie te breken, wordt hier gespeeld door een compenserend vector- en een hypermultiplet die nodig zijn om de verschillende `overbodige' symmetrie\"en te breken.\footnote{Na de eliminatie van de extra symmetrie\"en blijven er $n_V+1$ vectoren over, waarvan een bepaalde lineaire combinatie de rol van gravifoton zal spelen in de uiteindelijke super-Poincar\'e theorie.}  

De realisatie van deze stap op de hypermultiplets heeft drastische implicaties op de geometrie van de doelruimte. Om nu eerst de dilatatie symmetrie te breken, moeten we eerst en vooral de scalair die de stroming van het vectorveld $k$ parametriseert een constante waarde geven. Om nu echter supersymmetrie te waarborgen tijdens de procedure, moet deze scalair deel uitmaken van een compleet hypermultiplet. Het ligt voor de hand dat de andere drie scalairen die moeten worden ge\"elimineerd, corresponderen met de stromingen die afkomstig zijn van de vectorvelden $\vec k$ die de R-symmetrie transformaties genereren. Vanuit geometrisch standpunt gezien, komt het breken van deze symmetrie\"en overeen met het uitprojecteren van de vier richtingen langs de stroming van $k$ en $\vec k$. In Hoofdstuk~\ref{suptencalc} tonen we dan ook aan dat deze projectie een $4(n_H+1)$-dimensionale hypercomplexe vari\"eteit afbeeldt op een $4n_H$-dimensionale quaternionische ruimte of (in het geval met metriek) een hyperk\"ahler ruimte naar een quaternionische K\"ahler vari\"eteit transformeert. Het resultaat van deze projectie is dan de doelruimte van de $n_H$ super-Poincar\'e hypermultiplets. Bovendien bespreken we in dit deel van Hoofdstuk~\ref{suptencalc} hoe alle connecties, krommingstensoren, complexe structuren en vectorvelden die een symmetrie van de doelruimte genereren, opsplitsen tijdens deze projectie. 

Op het einde van Hoofdstuk~\ref{suptencalc} wordt tenslotte de conforme constructie voltooid in de context van een theorie met actie. Meer bepaald houden we tijdens het elimineren van de overbodige symmetrie\"en nu ook rekening met alle fermionische termen in de theorie. Dit resulteert in de volledige $\mathcal N=2$ super-Poincar\'e theorie gekoppeld aan een willekeurig aantal vector- en hypermultiplets samen met de supersymmetrie transformaties die deze theorie invariant laten. 
\section{Kaluza-Klein theorie}
In de Hoofdstukken~\ref{onshellsusy} en~\ref{suptencalc} hebben we theorie\"en zonder actie geconstrueerd via de studie van de $\mathcal N=2$ superalgebra. Zoals in de inleiding reeds werd vermeld, is het echter belangrijk om te weten hoe deze theorie\"en ingebed kunnen worden in snarentheorie. In het laatste Hoofdstuk~\ref{dimred} bespreken we daarom aan de hand van een eenvoudig voorbeeld hoe een theorie zonder actie kan volgen uit de dimensionale reductie van een supergravitatie die wel een Lagrangiaanse beschrijving kent. Zoals we reeds hebben uitgelegd is een dergelijke supergravitatie een effectieve theorie die een bepaalde sector van snarentheorie beschrijft, zodat we op deze manier een snarentheoretische oorsprong hebben gevonden voor (sommige) theorie\"en zonder actie. Meer bepaald bespreken we de constructie van een massieve tien-dimensionale supergravitatie uit de reductie van klassieke M-theorie, een supergravitatie die wel een actie bezit. 

Kaluza-Klein theorie, of het proces van dimensionale reductie, is de constructie van een lager-dimensionale theorie uit een hoger-dimensionale, via de veronderstelling dat \'e\'en of meerdere (ruim\-te\-lij\-ke) dimensies compact zijn. We kunnen deze methode illustreren met het volgende elementaire voorbeeld. Veronderstel dat we een re\"eel, massaloos, scalair veld $\hat \phi$ willen beschrijven in een vlakke ruimte $\mathbb R^{1+3}\times S^1$, d.w.z. een ruimte die het direct product is van een vier-dimensionale Minkowski ruimte-tijd (met co\"ordinaten $x^\mu$) en een cirkel (met co\"ordinaat $z$). Veronderstel verder dat het veld periodiek is omheen die cirkel, zodat we het kunnen ontbinden in Fouriermodes,
\begin{equation}
\hat \phi(x,z)= \sum_{n=-\infty}^{\infty} \phi_{|n|}(x)e^{\rmi \frac{n}{R}z}\,.
\end{equation}
Substitutie van deze reeks in de golfvergelijking levert dan
\begin{equation}
\Box_5 \hat \phi=0 \;\Rightarrow \Box \phi_n=\left(\frac{n}{R}\right)^2 \phi_n\,,
\end{equation}
waarbij $\Box_5$ de vijf-dimensionale en $\Box$ de vier-dimensionale d'Alembertiaan is. Met andere woorden, vanuit vier-dimensionaal oogpunt geeft het veld $\hat \phi$ aanleiding tot een oneindig aantal velden met massa's evenredig met $|n|/R$. Wanneer we nu slechts laag-energetische processen willen beschrijven met een typische energie $E$ waarvoor $E\ll R^{-1}$, dan kunnen we deze massieve modes verwaarlozen. Deze processen lijken met andere woorden vier-dimensionaal. Wanneer we echter deeltjes zouden kunnen versnellen tot een energie van de orde van $R^{-1}$, dan zouden we de massieve Kaluza-Klein deeltjes die corresponderen met $\phi_n$ kunnen waarnemen en bijgevolg de compacte dimensie ontdekken.

Deze constructie suggereert dat als we de cirkel reductie van pure vijf-dimensionale gravitatie willen uitvoeren, we de metriek moeten ontbinden in Fouriercomponenten en enkel de constante mode moeten weerhouden wanneer we een `effectieve' vier-dimensionale theorie willen beschrijven. Stoppen we deze constante mode in de bewegingsvergelijkingen, dan resulteert deze constructie in een vier-dimensionale theorie van gravitatie gekoppeld aan Maxwell-theorie (met ijkpotentiaal $A_\mu$) en een re\"eel scalair veld. De vier-dimensionale $\U(1)$ ijktransformatie $A_\mu \to A_\mu+ \partial_\mu\lambda$ heeft dan als vijf-dimensionale oorsprong de co\"ordinaat reparametrisaties van de vorm $z \to z-\lambda(x)$. Onder deze ijktransformatie zijn zowel de scalair als de vier-dimensionale metriek ongeladen. De reden hiervoor is dat deze velden niet afhangen van $z$. 
Dergelijke cirkel reducties zijn uiteraard ook mogelijk wanneer we vertrekken van gravitatie gekoppeld aan materievelden. In het algemeen zal echter geen enkel veld in de effectieve theorie geladen zijn onder de Abelse ijktheorie die volgt uit de reductie van de metriek. 

Wanneer de hoger-dimensionale theorie echter een globale symmetrie bezit, dan kunnen we dankzij deze symmetrie de hoger-dimensionale velden een welbepaalde afhankelijkheid van de compacte co\"ordinaat geven zodat de effectieve theorie toch nog onafhankelijk blijft van deze richting. Deze afhankelijkheid leidt dan in de lager-dimensionale theorie tot het feit dat de velden die transformeerden onder de oorspronkelijke globale symmetrie nu geladen zijn onder de $\U(1)$ ijksymmetrie afkomstig van de hoger-dimensionale metriek. We kunnen dit opnieuw illustreren met het eenvoudig voorbeeld van het vijf-dimensionaal, massaloos, scalair veld $\hat \phi$. Omdat we echter co\"ordinaat transformaties willen toelaten, veronderstellen we nu dat dit veld gekoppeld is aan gravitatie. Uit de vijf-dimensionale golfvergelijking $\Box_5 \hat \phi=0$ is duidelijk dat de transformatie $\delta \hat \phi=a$, met $a$ een constante, een symmetrie is van de bewegingsvergelijking. Veronderstellen we nu dat het veld $\hat \phi$ meerwaardig is langs de cirkel, i.e. $\hat \phi(z+2\pi R)=\hat \phi (z)+2\pi$, met $R$ de straal van de cirkel, dan kunnen we tijdens de reductie het veld toch laten afhangen van $z$. Kiezen we namelijk 
\begin{equation}\label{hoedphi}
\hat \phi=\phi (x)+z/R\,,
\end{equation}
dan hangt de vier-dimensionale bewegingsvergelijking nog steeds niet af van $z$, dankzij de hoger vermelde symmetrie. De transformatie $z \to z-\lambda$ impliceert nu echter dat het veld $\phi$ moet transformeren naar $\phi+\lambda/R$ om $\hat \phi$ nog steeds aan dezelfde uitdrukking~(\ref{hoedphi}) te laten voldoen. Vermits deze co\"ordinaat transformaties overeenkomen met een $\U(1)$ ijktransformatie in de effectieve theorie, is het duidelijk dat in deze veralgemeende reductie (ook wel Scherk-Schwarz reductie genoemd) $\phi$ een $\U(1)$ lading zal dragen. 

Pure gravitatie bezit een schaalsymmetrie van de bewegingsvergelijkingen. Gebruiken we deze symmetrie om Scherk-Schwarz reductie uit te voeren, dan bezit de resulterende theorie een potentiaal die niet afgeleid kan worden uit een actie. Deze constructie is te veralgemenen tot de Scherk-Schwarz reductie van elf-dimensionale supergravitatie (ook wel klassieke M-theorie genoemd). De resulterende tien-dimensionale (mIIA) theorie is een massieve supergravitatie\footnote{Dit is een supergravitatie waarbij sommige velden weggeijkt kunnen worden en zodoende een massa verkrijgen. `$IIA$' verwijst verder naar het feit dat de theorie niet-chiraal is.} die niet equivalent is met de enige andere massieve IIA supergravitatie geconstrueerd door Romans in~\cite{Romans:1986tz}, en die geen actie formulering bezit. De supersymmetrie-transformatie regels die de be\-we\-gings\-ver\-ge\-lij\-kingen invariant laten, werden afgeleid in~\cite{Gheerardyn:2002wp}.

Zoals reeds gezegd, moeten tijdens (veralgemeende) dimensionale reducties de hoger-di\-men\-sio\-na\-le velden op een welbepaalde manier van de compacte richting afhangen. In het geval van de constructie van de mIIA theorie moet bijvoorbeeld de metriek op de volgende manier een functie zijn van $z$:
\begin{equation}\label{deansatz}
\hat g_{\hat \mu \hat \nu}(x,z)=e^{2mz} \hat h_{\hat \mu \hat \nu}(x)\qquad {\rm met}\qquad \hat \mu\,,\hat \nu=0,\dots,10\,.
\end{equation}
Een gravitationele oplossing\footnote{I.e. een oplossing die enkel een niet-triviale metriek bevat.} van de elf-dimensionale theorie reduceert bij\-ge\-volg naar een oplossing van de tien-dimensionale supergravitatie wanneer er co\"ordinaten bestaan zodanig dat de metriek kan geschreven worden als in~(\ref{deansatz}). De co\"ordinaat-invariante manier om dit te zeggen is dat de metriek $\hat g$ een homothetische Killing vector $\hat k$, zoals in~(\ref{homKilvector}), moet bevatten.
Deze elf-dimensionale oplossing bezit bovendien evenveel supersymmetrie\"en als er parallelle spinoren 
\begin{equation}
\hat{ \mathfrak D} \hat \epsilon =0\,,
\end{equation}
bestaan. Deze spinoren genereren supersymmetrische invarianties van de tien-dimensionale op\-los\-sing wanneer ze, in co\"ordinaten overeenkomend met~(\ref{deansatz}), van $z$ afhangen als 
\begin{equation}
\hat \epsilon(x,z)=e^{\frac{m}2z}\hat \eta(x)\,.
\end{equation}
In Hoofdstuk~\ref{dimred} leiden we verder af dat dit equivalent is met de co\"ordinaat invariante uitdrukking
\begin{equation}
(\partial_{\hat \mu}\hat k_{\hat \nu}-\partial_{\hat \nu}\hat k_{\hat \mu})\Gamma^{\hat \mu}\Gamma^{ \hat \nu}\hat \epsilon=m\hat \epsilon\,,
\end{equation}
met $\Gamma$ de elf-dimensionale gamma-matrices.
Deze opmerkingen passen we dan vervolgens toe om supersymmetrische oplossingen te construeren in de tien-dimensionale theorie. Zo construeren we uit de reductie van elf-dimensionale vlakke ruimte veldconfiguraties die voldoen aan de be\-we\-gings\-ver\-ge\-lij\-kingen en die 16 supersymmetrie\"en bezitten. Verder bouwen we oplossingen met minder supersymmetrie uit de reductie van welbepaalde Ricci-vlakke kegels (nl. kegels met speciale holonomie)~\cite{Gheerardyn:2002wp}.
\section{Besluit}
In deze thesis hebben we enkele aspecten van on-shell supersymmetrie be\-stu\-deerd. We hebben eerst en vooral aangetoond dat supersymmetrie meer algemene theorie\"en toelaat dan deze die kunnen beschreven worden in termen van een Poincar\'e-invariante en supersymmetrische actie. In de context van $\mathcal N=2$ supersymmetrie leidde dit tot het inzicht dat dergelijke theorie\"en algemenere potentialen kunnen bevatten en andere doelruimtes kunnen beschrijven. Bovendien hebben we aangetoond dat we gelijkaardige theorie\"en via dimensionale reductie kunnen construeren. 
Dit werk is bijgevolg niet meer dan een nieuwe illustratie van de rijke structuur die zich verschuilt in elk systeem met een fermionische symmetrie.
\providecommand{\href}[2]{#2}\begingroup\raggedright\endgroup

\end{document}